\documentclass[
11pt, 
english, 
singlespacing, 
headsepline, 
]{MastersDoctoralThesis} 

\usepackage[utf8]{inputenc} 
\usepackage[T1]{fontenc} 

\usepackage{palatino} 

\usepackage[autostyle=true]{csquotes} 

\usepackage{pgf,tikz}
\usepackage{mathrsfs}
\usetikzlibrary{arrows}
\usepackage{upgreek}
\usepackage{color}

\makeatletter
\newcommand{\thickhline}{%
    \noalign {\ifnum 0=`}\fi \hrule height 1pt
    \futurelet \reserved@a \@xhline
}
\newcolumntype{"}{@{\hskip\tabcolsep\vrule width 1pt\hskip\tabcolsep}}
\makeatother

\usepackage{braket} 

\usepackage{bm,accents}

\newcommand{\mais}{{\textsc{\footnotesize +}}}
\newcommand{\menos}{{\textsc{\footnotesize $-$}}}

\usepackage[nottoc]{tocbibind} 

\thesistitle{Aspects of Black Hole Physics Beyond General Relativity:} 
\subtitle{extra dimensions, horizon wave function and applications} 
\supervisor{Dr. Rold\~ao da {Rocha} Jr.} 
\examiner{} 
\degree{Doutor em Física} 
\author{Rog\'erio Teixeira {Cavalcanti}} 
\addresses{} 

\subject{Theoretical Physics} 
\university{
{Federal University of ABC}} 
\department{
{Center of Natural and Human Sciences}} 
\group{
{Programa de Pós-Graduação em Física}} 
\faculty{\href{http://faculty.university.com}{Faculty Name}} 

\usepackage{titlesec, blindtext, color}
\definecolor{gray75}{gray}{0.75}
\usepackage{subcaption}
\usetikzlibrary{shapes,snakes}

\makeatletter
\def\thickhrulefill{\leavevmode \leaders \hrule height 1.2ex \hfill \kern \z@}
\def\@makechapterhead#1{
  \vspace*{10\p@}%
  {\parindent \z@ \centering \reset@font
        \textcolor{gray75}{\thickhrulefill}\quad 
        \scshape\bfseries\textit{\@chapapp{}  {\normalsize \thechapter}}  
        \quad \textcolor{gray75}{\thickhrulefill}
        \par\nobreak
        \vspace*{10\p@}%
        \interlinepenalty\@M
        \hrule
        \vspace*{10\p@}%
        \LARGE \bfseries #1 \par\nobreak
        \par
        \vspace*{10\p@}%
        \hrule
        \vskip 50\p@
  }}

\begin{document}

\frontmatter 

\pagestyle{plain} 


\newcommand{\bfx}{{\bf x}}
\newcommand{\bfy}{{\bf y}}
\newcommand{\bfk}{{\bf k}}
\newcommand{\bkp}{{\bf k'}}
\newcommand{\order}{{\cal O}}
\newcommand{\beqa}{\begin{eqnarray}}
\newcommand{\eeqa}{\end{eqnarray}}
\newcommand{\mpl}{m_{p}}
\newcommand{\lmk}{\left(}
\newcommand{\rmk}{\right)}
\newcommand{\lkk}{\left[}
\newcommand{\rkk}{\right]}
\newcommand{\lnk}{\left\{}
\newcommand{\rnk}{\right\}}
\newcommand{\Rbar}{\bar{R}}
\newcommand{\gbar}{\bar{g}}

\def\half{\textstyle{1\over2}}
\def\third{\textstyle{1\over3}}
\def\quarter{\textstyle{1\over4}}
\newcommand{\be}{\begin{equation}}
\newcommand{\ee}{\end{equation}}
\newcommand{\bea}{\begin{eqnarray}}
\newcommand{\eea}{\end{eqnarray}}
\newcommand{\ud}{\mathrm{d}}
\newcommand{\GN}{\!}
\newcommand{\ba}{\begin{eqnarray}}
\newcommand{\ea}{\end{eqnarray}}

\newcommand{\pro}[2]{\mbox{$\langle\, #1 \mid #2\,\rangle$}}
\newcommand{\expec}[1]{\mbox{$\langle\, #1\,\rangle$}}
\newcommand{\expecl}[1]{\mbox{$\left\langle\,
            \strut\displaystyle{#1}\,\right\rangle$}}
\newcommand{\bsy}{\boldsymbol} 
\newcommand{\re}{\Re{\it e}}
\newcommand{\im}{\Im{\it m}}
\renewcommand{\a}{\hat a}
\newcommand{\ac}{\hat a^{\dagger}}
\renewcommand{\b}{\hat b}
\newcommand{\bc}{\hat b^\dagger}
\renewcommand{\d}{\mbox{${\rm d}$}} 
\newcommand{\lp}{\ell_{\rm p}}
\newcommand{\gn}{G_{\rm N}}
\newcommand{\gd}{G_{D}}
\newcommand{\rh}{r_{\rm H}}
\newcommand{\Rh}{R_{\rm H}}
\newcommand{\rd}{r_{\rm HD}}
\newcommand{\RS}{R_{\rm S}}
\newcommand{\RSD}{R_{D}}
\newcommand{\Ah}{A_{\rm H}}
\newcommand{\M}{\mathcal{M}}
\newcommand{\Th}{T_{\rm H}}
\newcommand{\E}{\mathrm{E}}

\def\ben{\begin{eqnarray}}
\def\een{\end{eqnarray}}

\newcommand{\mf}{\mathbf}
\newcommand{\1}{\'{\i}}
\newcommand{\2}{\c c\~ao}
\newcommand{\3}{\c c\~oes}
\newcommand{\lie}{\pounds_{n}}
\newcommand{\el}{^}
\newcommand{\et}{espaço-tempo }

\newcommand{\beq}{\begin{eqnarray}}\newcommand{\benu}{\begin{enumerate}}\newcommand{\enu}{\end{enumerate}}
\newcommand{\eeq}{\end{eqnarray}}
\def\dual#1{\accentset{\boldsymbol{\neg}\vspace{-.13ex}}{#1}}
\def\dualsim#1{\accentset{\boldsymbol{\sim}\vspace{-.13ex}}{#1}}
\def\0{\bm0}
\def\vp{\bm\varphi}
\def\bpi{\bm\pi}
\def\j{\bm{j}}
\def\J{\bm{J}}
\def\K{\bm{K}}
\def\L{\bm{L}}
\def\x{\bm{x}}
\def\kb{\bm\kappa}
\def\I{\openone}
\def\s{\bm\sigma}
\def\p{\bm{p}}
\def\0{\bm0}
\def\vp{\bm\varphi}
\def\bpi{\bm\pi}
\def\j{\bm{j}}
\def\J{\bm{J}}
\def\K{\bm{K}}
\def\L{\bm{L}}
\def\x{\bm{x}}
\def\kb{\bm\kappa}
\def\I{\openone}
\def\s{\bm\sigma}




\begin{titlepage}
\begin{center}

{\scshape\LARGE \univname\par}\vspace{1.5cm} 
\textsc{\Large Doctoral Thesis}\\[0.5cm] 

\HRule \\[0.4cm] 
{\LARGE \bfseries \MakeUppercase \ttitle\par}
{\large \bfseries \MakeUppercase \stitle\par}\vspace{0.4cm} 
\HRule \\[1.5cm] 
 
\begin{minipage}[t]{0.5\textwidth}
\begin{flushleft} \large
\emph{Author:}\\
{\authorname} 
\end{flushleft}
\end{minipage}
\begin{minipage}[t]{0.4\textwidth}
\begin{flushright} \large
\emph{Supervisor:} \\
{\supname} 
\end{flushright}
\end{minipage}\\[2.8cm]
 
\large \textit{ Thesis submitted to the Federal University
of ABC (UFABC) in fulfillment of the requirements
for the degree of Doctor of Philosophy
in the field of Theoretical Physics.}\vspace{3.cm}\\[0.3cm] 

 \includegraphics[width=3.5cm]{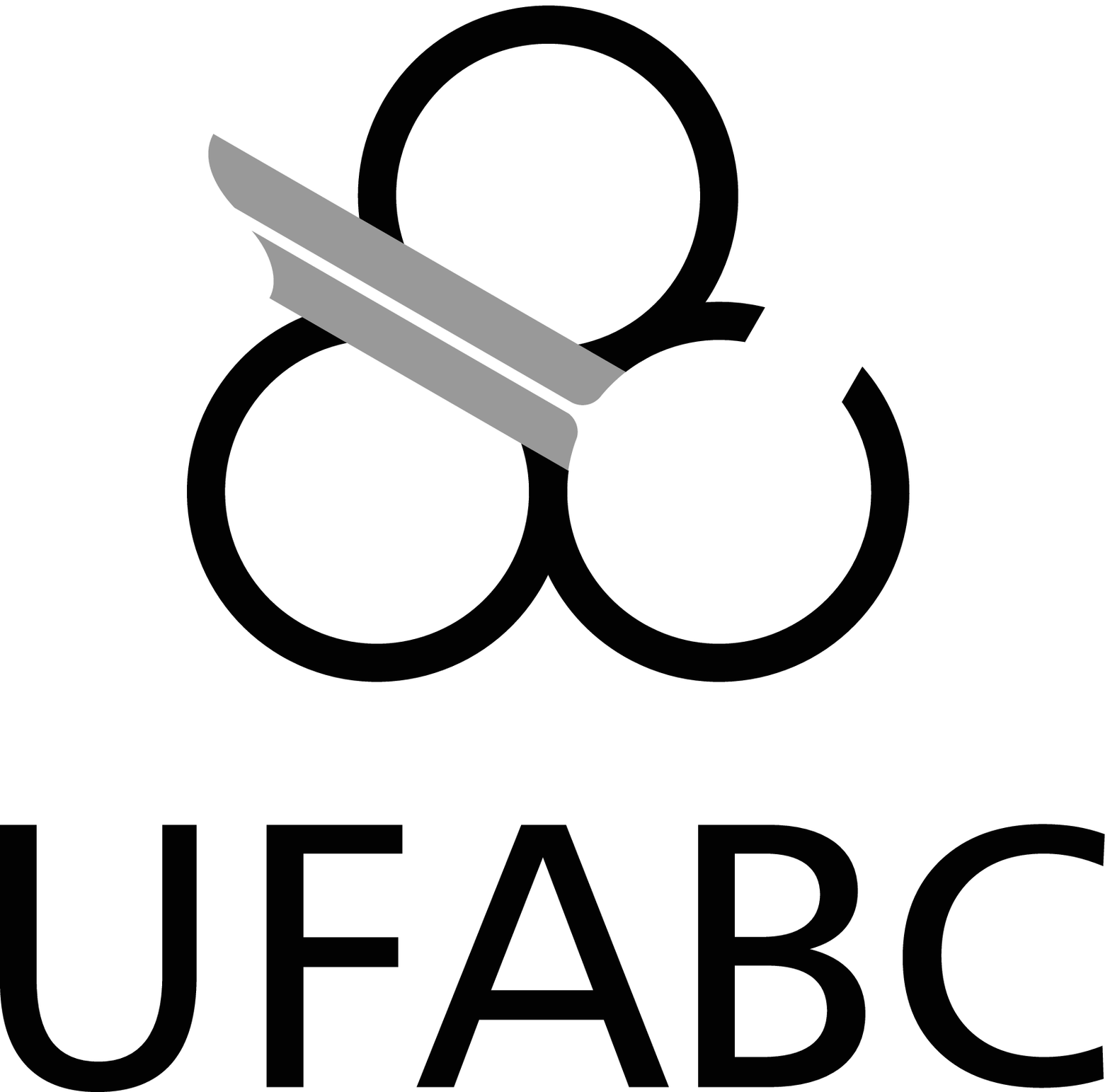} \\[1.8cm] 
 
{\large Santo Andr\'e \\ Brazil\\ 2017}\\[4cm] 

\vfill
\end{center}
\end{titlepage}

%

\includepdf{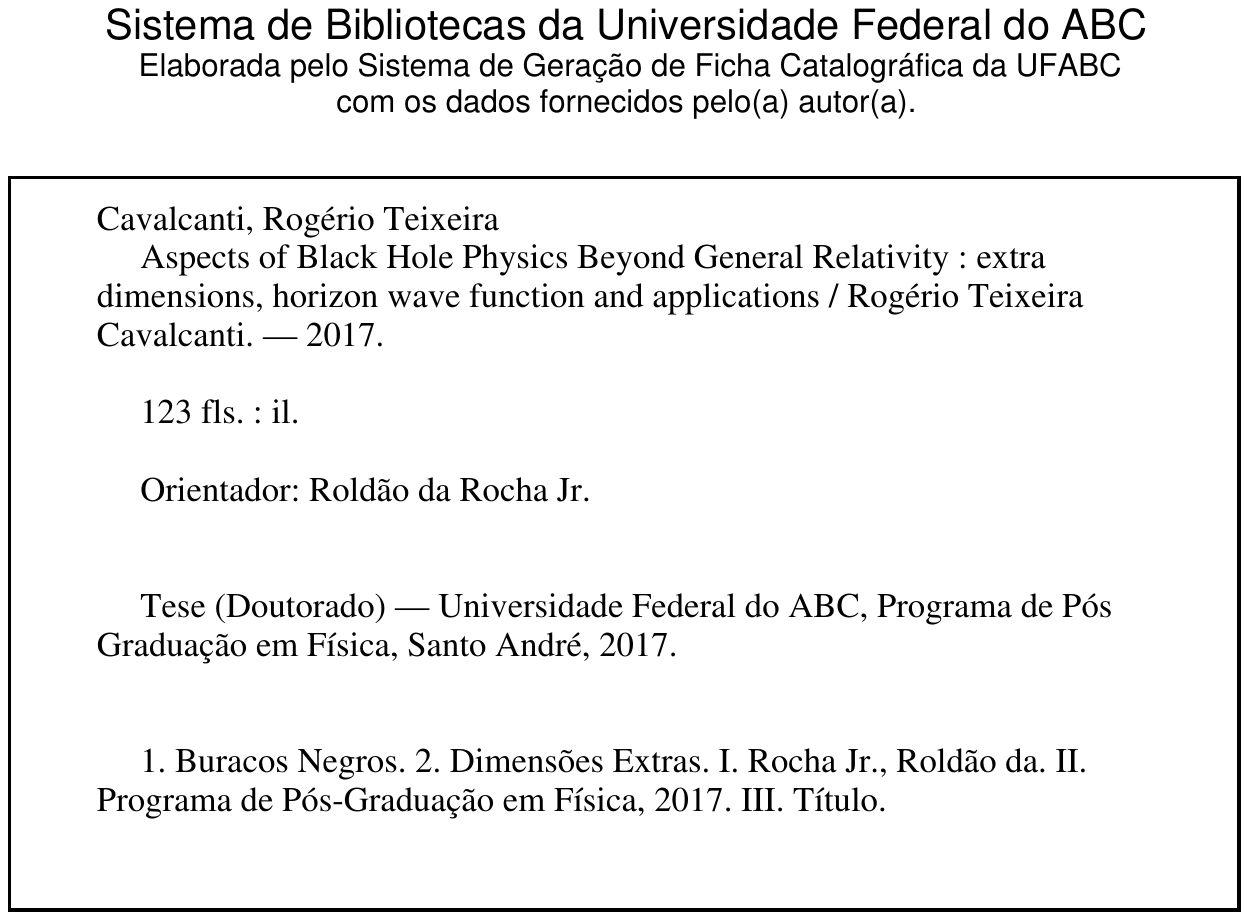}

\begin{acknowledgements}
\addchaptertocentry{\acknowledgementname} 

There are many people directly and indirectly related to this work who deserve more than a few words preceding the main text. My wife is surely the person who suffered more due to my constant absence and lack of time. I am very grateful for all the support she gave me, always helping me to become a better person. Thank you and I love you Juliana! 
My parents also supported me all this time and I'm very grateful for that. Thank you mom and dad (\textit{in memoriam}). 

I had the fortune of meeting some special people during my time as a graduate student. With them I had many discussions about physics, mathematics, science and life in general. I do not intend to mentioning everybody who deserves because I would surely miss a few of them, but José, Allan, André, Kelvyn, Armando, Bruno, Marina, Luis and Daniel are some of these who have helped me to better understand the meaning of friendship and science.

Over the year of 2015 I have been collaborating with professor Roberto Casadio, who hosted me at the Bologna university. Roberto was a great host and I have learnt a lot with him. Thank you!

Finally my advisor, professor Roldão da Rocha. I have learnt much more than how I should carry on a research and technical stuff with him. I am sure his other students have the same feeling. Sometimes he treats us as students, some times as friends, but always being very professional, kind and ready for helping. For me it is hard imagining a better advisor. Thank you very much professor Roldão! 

\end{acknowledgements}


\begin{abstract}
\addchaptertocentry{\abstractname} 

This work is devoted to investigate some consequences of black holes physics beyond the domain of general relativity, mainly in effective extra dimensional models. The investigation is carried along three gravitational effects, namely the Hawking radiation, the strong deflection of gravitational lensing and the formation of quantum black holes. A cosmological thick brane solution is also investigated. Effective theories and models provide a prominent approach for testing the limits of known theories and show what would be expected beyond that. Based on such idea we have used effective models for finding deviations of general relativity associated to each of the mentioned phenomena.

\end{abstract}

\begin{resumo}
\addchaptertocentry{Resumo} 

Neste trabalho foram investigadas algumas conseguências da física de buracos negros em teorias cujo domínio está além do domínio da relatividade geral, em especial em teorias efetivos com dimensões extras. A investigação foi em substancialmente conduzida baseando-se em três efeitos gravitacionais, a saber, a radiação Hawking, o regime de deflexão forte de lentes gravitacionais e a formação de buracos negros quânticos. Uma solução de modelo cosmológico imerso em uma brana espessa foi também investigada. Modelos e teorias efetivas fornecem meios para testar os limites de validade de teorias conhecidas e indicam o que deveríamos esperar além desses limites. Baseado nessa ideia foram usados alguns modelos efetivos para estudar efeitos não previstos pela relatividade geral, associados a cada um dos fenômenos mencionados.

\textbf{Palavras Chave:} Buracos Negros, Dimensões Extras, Lentes Gravitacionais, Buracos Negros Quânticos, Princípio da Incerteza Generalizado.
\end{resumo}




\tableofcontents* 
\listoffigures 
\listoftables 


\begin{abbreviations}{ll} 

\textbf{GR} & \textbf{G}eneral \textbf{R}elativity\\
\textbf{BH} & \textbf{B}lack \textbf{H}oles\\
\textbf{PBH} & \textbf{P}rimordial \textbf{B}lack \textbf{H}oles\\
\textbf{QBH} & \textbf{Q}uantum \textbf{B}lack \textbf{H}oles\\
\textbf{ABH} & \textbf{A}strophysical \textbf{B}lack \textbf{H}oles\\
\textbf{VSR} & \textbf{V}ery \textbf{S}pecial \textbf{R}elativity\\
\textbf{RS} & \textbf{R}andall-\textbf{S}undrum\\
\textbf{CFM} & \textbf{C}asadio-\textbf{F}abbri-\textbf{M}azzacurati\\
\textbf{MGD} & \textbf{M}inimal \textbf{G}eometric \textbf{D}eformation\\
\textbf{GL} & \textbf{G}ravitational \textbf{L}ensing\\
\textbf{SDL} & \textbf{S}trong \textbf{D}eflection \textbf{L}imit\\
\textbf{WDL} & \textbf{W}eak \textbf{D}eflection \textbf{L}imit\\
\textbf{GUP} & \textbf{G}eneralized \textbf{U}ncertainty \textbf{P}rinciple\\
\textbf{HUP} & \textbf{H}eisenberg \textbf{U}ncertainty \textbf{P}rinciple\\
\textbf{HWF} & \textbf{H}orizon \textbf{W}ave \textbf{F}unction\\

\end{abbreviations}


%
%
%


%
%
%
%




\mainmatter 

\pagestyle{thesis} 





\chapter{Introduction} 

\label{intro} 



The Einstein's theory of gravity, the so called general relativity, has been fascinating physicist and science enthusiasts for a century. It is not by chance, predictions of general relativity transcend our regular experience with physical phenomena so dramatically that one could have difficulties for separating physics to science fiction. The final answer is, as always, given by nature, through experiments or observations. Black holes are one of those odd predictions that have been challenging the imagination of physicists for decades. 

The physics of black holes was developed, mainly, after the 50s, carried by some of the most prominent physicists of the last century. Applying the machinery of Riemann geometry they discovered that black holes are associated with space-time singularities and are capable of changing the space-time causal structure. Even being highly non intuitive objects, they are pretty simple, in the sense that any asymptotically flat black hole solution predicted by general relativity is described by only three parameters, namely, mass, angular momentum and charge. In spite of its power and elegance, general relativity was not sufficient to explain some problems emerging on the interface between black hole physics and thermodynamics. An extension of the classical general relativity, called semi-classical gravity, was developed and employed to make both fields compatible. 

It is not the end, black holes are always inviting us to go deeper in fundamental physics.  The semi-classical approach has its limits. When quantum effects become more prominent, a new theory should be used to explain properly things like singularities and the physics around the Planck scale. This theory is not known up to the present date, but black holes might be used to guide us through the directions potentially interesting. In this sense, black holes are employed by phenomenological approaches looking beyond general relativity. This is a central point in the present work. In fact, we apply phenomenological theories/models, mainly using extra dimensions, to test deviations of general relativity. Some of those deviations can be tested by applying the effect gravitational lensing. Lensing effects have been being applied with great success by astrophysicists and cosmologists for a long time. However, only recently  an effective approach able to incorporate the strong gravitational field around black holes was proposed. Such approach is known as strong deflection limit and we have used it to calculate observables of extra dimensional black holes.  
 
Despite being usually related to astrophysical objects, black holes are also expected to be formed when a high concentration of energy is confined in a small region of the space-time. In order to better understanding  such phenomena was proposed the horizon quantum mechanics formalism. Some interesting results emerge  in this formalism, as the generalized uncertainty principle and minimum physics scales. Both phenomena appear in different candidates for the theory that merges gravity and quantum mechanics.

Summarizing, in the present thesis we investigate some effective approaches of black hole physics when one tries to push the borders of gravity and/or quantum mechanics.  

\section{Results and Thesis Organization}

Except by this introductory chapter and the next one, any chapter of this thesis contains at least one session filled with results published by us in indexed journals. They are listed accordingly: Ref. \cite{Cavalcanti:2015nna} was published in Adv. High Energy Phys. (Chapter \ref{cap1}), Ref. \cite{Bernardini:2014vba} in Gen. Rel. Grav. (Chapter \ref{ched}), Ref. \cite{Cavalcanti:2016mbe} in Class. Quant. Grav. (Chapter \ref{cap5}) and Ref. \cite{Casadio:2015jha} in Phys. Lett. B (Chapter \ref{cap6}). All the results were obtained and published during the doctorate degree time with my advisor and/or collaborators.  

There are more results published but not included here. Their topics are not very related to the theme of the present thesis and are listed as following:
\begin{itemize}
\item Ref. \cite{daRocha:2013qhu} was our first work, it deviates considerably of the present one and was further developed by another author and Roldão's student. 

\item Ref. \cite{Cavalcanti:2014uta} regards more formal aspects highly based on Clifford and DKP algebras, underlying dark spinors. 

\item In Ref. \cite{Cavalcanti:2014wia} I explicitly found the reciprocal Lounesto spinor classification. It has been useful for researchers in the field of non standard spinor fields.

\item Ref. \cite{Cavalcanti2015} contains results intimately related to Ref. \cite{Cavalcanti:2014uta}.

\item In Ref. \cite{Casadio:2016zhu} we employed the shear viscosity-to-entropy density ratio for estimating the post-Newtonian parameter. 

\item In Ref. \cite{daRocha:2016bil} new spinor solutions encompassing flag-pole and flag-dipole spinors in Kerr space-time have been derived.
\end{itemize}

In the next chapter our aim is to establish the basic notation, nomenclature, results and ideas that appear along the text. There we introduce a set of basic results of black hole physics in general relativity, from the possible asymptotically flat solutions to black hole thermodynamics. It provides a minimally smooth transition between results of classical black holes and our discussion of results found outside the domain of general relativity. In the Chapter \ref{cap1} we apply a recent method for calculating the Hawking temperature. The method is used to determine the temperature of a black hole emerged as an effective solution of string theory. Our result \cite{Cavalcanti:2015nna}, presented in Sec. \ref{elkokerrsen}, shows that the same temperature is obtained by considering the emission of a fermion beyond the standard model. Such fermion has its own dynamical equations, which makes the results highly non trivial.

The Chapter \ref{ched} is dedicated introduce some basic models and phenomenological aspects of large extra dimensions. We discuss some braneworld proposals and its effects on the limits of general relativity. The thick brane paradigm is also included in this chapter. It is used to introduce our results \cite{Bernardini:2014vba} in Sec. \ref{thickfrw}, which deal with the spherically symmetric thick brane version of the Friedmann-Robertson-Walker solution. We consider a scalar field minimally coupled with gravity, generating the warp factor, and analyse the solutions of the field equations for the cases of null, negative and positive curvature. We finish the chapter with a section introducing the extra dimensional black hole solutions used in the subsequent chapters.

The deflection of light by strong gravitational fields is investigated in the Chapter \ref{cap5}. We apply a different approach for calculating the angular change in the trajectory of light, the so called strong field deflection limit. In contrast to the conventional approach, this one is appropriate to deal with deflections happening nearby highly massive astrophysical objects. Such approach can be used for testing effects of black hole solutions not predicted by general relativity. In fact, in Sec. \ref{obs} we estimate the effects of the braneworld black holes introduced in the Chapter  \ref{ched} when compared to the standard Schwarzschild solution  \cite{Cavalcanti:2016mbe}.

A new approach for black holes emerging in scenarios where effects of both, gravity and quantum mechanics, are not negligible, is discussed in the Chapter \ref{cap6}. Such approach, known as horizon quantum mechanics or horizon wave function, provides an effective way for estimating the formation of mini black holes. We extend the horizon wave function formalism to an extra dimensional scenario in Sec. \ref{HDHWF}, showing that quantum effects could change the probability of those black hole formation \cite{Casadio:2015jha}. The results can change the number of events of black hole formation expected in the present colliders. Finally, in the Chapter \ref{cap7} we summarize the results and discuss some future perspectives.


\chapter{Preliminaries} 

\label{cap0} 

%

In this preliminary chapter we are going to discuss some basic and general results of general relativity (GR) as well as what we mean by beyond GR. Its purpose is to establish notations, conventions and nomenclatures used along the text, plus to introduce the basic setup of black holes in Einstein's gravity. By Einstein's gravity we mean the 4-dimensional (4D) theory of general relativity. This more than 100 years old theory has been survived against numerous observational tests and is the standard theory for gravitational interactions. Despite its remarkable success, GR by itself does not give enough clues on what should be expected when quantum effects are not negligible. In this case black hole physics could provide an effective approach to go further on the interface between gravity and quantum mechanics.

Let us start by some basic conventions. The fundamental geometric objects of GR are a manifold $\mathcal{M}$ and a metric $g$. The pair $(\mathcal{M}, g)$ is referred as the space-time. We shall use Greek letters representing the components of 4D fields defined in the tangent space $T_p\mathcal{M}$ of the point $p\in \mathcal{M}$. We adopt the indexes ranging in the set $\{0,1,2,3\}$. For higher dimensional space-times we use Latin capital letters for indexes in the set $\{0,1,2,3,5,6,...,n\}$. Hence, a line element of a 4D space-time is represented by
\begin{align}
ds^2=g_{\mu\nu}dx^\mu dx^\nu,
\end{align}
whereas a line element of a higher dimensional space-time reads
\begin{align}
ds^2=g_{_{AB}}dx^{_A} dx^{_B}.
\end{align}
The metric signature adopted is $(-,+,+,...)$, being, for example, spherically symmetric static 4D metrics  denoted by
\begin{align}
ds^2=-A(r)dt^2+B(r)dr^2+r^2d\Omega^2,
\end{align}
where $d\Omega^2= d\theta^2+\sin^2\theta d\phi^2$. Unless explicitly mentioned, the natural unity system ($c=\hbar=1$) will be used, where the Planck mass is related to the gravitational constant as $m_p={G}^{-1/2}$. In some situations, symmetric and anti-symmetric permutations of indexes are denoted by round and squared brackets, defining
\begin{align}
\nabla_{(\mu}V_{\nu)} \equiv \nabla_\mu V_{\nu}+\nabla_\nu V_{\mu} \;\;\text{ and }\;\; \nabla_{[\mu}V_{\nu]} \equiv \nabla_\mu V_\nu-\nabla_\nu V_\mu,
\end{align}
for an arbitrary field $V$.

Settled down the basic notational conventions, we can begin the discussion of GR. General relativity is a geometrized theory of gravity which generalizes the Einstein's special relativity and the Newtonian gravity. The non-vacuum dynamical field equations, also called Einstein equations, follow from the Einstein-Hilbert action plus the matter action
\begin{align}
S_{EH}+S_M=\int d^4x\sqrt{-g}\left(\frac{R}{16\pi G} + \mathcal{L}_M\right),
\end{align}
where $R$ is the Ricci scalar and $g$ is the metric determinant. The field equations are a system of coupled non linear partial differential equations for the metric components $g_{\mu\nu}$ and are obtained by varying the above action with respect to $g^{\mu\nu}$ . They connect geometry and matter in a very elegant way
\begin{align}\label{efeq}
G_{\mu\nu} \equiv R_{\mu\nu}-\frac{1}{2} R g_{\mu\nu}=8\pi G T_{\mu\nu}.
\end{align}
$T_{\mu\nu}=-\frac{2}{\sqrt{-g}}\frac{\delta(\sqrt{-g}\mathcal{L}_M)}{\delta g^{\mu\nu}}$ is the energy-momentum tensor.  It encompasses energy and momentum of matter fields, which act as a source for gravity. The left hand side is composed by the Ricci tensor $R_{\mu\nu}$ and the Ricci scalar, both measuring the curvature of space-time. The tensor $G_{\mu\nu}$ is called Einstein tensor.

The black holes we are going to introduce in the next section emerge as solutions of Eq. \eqref{efeq} for vanishing $T_{\mu\nu}$. Their properties have been pushing forward our knowledge about gravity for decades. They are also seen as prime candidates to reveal what might be expected beyond GR. A general discussion of such possibility, which have been influencing a considerable portion of present research on the interface gravity $\times$ quantum mechanics, can be found in Sec. \ref{grqm}. Before that we will briefly revise some features of black hole physics in GR.

\section{A Glimpse of Black Holes in General Relativity}

Theoretical predictions of black hole-like objects are more then 100 years older than the Einstein's general relativity. It dates back the 18th century, found in the work of John Michell and Pierre-Simon Laplace, when calculating the escape velocity of astronomical objects. If the escape velocity was higher than the velocity of light, the object would not radiate, thus it would be invisible for distant observers. Those objects were called dark stars. The modern concept of black hole begins with the first exact solution of the Einstein equation, provided by Karl Schwarzschild. However, black hole physics became a very active topic in theoretical physics only in the 60s, keeping such status up to the present date. The difference is that the present research of black holes physics is not restrict to the theoretical fields nor restricted to gravity. Many evidences supporting the existence of such objects were found in astronomical observations \cite{Reines:2016qun,Ghez:2008ms}, including the recent detection of gravitational waves by the LIGO collaboration \cite{Abbott:2016blz,TheLIGOScientific:2016wfe}. They also play a central hole in the gauge/gravity correspondence \cite{Maldacena:1997re} and its application to condensate matter  \cite{Pires2014}. In Chapter \ref{cap5}, for example, we use data from Sagittarius A$^*$, the supermassive black hole believed to reside in the center of our galaxy, to calculate the light deflection produced by black holes emerged beyond Einstein's gravity. 

There are three main physical process expected to form a black hole: primordial density fluctuations, high energy collisions and collapsing of highly of massive stars. The first process would form primordial black holes (PBH), the second would form quantum black holes (QBH) and the third is associated to astrophysical black holes (ABH). Only ABH are truly GR predictions, the other two require also quantum properties. Some features of each type are summarized below:

\begin{itemize}
\item[PBH] Those black holes are expected to form due primordial density inhomogeneities or topological defects from phase transition in the early universe. The high density of the early universe is a necessary but not sufficient condition for
PBH formation. Some kind of density fluctuation is also necessary in order to over dense
regions stop expanding and re-collapse into a PBH. Those fluctuations could be primordial
or arise spontaneously, as quantum effects during the inflation epoch. PBH is an active field of research. They were proposed to describe dark matter \cite{Carr:2016drx} and have been used to constrain models in cosmology \cite{Carr:2009jm} and inflation \cite{Green:1997sz}; 

\item[QBH] Quantum black holes are interesting tools for testing the merge of gravity and quantum mechanics. Those black holes should be formed when a large amount of energy lies in a very small region. In this case, if the energy is large enough, the causal structure of the space-time is changed by the horizon formation. Classical predictions expect such black hole formation happening close to the Planck scale, which is very far beyond the present experiments. However, some models predict this energy scale being lowered to a few TeV. We shall discuss more about quantum black hole in Sec. \ref{grqm} and Chapter \ref{cap6}.

\item[ABH] Astrophysical black holes are the best known type of black holes. The evidences of black holes found in astronomical observations are usually associated to ABH. Those black holes are formed after massive stars reach the endpoint of their thermonuclear burning
phase. Hence, nuclear reactions no longer supply thermal pressure and the gravitational collapse proceed (see Fig. \ref{fig:collapsing}). The first solution of the Einstein equations describing a collapsing star were proposed in 1939 by Oppenheimer and Snyder \cite{Oppenheimer:1939ue}. They calculated the collapse of a homogeneous sphere of pressureless
gas in general relativity, yielding a black hole.
\begin{figure}[!htb]
\centering
 \scalebox{.99}{\begin{tikzpicture}[line cap=round,line join=round,>=triangle 45,x=1.0cm,y=1.0cm]
\draw [rotate around={90:(-2,0.88)}] (-2,0.88) ellipse (1.65cm and 0.27cm);
\draw (-2,2.5)-- (3,2.5);
\draw (-2,-0.75)-- (3,-0.75);
\draw [shift={(3,2.5)}] plot[domain=4.71:5.76,variable=\t]({1*3.25*cos(\t r)+0*3.25*sin(\t r)},{0*3.25*cos(\t r)+1*3.25*sin(\t r)});
\draw [shift={(3,-0.75)}] plot[domain=0.52:1.57,variable=\t]({1*3.25*cos(\t r)+0*3.25*sin(\t r)},{0*3.25*cos(\t r)+1*3.25*sin(\t r)});
\draw [shift={(5.81,0.28)}] plot[domain=1.57:2.41,variable=\t]({1*0.9*cos(\t r)+0*0.9*sin(\t r)},{0*0.9*cos(\t r)+1*0.9*sin(\t r)});
\draw [shift={(5.81,1.48)}] plot[domain=3.87:4.71,variable=\t]({1*0.9*cos(\t r)+0*0.9*sin(\t r)},{0*0.9*cos(\t r)+1*0.9*sin(\t r)});
\draw (5.81,1.18)-- (9.7,1.18);
\draw (5.81,0.58)-- (9.7,0.58);
\draw [->] (-2.99,0.88) -- (0.3,0.88);
\draw [->] (-2.99,0.88) -- (-2.99,3.14);
\draw [line width=1.6pt,dash pattern=on 1pt off 2pt on 5pt off 4pt] (5.81,0.88)-- (9.7,0.88);
\draw [->] (6.72,0.88) -- (6.72,0.04);
\draw (6.14,0.04) node[anchor=north west] {Singularity};
\draw [dotted] (3,2.5)-- (5.81,2.5);
\begin{scriptsize}
\draw[color=black] (0.76,2.74) node {Star surface};
\draw[color=black] (7.9,1.36) node {Horizon};
\draw[color=black] (-1.08,1.1) node {Time};
\draw[color=black] (-2.4,3.04) node {Radius};
\draw[color=black] (4.5,2.72) node {Collapsing process};
\end{scriptsize}
\end{tikzpicture}}
\caption{Depicture of a spherically symmetric star collapsing into an ABH.}\label{fig:collapsing}
\end{figure}
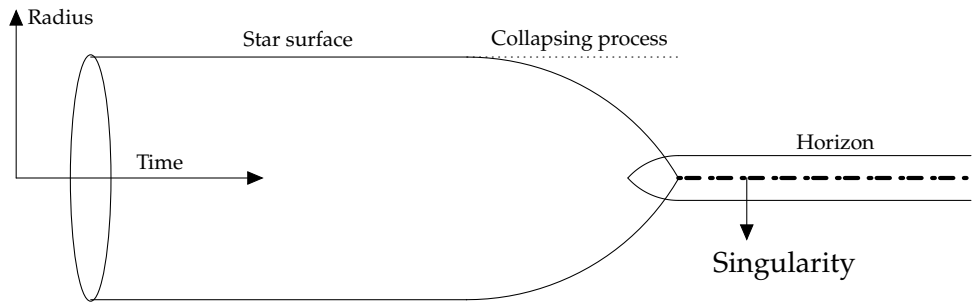
\end{itemize}

Now we are going to discuss some classical solutions and results of black hole physics in general relativity.

\subsection{Asymptotically Flat Solutions}\label{asyflat}

The first  discovered solution to the Einstein field equations, as well as the simplest black hole solution, is the famous Schwarzschild metric \cite{Schwarzschild:1916ae}, given by
\begin{align}\label{schsol}
ds^2=-\left(1+\frac{2M}{r}\right)dt^2+\left(1+\frac{2M}{r}\right)^{-1}dr^2+r^2d\Omega^2.
\end{align}
It was found in 1915, within months after Einstein achieved his final formulation of general relativity. Since the solution is static and spherically
symmetric, it was first proposed to describe the geometry outside a static and 
spherical star. It took several years to appear the interpretation of the Schwarzschild solution as describing a black hole.

One of the  main straightforward features of the Schwarzschild solution can be seen directly from Eq. \eqref{schsol}. The $g_{rr}$ coefficient becomes singular for $r=r_\mais=2M$ and for $r=0$. In general relativity there are two types of singularities that one must distinguish, namely, coordinate singularities and curvature singularities. The singularity at $r_\mais=2M$ can be removed by changing the coordinate system, thus it is called coordinate singularity. The one at $r = 0$, on the other hand, can not be removed by any coordinate system, being called curvature singularity. Curvature singulaties can be found by computing quantities that measure the gravitational field strength in a invariant way, by applying the so called diffeomorphism invariants or scalars, which give a measure of the density of matter. The Kretschmann scalar \cite{Carminati1991} $K=R^{\mu\nu\alpha\beta} R_{\mu\nu\alpha\beta}$ is one of those invariant quantities. For the Schwarzschild solution one finds $K\sim M^2/r^6$. Therefore $K$ diverges at $r = 0$ but is perfectly regular at $r_\mais=2M$, as mentioned earlier.
The Eddington-Finkelstein  coordinates \cite{Finkelstein:1958zz} is an example of coordinate system with no singularity at $r_\mais=2M$. In such coordinates the Schwarzschild solution is given by
\begin{align}
ds^2=-\left(1+\frac{r_\mais}{r}\right)dv^2+2dvdr+r^2d\Omega^2,
\end{align}
where $v=t+r^*$ and $r^*=r+2M\ln(r-2M)$.
Another way to avoid coordinate singularities is by using the Painlevé-Gullstrand coordinates \cite{Martel:2000rn},
\begin{align}
ds^2=-\left(1-\frac{r_\mais}{r} \right)dt_p^2+2\sqrt{\frac{r_\mais}{r}}dt_pdr+dr^2+r^2d\Omega^2.
\end{align}
We shall use the above Painlevé-Gullstrand coordinates in the next chapter 
to calculate the Hawking temperature.

A few years after Schwarzschild had found the solution carrying his name, Reissner and Nordström generalized it by considering charged spherically symmetric objects. Coupling to an electromagnetic field one finds the Einstein-Maxwell equations, which follow from the action
\begin{align}
S_{EM}=\int dx^4 \sqrt{-g}\left(\frac{R}{16\pi G}-\frac{1}{4}F_{\mu\nu}F^{\mu\nu}\right).
\end{align}
The term $F_{\mu\nu}F^{\mu\nu}$ results in the energy-momentum tensor 
\begin{align}\label{emTem}
T^{EM}_{\mu\nu}=\frac{1}{4\pi}\left(F_{\mu\lambda}F_\nu^{\;\;\lambda}-\frac{1}{4}g_{\mu\nu}F_{\alpha\beta}F^{\alpha\beta}\right),
\end{align}
corresponding to an electromagnetic field with field strength tensor $F_{\mu\nu}$. Having the only non-zero component of $F_{\mu\nu}$  as $F_{rt} = Q/r^2$ one finds the Reissner-Nordström solution
\begin{align}
ds^2=-\left(1+\frac{2M}{r}+\frac{Q^2}{r^2}\right)dt^2+\left(1+\frac{2M}{r}+\frac{Q^2}{r^2}\right)^{-1}dr^2+r^2d\Omega^2,
\end{align}
where $Q$ is the black hole charge. 
From the above metric we can see that, if $M > |Q|$, there are two values of $r$ for which the metric component $g_{rr}$ diverges:
\begin{align}
r_\pm=M\pm\sqrt{M^2-Q^2}.
\end{align}
The larger represents the event horizon $r_\mais$, whereas $r_\menos$ is known as inner horizon or Cauchy horizon. The Cauchy horizon delimits the region where initial data can have unique evolution. Beyond this horizon the evolution can be affected by unknown
boundary conditions at the singularity \cite{Frolov2011}.

The next solutions regard spinning and charged-spinning black holes, known as Kerr \cite{Kerr:1963ud} and Kerr-Newman \cite{Newman:1965tw} solutions respectively. In contrast to the static Schwarzschild solutions, found quickly after the Einstein's seminal work, it took almost 50 years before the analogous solution for a spinning black hole had been found. From the metric components one could guess the reason for this long delay between solutions:
\begin{align}\label{kerrnew}\nonumber
ds^{2} =&-\frac{\Delta-a^2\sin^2\theta}{\Sigma}dt^2+\frac{\Sigma}{\Delta}dr^2+\Sigma d\theta^2-\frac{2a\sin^2\theta}{\Sigma}\left[(r^2+a^2)-\Delta) \right]dtd\phi+\nonumber\\
&+\frac{\sin^2\theta}{\Sigma}\left[\left(r^2 +a^2 \right)^2-\Delta a^2\sin^2\theta \right]d\phi^2
\end{align}
where $\Delta=r^2-2Mr+a^2+Q^2$ and $\Sigma=r^2+a^2\cos^2\theta$. The above metric does not depend on $t$ or $\phi$. It implies that the space-time is stationary and
axisymmetric\footnote{However it is not static, as can be seen by making $t\mapsto -t$.}. The general metric \eqref{kerrnew} is the Kerr-Newman black hole, it describes a rotating solution coupled to \eqref{emTem}. The other solutions can be obtained from Eq. \eqref{kerrnew} by imposing some parameters to vanish. In fact, from Eq. \eqref{kerrnew} we have:
\begin{itemize}
\item $Q=0$ $\Rightarrow$ Kerr metric;
\item $a=0$ $\Rightarrow$ Reissner-Nordström metric;
\item $Q=0$ and $a=0$ $\Rightarrow$ Schwarzschild metric;
\item $Q=0$, $a=0$ and $M=0$ $\Rightarrow$ Minkowski metric.
\end{itemize}

\begin{figure}[!htb]
\centering
 \scalebox{.99}{\begin{tikzpicture}[line cap=round,line join=round,>=triangle 45,x=1.0cm,y=1.0cm]
\draw(4.1,-2.76) circle (2.9513386793114753cm);
\draw [rotate around={1.8245947789210832:(4.03,-2.755)}] (4.03,-2.755) ellipse (4.672492037195906cm and 2.964263287506549cm);
\draw [->] (6.686148269706775,-0.26535396196142624) -- (8.26,0.46);
\draw [->] (1.468265574763974,-1.4241954053726236) -- (0.14,0.02);
\draw [->] (7.78,-3.02) -- (9.68,-3.28);
\draw (-0.4,.44) node[anchor=north west] {\large $r_{\textsc{\footnotesize +}}$};
\draw (8.3,.75) node[anchor=north west] {\large $r_{_E}$};
\draw (9.76,-3.022) node[anchor=north west] {\large Ergosphere};
\draw (2.85,-2.6) node[anchor=north west] {{\large Black Hole}};
\end{tikzpicture}}
\caption{Depicture of Kerr or Kerr-Newman black hole and its ergosphere.}\label{fig:ergosphere}
\end{figure}
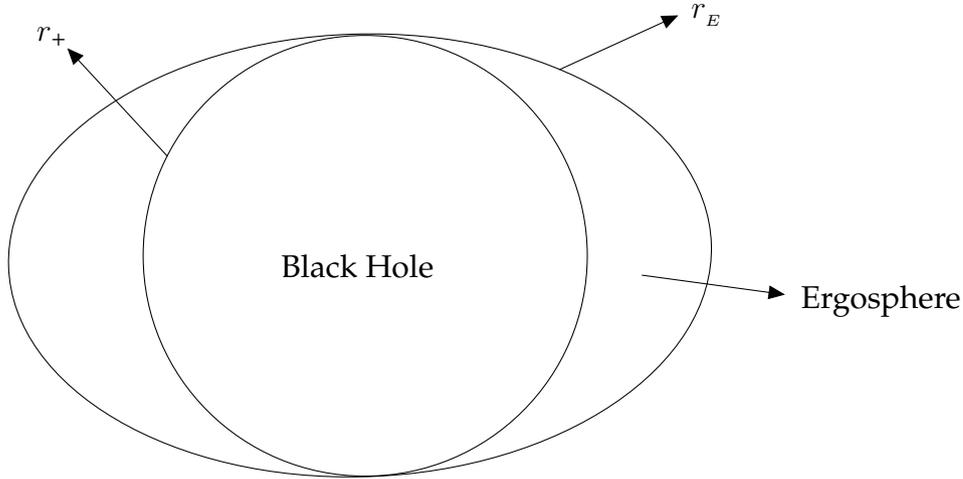

Once again, there are two singularities on Eq. \eqref{kerrnew}. The first one occurs when $\Sigma=0$ and the other when $\Delta=0$. The former is the actual curvature singularity, as can be seen in the Kretschmann scalar \cite{Henry:1999rm}
\begin{align}\nonumber
K=\frac{8}{\Sigma^6}&\left[6M^{2}(r^{6}-15a^{2}r^{4}cos^{2}\theta +15 a^{4} r^{2} cos^{4}\theta-a^{6}cos^{6}\theta)+\right. \\\nonumber
&- 12MQ^{2}r(r^4-10a^2r^{2}cos^{2}\theta+5a^{4}cos^{4}\theta)+\\
&+\left.Q^{4}(7r^{4}-34a^{2}r^2cos^{2}\theta+7a^{4}cos^{4}\theta)\right].
\end{align}
The latter is the coordinate singularity, which is fairly similar to the  Reissner-Nordström when $M^2>a^2+Q^2$. In this case there are two surfaces defined by
\begin{align}
r_\pm=M\pm\sqrt{M^2-a^2-Q^2}.
\end{align}
Again the surface defined by $r = r_\mais$ is the event horizon of the black hole and $r = r_\menos$ corresponds to the inner (or Cauchy) horizon. Moreover, the component $g_{tt}$ vanishes for $r_{_E}=M+\sqrt{M^2-a^2\cos^2\theta-Q^2}$,  which lies outside the event horizon. Except at its interception with the rotation axis, at $\theta=0$ and $\theta=\pi$, where they coincide. The region in between $r_\mais$ and $r_E$ is called ergosphere (see Fig. \ref{fig:ergosphere}). Penrose proposed a process to extract energy from the ergospheres of black holes, known as Penrose process \cite{Wald:1984rg}. From the Penrose process follows one of the black hole laws with tight analogy to thermodynamic laws, as seen in Sec. \ref{BHT}.

\subsection{Horizons}

In general relativity there are two types of horizons, the black hole horizon and the cosmological horizon. The black hole horizon is a future horizon, which means that it can be on the future cone of events, but not in the past cone. The cosmological horizon, on the other hand, can be on the past cone of events, but not in the future cone. In both cases the horizon defines a boundary in between regions causally disconnected in one direction. Here we are interested in black hole horizons and some of its definitions.

The naive definition of event horizon as the region from which nothing can escape is insufficient for the classical geometrical approach to GR. A mathematically clear definition is required. In fact, the formal definition of black holes and event horizons is the one used to prove some classical theorems of black hole physics. It is defined as follows \cite{Wald:1984rg}:

\begin{definition}  
Consider an asymptotically plane space-time $\mathcal{M}$. Give the null infinity future $\mathcal{I}^+$ and its causal past $J^-(\mathcal{I}^+)$, black holes are defined by
\begin{align}
\mathcal{B}\equiv \mathcal{M}-J^-(\mathcal{I}^+).
\end{align}
The event horizon is defined as the boundary of $\mathcal{B}$.
\end{definition}
  
The above definition can not be applied to non asymptotically plane space-times. Moreover, it is highly non-local, in the sense that it requires the whole space-time future. Those issues can be addressed by other characterizations of black hole horizons \cite{Hayward:1993wb,Booth:2005qc,Gourgoulhon:2008pu}. We shall define one of those  characterizations, called trapping surfaces, but first we need some preliminary definitions.

\begin{figure}[!htb]
    \centering
    \includegraphics[scale=.5]{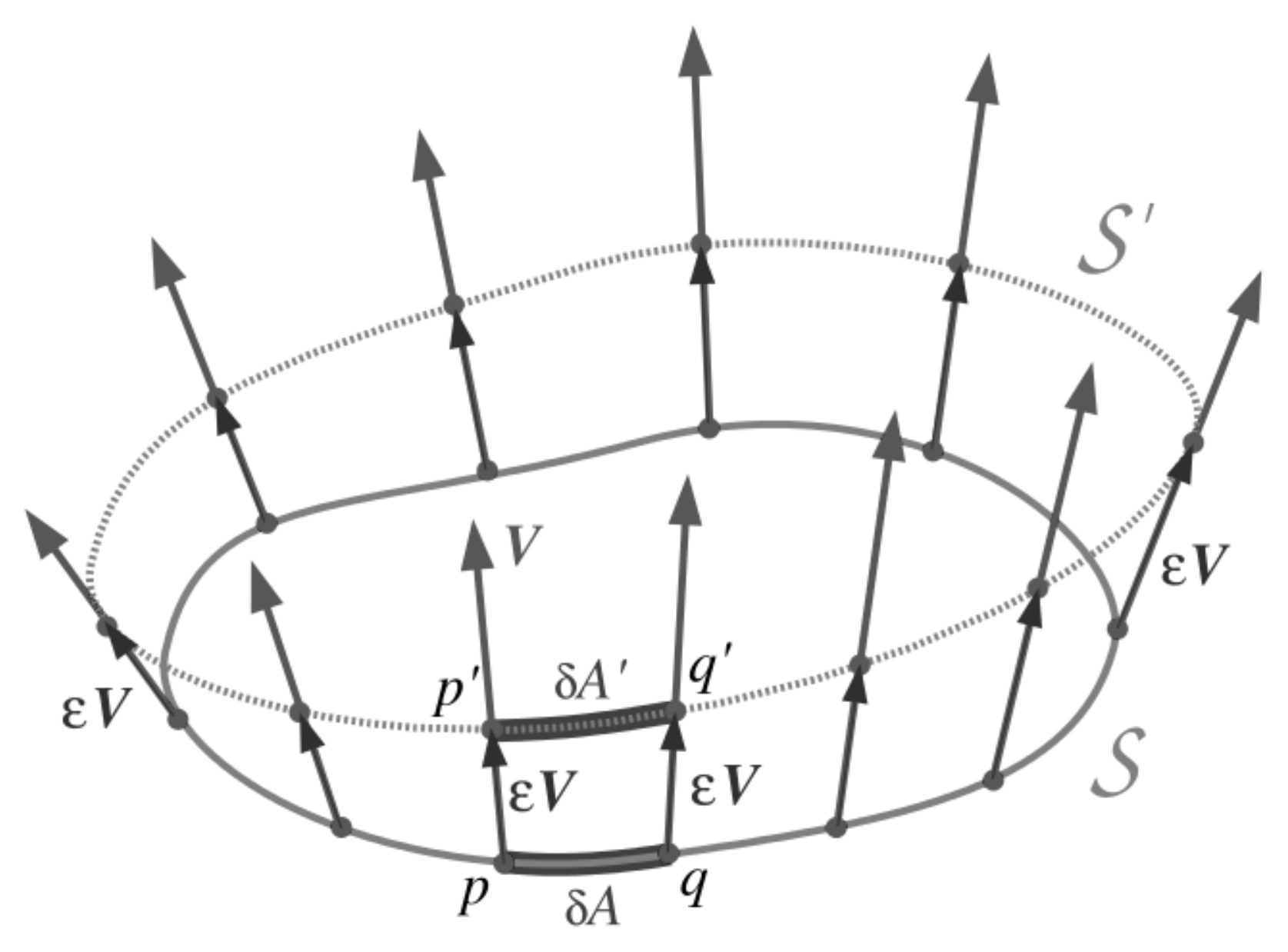}
    \caption{Representation of the displacement $p \mapsto p'=p+\varepsilon V$ producing $\delta A \mapsto \delta A'$. In this plot $\mathcal{S}$ appears as a closed line,
whereas it is actually 2-dimensional surface. Figure taken from \cite{Gourgoulhon:2008pu}.}
  \label{fig:trapp1}
  \end{figure}

Consider a space-like surface $\mathcal{S}$, closed and 2-dimensional, immersed in a space-time $(\mathcal{M}, g)$ and with induced metric $\bar{g}$. Consider also a vector field $V$ normal to $\mathcal{S}$ on each point $p\in \mathcal{S}$. Now make a small displacement in each  $p\in \mathcal{S}$ along $V$ such that $p \mapsto p'=p+\varepsilon V$ for $\varepsilon$ small (see Fig. \ref{fig:trapp1}). This process defines a new surface $\mathcal{S}'$.
  The displacement $p \mapsto p'=p+\varepsilon V$ produces an expansion of the surface $\mathcal{S}$ into $\mathcal{S}'$, which is quantitatively given by the variation of the area element $\delta A \mapsto \delta A'$ accordingly \cite{Gourgoulhon:2008pu},

\begin{align}
\theta^{(V)} \equiv \lim_{\varepsilon \to 0}\frac{1}{\varepsilon}\frac{\delta A'-\delta A}{\delta A}=\bar{g}^{\mu \nu}\nabla_\mu V_\nu.
\end{align}

\begin{figure}[!htb]
    \centering
    \includegraphics[scale=.5]{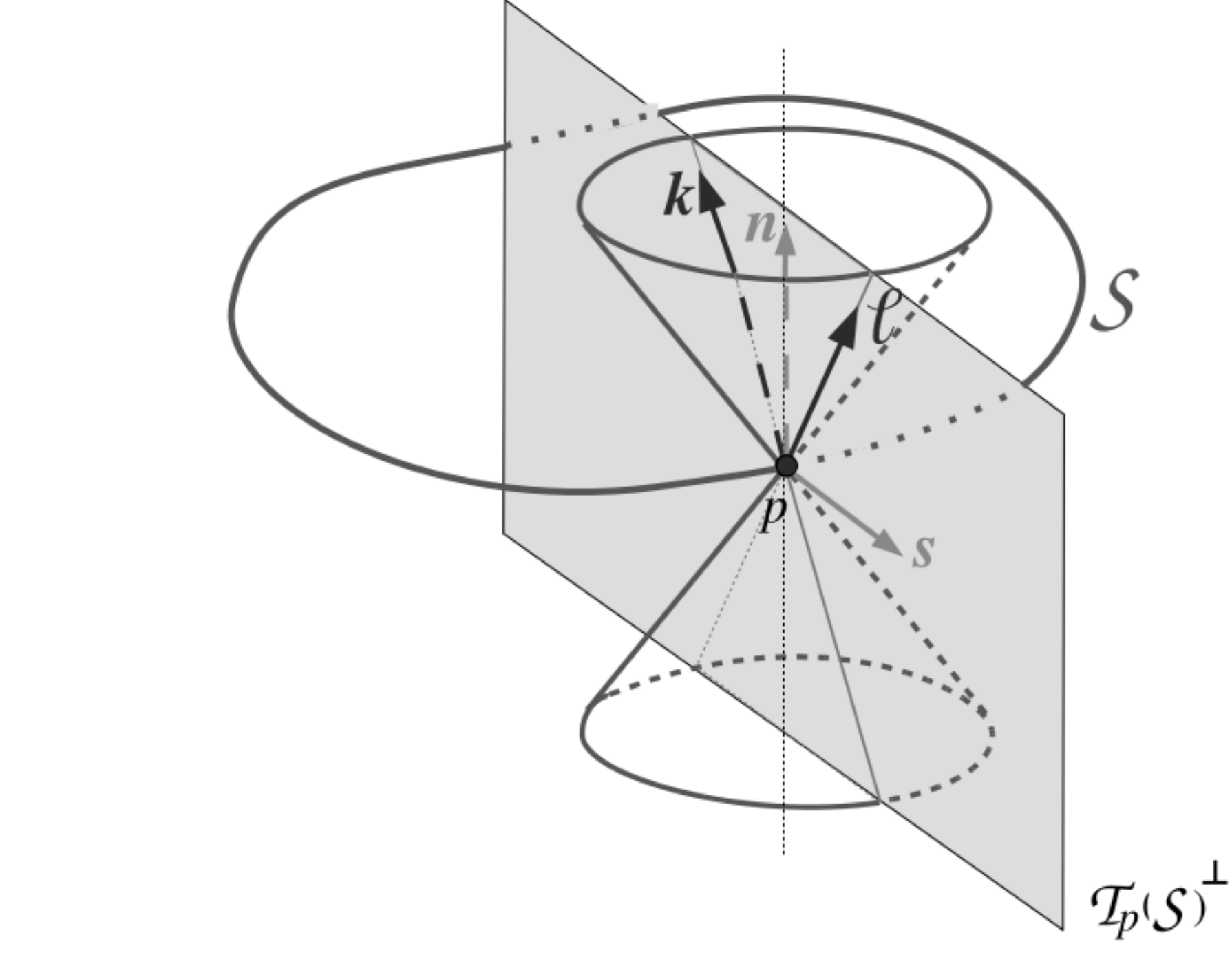}
    \caption{Intersection of the null cone of $p \in \mathcal{S}$ and the orthogonal tangent space $[T_p(\mathcal{S})^\bot]$ defining the directions $k$ and $l$. Figure taken from \cite{Gourgoulhon:2008pu}.}
  \label{scone}
  \end{figure}

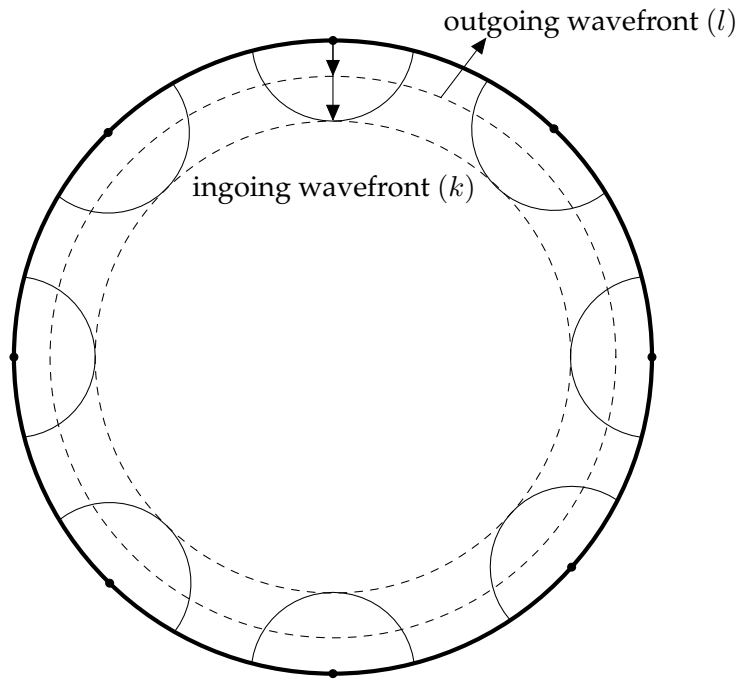
\begin{figure}[!htb]
\centering
 \scalebox{.99}{\begin{tikzpicture}[line cap=round,line join=round,>=triangle 45,x=1.0cm,y=1.0cm]
\draw [line width=1.6pt] (8.84,-2.98) circle (4.24cm);
\draw [dashed] (8.84,-2.98) circle (3.16cm);
\draw [dashed] (8.84,-2.98) circle (3.76cm);
\draw [->] (8.84,1.26) -- (8.84,0.18);
\draw [->] (8.84,1.26) -- (8.84,0.78);
\draw (10.18,1.82) node[anchor=north west] {$\text{outgoing wavefront } (l)$};
\draw (6.84,-0.4) node[anchor=north west] {$\text{ingoing wavefront } (k)$};
\draw [shift={(8.84,1.26)}] plot[domain=3.269297978120286:6.155479982649093,variable=\t]({1.*1.08*cos(\t r)+0.*1.08*sin(\t r)},{0.*1.08*cos(\t r)+1.*1.08*sin(\t r)});
\draw [shift={(11.777582639398368,0.07748397161871035)}] plot[domain=2.5055339971068458:5.388442304813938,variable=\t]({1.*1.0937659894116378*cos(\t r)+0.*1.0937659894116378*sin(\t r)},{0.*1.0937659894116378*cos(\t r)+1.*1.0937659894116378*sin(\t r)});
\draw [shift={(13.08,-2.98)}] plot[domain=1.6989558719525741:4.584229435227012,variable=\t]({1.*1.0838203132844921*cos(\t r)+0.*1.0838203132844921*sin(\t r)},{0.*1.0838203132844921*cos(\t r)+1.*1.0838203132844921*sin(\t r)});
\draw [shift={(12.012315679064193,-5.793185602189)}] plot[domain=0.9732935797091103:3.858951710917995,variable=\t]({1.*1.0822030978308996*cos(\t r)+0.*1.0822030978308996*sin(\t r)},{0.*1.0822030978308996*cos(\t r)+1.*1.0822030978308996*sin(\t r)});
\draw [shift={(8.84,-7.22)}] plot[domain=0.12875020115550911:3.0128424524342843,variable=\t]({1.*1.0887878089327299*cos(\t r)+0.*1.0887878089327299*sin(\t r)},{0.*1.0887878089327299*cos(\t r)+1.*1.0887878089327299*sin(\t r)});
\draw [shift={(5.872302810795055,-6.008262437964231)}] plot[domain=-0.6475169536075116:2.2385146135730527,variable=\t]({1.*1.080632657093991*cos(\t r)+0.*1.080632657093991*sin(\t r)},{0.*1.080632657093991*cos(\t r)+1.*1.080632657093991*sin(\t r)});
\draw [shift={(4.6,-2.98)}] plot[domain=-1.4429713934112156:1.4429713934112156,variable=\t]({1.*1.0810060157878016*cos(\t r)+0.*1.0810060157878016*sin(\t r)},{0.*1.0810060157878016*cos(\t r)+1.*1.0810060157878016*sin(\t r)});
\draw [shift={(5.851978907959317,0.028210423743660673)}] plot[domain=-2.231811614124939:0.654281306043026,variable=\t]({1.*1.0803746407693342*cos(\t r)+0.*1.0803746407693342*sin(\t r)},{0.*1.0803746407693342*cos(\t r)+1.*1.0803746407693342*sin(\t r)});
\draw [->] (10.274667640098265,0.49553287460424844) -- (10.88,1.3);
\begin{scriptsize}
\draw [fill=black] (8.84,1.26) circle (1.5pt);
\draw [fill=black] (8.84,-7.22) circle (1.5pt);
\draw [fill=black] (13.08,-2.98) circle (1.5pt);
\draw [fill=black] (4.6,-2.98) circle (1.5pt);
\draw [fill=black] (5.851978907959317,0.028210423743660673) circle (1.5pt);
\draw [fill=black] (12.012315679064193,-5.793185602189) circle (1.5pt);
\draw [fill=black] (11.777582639398368,0.07748397161871035) circle (1.5pt);
\draw [fill=black] (5.872302810795055,-6.008262437964231) circle (1.5pt);
\end{scriptsize}
\end{tikzpicture}}
\caption{Propagation of null wafefronts emitted on the tick circle, inside a trapping surface. Both, ingoing and outgoing wavefronts, converge to the surface interior}\label{fig:wavefront}
\end{figure}

As the surface $\mathcal{S}$ is space-like, it lies outside the light cone of any point $p \in \mathcal{S}$. Thus the tangent space orthogonal to  $\mathcal{S}$ at the point $p$ $[T_p(\mathcal{S})^\bot]$ intersects the light cone of $p$ along two future oriented null curves orthogonal to $\mathcal{S}$ (see Fig. \ref{scone}).
One of those curves goes ingoing  $\mathcal{S}$, whereas the other goes outgoing $\mathcal{S}$. The behaviour of geodesics in such directions determines the surface called trapping surface, as defined below. 
Denoting the ingoing null direction by $k$ and the outgoing by  $l$, the surface expansions along them are give by $\theta^{(k)}$ and $\theta^{(l)}$ respectively. It allows one to define:

\begin{definition}
The surface $\mathcal{S}$ is said to be a trapping surface if, and only if, its expansion along null geodesics given by $\theta^{(k)}$ and $\theta^{(l)}$ obeys
\begin{align}
\theta^{(l)}<0 \mbox{ and } \theta^{(k)}<0.
\end{align}
If $\theta^{(l)}=0 \mbox{ and } \theta^{(k)}<0$ then $\mathcal{S}$ is said to be a marginally trapping surface.
\end{definition}
It is worth emphasizing that the sign of $\theta^{(V)}$ says rather the surface is expanding or contracting along the geodesics. The condition $\theta^{(l)}<0 \mbox{ and } \theta^{(k)}<0$ means that both outgoing and ingoing null geodesics converge toward the surface interior (see Fig. \ref{fig:wavefront}). It is exactly the kind of behaviour expected by the "naive" definition of black hole horizon. If the space-times is flat or the gravitational field is not strong enough, it holds  $\theta^{(l)}>0 \mbox{ and } \theta^{(k)}<0$.

\subsection{General Properties}

The existence of trapping surfaces implies that, under general conditions which we shall describe below, the space-time contains a black hole. Before enunciating this theorem we need some more results and definitions. The first one regards the weak energy condition. As the Einstein theory of gravity is very unrestricted concerning the accepted types of matter, some energy conditions were proposed in order to avoid unphysical solutions coming from exotic matter fields.

\begin{definition}[Weak energy condition]
For every time-like vector field $\vec{X}$, the matter density measured by observers in their respective frames is always non-negative  
 $\rho = T_{ab} \, X^a \, X^b \ge 0.$
\end{definition}

Other required result deals with singularities. Those odd predictions of general relativity were extensively studied along the 60s and some results were established \cite{Wald:1984rg}.
  
\begin{theorem}[Penrose]
Assuming the weak energy condition, if $\mathcal{S}$ is a trapping surface then the space-time contains a singularity.
\end{theorem}

In spite of being widely accepted the next result keeps the status of conjecture. 

\begin{conjecture}[Cosmic censorship]
Naked singularities can not be formed by gravitational collapses of initially non singular states in asymptotically plane space-times.
\end{conjecture}

Here naked singularity means singularity without horizons covering it. Now we can enunciate a result connecting singularities, black holes and trapping surfaces:

\begin{theorem}[Hawking \& Ellis]
Assuming the cosmic censorship conjecture, if $\mathcal{S}$ is a trapping surface then the space-time contains a black hole such that $\mathcal{S} \subset \mathcal{B}$.
\end{theorem}

\begin{figure}[!htb]
\centering
 \scalebox{.99}{\begin{tikzpicture}[line cap=round,line join=round,>=triangle 45,x=1.0cm,y=1.0cm]
\draw (-3.8,0.28) node[anchor=north west] {Asymptotically Flat};
\draw (4.,3.24) node[anchor=north west] {Schwarzschild};
\draw (4.,1.24) node[anchor=north west] {Reissner-Nordström};
\draw (4.,-0.76) node[anchor=north west] {Kerr};
\draw (4.,-2.76) node[anchor=north west] {Kerr-Newman};
\draw (0.,0.)-- (1.5,2.);
\draw (1.5,2.)-- (3.5,3.);
\draw (0.,0.)-- (1.5,-2.);
\draw (1.5,-2.)-- (3.5,-3.);
\draw (1.5,-2.)-- (3.5,-1.);
\draw (1.5,2.)-- (3.5,1.);
\begin{scriptsize}
\draw [fill=black] (0.,0.) circle (2.5pt);
\draw [fill=black] (1.5,2.) circle (2.5pt);
\draw[color=black] (0.96,2.43) node {Static};
\draw [fill=black] (1.5,-2.) circle (2.5pt);
\draw[color=black] (0.68,-2.21) node {Stationary};
\draw [fill=black] (3.5,3.) circle (2.5pt);
\draw[color=black] (3.36,2.43) node {Vacuum};
\draw [fill=black] (3.5,1.) circle (2.5pt);
\draw[color=black] (3.04,0.55) node {Electrovacuum};
\draw [fill=black] (3.5,-1.) circle (2.5pt);
\draw[color=black] (3.18,-1.61) node {Vacuum};
\draw [fill=black] (3.5,-3.) circle (2.5pt);
\draw[color=black] (3.02,-3.51) node {Electrovacuum};
\end{scriptsize}
\end{tikzpicture}}
\caption{Possible asymptotically flat black holes in GR.}\label{fig:afbh}
\end{figure}
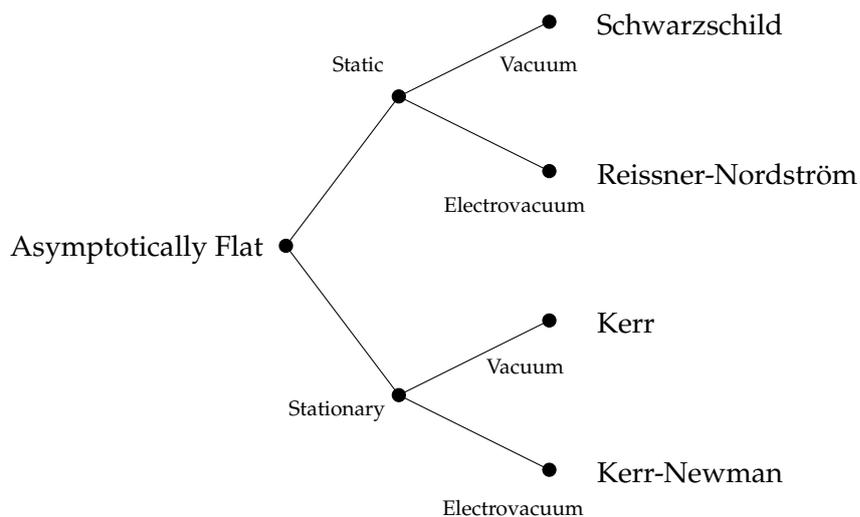

The next results we are going to enunciate touch upon black hole uniqueness. Those results highly restrict the possible types of black holes predicted by general relativity. In fact, in asymptotically flat space-times the only possible black hole solutions are the ones introduced in Sec. \ref{asyflat}. The first result states that the Schwarzschild solution is the only spherically symmetric solution of the vacuum field equations. This result extends to the Einstein-Maxwell equations: the only spherically symmetric solution to the  Einstein-Maxwell is the Reissner-Nordström metric. The next uniqueness result states that a stationary vacuum solution for a black hole in the
asymptotically flat space-time is characterized by two constants $M$ (mass) and $J$ (angular momentum) and coincides with the Kerr metric. Again the result extends to the Einstein-Maxwell equations and states that the only possible solution is the Kerr-Newman metric.  The possible asymptotically flat solutions are summarized in Figure \ref{fig:afbh}.

The above  black hole uniqueness results were followed by another very famous theorem for black holes in GR, known as no hair theorem:
\begin{theorem}[{No hair}]\label{nohair}
Stationary, asymptotically plane black hole solutions coupled  to  electromagnetic fields and non singular at the horizon are completely characterized by the parameters: mass $(M)$, 
electric charge $(Q)$ and angular momentum $(J)$.
\end{theorem}

Those remarkable uniqueness results of black holes in general relativity  are in general not preserved by theories beyond the Einstein's theory of gravity. In this work we are mainly interested in effective black hole solutions coming from gravity in higher dimensions. Some solutions are presented in the Chapter \ref{ched} and applied in the Chapters  \ref{cap5} and \ref{cap6}.

\section{Black Hole Mechanics and Thermodynamics}\label{BHT}

Black hole thermodynamics was a topic intensively discussed among the general relativity community all along the 70s and still is. It was motivated by an incompatibility between classical GR and thermodynamics, giving birth to one of the main widely accepted features of gravity beyond GR, the so called Hawking radiation. This section is devoted to introduce some consequences of the formal analogy between thermodynamics and classical black hole mechanics. It paves the way further discussions of BH physics beyond GR, particularly in the next chapter, where we deal with a powerful extension of the Hawking's original work.

Classically, black holes behave as perfect black bodies described by their mass $M$, charge $q$ and angular momentum $J$. Such scenario brings serious concerns related to the second law of thermodynamics, inasmuch as they do not have degrees of freedom to define entropy and their temperature is zero. In addition, if black holes do not have entropy, what would happen if one throws an entropic object into the black hole?   On the other hand, the laws that rules the mechanics of black holes are formally analogous to the laws of thermodynamics. Such analogy motivated, after the work of Bekenstein \cite{Bekenstein:1973ur} and Hawking \cite{Hawking:1974sw}, the field of research called black hole thermodynamics \cite{Wald:1984rg,Wald:1999vt}.

Now we shall introduce the laws of black hole mechanics. In order to enunciate the zeroth law one has to first define Killing horizons and surface gravity. A Killing horizon is a null surface defined by the region where the Killing vectors fields are null vectors. Killing vector fileds are solutions of the Killing field equation $\nabla_\mu \chi_\nu+\nabla_\nu \chi_\mu=0$, they are closely related to the isometries of the spacetime \cite{Wald:1984rg}. Any event horizon $\Sigma$ of asymptotically flat space-times is a Killing horizon for a given Killing field $\chi^\mu$ \cite{Wald:1984rg,Wald:1999vt}. Moreover,

\begin{itemize}
\item If the space-time is static, $\chi^\mu$ shall be a killing field $\chi^\mu=\xi^\mu$ representing the time translation symmetry;

\item If the space-time is stationary but not static, $\chi^\mu$ shall have axial symmetry, being represented by a Killing field related to the rotational symmetry $\psi^\mu$ and another one related to the time translation symmetry $\xi^\mu$. In this case the field $\chi^\mu$ is in general represented by
\begin{align}
\chi^\mu=\xi^\mu+\Omega_H\psi^\mu,
\end{align}
where $\Omega_H$ is the angular velocity at the event horizon.
\end{itemize}

Consider a Killing horizon $\mathcal{K}$, whose Killing normal field is denoted by $\chi^\mu$. 
There exist a function $\kappa$ on $\mathcal{K}$, called surface gravity of $\mathcal{K}$, defined by:
\begin{align}
\nabla^\mu(\chi^\alpha\chi_\alpha)=-2\kappa\chi^\mu
\end{align}
The surface gravity is a measure of the gravitational field strength on the event horizon. The zeroth law of black hole mechanics, due to Bardeen, Carter and Hawking is enunciated below \cite{Wald:1984rg,Wald:1999vt}:

\begin{proposition}[Zeroth law]
The surface gravity $\kappa$ is constant along the the horizon of stationary black holes.
\end{proposition}

The subsequent law follows from results enunciated in the previous sections. The first one follows from the no hair theorem:

\begin{proposition}[First law]
 For two stationary black holes differing only by small variations in
the parameters $M,J$ and $Q$, hold
\begin{align}
\delta M=\frac{1}{8\pi}\kappa \delta A+\Omega_H\delta J+\Phi \delta Q,
 \end{align} 
 where $A$ denotes the black hole area, $J$ the angular momentum and $\Phi$ is the electric potential at the horizon.
\end{proposition}

By assuming the cosmic censorship conjecture follows the second law of black hole mechanics:

\begin{proposition}[Second law]
The horizon area never decreases
\begin{align}
\delta A \geq 0
\end{align}
\end{proposition}

In 1969 Roger Penrose proposed a mechanism according to which would be possible to extract a limited amount of energy from the ergosphere of rotating black holes (see Fig. \ref{fig:ergosphere}). Such process reduces the black hole surface gravity. The final result became the third law of black hole mechanics\footnote{ See Ref. \cite{Wreszinski2009} for a precise discussion about the third law and the consequences of its relationship with thermodynamics.}

\begin{proposition}[Third law]
 It is impossible by any procedure to reduce the surface gravity $\kappa$ to zero in a  finite number of steps.
\end{proposition}

Collecting the above  laws one can see the intriguing analogy between them and the laws of thermodynamics, as shown in Table \ref{tabbhtd}.

\begin{table}[h]
\centering
\begin{tabular}{ccc}
\toprule
Law & Thermodynamics & Black hole mechanics \\ 
\toprule 
& & \\
$0^{\text{th}}$ &\begin{minipage}{0.4\linewidth} $T$ is constant throughout body
in thermal equilibrium \end{minipage}
&\begin{minipage}{0.4\linewidth} $\kappa$ is constant over the horizon
of stationary black hole \end{minipage}
\\ & & \\
$1^{\text{st}}$ & $\delta E=T\delta S+$ work terms & $\delta M=\frac{1}{8\pi}\kappa\delta A+\Omega_H\delta J$ \\ & & \\
$2^{\text{nd}}$ & $\delta S \geq 0$ & $\delta A \geq 0$ \\ & & \\
$3^{\text{rd}}$ & \begin{minipage}{0.35\linewidth}
Impossible to achieve $T=0$ by a physical
process
\end{minipage} & \begin{minipage}{0.35\linewidth}
Impossible to achieve $\kappa=0$ by a physical
process
\end{minipage} \\ 
\bottomrule
\end{tabular}
\caption{Correspondence between the laws of black hole mechanics and thermodynamics.}
\label{tabbhtd}  
\end{table}

 A main question at the time was: is this a mere formal analogy? Or there is something deeper behind? Besides $E$ and $M$ represent the same physical quantity, the black hole temperature, classically, is zero. Thus one can not straightforwardly establish a consistent connection between $T$ and $\kappa$. However, by using methods of semi-classical gravity Hawking has shown that black holes do radiate, producing a thermal bath surround it  whose temperature depends linearly of the surface gravity $\kappa$. It means that $T$ and $\kappa$, in fact, represent the same physical quantity.

Despite the Hawking's result, there is still an issue to deal with -- the black hole entropy.  Bekenstein came out with an attempt to make black hole physics compatible with the second law of thermodynamics by defining the generalized entropy $S'$, which is simply the regular entropy $S$ + the black hole entropy $S_{BH}$. His solution was arguing that the generalized entropy, not the regular one, should be taken into account in the second law \cite{Bekenstein:1973ur,Carlip:2008wv}.  Regarding all the physical constants, the generalized entropy is given by

\begin{eqnarray}
S'=S+S_{BH}=S+\frac{1}{4}k_B\frac{c^3A}{G\hbar}.
\end{eqnarray}
In natural and Planck units the black hole entropy reads
\begin{align}
S_{BH}=\frac{1}{4}\frac{A}{G}=m_p^2\frac{A}{4}=\frac{1}{4}\frac{A}{l_p^2}.
\end{align}
According to the generalized entropy, a decrease on the regular entropy (by falling objects into the black hole) should be followed by an increment of the black hole area. On the other hand, a decrease of the black hole area (by Hawking radiation emission) should be followed by an increment of the entropy outside the black hole. In any case should hold the generalized second law
\begin{align}
\delta S' \geq 0.
\end{align}

The pioneering results of Hawking and Bekenstein combined have given strong evidences that black holes do behave as thermodynamical objects. 
However, there still two question with no definitive answer in such interpretation. The first one regards the microscopic description of entropy. The second, the so called information paradox, deals with the possible non unitary evolution of radiating black holes as proposed by Hawking. Some proposes have been appeared in the last decades for both questions \cite{Carlip:2014pma, Harlow:2014yka} but there is no consensus for the final answer.

\subsection{Hawking Radiation}

Hawking radiation is a great achievement of semi-classical gravity. By semi-classical one means classical field theory from the gravitational side and quantum field theory from the matter side. The energy momentum-tensor is promoted to an operator and its expectation value  replaces the classical energy-momentum tensor. In this approach the field equations read
\begin{align}
G_{\mu\nu}=8\pi G \langle \,\hat{T}_{\mu\nu}\,\rangle.
\end{align}

From quantum field theory, the vacuum state is defined as a state without real propagating particles. Notwithstanding, due to quantum fluctuations it is populated by particles that are constantly created and annihilated, called virtual particles. In the absence of external fields the vacuum state is stable, in the sense that virtual particles do not exist enough time to become real. However, external fields can provide the necessary energy to turn virtual particles into real ones. This is the principle in which the Hawking radiation is based on. Here the gravitational field plays the role of the external field.  

Consider the particle creation in a static space-time. Particles propagating outside the horizon have positive energy \cite{Frolov2011}. Hence in a long living pair creation one of the particles must have negative energy and goes toward the black hole interior.  At the end the negative energy particle is absorbed by the black hole, which reduces its mass, whereas the positive energy particle escapes to infinity and is visible by  distant observers \cite{Hawking:1974sw}. That is the heuristic vision of the Hawking radiation, where the vacuum fluctuation happens at the horizon neighbourhood.

Let us sketch the Hawking temperature derivation for the emission massless spin-0 particles by Schwarszchild black holes. It is accomplished by using the Bogoliubov transformations \cite{Carlip:2014pma}, see Ref. \cite{Wald:1984rg} for a detailed derivation.
Given a quantum field $\hat{\phi}$, a solution basis for the field equations ${ f}_\omega$ and the creation [annihilation] operator $\hat{a}_\omega^\dagger$ [$\hat{a}_\omega$]. Then
%
\begin{align}
\hat{\phi}=\sum_\omega  \hat{a}_\omega f_\omega+ \hat{a}_\omega^\dagger f_\omega^*, \quad [\hat{a}_\omega, \hat{a}_{\omega'}^\dagger]=\delta_{\omega, \omega'}, \quad \hat{a}_\omega|0\rangle_a=0, \quad \hat{N}_\omega^a=\hat{a}_\omega^\dagger\hat{a}_\omega,
\end{align}
where $|0\rangle_a$ and $\hat{N}_\omega^a$ denote the vacuum state and the number operator written on the basis $f_\omega$. 
The same might be done by adopting another bases ${ h}_\omega$
\begin{align}
\hat{\phi}=\sum_\omega  \hat{b}_\omega h_\omega+ \hat{b}_\omega^\dagger h_\omega^*, \quad [\hat{b}_\omega, \hat{b}_{\omega'}^\dagger]=\delta_{\omega, \omega'}, \quad \hat{b}_\omega|0\rangle_b=0, \quad \hat{N}_\omega^b=\hat{b}_\omega^\dagger\hat{b}_\omega.
\end{align}
The basis are interchangeable by applying the Bogoliubov transformations:
\begin{align}
h_\omega= \sum_{\omega'}\alpha_{\omega \omega'}f_{\omega'}+\beta_{\omega \omega'}f_{\omega'}^*, \quad \hat{b}_\omega= \sum_{\omega'}\alpha_{\omega \omega'^*}\hat{a}_{\omega'}+\beta_{\omega \omega'}^*\hat{a}_{\omega'}^\dagger
\end{align}
whose components, to preserve unitarity, are constrained by
\begin{align}
\sum_{\omega'}\left(|\alpha_{\omega \omega'}|^2-|\beta_{\omega \omega'}|^2\right)=1.
\end{align}
Expressing the number operator on the basis ${ h}_\omega$ and calculating its vacuum expectation value on the basis ${ f}_\omega$ one finds
\begin{align}
_a\langle 0| \hat{N}_\omega^b|0\rangle_a=\, _a\langle 0|\hat{b}_\omega^\dagger \hat{b}_\omega|0\rangle_a=\sum_{\omega'}|\beta_{\omega \omega'}|^2.
\end{align}
It means that the vacuum state is frame dependent. 

By considering a star collapsing into the Schwarzschild black hole, Hawking  computed  the  Bogoliubov
coefficients of an initial vacuum state outside the star and a final vacuum state after the black hole formation. He showed that after the process, the Bogoliubov transformation connecting observers should obey 
\begin{align}
\sum_{\omega'}|\beta_{\omega \omega'}|^2=\frac{1}{e^{8\pi M \omega}-1}.
\end{align}
Comparing the above result to the Plack distribution follows the Hawking temperature
\begin{align}
T_H=\frac{1}{8\pi M }=\frac{\kappa}{2\pi}.
\end{align}

As we have seen, Hawking radiation plays an essential hole on black hole thermodynamics, however it is not restricted to this picture. Hawking radiation determines, for example, the lifetime of primordial black
holes and allows one to calculate the range of such black holes expected to be survived up to the present time \cite{Carr:2009jm}. In the next chapter we shall introduce a method for calculating the Hawking temperature which agrees with the heuristic description discussed above \cite{Kraus:1994fh,Parikh:1999mf,Vanzo:2011wq,Srinivasan:1998ty,Kerner:2008}. There, we are going to describe the method in details and apply it to spin-0 and spin-$1/2$ particles.

\section{Beyond General Relativity}\label{grqm}

The first thing we must do in this section is clarify what we mean by "beyond general relativity". There are  numerous ways to push forward the edges GR and go beyond that. For example, there are many theories of modified gravity, as $f(R)$, scalar-tensor, Brans-Dicke, Gauss-Bonnet, Lovelock, among others. See Ref. \cite{Clifton:2011jh} for a recent review. Some of them were proposed in order to fix observational issues not properly explained by GR, as dark matter and dark energy. As they are extensions of GR one could argue that all of them are theories beyond GR. However, in this work we do not deal with none those GR extensions. Actually, by beyond GR we mainly mean effective extra-dimensional models and microscopic black holes, both not incorporated by standard GR. What we mean by effective extra-dimensional models and microscopic black holes will become clear soon.

Physics laws depend on the scale at which the phenomenon happens. Quantum mechanics governs the physics of small lengths. Classical gravity, on the other hand, governs the physics of massive objects. However, when a large amount of mass is compressed in a very small region both theories have to be taken into account. Such situation should be described by an unknown theory generically called quantum gravity. Serious issues appear when one try to apply the standard quantum field theory tools to quantize gravity (see Ref. \cite{Wald:1984rg} for a general discussion on the issues and proposals for quantizing gravity). It makes finding the theory of quantum gravity the main unsolved problem of current theoretical physics. There are two main candidates to occupy the place of such theory, namely M-theory/string theory and loop quantum gravity. Even though both have been developed for decades there is no complete theory of quantum gravity up to this date. Moreover, the predictions of both theories are far from current experimental capacity. The lack of experimental quantum gravity signal is a major obstacle for further developments and turns the situation even more complicated.

Facing the foregoing situation some alternative approaches have emerged, employing effective theories to describe quantum gravity effects and looking for possible experimental signatures. Those effective theories and/or models deviate to GR by implementing  features predicted by quantum gravity proponents. Effective theories do not intend to be the final theory, but could enlighten the route and give some hints on what expect from the full theory.  Furthermore, the predictions arise without employing the heavy machinery of string theory and loop quantum gravity. Effective extra dimensional models fit in the above description. We shall introduce models of extra dimensions in the Chapter \ref{ched}.

The borders separating gravity, quantum mechanics and quantum gravity are somewhat fuzzy, nevertheless it can be enlightened by some considerations regarding mass and length scales. One of the basic distinguished characteristics arose in quantum mechanics is the uncertainty principle $\Delta x > \hslash/\Delta p$. Since the momentum of particles of mass $M$ is bounded by $Mc$, one
cannot localize a particle of mass $M$ in scales smaller then $\hslash/Mc$, which is the Compton wavelength $\lambda_M$ of the particle.  $R < \lambda_M$ might be regarded as the quantum mechanics domain, in the sense that the classical physics breaks
down  there. A general assumption of GR is that if massive spherically symmetric objects are smaller than its Schwarzschild radius, then it will form a black hole (see Chapter \ref{cap6}). Hence the Schwarzschild radius $r_\mais=2GM/c^2$ defines the region under influence of general relativity. There are no stable classical configurations here. The Compton and Schwarzschild boundaries intersect at the Planck scales,
\begin{align}
r_p =\sqrt{\frac{\hslash G}{c^3}}\sim 10^{-33}cm \quad \text{and} \quad m_p=\sqrt{\frac{\hslash c}{G}} \sim 10^{-5}g. 
\end{align}
Quantum gravity effects are expect to be important below the Planck scale as well as for densities higher than the Planck density 
\begin{align}
\frac{M}{R^3} > \frac{m_p}{r_p^3},
\end{align}
which comes from curvature singularities associated to the big bang or the center of black holes \cite{Proceedings:2016xkf}, resulting in
\begin{align}
R<\left(\frac{M}{m_P}\right)^{1/3}r_p.
\end{align}
\begin{figure}[!h]
\centering
 \scalebox{.86}{\begin{tikzpicture}[line cap=round,line join=round,>=triangle 45,x=1.0cm,y=1.0cm]
\fill[fill=black,fill opacity=0.15] (10.9,0.12) -- (3.8,-2.7) -- (-1.5,-2.7) -- (-1.5,-4.02) -- (10.9,-4.04) -- cycle;
\draw [->] (-1.5,-4.02) -- (-1.5,3.68);
\draw [->] (-1.5,-4.02) -- (10.9,-4.04);
\draw (-1.5,-2.7)-- (3.8,-2.7);
\draw (3.8,-2.7)-- (-0.58,2.3);
\draw (3.8,-2.7)-- (9.,2.3);
\draw (3.8,-2.7)-- (10.9,0.12);
\draw [dotted] (3.8,-2.7)-- (10.9,-2.7);
\draw [dotted] (3.8,-2.7)-- (3.8,3.18);
\draw (-1.16,-1.1) node[anchor=north west] {quantum mechanics };
\draw (7.84,-1.92) node[anchor=north west] {quantum gravity};
\draw (7.46,0.5) node[anchor=north west] {general relativity};
\draw (2.46,1.88) node[anchor=north west] {classical domain};
\draw (5.18,-3.18) node[anchor=north west] {Planck scale};
\draw [->] (5.02,-3.28) -- (3.8,-2.7);
\draw (0.22,3.9) node[anchor=north west] {Compton wavelength};
\draw (6.76,3.94) node[anchor=north west] {horizon radius};
\draw [->] (1.72,3.44) -- (0.12913643729461066,1.4904835190700787);
\draw [->] (7.92,3.42) -- (8.102690238278248,1.4372021521906229);
\draw (3.7,-4.2) node[anchor=north west] {-5};
\draw (8.64,-4.) node[anchor=north west] {$\log [M(\text{g})]$};
\draw (-2.42,4.46) node[anchor=north west] {$\log [R(\text{cm})]$};
\draw (-2.18,-2.5) node[anchor=north west] {-33};
\end{tikzpicture}}
\caption{$(M,R)$ diagram showing different physics domains. In logo scale, the mass correspondig to -5 ($10^{-5}$ g) is the Planck mass and length equals to -33 ($10^{-33}$ cm) is the Planck length.}\label{fig:scales}
\end{figure}
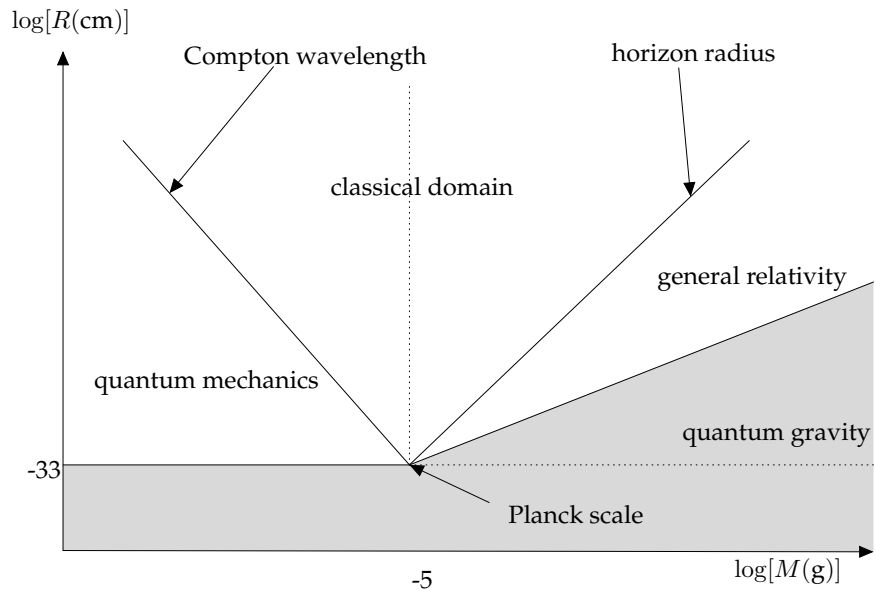

According to the above boundaries, the regions under domain of the different theories can be qualitatively represented in the $(M,R)$ diagram of Fig. \ref{fig:scales}. One exiting expectation of effective extra dimensions is that in those models the clash of quantum mechanics and gravity occur in scales closer to the current experimental capacity. It could give signals of quantum gravity in the
TeV region by producing small black holes in high energy collisions of
particles.  We shall discuss such possibility in the subsequent section and how extra dimensional models might lower the fundamental scale is discussed in Chapter \ref{ched}.

\subsection{Why Black Holes?}

Black holes are interesting by themselves because of their many unexpected features. However, those objects are even more interesting when one searches for hints on what happens in the previously described scenario, namely when gravity meets quantum mechanics. Black holes have properties that one expects to be satisfactorily explained only by a full quantum gravity theory, as singularities and microscopic entropy. Hence they are excellent candidates to test the limits of GR and alternative theories. In the Chapter \ref{cap5} we use the deflection of light by strong gravitational fields to compare observable parameters of black holes beyond GR to the ones for the Schwarzschild solution. Unfortunately, the current precision of astronomical observations is not good enough to rule out alternative theories yet, but it might be possible soon. Another possibility, as anticipated in the previous section, is the black hole formation in particle colliders. Such collisions would form  black holes of two types, semi-classical or quantum gravity black holes. In the former case they would evaporate via Hawking radiation of various particles until it vanish or preserve remnants with the mass of order of the Planck scale. The terminal stage of BH
evaporation is not known. Since black holes at colliders might be tiny and the Hawking radiation is inversely proportional to the black hole mass, the temperature should be  very hot and
therefore the evaporation would happen very fast. 



\chapter{Elko Dark Spinors Emission Through Hawking Radiation} 

\label{cap1} 


In spite of its relevance on semi-classical gravity and beyond, it took almost 20 years to have an approach for calculating the Hawking radiation, which encapsulates the original heuristic view proposed by Hawking himself \cite{Hawking:1974sw}. As discussed in the previous chapter, the Hawking radiation regards vacuum fluctuations at the horizon neighbourhood, whose virtual particles eventually become real ones.  This approach is  called tunnelling method \cite{Kraus:1994fh}. The latter name comes from the description of the phenomenon by considering classically forbidden trajectories. The method has been recently calling the attention,  due to its computational simplicity and versatility, being applied to different types of black holes, including non asymptotically flat, extra dimensional and string theory inspired solutions \cite{Vanzo:2011wq}. The tunnelling method was extended to encompass fermions emission, dynamical black holes and placed at solid foundation by Kerner \cite{Kerner:2008} and Acquaviva \cite{Acquaviva:2012}, respectively.

In this chapter, we are going to apply the tunnelling method  to calculate the Hawking radiation associated to a fermion beyond the standard model, emitted by an exact classical solution in the low energy effective field theory describing heterotic string theory: the Kerr-Sen dilaton-axion black holes \cite{Sen:1992ua}. The solution presents charge, magnetic dipole moment, and angular momentum,    involving the antisymmetric tensor field coupled to the Chern-Simons 3-form.  A myriad  of black hole solutions beyond general relativity, including the Kerr-Sen dilaton-axion itself, has been considered for tunnelling methods of fermions and bosons, as rotating and accelerating black holes, topological, BTZ, Reissner-Nordstr\"om, Kerr-Newman, and Taub-NUT-AdS black holes, including also the tunnelling of higher spin fermions  as well \cite{Yale:2008kx}.
 We shall study a similar method for a spin-1/2 fermion of mass dimension one. 
 This fermion %
is described by Elko dark spinors, or simply Elko, which is an acronym for the German expression \textit{Eigenspinoren des LadungsKonjugationsOperators}, meaning eigenspinor of the charge conjugator operator. It was proposed as a candidate to describe dark matter \cite{Ahluwalia:2004ab}.

This chapter is presented as follows: the Hamilton-Jacobi method -- which is the current main approach for the tunneling method -- will be introduced in the next section. We start by discussing the method for bosons and how to extend it for fermions. The Hamilton-Jacobi method applies the WKB approximation to compute  the tunnelling rate and its resulting tunnelling probability. In the subsequent section, we briefly revisit the Kerr-Sen dilaton-axion black hole. 
 It will be followed by a section regarding the dark spinors  framework. We thus shall  
calculate in Section \ref{ketunn} the probabilities of emission and absorption 
of Elko dark particles  across these black holes.  Finally, the associated Hawking temperature shall be obtained, corroborating the universal character of the Hawking effect and further extending it to particles beyond the standard model. We published the results of the last section in the reference \cite{Cavalcanti:2015nna}. 

\section{The Hamilton-Jacobi Method}

Black hole tunnelling procedures have been placed as prominent   methods to calculate the  Hawking temperature  \cite{Kraus:1994by,Parikh:1999mf,Srinivasan:1998ty,Angheben:2005rm,Arzano:2005rs,Jiang:2005ba,Li:2008zra,Vanzo:2011wq}.
Various types of black holes have been investigated in the context of tunnelling of fermions and bosons as well \cite{Kraus:1994by,Vanzo:2011wq,Parikh:1999mf,Kerner:2007rr}. Tunnelling procedures rely on classically forbidden paths that particles go through, from the inside to the outside of the horizon. The first approach for the tunnelling method is known as the null geodesic method. Besides its great achievements \cite{Kraus:1994by,Parikh:1999mf}, there were some issues to be adressed \cite{Vanzo:2011wq}. For example, the fact that it strongly relies on a very specific choice of coordinates.
Complementarily to the first results, the Hamilton-Jacobi method was employed  \cite{Srinivasan:1998ty} and  further generalized, by applying the WKB approximation to the Dirac equation \cite{Kerner:2007rr} and by providing Hawking radiation due to dark spinors for black strings \cite{daRocha:2014dla}. 
Mass dimension 3/2 fermions tunnelling  has been studied in the charged dilatonic black hole, rotating Einstein-Maxwell-Dilaton-Axion black hole and rotating Kaluza-Klein black hole likewise \cite{Chen:2008vi}. Ref. \cite{Vanzo:2011wq} is a clear and comprehensive review of the field.

 The Hamilton-Jacobi is based upon the particle description of Hawking radiation, under the assumption that the emitted particle action does satisfy the relativistic Hamilton–Jacobi equation. This method applies to any well-behaved coordinate system across the horizon. As mentioned, the tunnelling method relies on allowing particles to travel
along classically forbidden trajectories, from just behind the horizon onward to
infinity. As an indication of the classically forbidden trajectory, the classical action becomes complex. The imaginary part encloses the emission rate of the black hole. Inasmuch as the forbidden motion happens at the horizon crossing, the contribution to the rate emission should comes only from an infinitesimal part of the whole trajectory, nearby the horizon. It suggests the natural choice of performing near horizon approximations. Finally, divergences appearing on the imaginary action are regularized according to Feynman’s $i\epsilon-$prescription.

Since a black hole has a well-defined
temperature, in principle it should radiate not only spin-0 particles. Besides being a plausible argument, the result is not trivial. It is achieved by an extension of the Hamilton-Jacobi method in Kerner's seminal work \cite{Kerner:2007rr,Kerner:2008}.  When   spin-1/2 fermions are taken into account,  the basic assumption is that the effect of the spin of each type of fermion cancels out, due to particles emission with the spin in opposite directions. Hence, the change in the black hole angular momentum is completely negligible in tunnelling processes.

\subsection{Bosonic Emission}

The Hamilton-Jacobi method for calculation the Hawking temperature of spin-0 particles can be summarized in a sequence of steps, as following \cite{Vanzo:2011wq}:

\begin{enumerate}
\item Assume that at the region of interest (nearby the horizon), the particle dynamics is described by the relativistic Hamilton-Jacobi equation,
\begin{align}
g^{\mu \nu}\partial_\mu I \partial_\nu I+m^2=0
\end{align}
 where $m^2$ is the invariant mass and $I$ is the classical action. Alternatively, one can   apply the WKB approximation to the Klein-Gordon equation. The leading order gives  the above equation. In fact, by applying the WKB approximation, the spin-0 field can be expressed as $\phi=\exp[\frac{i}{\hslash}I+O(\hslash^0)]$. Substituting it in the field equation  $g^{\mu \nu}\partial_\mu \partial_\nu \phi-\frac{m^2}{\hslash^2}\phi=0$ results $$-\left[g^{\mu \nu}\partial_\mu I \partial_\nu I+m^2\right]+O(\hslash)=0;$$

\item Reconstruct the action $I$ by using the problem symmetries and its partial derivatives $\partial_\mu I$ by means of
\begin{align}\label{acttunn}
I=\int \partial_\mu I dx^\mu.
\end{align}
The integration is taken along an oriented null path with at least one point on the horizon;

\item Take only the piece of the trajectory around the horizon crossing, where the action becomes complex;

\item Perform a near horizon approximation on Eq. \eqref{acttunn} and regularize the divergences by using the Feynman's $i\epsilon-$prescription;

\item At the end, the imaginary part of the action reads

\begin{align}
\mathtt{Im} I=\frac{\pi \omega}{\kappa},
\end{align}
where $\omega$ is the particle energy and $\kappa$ is the black hole surface gravity;

\item The WKB approximation predicts the tunnelling probability rate $\Gamma$ as
\begin{align}\label{probemabs}
\Gamma=\frac{\Gamma_{{emission}}}{\Gamma_{{absorption}}} \sim e^{-4 \mathtt{Im}I}.
\end{align}
On the other hand, from statistical mechanics, the probability of finding (tunnelling rate), chosen at random from the ensemble, a particle of energy $\omega$ should be equals to the Boltzmann factor
 $e^{-\omega/T}$,  following 
 \begin{align}
T =\frac{\kappa}{2\pi}.
 \end{align}
\end{enumerate}

The Hamilton-Jacobi method, comparing to the standard derivations of the Hawking equation, is fairly simple and straightforward. As an  illustrative example, let us calculate the well known result of the Schwarzschild Hawking temperature. We mentioned that the space-time metric must be regular on the horizon, thus the Schwarzschild metric is not suitable for the bosonic tunneling method. For that reason, we are going to use the Painlevé-Gullstrand metric, briefly introduced in Sec. \ref{asyflat}.
 The Painlevé-Gullstrand coordinates are derived from the Schwarzschild coordinates by means of a new coordinate $t_p$ defined as
\begin{align}
t= t_p-h(r)
\end{align}
where $h(r)$ is a function to be determined. The above transformation results
\begin{align}
dt^2= dt_p^2-2h'(r)dt_pdr+h'(r)^2dr^2.
\end{align}
Written in the new coordinates the Schwarzschild metric is given by
\begin{align}\label{pp}\nonumber
ds^2=&-\left(1-\frac{r_\mais}{r} \right)dt_p^2+2h'(r)\left(1-\frac{r_\mais}{r} \right)dt_pdr+\left[\frac{1}{1-\frac{r_\mais}{r}}-\left(1-\frac{r_\mais}{r} \right)h'(r)^2\right]dr^2+\\
&+r^2\left(d\theta^2+\sin^2\theta d\phi^2 \right),
\end{align}
where $r_\mais$ denotes the horizon radius. Now the function $h'(r)$ is chosen in order to have $g_{rr}=1$,  giving 
\begin{align}\label{fp}
h'(r)=\frac{\sqrt{r_\mais r}}{r-r_\mais},
\end{align}
and
\begin{align}
h(r)=\int \frac{\sqrt{r_\mais r}}{r-r_\mais}dr=2\sqrt{r_\mais r}-r_\mais\ln\left(\frac{\sqrt{r}+\sqrt{r_\mais}}{\sqrt{r}-\sqrt{r_\mais}}\right). 
\end{align}
Hence, by substituting Eq. \eqref{fp} into Eq. \eqref{pp}, the Schwarzschild solution in Painlevé-Gullstrand coordinates yields
%
\begin{align}\label{painleve}
ds^2=-\left(1-\frac{r_\mais}{r} \right)dt_p^2+2\sqrt{\frac{r_\mais}{r}}dt_pdr+dr^2+r^2\left(d\theta^2+\sin\theta d\phi^2 \right).
\end{align}

The Hamilton-Jacobi equation, adopting the Painlevé-Gullstrand metric, reads
\begin{align}
-\left(\partial_{t_p}I\right)^2+2\sqrt{\frac{r_\mais}{r}}\partial_{t_p}I\partial_rI +\left(1-\frac{r_\mais}{r} \right)\left(\partial_rI\right)^2+m^2=0.
\end{align}
Being time independent, the space-time has a Killing vector associated to the particle energy by  $\omega=-\partial_{t_p}I$, hence
\begin{align}
-\omega^2-2\sqrt{\frac{r_\mais}{r}}\omega \partial_rI +\left(1-\frac{r_\mais}{r} \right)\left(\partial_rI\right)^2+m^2=0.
\end{align}
Consequently,
\begin{align}
\partial_rI=\frac{\sqrt{r_\mais r}}{r-r_\mais}\omega\pm\frac{r}{r-r_\mais}\sqrt{\omega^2-m^2\left(1-\frac{r_\mais}{r}\right)},
\end{align}
where the sign $-[+]$ represents the particle going inward [outward] the horizon.
On the other hand, constrained to null radial motion, from Eq. \eqref{painleve} one has
\begin{align}
0&=-\left(1-\frac{r_\mais}{r} \right)d t_p^2+2\sqrt{\frac{r_\mais}{r}}d t_pd r+d r^2.
\end{align}
Close to the horizon, by means of using $r \sim r_\mais$, one finds
\begin{align}\label{drdt}
d t_p=-\frac{1}{2}d r.
\end{align}
 Taking that into account, the reconstruction of the imaginary part of the action follows,
\begin{align}
\mathtt{Im}I_{\pm}&=\mathtt{Im}\int \partial_\mu I dx^\mu\\
&=\mathtt{Im}\int \partial_r I dr+\partial_{t_p}Idt_p\\ \label{usingdrdt}
&=\mathtt{Im}\int dr\left( \partial_r I +\frac{\omega}{2}\right)\\
&=\mathtt{Im}\int dr \partial_r I \\
&=\pm \mathtt{Im}\left[\int dr\frac{r}{r-r_\mais}\sqrt{\omega^2-m^2\left(1-\frac{r_\mais}{r}\right)} \right].
\end{align}
By using the horizon approximation $r \sim r_\mais$ and regularizing the divergence according to the Feynman prescription one finds,
\begin{align}
\mathtt{Im}I_{\pm}&=\pm \mathtt{Im}\int \frac{\omega r}{r-r_\mais-i\epsilon}\\
&=\pm \pi r_\mais \omega\\
&=\pm 2\pi M \omega.
\end{align}
The resulting tunnelling rate is
\begin{align}
\Gamma=\frac{\Gamma_{+}}{\Gamma_{-}}=\frac{e^{-2\mathtt{Im}I_+}}{e^{-2\mathtt{Im}I_-}}=e^{-8\pi M \omega}
\end{align}
and the Boltzmann factor finally gives the Hawking temperature
\begin{align}\label{tempbos}
T_H=\frac{1}{8\pi M }=\frac{\kappa}{2\pi}.
\end{align}

\subsection{Fermionic Emission}\label{femem}

Despite the remarkable results of the Hamilton-Jacobi method for spin-0 particles,  in this chapter, we are interested on spin-$1/2$ fermions emission. Those  particles are expected to be emitted as Hawking radiation due to the fact, as mentioned earlier, that black holes are surrounded by a thermal bath of finite temperature, from where all sort of particles could emerge  \cite{Page:1976df,Page:1976ki,Kerner:2007rr}. 
One key point of the Hamilton-Jacobi method is the use of the classical action. For fermions, the Dirac equation for curved space-times replaces the Hamilton-Jacobi equation, which is the natural choice given that the Dirac equation is the proper dynamical equation for fermions. In addition, it should be noticed that the action for spinning particles  $I_f$ has the form \cite{Barducci:1976qu,Vanzo:2011wq}
\begin{align}
I_f=I+\mbox{ (spin corrections), }
\end{align}
where $I$ denotes the kinetic term, as the classical action of the previous section. The spin corrections come from the coupling of the spin with the manifold spin connection. Given that the expectation of finding particles with a determined spin or its opposite is statistically equivalent, the emission of such particles should not substantially change  the black hole angular momentum. That is the starting point of the approach proposed by Kerner \cite{Kerner:2008}. The next step is to apply the WKB approximation and find the imaginary part of the classical action, which again encodes the Hawking temperature.

The procedure for the fermionic case can be summarized as follows:

\begin{enumerate}
\item Use the metric components to find the curved space-time version of the Dirac equation, at first keeping $\hslash$;

\item Choose a spin up or spin down spinor and apply the WKB approximation. The spin up, for example, results

\begin{align}
\Psi_{\uparrow}(t,r,\theta,\phi)=\begin{pmatrix}
A(t,r,\theta,\phi)\\
0\\
B(t,r,\theta,\phi)\\
0
\end{pmatrix}\exp\left[\frac{i}{\hslash}I_{\uparrow}(t,r,\theta,\phi)\right];
\end{align}

\item Substitute the above spinor in the Dirac equation and find the imaginary, radial part;

\item Analogously to the spin-0 case, the imaginary part encodes the tunnelling probability which, by setting it equals to the Boltzmann factor gives the temperature. 
\end{enumerate}

Once again, as a matter of clarification, we are going to calculate the Hawking temperature associated to Dirac fermions. Now adopting a general spherically symmetric space-time, using the Painlevé-Gullstrand coordinates. A general spherically symmetric metric is expressed as
\begin{align}\label{gsphym}
dr^2=-f(r)dt^2+\frac{1}{g(r)}dr^2+r^2d\Omega^2,
\end{align}
where $d\Omega^2=d\theta^2+\sin^2\theta d\phi^2$. Analogously to the Schwarzschild case, the Painlevé-Gullstrand metric for such space-time is obtained defining the coordinate $t_p$ as $t=t_p-h(r)$. It gives

\begin{align}\label{ppf}
ds^2=&-f(r)dt_p^2+2f(r)h'(r)dt_pdr+\left[\frac{1}{g(r)}-f(r)h'(r)^2\right]dr^2+
r^2d\Omega^2.
\end{align}
Again the function $h'(r)$ is chosen in order to have $g_{rr}=1$, following 
\begin{align}\label{fpf}
h'(r)=\sqrt{\frac{1-g(r)}{f(r)g(r)}}.
\end{align}
Hence, the metric \eqref{gsphym} in Painlevé-Gullstrand coordinates reads
%
\begin{align}\label{ppff}
ds^2=&-f(r)dt_p^2+2\sqrt{\frac{f(r)[1-g(r)]}{g(r)}}dt_pdr+dr^2+
r^2d\Omega^2.
\end{align}
That is the metric we are going to use in this section.

Going to the first step of the fermionic derivation of Hawking temperature, the curved space-time version of the Dirac equation is given by
\begin{align}\label{dicurv}
\left(\;i{\hslash}\gamma^a \textbf{e}^{\mu}_{\;\;a} D_\mu+{m}\;\right)\Psi_{\uparrow}(t,r,\theta,\phi)=0,
\end{align}
where
\begin{align}
D_\mu=\partial _{\mu }+\frac{i}{2}{}{} \Gamma _{\mu }^{\alpha \beta }\Sigma_{\alpha\beta},
\end{align}
 with $\Sigma_{\alpha\beta}=\frac{i}{4}\left[ \gamma
^{\alpha },\gamma ^{\beta }\right]$. The matrices $\gamma ^{\sigma}$ are the usual Clifford algebra generators for the Minkowski space-time and $\textbf{e}^{\mu}_{\;\;a}$ are the  vierbein fields
\begin{align}
g^{\mu\nu}=\textbf{e}^{\mu}_{\;\;a}\textbf{e}^{\nu}_{\;\;b}\eta^{ab}.
\end{align}
 Here, lowercase latin indexes denote the vierbein flat space index, whereas Greek indexes are the curved space-time indexes. To avoid confusion, when we make explicitly use of the  vierbein components, we are going label them as the space-time coordinates $(t,r,\theta,\phi)$ for curved space-time and numbers for the flat one.

Before proceeding, notice that by applying the operator ${\hslash}D_\mu$ to $\Psi_\uparrow$ most of the resulting terms are higher order in $\hslash$. In fact,

\begin{align}
{\hslash}D_\mu\Psi_{\uparrow}(t,r,\theta,\phi)=&\,{\hslash}\begin{pmatrix}
\partial_\mu A\\
0\\
\partial_\mu B\\
0
\end{pmatrix}
e^{\frac{i}{\hslash}I_{\uparrow}}+{i}\partial_\mu I_\uparrow \Psi_{\uparrow}-\frac{\hslash}{8}\Gamma^{\alpha\beta}_\mu\left[\gamma_\alpha,\gamma_\beta\right]\Psi_{\uparrow}\\ \label{leadingdirac}
=&\,{i}\partial_\mu I_\uparrow \Psi_{\uparrow}+\mathcal{O}(\hslash).
\end{align}
Accordingly, we have to consider only the action derivative term. Now we can setup $\hslash=1$. 

The vierbein fields of the general spherically symmetric Painlevé-Gullstrand metric, required to find $\gamma^a\textbf{e}^{\mu}_{\;\;a}$, are given by
\begin{align}
\textbf{e}^{t_p}_{\;\;0}=&\frac{1}{\sqrt{f(r)}}\,,\qquad\textbf{e}^{t_p}_{\;\;1}=\frac{1}{\sqrt{f(r)[1-g(r)]}}\,,\\
\textbf{e}^r_{\;\;1}=&\frac{1}{\sqrt{g(r)}}\,,\qquad
\textbf{e}^{\theta}_{\;\;2}=\frac{1}{r}\;,\qquad
\textbf{e}^{\phi}_{\;\;3}=\frac{1}{r\sin \theta}\,.
\end{align}
Hence, the $\gamma^\sigma$ matrices are chosen accordingly
\begin{align}\label{v1}
\textbf{e}^{t_p}_{\;\;0}\gamma^0+\textbf{e}^{t_p}_{\;\;1}\gamma^1
=&\frac{1}{\sqrt{f(r)}}\left(\begin{array}{cc}
0 & \mathbb{I}_2 \\ 
-\mathbb{I}_2&0 
\end{array}\right)+\frac{1}{\sqrt{f(r)[1-g(r)]}}\left(\begin{array}{cc}
0 & \sigma^3 \\ 
\sigma^3 & 0
\end{array}\right)\,,\\  \label{v2}
\textbf{e}^r_{\;\;1}\gamma^1=&
\frac{1}{\sqrt{g(r)}}\left(\begin{array}{cc}
0 & \sigma^3 \\ 
\sigma^3 & 0
\end{array}\right)\,,\quad
\textbf{e}^\theta_{\;\;2}\gamma^2=
\frac{1}{r}\left(\begin{array}{cc}
0 & \sigma^1 \\ 
\sigma^1 & 0
\end{array}\right)\,, \\ \label{v3}
\textbf{e}^\phi_{\;\;3}\gamma^3=&
\frac{1}{r\sin\theta}\left(\begin{array}{cc}
0 & \sigma^2 \\ 
-\sigma^2 & 0
\end{array}\right)\,.
\end{align}
Substituting Eqs. \eqref{v1}, \eqref{v2} and \eqref{v3} into Eq. \eqref{dicurv} one finds -- by taking into account the leading order of Eq. \eqref{leadingdirac} -- the system of equations

\begin{align}
-B\left( \frac{1}{\sqrt{f(r)}}(1+\sqrt{1-g(r)})\partial_t I_\uparrow+\sqrt{g(r)}\partial_r{I}_\uparrow\right) +mA&=0,\\
A\left(\frac{1}{\sqrt{f(r)}}(1-\sqrt{1-g(r)})\partial_t I_\uparrow- \sqrt{g(r)}\partial_r{I}_\uparrow\right) +mB&=0,\\ 
-\frac{B}{r}\left( \partial_\theta I_\uparrow+\frac{i}{\sin\theta}\partial_\varphi I_\uparrow\right) &=0,\\ 
-\frac{A}{r}\left( \partial_\theta I_\uparrow+\frac{i}{\sin\theta}\partial_\varphi I_\uparrow\right) &=0.
\end{align}
The spherical symmetry of the space-time motivates the ansatz 
\begin{align}\label{ansatz1}
I_\uparrow=-\omega t+W(r)+J(\theta,\phi).
\end{align}
Replacing it in the above system results
\begin{align}
B\left( \frac{\omega}{\sqrt{f(r)}}(1+\sqrt{1-g(r)})- \sqrt{g(r)}W'(r)\right) +mA&=0,\\
-A\left( \frac{\omega}{\sqrt{f(r)}}(1-\sqrt{1-g(r)})+ \sqrt{g(r)}W'(r)\right) +mB&=0, \\ \label{j1}
-\frac{B}{r}\left( \partial_\theta J(\theta,\phi)+\frac{i}{\sin\theta}\partial_\varphi J(\theta,\phi)\right) &=0,\\ \label{j2}
-\frac{A}{r}\left( \partial_\theta J(\theta,\phi)+\frac{i}{\sin\theta}\partial_\varphi J(\theta,\phi)\right) &=0.
\end{align}
We can discard the angular equations as they do not contribute to the motion across the horizon. Alternatively, note that  Eqs. \eqref{j1} and \eqref{j2} are the same, regardless of $A$ and $B$. It means that the inward and outward equations are the same. Consequently, the contribution from $J(\theta,\phi)$ cancels out upon dividing the outcoming probability by the incoming probability, as in
Eq. \eqref{probemabs} \cite{Kerner:2007rr}.  The remaining equations can be written as
\begin{align}
\Xi\begin{pmatrix}
A\\
B
\end{pmatrix}=\begin{pmatrix}
0\\
0
\end{pmatrix},
\end{align}
where
\begin{align}\nonumber
\Xi=\begin{pmatrix}
m&\frac{\omega}{\sqrt{f(r)}}(1+\sqrt{1-g(r)})- \sqrt{g(r)}W'(r)\\
-\frac{\omega}{\sqrt{f(r)}}(1-\sqrt{1-g(r)})- \sqrt{g(r)}W'(r)&m
\end{pmatrix}.
\end{align}
It has non-trivial solution if
\begin{align}
\det\Xi=\frac{\omega^2}{f(r)}g(r)-g(r)W'(r)^2+2\omega W'(r)\sqrt{\frac{g(r)}{f(r)}}\sqrt{1-g(r)}+m^2=0.
\end{align}
Solving the above quadratic equation results
\begin{align}
W'_{\pm}(r)=\omega\sqrt{\frac{1-g(r)}{f(r)g(r)}}\pm\frac{\sqrt{\omega^2+m^2f(r)}}{\sqrt{f(r)g(r)}},
\end{align}
%
%
where $W_+ [W_-]$ corresponds to outward [inward] solution. Remembering that the overall tunnelling probability is 
\begin{align}
\Gamma=\frac{\Gamma_{+}}{\Gamma_{-}}=\frac{e^{-2\mathtt{Im}I_+}}{e^{-2\mathtt{Im}I_-}}=e^{-2\mathtt{Im}(I_+-I_-)},
\end{align}
in the present case we have
\begin{align}
\Gamma=\frac{e^{-2\mathtt{Im}W_+}}{e^{-2\mathtt{Im}W_-}}=e^{-2\mathtt{Im}(W_+-W_-)}.
\end{align}
Hence we have to calculate only the difference 
\begin{align}\nonumber
W(r) \equiv W_+(r)-W_-(r)=&2\int\frac{\sqrt{\omega^2+m^2f(r)}}{\sqrt{f(r)g(r)}}\\ \label{Wm}
=&2\int\frac{\omega}{\sqrt{f(r)g(r)}}\sqrt{1+\frac{m^2}{\omega^2}f(r)}.
\end{align}
The functions $f(r)$ and $g(r)$ have both a first order zero on the horizon \cite{Wheeler1995}. As that is the point of interest for us, we perform a Taylor expansion of $f(r)$ and $g(r)$ at the horizon neighbourhood,
\begin{align}
f(r)=& f'(r_+)(r-r_+)+\mathcal{O}(r^2)\\
g(r)=& g'(r_+)(r-r_+)+\mathcal{O}(r^2),
\end{align}
thus
\begin{align}\label{taylorfg}
\sqrt{f(r)g(r)} \approx& \sqrt{f'(r_+)g'(r_+)}(r-r_+).
\end{align}
Near the horizon, as $f(r)\to 0$, the term with $m$ on the equation \eqref{Wm} should  vanish. Taking it into account and substituting Eq. \eqref{taylorfg} on Eq. \eqref{Wm} follows the imaginary part of the action
\begin{align}
W(r) =\frac{2\omega}{ \sqrt{f'(r_+)g'(r_+)}}\int\frac{1}{(r-r_+)}=i\frac{2\omega\pi}{ \sqrt{f'(r_+)g'(r_+)}}.
\end{align}
 It  yields
\begin{align}
2\mathtt{Im}(I_{+}-I_-)=\frac{4\omega\pi}{ \sqrt{f'(r_+)g'(r_+)}}.
\end{align}
Finally, from $2\mathtt{Im}(I_{+}-I_-)=\omega/T$,  the Hawking temperature is acquired:
\begin{align}\label{th}
T_H&=\frac{ \sqrt{f'(r_+)g'(r_+)}}{4\pi}.
\end{align}

It is worth mentioning that we can turn back to the Schwarzschild case taking
\begin{align}
f'(r_+)=g'(r_+)=\frac{1}{r_+},
\end{align}
which gives the right temperature we got on Eq.\eqref{tempbos}
\begin{align}\label{th}
T_H&=\frac{ 1}{4\pi r_+}=\frac{ 1}{8\pi M}.
\end{align}

In the next section, we are going to introduce a different black hole solution, coming the effective string theory, whose metric will be used for calculating the Hawking radiation associated with the Dirac spinors as well as the class of spinors discussed in Section \ref{secelko}.

\section{The Kerr-Sen Dilaton-Axion Black Hole}

The Kerr-Sen dilaton-axion black hole is a solution of effective field equations coming from string theory found as a deviation from the Kerr solution \cite{Sen:1992ua}. In this section, we are going to briefly introduce such solution in order to apply it in Section \ref{ketunn}.
The low energy effective action of the heterotic
string theory is ruled by an action that, up to higher
derivative terms and other fields which are set to zero for the
particular class of backgrounds  considered \cite{Sen:1992ua}, is given 
\begin{small}
\begin{eqnarray}
S&=&-\!\int d^4 x\sqrt{-\det G} e^{-\Phi} \left(-R+\frac{1}{12}H_{\mu\nu\rho}H^{\mu\nu\rho}-G^{\mu\nu}\partial_\mu\Phi\partial_\nu\Phi
+\frac18 F_{\mu\nu}F^{\mu\nu}\right),
\label{acao}
\end{eqnarray}%
\end{small}%
where  $G_{\mu\nu}$ is a metric related to the Einstein metric
by  $
e^{-\Phi} G_{\mu\nu}$. Here $\Phi$ denotes the dilaton field, $R$ stands for the scalar curvature,
$F_{\mu\nu}=\partial_{[\mu} A_{\nu]}$ is the Maxwell  field strength,  
$H_{\mu\nu\rho}=\partial_\mu B_{\nu\rho} + \partial_\rho B_{\mu\nu} + \partial_\nu B_{\rho\mu} - 
\Omega_{\mu\nu\rho}$ and $\Omega_{\mu\nu\rho}-\frac14 A_{(\mu} F_{\nu\rho)}$ is the 
Chern-Simons 3-form . 

The Kerr-Sen dilaton-axion black hole metric is a solution of the field equations derived from (\ref{acao}). In Boyer-Lindquist coordinates  it reads:
\begin{small}
\begin{align}\nonumber
ds^{2} =&-\frac{\Delta-a^2\sin^2\theta}{\Sigma}dt^2+\frac{\Sigma}{\Delta}dr^2+\Sigma d\theta^2+\frac{\sin^2\theta}{\Sigma}\left[\left(r^2-2\beta r +a^2 \right)^2-\Delta a^2\sin^2\theta \right]d\phi^2+\nonumber\\
&-\frac{2a\sin^2\theta}{\Sigma}\left[(r^2-2\beta r+a^2)-\Delta) \right]dtd\phi\label{metricaaa}
\end{align}
\end{small}
where
\begin{align}
\!\!\!\!\Sigma=r^2-2\beta r +a^2\cos^2\theta,\quad \Delta=r^2 - 2\eta r + a^2 = (r - r_\mais )(r - r_\menos ),\quad \beta = \eta \sinh^2 \frac{\alpha}{2}
\end{align} and 
$r_\mais [r_\menos]$ denote the outer [inner] horizons.

The metric (\ref{metricaaa}) describes a black hole solution with charge $Q$, mass $M$,  magnetic dipole moment $\mu$ and 
angular momentum $J$, given respectively by
\begin{eqnarray}
Q={\eta \over\sqrt 2}\sinh\alpha, && M={\eta\over 2} (1+\cosh\alpha), \\
\mu ={1\over\sqrt 2} \eta a\sinh\alpha\,, &&J=
{\eta a\over 2} (1+\cosh\alpha)\,.
\end{eqnarray}

When the charge $Q$ vanishes, or equivalently, when the parameter $\alpha$ vanishes, the Kerr-Sen dilaton-axion solution reduces to the Kerr solution. Moreover, the  parameters can be expressed in terms of genuinely physical quantities as
$$
\eta=M-{Q^2\over 2M}, ~~~~~~\alpha =\arcsin\!{\rm h}\left({2\sqrt 2 QM\over 2M^2 -Q^2}\right), ~~~~~~
a={J\over M}.
$$
The coordinate singularities read
$
r_\pm = M-{Q^2\over 2M}\pm\sqrt{-{J^2\over M^2}+\left(M-{Q^2\over 2M}\right)^2
}
$ which vanishes unless $
|J| < M^2-\frac{Q^2}{2}
$.

By performing the transformation  $\phi = \varphi -\Omega t$, where
$\Omega=\frac{a\left(a^2-2\beta r +a^2-\Delta\right)}{\left(r^2-2\beta r+a^2\right)^2-\Delta a^2\sin^2\theta}$, 
the metric (\ref{metricaaa}) takes the form
\begin{align}\label{metrrr}
\!\!\!\!\!ds^{2} =&\!-\frac{\Delta\Sigma}{\left(r^2\!-\!2\beta r\!-\!a^2 \right)^2\!-\!\Delta a^2\sin^2\theta}dt^2\!+\!\frac{\Sigma}{\Delta}dr^2\!+\!\Sigma d\theta^2\!+\\ &+\frac{\sin^2\theta}{\Sigma}\left[\left(r^2\!-\!2\beta r \!-\!a^2 \right)^2\!-\!\Delta a^2\sin^2\theta \right]d\varphi^2\,.
\end{align}To study the Hawking radiation at the event horizon,  the near-horizon approximation is performed on the metric coefficients \cite{Chen:2009zzk}. It gives
\begin{align}
ds^2=-F(r_\mais)dt^2+\frac{1}{G(r_\mais)}dr^2+\Sigma(r_\mais)d\theta^2+\frac{H(r_\mais)}{\Sigma(r_\mais)}d\varphi^2
\end{align}
where
\begin{align}
H(r_\mais)&= \sin^2\theta\left(r_\mais^2-2\beta r_\mais +a^2 \right)^2,\qquad 
F(r_\mais)= \frac{2(r_\mais-\eta)(r-r_\mais)\Sigma(r_\mais)}{\left(r_\mais^2-2\beta r_\mais+a^2\right)^2}\\
G(r_\mais)&= \frac{2(r_\mais-\eta)(r-r_\mais)}{\Sigma(r_\mais)}, \quad\qquad \qquad \Sigma(r_\mais)=\Sigma(r=r_\mais)\,.
\end{align}

\section{Elko Dark Spinors}\label{secelko}

Elko dark spinors are dual-helicity eigenspinors of the charge conjugation operator \cite{Ahluwalia:2004ab,Ahluwalia:2016rwl}. Furthermore,  they are  spin-1/2  fermions of mass dimension one, with  novel features that make them capable to  incorporate both the Very Special Relativity (VSR) paradigm and the dark matter description as well \cite{Ahluwalia:2004ab,Ahluwalia:2008xi,daRocha:2013qhu,Cavalcanti:2014wia,
Bernardini:2012sc,daSilva:2012wp,Villalobos:2015xca,Lee:2015tcc,Lee:2014opa}. 
 Due to its very small coupling with the standard model fields, excluding  the Higgs field, dark spinors 
are self-interacting dark matter  candidates \cite{Ahluwalia:2004ab}. 
In fact, the Lagrangian of such field contains a quartic self interaction term and the interaction term of the new field with spin-zero bosonic fields. 
 Elko  is a representative of the type-5 spinor field class in Lounesto's spinors classification \cite{Lounesto2001,Vaz:2016qyw}. However, is not the most general in the type-5 class, as shown in \cite{Cavalcanti:2014wia}. Some attempts to detect Elko  at the LHC have been proposed, and   important applications to cosmology  have been widely investigated as well \cite{daRocha:2014dla,Ahluwalia:2004ab,Ahluwalia:2008xi,Boehmer:2010ma,daRocha:2011yr,S.:2014dja}.

In order to analyze the tunnelling of Elko dark particles  across the Kerr-Sen black hole event horizon, we will study the role that Elko dark particles  play in this background. The essential prominent Elko
 particles features  are in short 
revisited \cite{Ahluwalia:2004ab}. Elko dark spinors  $\lambda(p^\mu)$ are eigenspinors of the charge
conjugation operator $C$, namely, $C\lambda(p^\mu)=\pm \lambda(p^\mu)$. 
The plus [minus] sign regards {self-conjugate},  [{anti self-conjugate}]  spinors , denoted by $\lambda^{S}(p^\mu)$ [$\lambda^{A}(p^\mu)$]. For rest spinors  $\lambda(k^\mu)$ the boosted spinors read $
	\lambda(p^\mu) = e^{i \kb\cdot\vp} \lambda(k^\mu),$ where $
		k^\mu = (m,\lim_{p\rightarrow 0}{\p}/{\vert\p\vert})$ and  $e^{i \kb\cdot\vp}$ denotes the boost operator. 
The $\phi_{}(k^\mu)$ are defined to be  eigenspinors of  the helicity operator, as $
\s\cdot\hat \p\, \phi_{}^\pm(k^\mu) = \pm \phi_{}^\pm(k^\mu)$, 
where \cite{Ahluwalia:2004ab} 
\begin{eqnarray}
\phi^+_{}(k^\mu) &=& \sqrt{m} \left(
									\begin{array}{c}
									\cos\left(\frac{\theta}{2}\right)e^{- i \varphi/2}\\
									\sin\left(\frac{\theta}{2}\right)e^{+i \varphi/2}
											\end{array}\right)\equiv\left(\begin{array}{c}
\alpha\\\beta			\end{array}
									\right)\,, \label{phim}\\ 
\phi^-_{}(k^\mu) &=& \sqrt{m} \left(
									\begin{array}{c}
									-\sin\left(\frac{\theta}{2}\right)e^{- i \varphi/2}\\
									\cos\left(\frac{\theta}{2}\right)e^{+i \varphi/2}
											\end{array}
									\right)=\left(\begin{array}{c}
-\beta^*\\\alpha^*			\end{array}
									\right)\,, 	\label{phime}					\end{eqnarray}
$\alpha,\beta:\mathbb{R}^{1,3} \to \mathbb{C}$ are scalar fields.		
The spinors   $\lambda(k^\mu)$ are constructed   as 
\begin{eqnarray}
 \lambda^S_\pm(k^\mu) & =& \left(
					\begin{array}{c}
					\sigma_2\left(\phi_{}^\pm(k^\mu)\right)^\ast\\
								\phi_{}^\pm(k^\mu)
					\end{array}
					\right)\,,\quad\quad  
				\lambda^A_\pm(k^\mu) = \pm\left(
							\begin{array}{c}
								-\sigma_2\left(\phi_{}^\mp(k^\mu)\right)^\ast\\
								\phi_{}^\mp(k^\mu)
												\end{array}
												\right)\,,
			\label{ppm}
\end{eqnarray}
and have dual helicity, as $ -i\sigma_2(\phi^\pm)^\ast$ has helicity dual to that of 
$\phi^\pm$. The boosted terms  \begin{eqnarray}
		\lambda^A_\pm(p^\mu) &=& \sqrt{\frac{E+m}{2 m} }\left( 1\pm\frac{p^\mu}{E+m}\right)\lambda^A_\pm,\\
		\lambda^S_\pm(p^\mu) &=& \sqrt{\frac{E+m}{2 m} }\left( 1\mp\frac{p^\mu}{E+m}\right)\lambda^S_\pm, 		\label{jj}
		\label{jj1}
\end{eqnarray}
 are the expansion coefficients of  a mass dimension one quantum field. The  Dirac operator does not annihilate the $\lambda(p^\mu)$, but instead the equations of motion read  \cite{Ahluwalia:2004ab,Ahluwalia:2016rwl}:
\begin{align}\label{elkodyn}
\gamma^\mu \partial_\mu \lambda^S_\pm&=\pm i\frac{m}{\hslash} \lambda^S_\mp\\
\gamma^\mu \partial_\mu \lambda^A_\mp&=\pm i\frac{m}{\hslash} \lambda^A_\pm\label{4}
\end{align}

A mass dimension one quantum field can be thus constructed as \cite{Ahluwalia:2016rwl} 
\begin{eqnarray}
\!\!\!\!\!\!\mathfrak{f}(x) =  \int \frac{\text{d}^3p}{(2\pi)^3}  \frac{1}{\sqrt{2 m E(\p)}} \sum_\rho \Big[ b^\dagger_\rho(\p)\lambda^A(\p) e^{i p_\mu x^\mu}+a_\rho(\p)\lambda^S(\p) e^{- i p_\mu x^\mu}{\Big]}\,.
\end{eqnarray}
The creation and annihilation operators $a_\rho(\p), a^\dagger_\rho(\p)$ satisfy the  Fermi statistics  \cite{Ahluwalia:2016rwl}, with similar anticommutators for $b_\rho(\p)$ and 
$b^\dagger_\rho(\p)$.

\section{Tunneling From Kerr-Sen Dilaton-Axion}\label{ketunn}

Following the procedure for calculating the Hawking temperature associated to fermions emission, as introduced in the Section \ref{femem}, we are now able to calculate the temperature of the Kerr-Sen dilaton-axion black hole. Before performing the calculations for Elko dark spinors, which is fairly more subtle, we are going to calculate the temperature from Dirac spinors emission. The results can then be compared at the end of the Section \ref{elkokerrsen}, when we finally find the temperature of Elko emission by Kerr-Sen dilaton-axion black holes.

\subsection{Dirac Spinors}

One starts by listing the vierbein associated to Eq. (\ref{metrrr})\footnote{As there are no off diagonal terms, we are representing the vierbein by a single index.} 

\begin{align}
\textbf{e}^t=\frac{1}{\sqrt{F(r_\mais)}}\,,\quad\textbf{e}^r=\sqrt{G(r_\mais)}\,,\\
\textbf{e}^{\theta}=\frac{1}{\sqrt{\Sigma(r_\mais)}}\;,\quad\textbf{e}^{\varphi}=\sqrt{\frac{\Sigma(r_\mais)}{H(r_\mais)}}.
\end{align}
 Contracting with the chosen $\gamma^\sigma$ matrices  yields 
\begin{align}
\textbf{e}^t\gamma^0=&
\frac{1}{\sqrt{F(r_\mais)}}\left(\begin{array}{cc}
i\mathbb{I}_2 & 0 \\ 
0 & -i\mathbb{I}_2
\end{array}\right)\,,\quad  \textbf{e}^\theta\gamma^2=
\frac{1}{\sqrt{\Sigma(r_\mais)}}\left(\begin{array}{cc}
0 & \sigma^1 \\ 
\sigma^1 & 0
\end{array}\right)\,, \\
\textbf{e}^r\gamma^3
=&{\sqrt{G(r_\mais)}}\left(\begin{array}{cc}
0 & \sigma^3 \\ 
\sigma^3 & 0
\end{array}\right)\,,\quad \textbf{e}^\varphi\gamma^2=
\sqrt{\frac{\Sigma(r_\mais)}{H(r_\mais)}}\left(\begin{array}{cc}
0 & \sigma^2 \\ 
\sigma^2 & 0
\end{array}\right)\,.
\end{align}
The leading order of the Dirac equation, as in Eq. \eqref{leadingdirac}, gives the equations
\begin{align}
-\left( B\sqrt{G(r_\mais)}\partial_r{I}+\frac{iA}{\sqrt{F(r_\mais)}}\partial_t I\right) +mA&=0,\\
\left( \frac{iB}{\sqrt{F(r_\mais)}}\partial_t I-A\sqrt{G(r_\mais)}\partial_r I\right) +mB&=0,\\
-B\left( \frac{1}{\sqrt{\Sigma(r_\mais)}}\partial_\theta I+i\sqrt{\frac{\Sigma(r_\mais)}{H(r_\mais)}}\partial_\varphi I\right) &=0,\\
-A\left( \frac{1}{\sqrt{\Sigma(r_\mais)}}\partial_\theta I+i\sqrt{\frac{\Sigma(r_\mais)}{H(r_\mais)}}\partial_\varphi I\right) &=0.
\end{align}

We apply the ansatz $I(t,r,\theta,\varphi)=-(\omega-\Omega)t+W(r)+\Phi(\varphi)+\Theta(\theta)$ and disregard the angular terms, following\footnote{Here $\Omega$ denotes the particle angular velocity.}
\begin{align}
-\left( B\sqrt{G(r_\mais)}W'(r)-\frac{iA}{\sqrt{F(r_\mais)}}(\omega-\Omega)\right) +mA&=0,\\
-\left( \frac{iB}{\sqrt{F(r_\mais)}}(\omega-\Omega)+A\sqrt{G(r_\mais)}W'(r)\right) +mB&=0.
\end{align}
The above equations can be written as
\begin{align}
\begin{pmatrix}
i\frac{\omega-\Omega}{\sqrt{F(r_\mais)}}+m&-\sqrt{G(r_\mais)}W'(r)\\
-\sqrt{G(r_\mais)}W'(r)&-i\frac{\omega-\Omega}{\sqrt{F(r_\mais)}}+m
\end{pmatrix}\begin{pmatrix}
A\\
B
\end{pmatrix}=\begin{pmatrix}
0\\
0
\end{pmatrix},
\end{align}
which has non-trivial solution only if
\begin{align}
\det\begin{pmatrix}
i\frac{\omega-\Omega}{\sqrt{F(r_\mais)}}+m&-\sqrt{G(r_\mais)}W'(r)\\
-\sqrt{G(r_\mais)}W'(r)&-i\frac{\omega-\Omega}{\sqrt{F(r_\mais)}}+m
\end{pmatrix}=\frac{(\omega-\Omega)^2}{F(r_\mais)}+m^2-G(r_\mais)W'(r)^2=0.
\end{align}
Hence it follows
\begin{align}
W_\pm(r)=&\pm\int\sqrt{\frac{(\omega-\Omega)^2-F(r_\mais)m^2}{F(r_\mais)G(r_\mais)}}dr.
\end{align}
For a vanishing mass term one finds
\begin{align}
\displaystyle W(r)_\pm=&\pm\int\sqrt{\frac{(\omega-\Omega)^2}{F(r_\mais)G(r_\mais)}}dr,\\
\displaystyle=&\pm\int\frac{\omega-\Omega}{\sqrt{\left[\frac{2(r_\mais-\eta)(r-r_\mais)\Sigma(r_\mais)}{\left(r_\mais^2-2\beta r_\mais+a^2\right)^2}\right]\left[\frac{2(r_\mais-\eta)(r-r_\mais)}{\Sigma(r_\mais)}\right]}}dr,\\
=&\pm \frac{r_\mais^2-2\beta r_\mais+a^2}{2(r_\mais-\eta)}(\omega-\Omega)\int \frac{1}{r-r_\mais} dr.\\
\end{align}
Accordingly
\begin{align}
W(r)_\pm=\pm i\pi \frac{r_\mais^2-2\beta r_\mais+a^2}{2(r_\mais+\eta)}(\omega-\Omega)\,,
\end{align}
%
%
with the imaginary part
\begin{align}
\mathtt{Im}\,I_{\pm}=\pm \pi \frac{r_\mais^2-2\beta r_\mais+a^2}{2(r_\mais+\eta)}(\omega-\Omega)\,.
\end{align}
Thus, the resulting tunnelling probability is given by
\begin{align}
\Gamma&=\frac{\Gamma_+}{\Gamma_-}=\frac{e^{-2\mathtt{Im}I_+}}{e^{-2\mathtt{Im}I_-}}=e^{-2(\mathtt{Im}I_+-\mathtt{Im}I_-)}\,,
\end{align}
and
\begin{align}
-2(\mathtt{Im}I_+-\mathtt{Im}I_-)=-2\pi \frac{r_\mais^2-2\beta r_\mais+a^2}{(r_\mais+\eta)}(\omega-\Omega)\,.
\end{align}
Finally,  the Hawking temperature of the Kerr-Sen dilaton-axion black hole is obtained
\begin{align}\label{th}
T_H&=\frac{1}{2\pi} \frac{(r_\mais+\eta)}{r_\mais^2-2\beta r_\mais+a^2}.
\end{align}
%
This result was obtained by Chen and Zu in the reference \cite{Chen:2009zzk}.
\subsection[Elko Dark Spinors]{Elko Dark Spinors\protect\footnote{We have published the results of this section in Ref. \cite{Cavalcanti:2015nna}.}}\label{elkokerrsen}

Just like in the previous section, the vierbein associated to the metric (\ref{metrrr}) are 
%
\begin{align}
\textbf{e}^t=\frac{1}{\sqrt{F(r_\mais)}}\,,\quad\textbf{e}^r=\sqrt{G(r_\mais)}\,,\nonumber\\
\textbf{e}^{\theta}=\frac{1}{\sqrt{\Sigma(r_\mais)}}\;,\quad\textbf{e}^{\varphi}=\sqrt{\frac{\Sigma(r_\mais)}{H(r_\mais)}}.\nonumber
\end{align}
Elko dark spinors are better described using the Weyl representation of the $\gamma$ matrices, it yields to the algebra generators

\begin{align}
\gamma^t&=\textbf{e}^t\gamma^0=
\frac{1}{\sqrt{F(r_\mais)}}\left(\begin{array}{cc}
0 & \mathbb{I}_2 \\ 
\mathbb{I}_2 & 0
\end{array}\right)\,,\quad  \gamma^\theta=\textbf{e}^\theta\gamma^2=
\frac{1}{\sqrt{\Sigma(r_\mais)}}\left(\begin{array}{cc}
0 & \sigma^2 \\ 
\sigma^2 & 0
\end{array}\right)\,, \\
\gamma^r&=\textbf{e}^r\gamma^1
={\sqrt{G(r_\mais)}}\left(\begin{array}{cc}
0 & \sigma^1 \\ 
\sigma^1 & 0
\end{array}\right)\,,\quad \gamma^\varphi=\textbf{e}^\varphi\gamma^3=
\sqrt{\frac{\Sigma(r_\mais)}{H(r_\mais)}}\left(\begin{array}{cc}
0 & \sigma^3 \\ 
\sigma^3 & 0
\end{array}\right)\,.
\end{align}

Elko dark spinors can be written as 
\begin{eqnarray}\label{elkof1}
\lambda^\mathrm{S}_+ &=&\!\left(
\begin{array}{c}
- i\beta^*  \\
i\alpha^* \\
\alpha \\
\beta
\end{array}%
\right) \exp \left( \frac{{}i{} }{\hbar }\tilde{I}\right)\,,\qquad 
\label{elkof2}\lambda^\mathrm{S}_- =\left(
\begin{array}{c}
- i\alpha  \\
-i\beta \\
-\beta^* \\
\alpha^*
\end{array}%
\right) \exp \left( \frac{{}i{} }{\hbar }\tilde{I}\right)\,,\\
\label{elkof3}\lambda^\mathrm{A}_+   &=&\left(
\begin{array}{c}
i\alpha  \\
i\beta \\
-\beta^* \\
\alpha^*
\end{array}%
\right) \exp \left( \frac{{}i{} }{\hbar }\tilde{I}\right)\,,\qquad\lambda^\mathrm{A}_-  =\left(
\begin{array}{c}
- i\beta^*  \\
i\alpha^* \\
-\alpha \\
-\beta
\end{array}%
\right) \exp \left( \frac{{}i{} }{\hbar }\tilde{I}\right)\,,
\end{eqnarray}
where   $\tilde{I}=\tilde{I}(t,r,\theta,z)$ represents the classical action.  We will use the above forms     in each one of the Eqs. (\ref{elkodyn}-\ref{4}), and then solve this coupled system of  equations. 
By identifying  $\lambda$ [$\mathring{\lambda}$] to the Elko spinor on the left [right] hand side of Eqs. \eqref{elkodyn} and \eqref{4}, then those equations read $\gamma^\mu D_\mu  \lambda=i\frac{m}{\hslash} \mathring{\lambda}$. Using the WKB approximation, where $\tilde{I}=I+\mathcal{O}(\hslash)$, it yields    
\begin{align}\label{elkoelko}
\gamma^\mu D_\mu \lambda=&i\frac{m}{\hslash} \mathring{\lambda}\\
\gamma^\mu \left(\partial_\mu+\frac{i}{2}{}{} \Gamma _{\mu }^{\alpha \beta }\Sigma_{\alpha\beta}\right) \lambda=&i\frac{m}{\hslash} \mathring{\lambda}\\
\gamma^\mu \left(\frac{\partial_\mu\tilde{I}}{\hslash}+\frac{i}{2}{}{} \Gamma _{\mu }^{\alpha \beta }\Sigma_{\alpha\beta}\right) \lambda=&i\frac{m}{\hslash} \mathring{\lambda}\\
\gamma^\mu \left(\partial_\mu{\tilde{I}}+{\hslash}\frac{i}{2}{}{} \Gamma _{\mu }^{\alpha \beta }\Sigma_{\alpha\beta}\right) \lambda=&i{m} \mathring{\lambda}\\
\gamma^\mu \partial_\mu{I}\lambda=&i{m} \mathring{\lambda}+\mathcal{O}(\hslash)\,.
\end{align}
Taking merely the leading order terms in the above equation, from a general form 
$\lambda=(a,b,c,d)^\intercal$, $  \mathring{\lambda}=(\mathring{a}, 
\mathring{b}, \mathring{c}, 
\mathring{d})^\intercal\,,$
we have general Elko dynamic equations governed by 
\begin{align*}
\left( \sqrt{G(r_\mais)}\partial_r{I}-i\,\frac{1}{\sqrt{\Sigma(r_\mais)}}\partial_\theta I\right) d+\left( \sqrt{\frac{\Sigma(r_\mais)}{H(r_\mais)}}\partial_\varphi I+\frac{1}{\sqrt{F(r_\mais)}}\partial_t I\right) c=i\,\mathring{a}\,m,\\
\left( \frac{1}{\sqrt{F(r_\mais)}}\partial_t I-\sqrt{\frac{\Sigma(r_\mais)}{H(r_\mais)}}\partial_\varphi I\right) d+\left( i\,\frac{1}{\sqrt{\Sigma(r_\mais)}}\partial_\theta I+\sqrt{G(r_\mais)}\partial_r I\right) c=i\,\mathring{b}\,m,\\
\left( i\,\frac{1}{\sqrt{\Sigma(r_\mais)}}\partial_\theta I-\sqrt{G(r_\mais)}\partial_r I\right) b+\left( \frac{1}{\sqrt{F(r_\mais)}}\partial_t I-\sqrt{\frac{\Sigma(r_\mais)}{H(r_\mais)}}\partial_\varphi I\right) a=i\,\mathring{c}\,m,\\
\left( \frac{1}{\sqrt{F(r_\mais)}}\partial_t I+\sqrt{\frac{\Sigma(r_\mais)}{H(r_\mais)}}\partial_\varphi I\right) b-\left( i\,\frac{1}{\sqrt{\Sigma(r_\mais)}}\partial_\theta I+\sqrt{G(r_\mais)}\partial_r I\right) a=i\,\mathring{d}\,m.
\end{align*}
The ansatz $I(t,r,\theta,\varphi)=-(\omega-\Omega)t+W(r)+\Phi(\varphi)+\Theta(\theta)$ is again used by considering symmetries of the space-time solution \cite{Chen:2009zzk}, where $\omega$ and $\Omega$ denote the energy and angular velocity respectively. The angular function $\Phi(\varphi) +\Theta(\theta)$ must have the same solution for both the incoming and outgoing cases as well. It implies that the contribution of such function vanishes after dividing the outgoing probability by the incoming one, just like to $J(\theta,\phi)$ of Eq. \eqref{ansatz1}. Hence the angular function can be neglected hereupon. Using the ansatz on the above system of equations and disregarding the angular terms we have, 
\begin{align}
 \sqrt{G(r_\mais)} W'd-\frac{(\omega-\Omega)}{\sqrt{F(r_\mais)}} c=&i\,\mathring{a}\,m,\\
- \sqrt{G(r_\mais)}W' a-\frac{(\omega-\Omega)}{\sqrt{F(r_\mais)}} b=&i\,\mathring{d}\,m,\\
 \sqrt{G(r_\mais)}W' c-\frac{(\omega-\Omega)}{\sqrt{F(r_\mais)}} d=&i\,\mathring{b}\,m,\\
 \sqrt{G(r_\mais)}W' b+\frac{(\omega-\Omega)}{\sqrt{F(r_\mais)}} a=&-i\,\mathring{c}\,m.
\end{align}
Moreover, the parameters $a,b,c,d$  are not independent. In fact, Eqs. (\ref{elkof1}) and   (\ref{elkof3}) assert that for the self-conjugate spinors  $\lambda^S$ we have $a=-id^*$ and $b=ic^*$, whereas for the anti-self-conjugate spinors $\lambda^A$ it reads $a=id^*$ and $b=-ic^*$ \cite{Cavalcanti:2014wia}. Thus, by corresponding the $\lambda^S$  [$\lambda^A$] spinors to the upper [lower] sign below follows,
\begin{align}\label{s1}
 \sqrt{G(r_\mais)} W'd-\frac{(\omega-\Omega)}{\sqrt{F(r_\mais)}} c=\pm\,\mathring{d}^*\,m,\\ \label{s2}
\pm \sqrt{G(r_\mais)}W' d^* \mp \frac{(\omega-\Omega)}{\sqrt{F(r_\mais)}} c^* =\,\mathring{d}\,m,\\ \label{s3}
\sqrt{G(r_\mais)}W' c-\frac{(\omega-\Omega)}{\sqrt{F(r_\mais)}} d =\mp\,\mathring{c}^*\,m,\\ \label{s4}
\pm  \sqrt{G(r_\mais)}W' c^*\mp \frac{(\omega-\Omega)}{\sqrt{F(r_\mais)}} d^*=-\,\mathring{c}\,m.
\end{align}
We still have a system of 4 coupled equations not straightforward to be solved for each 4 types of Elko spinors. It means that we do not know \textit{a priori} if all of then are emitted at the same temperature, a feature that would be weird. However, we are going to show that the situation can be simplified using properties of Elko components. In fact, labelling $A=\sqrt{G(r_\mais)}W'$, $B=\frac{(\omega-\Omega)}{\sqrt{F(r_\mais)}}$ and replacing explicitly the the components of each of the Elko spinors we find

\begin{align}
\lambda^S_+:& \left\{\begin{array}{ccc}
A \beta+B\alpha & = & \alpha m \\ 
A\beta^*-B\alpha^* & = & \alpha^* m\\
A \alpha-B\beta & = & -\beta m \\ 
A \alpha^*-B\beta^* & = & \beta^* m 
\end{array} \right.\\
\lambda^A_-:& \left\{\begin{array}{ccc}
-A \beta-B\alpha & = & -\alpha m \\ 
A\beta^*-B\alpha^* & = & \alpha^* m\\
-A\alpha+B\beta & = & \beta m \\ 
A \alpha^*-B\beta^* & = & \beta^* m 
\end{array} \right.\\
\lambda^S_-:& \left\{\begin{array}{ccc}
A \alpha^*-B\beta & = & \beta^* m \\ 
A\alpha +B\beta & = & \beta m \\ 
-A\beta^*-B\alpha^* & = & \alpha^* m\\
-A \beta-B\alpha & = & -\alpha m
\end{array} \right.
\\
\lambda^A_+:& \left\{\begin{array}{ccc}
A \alpha^*-B\beta & = & \beta^* m \\ 
-A\alpha -B\beta & = & -\beta m \\ 
-A\beta^*-B\alpha^* & = & \alpha^* m\\
A \beta+B\alpha & = & \alpha m
\end{array} \right.
\end{align}

From the systems above, we see that the equations for $\lambda^S_+$ are completely equivalent to the equations for $\lambda^A_-$. The same happens to the pair $\lambda^S_-,\lambda^A_+$. Thus we have to deal only with $\lambda^S_+$ and $\lambda^S_-$. Furthermore, there are more equivalences:  Eq. \eqref{s1} for $\lambda^S_+$ is equivalent to Eq. \eqref{s4} for $\lambda^S_-$ and Eq. \eqref{s1} for $\lambda^S_-$ is equivalent to Eq. \eqref{s4} for $\lambda^S_+$. At the end, rather than have four systems of four equations we have the three pairs of equations below:

\begin{align}\label{ss1}
\mbox{Eqs. for } \lambda^S_+ \mbox{ and }\lambda^S_-:& \left\{\begin{array}{ccc}
\sqrt{G(r_\mais)} W'\alpha^*-\frac{(\omega-\Omega)}{\sqrt{F(r_\mais)}}\beta^* & = & \beta^* m, \\ 
\sqrt{G(r_\mais)} W'\beta+\frac{(\omega-\Omega)}{\sqrt{F(r_\mais)}}\alpha & = & \alpha m,
\end{array} \right.\\\label{ss2}
\mbox{Eqs. for } \lambda^S_+:& \left\{\begin{array}{ccc}
\sqrt{G(r_\mais)} W'\alpha-\frac{(\omega-\Omega)}{\sqrt{F(r_\mais)}}\beta & = & -\beta m, \\ 
\sqrt{G(r_\mais)} W'\beta^*-\frac{(\omega-\Omega)}{\sqrt{F(r_\mais)}}\alpha^* & = & \alpha^* m,
\end{array} \right.\\\label{ss3}
\mbox{Eqs. for } \lambda^S_-:& \left\{\begin{array}{ccc}
\sqrt{G(r_\mais)} W'\alpha+\frac{(\omega-\Omega)}{\sqrt{F(r_\mais)}}\beta & = & \beta m, \\ 
\sqrt{G(r_\mais)} W'\beta^*+\frac{(\omega-\Omega)}{\sqrt{F(r_\mais)}}\alpha^* & = & -\alpha^* m.
\end{array} \right.
\end{align}

Combining either Eqs. \eqref{ss1} and \eqref{ss2} or \eqref{ss1} and \eqref{ss3} implies the equations for $\lambda^S_+$ or $\lambda^S_-$ respectively. Hence, for each $\lambda^S$ there is a system of  coupled equations for the dark spinor components $\alpha$ and  $\beta$ and also another coupled system for $\alpha^*$ and  $\beta^*$, which are going to be solved separately. We denote now the first equation of each one of the systems below to be the equations related to $(\alpha, \beta)$, whereas the second ones regard $(\alpha^*, \beta^*)$. We can determine the above functions as:
\begin{align}\label{sss1}
\lambda^S_+:& \left\{\begin{array}{cccc}
&W_1(r) & = & \pm \int\sqrt{\frac{m^2F(r_\mais)-(\omega-\Omega)^2}{F(r_\mais)G(r_\mais)}}dr \\ 
&W_2(r) & = & \pm \int\sqrt{\frac{(m\sqrt{F(r_\mais)}+\omega-\Omega)^2}{F(r_\mais)G(r_\mais)}}dr
\end{array} \right.\\\label{sss2}
\lambda^S_-:& \left\{\begin{array}{ccccc}
&W_3(r) & = & \pm \int\sqrt{\frac{(m\sqrt{F(r_\mais)}-\omega+\Omega)^2}{F(r_\mais)G(r_\mais)}}dr \\ 
&W_4(r) & = & iW_2(r)
\end{array} 
\right.
\end{align}
For massless particles the solutions for $\lambda^S_\pm$ are equivalent and $W_2(r)=-iW_1(r)$. In such case $W_2(r)$ reads 
\begin{align}
W_2(r)=&\pm\int\frac{\omega-\Omega}{\sqrt{F(r_\mais)G(r_\mais)}}dr\\
=&\pm i\pi \frac{r_\mais^2-2\beta r_\mais+a^2}{2(r_\mais+\eta)}(\omega-\Omega)\,.
\end{align}
Actually, since $F\to 0$ near the black hole horizon, the results hold both 
for massive and their massless limit particles as well. 
It is worth emphasizing that it implies a real solution for $W_1(r)$, and thus a null contribution for the tunnelling effect.

The above expression for $W_2(r)$ is exactly the same as obtained in the previous section for Dirac spinors. Thus, the associated Hawking temperature is also the same, given the temperature provided by
\begin{align}\label{th}
T_H&=\frac{1}{2\pi} \frac{(r_\mais+\eta)}{r_\mais^2-2\beta r_\mais+a^2}.
\end{align}

Thus, the temperature of the Kerr-Sen dilaton-axion black hole was computed and demonstrated to  confirm the universal character of the Hawking effect, even for mass dimension one fermions of spin-1/2 that are beyond the standard model. 
 For future works, corrections of higher order in $\hbar$ to the Hawking temperature (\ref{th}) of type $I = I_0 + \sum_{n\geq 1} \hbar^n I_n$  \cite{Banerjee:2009wb} can be still implemented in the context of mass dimension one spin-1/2 fermions.  

\chapter{Extra Dimensions and Braneworld FRW} 

\label{ched} 

The first proposal of an extra dimensional space-time appeared through the work of Kaluza and Klein, around the 1920s. Kaluza aimed a classical extension of general relativity to five dimensions and 
derived the 4D Einstein field equations, the Maxwell equations, and an equation for the scalar field. Later, in 1926, Klein provided a quantum interpretation of Kaluza's model. Despite of the lack of new predictions from their program, pieces of their formalism were further developed and it is still alive. After the 1960s, extra spatial dimensions were bring back by string theorists. This time playing an essential role on the consistency of the theory. The not so exciting fact is that the extra dimensions were supposed to be compactified at a scale close to the Planck scale, pushing the new physics very far away from the energy reachable by current experiments. This scenario gave birth to the modern extra dimensions models, also called large extra dimensions paradigm. Its potential capacity to bring new physics to a reachable energy scale made those models very popular around the turn of the last century. The rich phenomenology emerging from large extra dimensions keeps it as a serious candidate among present phenomenological models.

There are plenty of papers and models of extra dimensions. On the HEP-Inspire database, for example, there are 2,708 papers whose title has ``extra dimensions" or ``braneworld". Far from a comprehensive cover of the field, this chapter was written with four goals in mind. The first one is to introduce some basic ideas of extra dimensions, giving an overview of the main models. It allows the discussion of the second goal, which is to present the content of the paper \cite{Bernardini:2014vba} in the Section \ref{thickfrw}. The third goal is introduce the two braneworld black holes studied in the next chapter, in the context of gravitational lensing in the strong field deflection regime. The last one is to support the results of the Chapter \ref{cap6}, where we analyse the consequences coming from the horizon have function formalism by assuming the existence of extra dimensions. 

\section{Overview}



Since the wide success of the Maxwell theory in the 19th century, unifying physical theories has become a hot topic of research in physics. In such context, the existence of extra dimensions has been shown fruitful and quite recurrent nowadays. The first attempt of an unifying theory which includes Newtonian gravity and extra dimensions was proposed by Nordström in 1914 \cite{Nordstrom:1988fi}. His aim was to set up a theory unifying gravity and electromagnetism, the two known interactions at the time, assuming a 3D space embedded in a 4D one. 

	The establishment of general relativity as the theory of gravity at the beginning of the 20th century turned the Nordström attempt obsolete. Any theory intending to unify gravity and electromagnetism should incorporate general relativity. That was the underlying idea of the Kaluza-Klein program \cite{kaluza,Klein:1926tv}.  For the first time the extra dimensions were assumed to be compact, which as we are going to discuss in the Section \ref{kkm}, have some interesting consequences. 

Going forward to the 80s, some ideas of a paper entitled “Do We Live Inside A Domain Wall?”, by Rubakov and Shaposhnikov \cite{Rubakov:1983bb}, have motivated a class of extra dimensional models currently known as thick branes. Assuming the existence of extra dimensions, it suggests that the known matter fields are localized in a topological defect placed in a 5D space-time. Such feature is largely assumed by current models, where the brane plays the role of the topological defect. We are going to discuss a model of thick branes in the Section \ref{thick_b}.

Also in the 80s, string theory had an intense development. This theory, among the quantum gravity candidates, is the one which has inspired more phenomenological models since then. From the extra dimensional models perspective, one of the main objects predicted by string theory is 	the so called Dirichlet branes, or D-branes for short. D-branes are extended objects where open strings are attached \cite{Polchinski:1995mt,Witten:1998cd,Bandos:1997gd,Cardona:2014ora,Duff:2015jka}. They provide a natural justification for matter field localization, like on the topological defects proposed by Rubakov and Shaposhnikov. 
In the former case, however, the width of the brane can be considered tiny enough to be negligible. It justifies the emerging brane models being called thin branes models, also known as braneworld models or braneworld scenarios. According to the current braneworld terminology, the place where the known fields are confined is called brane, while the whole extra dimensional space is called bulk.

\subsection{Braneworld}

An important class of braneworld models, including the famous Randall-Sundrum (RS) model (see Section \ref{warp_ed}), is highly inspired by the work of Horava and Witten \cite{SatheeshKumar:2006ac, Horava:1995qa, Horava:1996ma}. The original model purposes a 11D string theory scenario where the matter fields are confined in a 10D manifold. The model supposes the existence of two branes of opposite tension, which generates curvature on the bulk.  The bulk curvature warps the brane metric tensor, being the braneworlds with such feature called warped braneworlds.

Independently of its specific properties, a question that must be addressed by any physically accepted braneworld model regards the fact that no extra dimensions have been observed on nature until the present date. Any realistic model has to prevent the effects of the extra dimensions to be observed at low energy scales. A way of solving this problem is to suppose that the extra dimensions are small\footnote{In some models the extra dimensions might be of order of millimetres, therefore they are large compared to the ones from string theory.} and compact, as in string theory. This solution is also applied to compact flat extra dimension (Section \ref{compact_ed}). A pretty similar argument is used on thick brane models (Section \ref{thick_b}), where the width of the topological defect has to be very small. The third one regards the warped braneworld models (Section \ref{warp_ed}), where the bulk curvature prevents gravitons to access the whole bulk, thus keeping gravity fairly as we know.

One of the interesting features of braneworld models is that they provide an original solution to the hierarchy problem of the standard model. The hierarchy problem deals with the fact, unexplained using only the standar model physics, that the Planck scale and the electroweak scale are different by a factor of $10^{16}$. The Planck mass, which determines the strength of the gravitational interaction, is proportional to G$^{-1/2}$, where G is the Newton's constant.  The weak scale, on the other hand, is given by the vaccum expected value (VEV) of the Higgs field. However, considering its couplings to gauge bosons, the Higgs mass under radiative corrections shows a quadratic sensitivity to the ultraviolet cutoff. Resulting in larger Higgs mass, and therefore a larger electroweak symmetry break scale, when the energy goes toward the Plack scale. It imposes a huge and very unnatural fine tunning on the Higgs bare mass in the standard model Lagrangian.  As we shall discuss in Sec. \ref{ThinBW}, braneworld models solve the hireachy problem by lowering the Planck scale to the order of the electroweak scale, arguing that the weakness of the gravitational interaction is an direct effect of the existence of extra espatial dimensions. 

Besides their role dealing with the hierarchy problem, braneworld models are very versatile. Their range of applications goes, for exemple, from  cosmology \cite{Binetruy:1999ut, Binetruy:1999hy, Dvali:1998pa} and particle physics \cite{Davoudiasl:2000wi, Nath:2006ut, Randall:1999ee} to black hole physics \cite{Dadhich:2000am, Horowitz:2012nnc}. Refs. \cite{Maartens:2010ar} and \cite{Dzhunushaliev:2009va} are comprehensive reviews on thin and thick brane respectively.

\subsection{Braneworld Cosmology}

Applications of braneworld on cosmology has been investigated in several suitable contexts.
Classes of exact solutions with a constant 5D radius on a cosmologically evolving brane were provided in \cite{Binetruy:1999ut}, allowing unconventional cosmological equations with the matter content of the brane dominating that of the bulk.
This framework is in full compliance to standard cosmology, as the present values of the Hubble parameter and of the cosmological background radiation temperature fits their respective values at the time of nucleosynthesis.
Moreover, braneworld cosmology in thin branes has been studied for any equation of state describing the matter on the brane, where standard cosmological evolution can be obtained after an early non-conventional phase in typical Randall-Sundrum \cite{Binetruy:1999hy} scenarios, where the brane tension compensates the bulk cosmological constant. 
The accelerated Universe could be the result of the gravitational leakage into extra dimensions on Hubble distances rather than the consequence of non-zero cosmological constant \cite{Deffayet2001}.
 
Subsequent to the braneworld cosmology on thin branes, the thick braneworld paradigm has exhibited a fine structure \cite{Ahmed2012,Ahmed2013, Ahmed2014} that supports the above discussed phenomenology.
In spite of their success, thick braneworld models do encompass neither anisotropy on the brane nor the important framework of asymmetric branes, as well as spherically symmetric thick brane worlds.

The study of the dynamics of a scalar field with an arbitrary potential trapped in braneworld model can be further performed \cite{Ahmed2013,Ahmed2014,Kobayashi2001,Wang:2002pka}.
Homogeneous and anisotropic branes filled also with a perfect fluid are the mostly approached models.
In particular, by taking into account the effect of a positive dark radiation term on the brane \cite{Escobar:2013js}, the effect of the projection of the 5D Weyl tensor onto the brane in the form of a negative dark radiation term is considered \cite{Escobar2012}. 

All the above-mentioned reasons motivate the investigation of both spherically symmetric and anisotropic brane models.

\section{Kaluza-Klein Modes}\label{kkm}

Let us now to sketch the Kaluza-Klein (KK) theory. Their original program, as already mentioned, was proposed as an attempt to unify GR and electromagnetism. Its basic set up can be described as follows: departing from the metric tensor\footnote{Here we are not regarding the scale and units, as we are going to do it in the next section.},
\begin{align}
{^{(5)}g}_{AB}&=  \delta_A^\mu\delta_B^\nu\left(g_{\mu \nu}+A_{\mu}A_{\nu}\right)+\delta_A^5\delta_B^5\\
{^{(5)}g}_{\mu 5}&=  A_{\mu}
\end{align}
and assuming that ${^{(5)}g}_{AB}$ and $ A_{\mu}$ does not depend on the extra dimension $y$, they showed that the 5D Ricci tensor decomposes as
\begin{align}
^{(5)}{R}=R+\frac{1}{4} F_{\alpha \beta}F^{\alpha \beta},
\end{align}
where $F_{\mu\nu}=\partial_\mu A_\nu-\partial_\nu A_\mu$.
Putting this Ricci tensor in the 5D Einstein-Hilbert action, here denoted by $S_{KK}$, follows
\begin{eqnarray}
S_{KK}&=&\int dy d^4x\sqrt{-{}^{(5)}g} \;\;{}^{(5)}{R}\\
&=&\int dy  \int d^4x\sqrt{-g}\left[R+\frac{1}{4} F_{\alpha \beta}F^{\alpha \beta} \right]\\ \label{ehem}
&\equiv & \Delta \int d^4x\sqrt{-g}\left[R+\frac{1}{4} F_{\alpha \beta}F^{\alpha \beta} \right]\\
&= & \Delta(S_{EH}+S_{MW}).
\end{eqnarray}
In order to consistently identify the equation \eqref{ehem} as a sum of the 4D Einstein-Hilbert and Maxwell actions, the length of the extra dimension $\Delta=\int dy$ must be finite. It led them to postulate that the extra dimension is compact.

In spite of having no new consequences arising from the above action decomposition, the interesting property of the KK theory, that has survived almost a century, is its prediction of additional modes of known fields. We are going to describe those modes in the following.
Consider a 5D space-time with the extra dimension compactified through the identification $y\leftrightarrow y+2n\pi L$, where $n = 0, 1, 2,...$. Such compactification means that the extra dimension is the circle $S^1$. Applying it to a scalar field one has 
\begin{align}
\varphi(x^\mu,y)=\varphi(x^\mu,y+2n\pi L).
\end{align}
 The $2\pi L$ periodicity in
$y$ results that the field $\varphi(x^\mu,y)$ can be expressed as a Fourier series,
\begin{align}\label{kkmodes}
\varphi(x^\mu,y)= \sum_{k=-\infty}^{\infty}\varphi_{k}(x^\mu)e^{iky/L},
\end{align}
where the expansion coefficients depend only on the $x^\mu$ coordinates, being the $k \neq 0$ modes often referred to as Kaluza-Klein excitations or Kaluza-Klein tower. In order to interpret the consequences of the above assumption, let us apply the 5D version of the Klein-Gordon equation to a massless field $\varphi(x^\mu, y)$ ,
\begin{align}
^{(5)}\square \varphi(x^\mu, y) \equiv \left(\square-\partial^2_y \right)\varphi(x^\mu, y)=0.
\end{align}
Substituting the Fourier decomposition \eqref{kkmodes} into the previous equation  follows that each mode $\varphi_k$ satisfies the 4D Klein-Gordon equation
\begin{align}
\left(\square-\partial^2_y \right)\varphi(x^\mu, y)=\left(\square+\frac{k^2}{L^2}\right)\varphi_{k}(x^\mu)=0.
\end{align}
From the above equation one sees that the zero mode $\varphi_0(x^\mu)$ remains massless, whereas all other modes become effective 4D massive fields, with $|k|/L$ playing the role of the mass term,
\begin{align}
m_k=\frac{|k|}{L}.
\end{align}

The KK tower is expected to exist in every theory with compact extra dimensions, with energy scale varying with the inverse of the length of the extra dimension. In fact, the excited modes are one of the main features of the braneworld models that have been tested at the LHC.

\section{Thin Braneworld Models}\label{ThinBW}

One of the main properties of braneworld models is that the
4D Planck scale $m_p\equiv m_4$ no longer plays the role of fundamental scale, which is played by $m_D\equiv m_{4+d}$, where
$d$ is the number of extra dimensions. It might allow the possibility of the fundamental scale being around TeV. That is the reason for its huge popularity among researchers of high energy phenomenology.  In fact, the fundamental scale suppression is usually seen from the extra dimensional gravitational potential. The weakness of
gravity is interpreted as being due to the fact that it ``spreads" into extra dimensions and only part of it is felt in our 4D space-time. 

The extra dimensional field equations reads \cite{Maartens:2010ar}
\begin{align}
^{(4+d)}R_{AB}-\frac{1}{2}{}^{(4+d)}R{}^{(4+d)}g_{AB}=-\Lambda_{4+d}{}^{(4+d)}g_{AB}+\kappa^2_{4+d}{}^{(4+d)}T_{AB},
\end{align}
where the gravitational coupling constant $\kappa_{4+d}$ is related to extra dimensional Newton's constant and the fundamental scale by
\begin{align}
\kappa^2_{4+d}=8\pi G_{4+d}=\frac{8\pi}{m_{4+d}^{2+d}}.
\end{align}
The static weak field limit of the field equations leads to the $(4+d)$-dimensional Poisson equation,
whose solution is the gravitational potential,
\begin{align}
V(r) \propto \frac{\kappa^2_{4+d}}{r^{1+d}}.
\end{align}

However, on scales such that $r\gg L$,  the extra dimensional
potential behaves like a 4D potential. It means that $V(r)^{-1} \sim V_{d}\,r$, where $V_d$ is the volume of the $d$-extra dimensional space. The interesting result is that the usual Planck scale becomes part of an
effective coupling constant, describing gravity effectively on scales much larger than the extra dimensions. The Planck scale and the fundamental scales are related by
\begin{align}\label{scalesupression}
m_p^2 \sim m_{4+d}^{2+d}V_{d}.
\end{align}
Setting the number of extra dimensions and specifying its geometry, the volume can be calculated. Thus the above equation gives the true fundamental scale. This is the basic setup of the fundamental scale suppression applied to the ADD model, introduced in the next section. Apart from bringing the fundamental scale close to the experimental scale, the scale suppressing is also a solution to the hierarchy problem of the standard model. 
\subsection{Compact Flat Extra Dimensions}\label{compact_ed}

The main model to apply compact flat extra dimensions is the so called ADD model, proposed by Arkani-Hamed, Dimopoulos e Dvali  \cite{ArkaniHamed:1998rs}.  The model assumes a non specified \textit{a priori} number $d$ of extra dimensions. Thus, using the fundamental scale suppression previously introduced, one can estimate the least number of extra dimensions such that the 4D gravity is preserved. As in any braneworld model, the idea is to assume that 4D gravity is an effective theory of a fundamental theory with extra dimensions. The basic assumptions of the model are summarized below:
\begin{itemize}
\item The length of the brane is  neglected;
\item All extra dimensions are assumed to have equal size $L$;
\item It is postulated that only gravity propagates in the bulk;
\item A compactification for the extra dimensions is chosen \textit{a priori}.
\end{itemize}
For the sake of simplicity, it is usually assumed the toroidal compactification. It leads to the volume $V_d\sim (2\pi L)^d$.  Using this volume  on the equation \eqref{scalesupression} reads
\begin{eqnarray}\label{lmp}
L \sim \frac{1}{{m}_{4+d}}\left(\frac{{m}_p}{{m}_{4+d}}\right)^{2/d}.
\end{eqnarray}
If one assumes that ${m}_{4+d} \sim  10 $ TeV,  fixing  ${m}_p \sim 10^{16}$ TeV  the compactification radius, depending on $d$, can be estimated
\begin{eqnarray}
L \sim 10^{-1}\text{TeV}^{-1}\left(10^{15}\right)^{2/d}=10^{30/d-1}\;\text{TeV}^{-1},
\end{eqnarray}
or
\begin{eqnarray}
L \sim  10^{30/d}10^{-17}cm.
\end{eqnarray}

\begin{enumerate}
\item $d=1 \Rightarrow L \sim 10^{13}\,cm$, which is definitely inconsistent
with the Newton law well-established at such distances. 
\item  $d=2 \Rightarrow L \sim 0.1 \,mm$. That is the present experimental bound for gravitational experiments \cite{Csaki:2004ay, Cheng:2010pt, Adelberger:2003zx}.
\end{enumerate}
Alternatively, one could assume only one extra dimension whose length is $L \sim 0.1 \,mm$ and from the Eq. \eqref{lmp} find $m_{5}\sim 10^6$ TeV, which is far beyond the standard model energy scale. Thus not a particularly interesting result.

Motivated by the matching of the fundamental and standard model scales, the original assumption made for $m_{4+d}$  was $m_{4+d}\sim$ TeV. In this case the possibilities $d<3$ are ruled out by deviations from the inverse square law, whereas $d=3$ gives $L\sim 10^{-6}mm$.

Several simplifying assumptions were made above, one of them or more could be easily lifted. This would change technical details and/or results, leaving
conceptual foundations of the approach intact. However, a weakness of such model is that the assumptions were driven by simplicity or fine tuning.

\subsection{Warped Extra Dimensions}\label{warp_ed}

The Randall-Sundrum (RS) braneworld is the seminal warped extra dimensional model. It is one of the most studied braneworld model \cite{Csaki:2004ay}. Many  generalizations have been developed since it was proposed in 1999 \cite{Randall:1999ee,Randall1999}. There are two RS models, RS-1 and RS-2, with one and two branes respectively. Unlike the ADD model, there is no need to assume a compact extra dimension here. The model structure makes the extra dimension, even being infinity, have an finite effective length, related to the bulk curvature. This is, in fact, the picture of the model known as RS-1 \cite{Randall1999,Maartens:2010ar}.  The RS  models can be summarized as follows: one starts by assuming branes in a 5D bulk. The gravitational effects of the brane tensions are balanced by a bulk
cosmological constant, which acts to ``squeeze” the gravitational field closer to the brane and makes the brane cosmological constant to vanish. Thus our Universe will be seem as static and flat for an observer on the brane. The price to pay for this fine tuning is a highly curved 5D background.

One starts to build up the RS-2 model departing from the 5D Einstein-Hilbert action

\begin{equation}\label{5daction}
^{(5)}S_{EH}= \int d^4xdy \sqrt{-{}^{(5)}{g}}\left(m_5^3{}^{(5)}{R}-\Lambda_5\right).
\end{equation}
Besides the bulk being curved, since we want the Lorentz invariance to be preserved on the brane, we have to assume that the induced metric is proportional to the
 Minkowski metric, while the components of the 5D metric depend only on the fifth coordinate $y$. The ansatz for a general metric with these properties is given by \cite{Randall:1999ee,Randall1999}

\begin{equation} \label{ans}
g_{AB}(y)=e^{-2\Phi(y)}\eta_{\mu \nu}\delta^\mu_A \delta^\nu_B+ \delta^5_A\delta^5_B.
\end{equation}


The explicit form of the function $\Phi(w)$ is given by the field equations, which also impose a negative 5D cosmological constant. Thus the bulk must be anti de-Sitter. After placing the two opposite tension branes at $y=0$ and $y=L$ and using the field equations one finds
\begin{equation}
\Lambda_5=-\frac{1}{6}\frac{\sigma^2}{{m}_5^3},
\end{equation}
where $\sigma$ is the brane tension and ${m}_5$ the fundamental scale. In addition,
\begin{align}
\Phi(y)=\sqrt{-\frac{\Lambda_5}{6m_5^3}}|y| \equiv k|y|
\end{align}
and the metric is then given by
\begin{equation} \label{ans2}
g_{AB}(w)=e^{-2k|y|}\eta_{\mu \nu}\delta^\mu_A \delta^\nu_B+ \delta^5_A\delta^5_B.
\end{equation}
The extra dimension has length $L$ of the order of the Planck scale, thus the KK expansion can be applied. It also has $\mathbb{Z}_2$ symmetry, meaning the identification $y\leftrightarrow -y$ and $y+L \leftrightarrow y-L$.

One of the branes, the positive-tension brane, carry fundamental scale  and is called Planck brane. The standard model fields
are assumed to be confined on the negative tension brane, the TeV brane. Due to the negative exponential warping
factor, the effective Planck scale on the TeV brane (the one at $y=L$) is given by
\begin{align}
m_p^2=m_5^3\ell\left[e^{2L/\ell}-1\right]
\end{align}
where $\ell$ is the anti de-Sitter curvature radius. It gives a new approach to deal with the hierarchy problem. The RS-1 model is very similar to the RS-2, the main difference is the effective presence of only one brane, due to the assumption $L\to \infty$. %

\subsection{Effective Field Equations}\label{brane}


Numerous generalizations have appeared after the seminal paper by Lisa Randall and Raman Sundrum. In such scenario an approach to find effective 4D field equations from the original 5D would be very welcome. That is exactly the proposal of the formalism known as Shiromizu–Maeda–Sasaki (SMS) \cite{Shiromizu:1999wj}. By using the Gauss-Codazzi equations and the $\mathbb{Z}_2$ symmetry they figure out the corrections to the 4D field equation that emerge as consequence of the bulk curvature. Such formalism is useful, for example,  if one aims to find braneworld black hole solutions or any non flat solution on the brane. A sketch of the formalism is given as below. For details we refer to the original paper \cite{Shiromizu:1999wj} or braneworld reviews  \cite{Maartens:2010ar}.

At first let us write down the 5D field equations
\begin{align}\label{ein5}
{}^{(5)}{G}_{_{AB}}=-\Lambda_5{}^{(5)}{g}_{_{AB}}+{\kappa}_5^2\left[{}^{(5)}{T}_{_{AB}}+\left({T}_{_{AB}}-\sigma\right)\delta(y)\right].
\end{align}
where ${T}_{_{AB}}$ and $\sigma$ are the energy-momentum tensor and the brane tension (energy density) respectively, both localized on the brane. Now take $y$ as a coordinate normal to the brane, which we assumed to be placed at $y=0$. Thus, from the unit normal vector $n^{_{A}}$  we have $n_{_{A}}dx^{_{A}}=dx^4=dy$. The 5D metric given in terms of the induced metric on surfaces normal to $n^A$ reads
\begin{align}
^{(5)}{g}_{_{AB}}={g}_{_{AB}}+n_{_{A}}n_{_{B}}, \qquad  ds^2=g_{\mu \nu}(x^\alpha,y)dx^\mu dx^\nu+dy^2.
\end{align}
The Gauss equations, which relate extrinsic and intrinsic curvatures, in the present case give the Riemann tensor projected on the brane\footnote{$K_{_{A[C}}K_{_{D]B}}\equiv K_{_{AC}}K_{_{DB}}-K_{_{AD}}K_{_{CB}}$.},
\begin{align}\label{gauss}
{R}_{_{ABCD}}={}^{(5)}{R}_{_{EFGH}}{g}_{_{A}}^{_{\;\;E}}{g}_{_{B}}^{_{\;\;\;F}}{g}_{_{C}}^{_{\;\;G}}{g}_{_{D}}^{_{\;\;H}}+2K_{_{A[C}}K_{_{D]B}}.
\end{align}
The Codazzi equation, in contrast, relates the extrinsic curvature tensor to the bulk Ricci tensor as\footnote{$K=K^A_{\;\;A}$.} 
\begin{align}\label{codazzi}
\nabla_{_B}K_{^{\;\;A}}^{_{B}}-\nabla_{_A}K={}^{(5)}{R}_{_{BC}}{g}_{_{A}}^{_{\;\;B}}n^{_{C}},
\end{align}
with the extrinsic curvature tensor defined by $K_{_{AB}}=\frac{1}{2}\lie g_{_{AB}}=g_{_{A}}^{\;\;_C}{\nabla}_{_C}n_{_B}$, such that $K_{AB}n^B=0$. Now the bulk Riemann tensor is decomposed in terms of the bulk  Ricci, Weyl and metric tensors, as well as the Ricci scalar 
\begin{align}
{}^{(5)}{R}_{_{ABCD}}={}^{(5)}{C}_{_{ABCD}}+\frac{2}{3}\left({}^{(5)}{g}_{_{A[C}}{}^{(5)}{R}_{_{D]B}}-{}^{(5)}{g}_{_{B[C}}{}^{(5)}{R}_{_{D]A}}\right)-\frac{1}{6}{}^{(5)}{g}_{_{A[C}}{}^{(5)}{g}_{_{D]B}}{}^{(5)}{R}.
\end{align}
By contracting the Eq. \eqref{gauss} and replacing in Eq. \eqref{ein5} follows the effective field equations on the brane
\begin{align}\label{efffieldeq}\nonumber
G_{\mu \nu}&=-\frac{1}{2}\Lambda_5g_{\mu \nu}+\frac{2}{3}{\kappa}_5^2\left[{}^{(5)}{T}_{_{AB}}g_\mu^{\;_A}g_\nu^{\;_B}+\left({}^{(5)}{T}_{_{AB}}n^{_A}n^{_B}-\frac{1}{4}{}^{(5)}{T}\right)g_{\mu \nu} \right]\\ \nonumber
&+KK_{\mu \nu}-K_\mu^\alpha K_{\alpha \nu}+\frac{1}{2}\left(K^{\alpha\beta} K_{\alpha \beta}-K^2\right)g_{\mu\nu}-\mathcal{E}_{\mu\nu}.
\end{align}
The tensor $\mathcal{E}_{\mu\nu}={}^{(5)}{C}_{_{ABCD}}n^{_{C}}n^{_{D}}g_\mu^{\;_A}g_\nu^{\;_B}$ depicts the projection of the Weyl tensor orthogonal to $n^A$. The extrinsic curvature is related to the energy-momentum tensor on the brane and follows from the Israel junction conditions \cite{Israel:1966rt}
\begin{align}
[g_{\mu \nu}]=0, \qquad [K_{\mu \nu}]=-{\kappa}_5^2\left[ T_{\mu \nu}+\frac{1}{3}\left(\sigma- T\right) g_{\mu \nu}\right],
\end{align}
where the bracket denotes $[X(y)]\equiv\lim_{\epsilon\to 0}X(y+\epsilon)-X(y-\epsilon)\equiv X^+-X^-$. Due to the bulk $\mathbb{Z}_2$ symmetry, normal vectors pointing to opposite sides on the brane, when placed at the same point are, up to the sign, equals. This fact on the Eq. \eqref{codazzi} results that the extrinsic curvature, also on opposite sides, are equals up to the sign $K_{\mu \nu}^-=-K_{\mu \nu}^+$, hence $[K_{\mu \nu}]=2K_{\mu \nu}^+\equiv2K_{\mu \nu}$. Thus the extrinsic curvature is completely determined by  the energy-momentum tensor and the brane induced metric
\begin{align}\label{cext}
K_{\mu \nu}=-\frac{{\kappa}_5^2}{2}\left[ T_{\mu \nu}+\frac{1}{3}\left(\sigma- T\right)g_{\mu \nu}\right].
\end{align}
It gives to the contractions of $K_{\mu \nu}$ in the equation \eqref{efffieldeq} the meaning of a second order terms of the energy-momentum tensor on the brane. By combining the equations \eqref{efffieldeq} and \eqref{cext} one finds \cite{Shiromizu:1999wj,Maartens:2010ar},
\begin{align}\label{eff5d}
G_{\mu\nu}=-\Lambda_4g_{\mu\nu}+k_4^2T_{\mu\nu}+6\frac{\kappa_4^2}{\sigma}\pi_{\mu\nu}+4\frac{\kappa_4^2}{\sigma}\tau_{\mu\nu}-\mathcal{E}_{\mu\nu}.
\end{align}
The term $\pi_{\mu\nu}$ contains the higher order terms of the brane energy-momentum tensor and $\tau_{\mu\nu}$ is the bulk energy-momentum tensor. The projection of the bulk Weyl tensor on the brane $\mathcal{E}_{\mu\nu}$ encodes corrections from 5D
graviton effects. For an observer on the brane the energy-momentum corrections in $\pi_{\mu\nu}$ are local, whereas $\mathcal{E}_{\mu\nu}$ contains the Kaluza-Klein corrections and acts as a non-local source. These nonlocal corrections cannot be determined purely from data on the brane. 

By identifying the equations \eqref{efffieldeq} and \eqref{eff5d} follows the expression correlating the 4D and 5D cosmological and coupling constants being 
\begin{align}
\kappa_4^2&=\frac{1}{6}\sigma\kappa_5^4,\\
\Lambda_4&=\frac{\kappa_5^2}{2}\Big(\Lambda_{5}+\frac{1}{6}{\kappa}_5^{2}\sigma^{2}\Big).\label{la4la5}
\end{align}

The effective field equations are even more intricate than the Einstein equations due to the additional terms on Eq. \eqref{eff5d}. However, a practical way of finding the brane metric on the brane neighbourhood is to expand the metric around the brane. By performing a Taylor series around $y=0$ using the Lie derivative one finds
\begin{align}
g_{\mu\nu}(x^\alpha,y)=g_{\mu\nu}+\sum_{n=1}^\infty\lie^{(n)} g_{\mu\nu}|y|^n,
\end{align}
with $g_{\mu\nu}(x^\alpha,0) \equiv g_{\mu\nu}$. Explicit calculations leads to
\begin{footnotesize}
 \begin{align}
\hspace*{-0.4cm}{\;\;\;\;}g_{\mu\nu}(x^\alpha,y)&\!\!=\!\! g_{\mu\nu}-\kappa_5^2\left[
T_{\mu\nu}\!+\!\frac{1}{3}(\sigma-T)g_{\mu\nu}\right]|y|  +\label{tay} \\ 
&+\left[\frac{1}{4}\kappa_5^4\!\left(
T_{\mu\alpha}T^\alpha_\nu -{\cal E}_{\mu\nu} +\frac{2}{3} (\sigma-T)T_{\mu\nu}
\right)\! +\!\left( \frac{1}{36}
\kappa_5^4(\sigma-T)^2-\frac{\Lambda_5}{6}
\right)\!g_{\mu\nu}\right] y^2 + {\;}\nonumber\\
& +\left.\Bigg[2K_{\mu\beta}K^{\beta}_{\;\,\alpha}K^{\alpha}_{\;\,\nu} - ({\cal E}_{\mu\alpha}K^{\alpha}_{\;\,\nu}+K_{\mu\alpha}{\cal E}^{\alpha}_{\;\,\nu})-\frac{1}{3}\Lambda_5K_{\mu\nu}-\nabla^\alpha{\cal B}_{\alpha(\mu\nu)} + \frac{1}{6}
\Lambda_5\left(K_{\mu\nu}-g_{\mu\nu}K\right)+
 \right.\nonumber\\
&\left. +K^{\alpha\beta}R_{\mu\alpha\nu\beta}+ 3K^\alpha{}_{(\mu}{\cal
E}_{\nu)\alpha}-K{\cal E}_{\mu\nu}+\left(K_{\mu\alpha}K_{\nu\beta}
-K_{\alpha\beta}K_{\mu\nu}\right)K^{\alpha\beta}-\frac{\Lambda_5}{3}K_{\mu\nu}\Bigg]\;\frac{|y|^3}{3!} + \cdots \right.  \nonumber
 \end{align}
\end{footnotesize}%
where ${\cal B}_{\mu\nu\alpha}=g_{\mu}^{\;\;\rho}g_{\nu}^{\;\;\sigma}C_{\rho\sigma\alpha\beta}n^\beta$. Such expansion was analysed on \cite{Maartens:2010ar} up to the second order. In \cite{daRocha:2012pt} the general case was obtained up to the forth order, including a variable brane tension. Moreover, such expansion is the fundamental 
of the black strings study near the brane. In fact, 
bounds on ultra high energy cosmic rays have been 
obtained in \cite{Anjos:2015coa}, and this expansion 
was also used in Refs. \cite{Bazeia:2013bqa,Casadio:2013uma}
to study the bulk regularity and black strings associated to the McVittie solution and the minimal geometric deformation, respectively, as well as in the context of dark dust and dark radiation \cite{Herrera-Aguilar:2015koa}. In the case of variable brane tension there is no correction to the first and the second order, it happens only beyond the third order, whose correction reads
\begin{align}
-\frac{2\kappa_5^2}{3}\left((\nabla^\alpha\nabla_\alpha\sigma)g_{\mu\nu}-(\nabla_{(\nu}\nabla_{\mu)}\sigma)
\right).\label{magnetico}
\end{align}
The fourth order correction, though, is given by 
\begin{footnotesize}
\begin{align}
&6 \left[(\Box\sigma)K_{(\mu\tau}{E}_{\nu)}^\tau - \nabla^\alpha((\nabla_{(\mu}\sigma)\, {E}_{\nu)\alpha})\right]
+2\left(K + \frac{7}{3}\kappa_5^2\right)\left[(\Box\sigma)K\,K_{\mu\nu}-\nabla^\alpha((\nabla_{(\mu}\sigma)\, K\,K_{\alpha\vert\nu)})\right]\nonumber\\
&- \frac{1}{3}\kappa_5^2\,\left[\Box(\Box\sigma)g_{\mu\nu}-\nabla_{(\nu}\nabla_{\mu)}(\Box\sigma)\right]+\left(\frac{1}{3}\kappa_5^2+2 K\right)[(\Box\sigma){E}_{(\mu\nu)} - \nabla^\alpha\left((\nabla_{(\mu}\sigma)\, {E}_{\nu)\alpha}\right)]\nonumber\\
&+ \frac{1}{3}\kappa_5^2 \left[(\Box\sigma)
(R_{\mu\nu}+K_{(\mu\vert\tau}K_{\nu)\beta} 
K^{\tau\beta} - K^2\,K_{(\mu\nu)})-
\nabla^\alpha((\nabla_{(\mu}\sigma)
(R_{\alpha\vert\nu)}  - 
K_{\alpha\tau}K^{\tau}_{\nu)} - K
\,K_{\alpha\nu)}))\right]
\nonumber\\
&-2 K^{\tau\beta}\left[(\Box\sigma)R_{(\mu\vert\tau\vert\nu)\beta} - \nabla^\alpha\left((\nabla_{(\mu}\sigma)\, R_{\alpha\tau\vert\nu)\beta}\right)\right]
+\left(2\,K^2-\frac{1}{3}\Lambda_5 \right)[(\Box\sigma)g_{\mu\nu}-\nabla_{(\nu}\nabla_{\mu)}\sigma] \label{magnetico1}
\end{align}
\end{footnotesize}%
The above expansion is particularly useful for black hole like solutions, in which one construct a Taylor expansion of the metric, near the horizon, along the extra dimension \cite{daRocha:2012pt,Bazeia:2014tua}.

\section{Thick Branes}\label{thick_b}

An alternative approach for the thin braneworld models previously introduced consists in the so called thick brane models. In those models the brane itself comes from topological defects. An advantage of this approach is that the brane does not appear ad hoc, but emerges from the field equations. Similarly to the warped extra dimension models, when the 4D space-time is assumed to be flat the 5D metric is given by
\begin{equation}\label{tmetric}
ds^2=e^{2A(y)}\eta_{\mu\nu}dx^\mu dx^\nu-dy^2
\end{equation}  
where again $e^{2A(y)}$ is called warp factor. However, unlike the RS models, here the warp factor is a smooth function of the extra dimensional coordinate. A comparison of both warp factors is shown in Fig. \ref{fig_1}.
\begin{figure}[h!]
\centering
\includegraphics[scale=0.2]{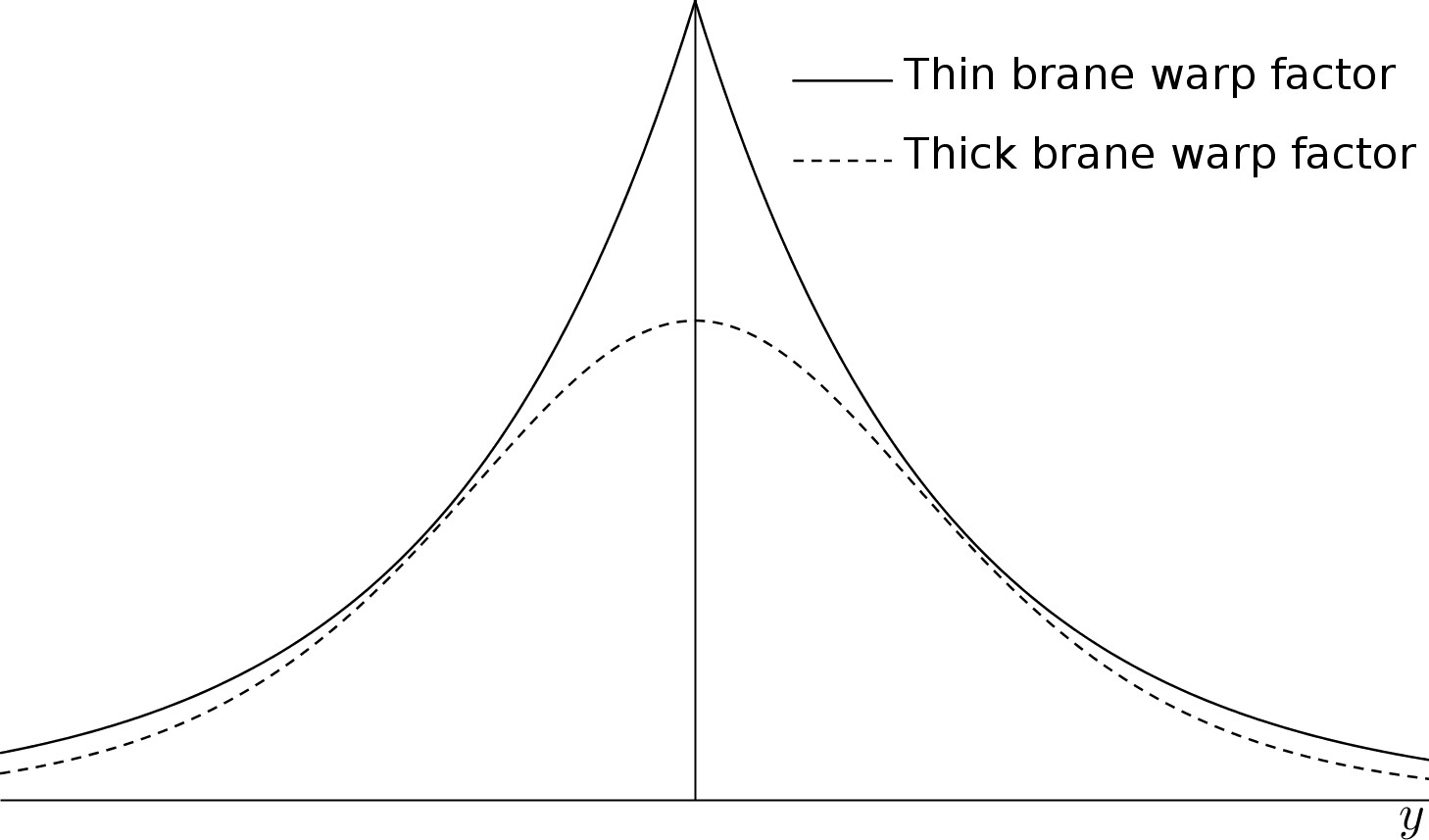}
\caption{Typical warp factor profile for thin and thick brane models.}
\label{fig_1}
\end{figure}

Our aim here is to apply the thick brane approach to the 5D Friedman-Robertson-Walker (FRW) solution, in the Section \ref{thickfrw}. Let us first introduce some basic underlying features of thick branes though.  
\subsection{Topological Defects and Domain Wall Branes}

Dynamically generated thick branes are described by topological defects. A topological defect is defined as a stable solution of the field equations which cannot be continuously deformed into the vacuum solution. In other words, are solutions localised with non-dispersive energy density which travel undistorted in shape \cite{Rajaraman:1982is, Manton:2004,1994csot.book.....V}. They encompass a popular branch of nonlinear science and have a wide range of applications \cite{Presnov2014,Thom:1968,Bazeia:2012qh,Bazeia:2016pvk,Bazeia:2016est,
Bazeia:1995en,Bazeia:2002xg, Liu:2013kxz,Duan:2007hz,Landim:2013dja,Alencar:2010hs,Mendes:2010zza}. In particular we could mention cosmology, high energy physics and condensed matter physics, where they can be used to describe
phase transitions in the early universe and contribute to pattern formation in nature \cite{Sornborger:2000ka,Gorbunov:2011zz,Basar:2008im}. Here we are interested on topological defects generated by real scalar fields \cite{HoffdaSilva:2012em,Bazeia:2008zx,Bazeia:2010vb,Cruz:2013uwa,Cruz:2014eca}.

Consider the action for a real scalar field in 1+1 dimensions

\begin{equation}\label{scalaraction}
S=\frac{1}{2}\int d^2x[\partial^\nu \phi \partial_\nu \phi - 2V(\phi)], \qquad \nu=0,1
\end{equation}
with potential given by
\begin{align}\label{pottop}
V(\phi)=\frac{\lambda}{4}(\phi^2-v^2)^2, \qquad v=\frac{\mu}{\sqrt{\lambda}}.
\end{align}
The above action is invariant under the transformation $\phi\mapsto-\phi$, a $\mathbb{Z}_2$ symmetry. However, at low energy scales, this symmetry is broken due to the discreteness of the vacuum states.

\begin{figure}[h!]
\centering
\includegraphics[scale=0.2]{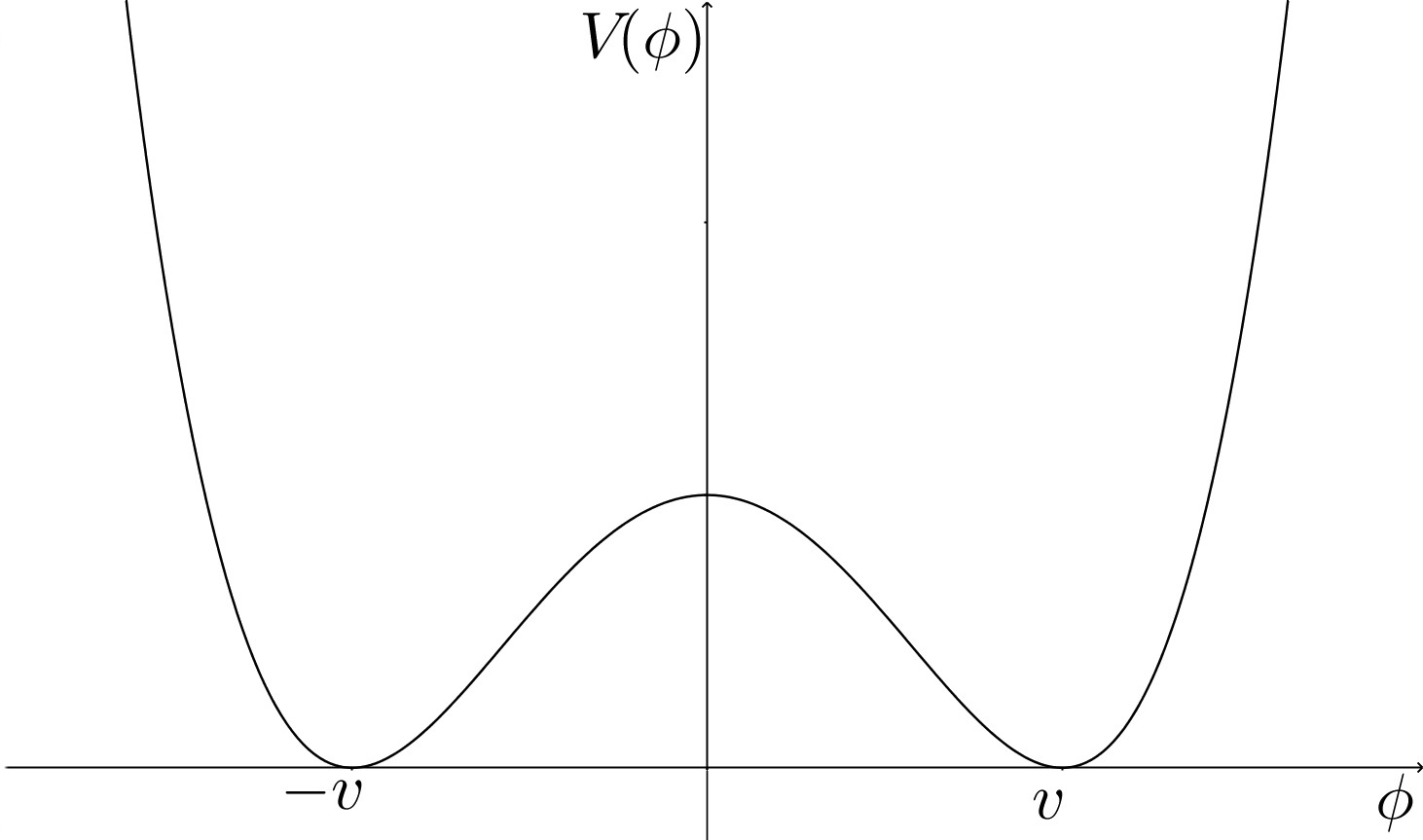}
\caption{Potential profile of the Eq. \eqref{pottop}. The $\mathbb{Z}_2$ symmetry is broke at low scaled}
\label{fig_1}
\end{figure}

In addition to the trivial vacuum solutions, the two solutions to the field equations coming from the Eq. \eqref{scalaraction} are called kink and anti-kink. Their asymptotically behaviour are given by $\displaystyle \lim_{x\to \pm\infty}\phi(x)=\pm v$ and $\displaystyle \lim_{x\to \pm\infty}\phi(x)=\mp v$ respectively. Thus, kink and anti-kink are solutions that interpolate the vacuum solutions $\pm v$. Their asymptotically behaviour distinguish the solutions. One interesting fact is that there is no way to connect the solutions with a continuous and energetic allowed transformation. Any intermediate solution would have a non vacuum value  when $x\to \infty$, thus with infinite energy.\footnote{ The precise way to say that the solution classes are disjoint and inequivalent is by using the concept of homotopy. A homotopy between fields $f,g:X\to Y$ is a continuous map defined by
$$H:[0,1]\times X \to Y, \;\; \mbox{ such that } H(0,x)=f(x)\;\mbox{ and }\;H(1,x)=g(x).$$
Intuitively, two fields are homotopically equivalent if one can be continuously deformed into the other.}   

The finiteness of the solution can be guessed from the scalar field energy
\begin{align}
E=\int dx\left[ \frac{1}{2}\left(\frac{d\phi}{dx}\right)^2+\frac{\lambda}{4}(\phi^2-v^2)^2\right],
\end{align}
where both terms on the integrand have to vanish quick enough when $x \to \pm \infty$.


\begin{figure}[h!]
\centering
\includegraphics[scale=0.2]{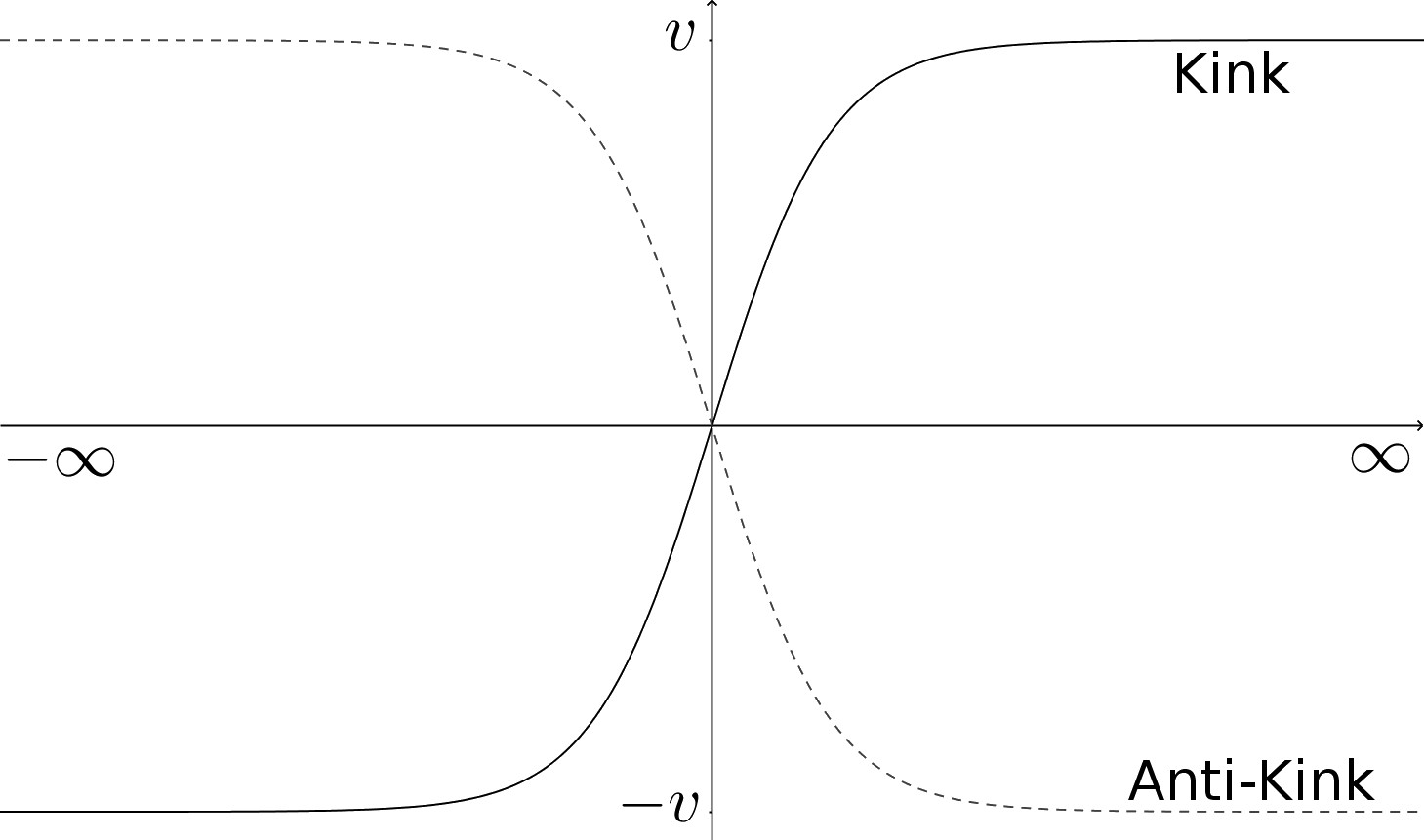}
\caption{Kink and anti-kink profile.}
\label{fig_1}
\end{figure}

Let us now look at the field equations. By varying the action \eqref{scalaraction} one finds 
\begin{align}
\phi''=\frac{\partial V}{\partial \phi}.
\end{align}
Multiplying both sides by $\phi'$ and integrating on $x$ it follows
\begin{align}
\frac{1}{2}\phi'^{2}- V=\epsilon_0,
\end{align}
where $\epsilon_0$ is the constant of integration. From the contour conditions, the field and potential should vanish when $x \to \pm \infty$, thus $\epsilon=0$. As mentioned, besides the vacuum solutions $v$ and $-v$, the two inequivalent solutions are the kink [positive sign] and anti-kink [negative sign], explicitly given by 
\begin{align}
\phi_k(x)=\pm\frac{\mu}{\sqrt{\lambda}}\tanh\left(\frac{\mu}{\sqrt{2}}x\right).
\end{align}
Those are simple solutions used for guiding when we have to face more complicated field equations.

Models of 3+1 or 4+1-dimension can also have the same solutions for the field equations, depending on only one coordinate. Those solutions can be represented by surfaces that concentrate the energy and separate vacuum solutions, called domain walls. The existence of domain walls is expected in models with discrete vacuum states \cite{Rajaraman:1982is,Dzhunushaliev:2009va}.

\subsection[Thick Brane FRW Solution]{Thick Brane FRW Solution\protect\footnote{We have published the results of this section in Ref. \cite{Bernardini:2014vba}.}}\label{thickfrw}
%
%

Our aim here is to provide  an approach for spherically symmetric thick brane cosmology.
By exploring the framework of isotropic thick branes \cite{Ahmed2012,Ahmed2013, Ahmed2014}, one can realize that the separability of the warp factor is fundamental in order 
to explicitly describe the time-dependent solutions.
It is noway obvious that, for spherically symmetric thick braneworlds, the  warp factors to be considered in this section -- that are dependent on time, extra dimension, and radial coordinate on the brane -- should be separable in the context of solving the equations of motion. 
Likewise, it suggests that it might be hopeful to find time-dependent soliton solutions leading to non-separable forms of the warp factor \cite{Giovannini2007,George2008,Kadosh2012}.
Separable solutions are normally discussed in the framework of thin braneworld models that are rather unnatural in case of thick defects, since the brane thickness  must fulfill  the limits $2.0\times 10^{-19} {\rm m}\lesssim \Delta\lesssim 44\mu{\rm m}$ \cite{Kapner2006}, having thus a minimal thickness \cite{HoffdaSilva:2012em}.
In fact, thick brane cosmology has been widely 
discussed in \cite{Giovannini2007,George2008,Kadosh2012,German:2012rv}, further regarding other type of warp factors \cite{Gogberashvili2012,Carrillo-Gonzalez:2013exa,Bernardini:2013tba} and tachyonic solutions, with a decaying warp factor that enables localization of 4D gravity as well as other matter fields \cite{German:2012rv}. Some applications in the thin brane limit have been provided, e. g., in \cite{daRocha:2012pt}.

Departing from a general 5D spherically symmetric warped spacetime, our purpose is to solve the coupled system of gravitational field equations and the equations of motion for a scalar field.
The procedure introduced in the following results into an explicit formula for both the extra dimension-dependent part of the warp factor and the spherically symmetric time-dependent component.
The warp factors for flat, closed, and open spacetimes are obtained and discussed, and the properties of Hubble type parameter are also investigated.
Our analysis results into deploying the fundamentals of thick braneworld cosmology with time dependent spherically symmetric warp factors, exclusively departing from the field equations. 

 
To provide a generalization of the successful achievements on braneworld cosmology in the thin brane paradigm \cite{Binetruy:1999ut, Binetruy:1999hy} as well as in the thick brane scenario \cite{Ahmed2013,Ahmed2014}, one considers 5D spacetimes for which the metric assumes
the following form:
\begin{align}
ds^2&=a^2(t,r,y)g_{\mu\nu}dx^\mu dx^\nu+dy^2, \label{metric}
\end{align}
where $g_{\mu\nu}$ is the FRW metric given by 
\begin{eqnarray}
g_{\mu\nu}dx^\mu dx^\nu = -dt^2 + \left[ \frac{dr^2}{1-kr^2} + r^2(d\theta^2 +\sin^2\theta d\varphi^2)\right], \nonumber
\end{eqnarray}
with $k$ denoting the curvature parameter assuming the values $-1$, 0 and $1$, leading respectively to an open, a flat or a closed Universe. 
The function $a(t,r,y)$ plays simultaneously the role of conformal scale factor and brane warp factor. The 4D solutions are sourced by the bulk scalar field. 

The action for scalar field in the presence of 5D gravity is given by\footnote{All the tensor and scalars here are the 5D ones, no effective 4D quantities are used in the present section. Thus there is no reason to use ${}^{(5)}$ to distinguish 4D and 5D quantities.}
\begin{equation}
S_{5D}=\int d^4xdy \sqrt{-{ g}}\left(-\frac{1}{2}{ g}^{MN}\nabla_{M}\phi\nabla_{N}\phi-V(\phi)+2m_5^{3}R\right).
\label{action}
\end{equation}
For the above prescribed scenario, one assumes that the scalar field, $\phi$, depends exclusively on time and upon the extra coordinate $y$.

The field equations and the equation of motion for $\phi$ resulting from the above action \eqref{action} are provided by 
\begin{eqnarray}
\nabla^2\phi-\frac{dV}{d\phi}&=&0, \label{eineq2}\\
R_{MN}-\frac{1}{2}{ g}_{MN}R&=&\frac{1}{4m_5^{3}}T_{MN},\label{eineq1}
\end{eqnarray}
where $\nabla^2$ is the 5D Laplacian operator, and the energy-momentum tensor, ${}^{(5)}T_{MN}$, for the
scalar field $\phi(t,y)$ reads
\begin{equation}
T_{MN}=-{ g}_{MN}\left(\frac{1}{2}(\nabla\phi)^{2}+V(\phi)\right)+\nabla_{M}\phi\nabla_{N}\phi\,.
\nonumber\label{emt}
\end{equation}
In particular, the energy-density ($T_{00}$) is implied by $\phi(t,y)$ and localized near $y = 0$. Moreover, the  equation of motion for the scalar field is expressed by
\beq
\phi''-\frac{1}{a^2}\ddot\phi+\frac{4{a'}}{a}\phi' -\frac{2\dot a}{a^3}\dot\phi=\frac{dV}{d\phi}\,, \nonumber
\eeq
where one denotes 
$\frac{\partial f}{\partial t} = \dot{f},\; \frac{\partial f}{\partial r} = \bar{f},$ and $ \frac{\partial f}{\partial y} = {f'}
$, for any scalar function $f$ hereupon. 
By assuming a static scalar field scenario, the components of the Einstein tensor
are given by the following expressions:
\begin{footnotesize}
\begin{eqnarray}\nonumber
G_{00}&=&a^{2}\left\{3\left[ \frac{\dot{a}^{2}}{a^{4}}-\left( \frac{a''}{a}+\frac{a'^{2}}{a^{2}}\right)+\frac{k}{a^{2}}\right] +(1-kr^{2})\left[\frac{\bar{a}^{2}}{a^{4}}-\frac{2\bar{\bar{a}}}{a^{3}}-\frac{6\bar{a}}{a^{3}r} \right]+\frac{2\bar{a}}{a^{3}r} \right\},\\\nonumber 
G_{11}&=&-g_{11}\left\{\left[ \frac{1}{a^{2}}\left(\frac{2\ddot{a}}{a}-\frac{\dot{a}^{2}}{a^{2}}\right)-3\left( \frac{a''}{a}+\frac{a'^{2}}{a^{2}}\right)+\frac{k}{a^{2}}\right] -(1-kr^{2})\left[\frac{3\bar{a}^{2}}{a^{4}}+\frac{4\bar{a}}{a^{3}r} \right] \right\},\\\nonumber
G_{22}&=&-g_{22}\left\{\left[ \frac{1}{a^{2}}\left(\frac{2\ddot{a}}{a}-\frac{\dot{a}^{2}}{a^{2}}\right)-3\left( \frac{a''}{a}+\frac{a'^{2}}{a^{2}}\right)+\frac{k}{a^{2}}\right] -(1-kr^{2})\left[\frac{2\bar{\bar{a}}}{a^{3}}-\frac{\bar{a}^{2}}{a^{4}}+\frac{4\bar{a}}{a^{3}r} \right]+\frac{2\bar{a}}{a^{3}r} \right\},\\ \nonumber
G_{33}&=&{g_{33}}G_{22}/{g_{22}},\\ \nonumber
G_{55}&=&3\left(\frac{2a'^{2}}{a^{2}}- \frac{\ddot{a}}{a^3}-\frac{k}{a^{2}}\right)+3(1-kr^{2})\left(\frac{\bar{\bar{a}}}{a^{3}}+\frac{3\bar{a}}{a^{3}r} \right)-\frac{3\bar{a}}{a^{3}r}, \\ \nonumber
G_{01}&=&2\left(2\frac{\bar{a}\dot{a}}{a^{2}}-\frac{\dot{\bar{a}}}{a} \right),\\ \nonumber
G_{05}&=&3\left(\frac{\dot{a}{a'}}{a^{2}}-\frac{\dot{{a}}'}{a} \right),\\ \nonumber
G_{15}&=&3\left(\frac{\bar{a}{a'}}{a^{2}}-\frac{{\bar{a}}'}{a} \right).
\end{eqnarray}
\end{footnotesize}

The Einstein equations $G_{MN}=T_{MN}$, adopting $4M_*^{3} = 1$, can be used to find the 
form of the warp factor. 
By separating the variables $a(t,r,y)=A(t,r)B(y)$, the Einstein equation $G_{01}=T_{01}=0$ yields 
\begin{eqnarray}
0&=&
2\frac{\bar{A}}{A}-\frac{\dot{\bar{A}}}{\dot{A}}=\partial_r\ln A^{2}- \partial_r \ln \dot{A}\nonumber\\
&\Rightarrow & \ln \frac{A^{2}}{\dot{A}}=T(t) \Leftrightarrow \dot{A}=A^{2}e^{-T}\,,\label{dota}\end{eqnarray}
or
\begin{eqnarray}
0&=&2\frac{\dot{A}}{A}-\frac{\dot{\bar{A}}}{\bar{A}}=\partial_t\ln \frac{A^{2}}{\bar{A}}\nonumber\\\label{bara}
&\Rightarrow & \ln \frac{A^{2}}{\bar{A}}=R(r) \Leftrightarrow \bar{A}=A^{2}e^{-R}\,,
\end{eqnarray}
implying that 
\begin{eqnarray}
 \label{dotbar}
 \dot{A}=\bar{A}e^{R-T}\,.
\end{eqnarray}
The expressions $\dot{A}^{2}=A^{4}e^{-2T}$ and $\bar{A}^{2}=A^{4}e^{-2R}$ follow from \eqref{dota}, \eqref{bara} and \eqref{dotbar}, and they imply that 
\begin{eqnarray}
\dot{\bar{A}}&=&
=2A^{3}e^{-(T+R)}\,,\nonumber\\
\ddot{A}&=&
=A^{2}e^{-T}\left(2Ae^{-T}-\dot{T}\right)\,,\label{aa0}\\
\bar{\bar{A}}&=&
A^{2}e^{-R}\left(2Ae^{-R}-\bar{R}\right)\,.\label{aa1}
\end{eqnarray}
One of the off-diagonal Einstein equations $G_{05}=T_{05}=\dot{\phi}\phi'$ yields
\begin{eqnarray}
\dot{\phi}\phi'&=&3
\left(\frac{\dot{A}B'}{AB}-\frac{\dot{A}B'}{AB} \right)=0 \;\;\;\Rightarrow\;\;\; \dot{\phi}=0\,,\nonumber
\end{eqnarray}
that means that the scalar field $\phi$ is time independent. Hence the components of energy-momentum tensor become
\begin{eqnarray}
T_{\mu \nu}&=&-g_{\mu \nu}\left(\frac{1}{2}\phi'^{2} + V(\phi) \right)\,, \qquad\quad
T_{55}=\frac{1}{2}{\phi'}^{2} - V(\phi)\,. 
\nonumber\end{eqnarray}
The explicit form of the components of the field equation for the metric ansatz \eqref{metric}
can be written for $k = 0,~\pm1$, by 
denoting the spatial curvature of the 4D homogeneous and isotropic
space-time for Minkowski, de Sitter and anti-de Sitter space, respectively.
The diagonal components (remembering that the component $_{22}$ equals $_{33}$) are respectively expressed as:
\begin{footnotesize}
\begin{eqnarray}\nonumber
\frac{1}{2}\phi'^2\!+\!V(\phi) &=&\frac{1}{B^{2}}\!\left\{\! -3\left(B{B'}\right)' \!+\!\frac{1}{A^{2}}\left[(1-kr^{2})\left(\frac{\bar{A}^{2}}{A^{2}}-4\frac{\bar{\bar{A}}}{A}-6\frac{\bar{A}}{Ar} \right)\!+\!2\frac{\bar{A}}{Ar}\!+\!3\frac{\dot{A}^{2}}{A^{2}}\!+\!3{k}\right]\right\},\nonumber\\
\frac{1}{2}\phi'^2\!+\!V(\phi) &=&\frac{1}{B^{2}}\!\left\{\! -3\left(B{B'}\right)' \!+\!\frac{1}{A^{2}}\left[(1-kr^{2})\left(3\frac{\bar{A}^{2}}{A^{2}}\!+\!4\frac{\bar{A}}{Ar} \right)\!+\!2\frac{\ddot{A}}{A}-\frac{\dot{A}^{2}}{A^{2}}\!+\!{k}\right]\right\},\nonumber\\
\frac{1}{2}\phi'\!^2+\!V(\phi) &=&\frac{1}{B^{2}}\!\left\{\! -3\left(B{B'}\right)' \!+\!\frac{1}{A^{2}}\left[(1-kr^{2})\left(2\frac{\bar{\bar{A}}}{A}-\frac{\bar{A}^{2}}{A^{2}}\!+\!4\frac{\bar{A}}{Ar} \right)\!+\!2\frac{\ddot{A}}{A}-\frac{\dot{A}^{2}}{A^{2}}\!+\!{k}\!+\!2\frac{\bar{A}}{Ar}\right]\right\},\nonumber\\
\frac{1}{2}\phi'^2-V(\phi)
 &=&-\frac{1}{B^{2}}\left\{ -6{B'^{2}} \!+\!\frac{3}{A^{2}}\left[(kr^{2}-1)\left(\frac{\bar{\bar{A}}}{A}\!+\!3\frac{\bar{A}}{Ar} \right)\!+\!\frac{\ddot{A}}{A} \!+\!{k}\!+\!\frac{\bar{A}}{Ar} \right]\right\}.\label{by}
 \end{eqnarray}
\end{footnotesize}
 Therefore, the equations for $\phi$ or $y$
 can be expressed as
\begin{eqnarray}
\phi''(y)+4\frac{{a'}}{a}{\phi'}(y)&=&\frac{dV}{d\phi},\nonumber\\B^2(y)\left[ 3\left( \frac{B''(y)}{B(y)}+{\frac{B'^2(y)}{B^2(y)}}\right)+\frac{1}{2}\phi'(y)+V(\phi)\right]&=&c_0,\nonumber\\
B^2(y)\left[ 6{\frac{B'^{2}(y)}{B^2(y)}}-\frac{1}{2}\phi'(y)+V(\phi)\right]&=&c_5\,,\label{c0c5}\end{eqnarray} for $c_0$ and $c_5$ separation constants, whereas the ones for $t$ and $r$ are summarized 
respectively by
\begin{eqnarray}
&&\dot{A}(t,r)=A^2(t,r)e^{-T(t)},\nonumber\\
&&\bar{A}(t,r)=A^2(t,r)e^{-R(r)},\label{a2a}\\\label{e00}
_{00}:&& \frac{1}{A^{2}}\left[(1-kr^{2})\left(\frac{\bar{A}^{2}}{A^{2}}-\frac{4\bar{\bar{A}}}{A}-\frac{6\bar{A}}{Ar} \right)+2\frac{\bar{A}}{Ar}\right]+\frac{3\dot{A}^2}{A^{4}}+\frac{3k}{A^2}=c_0,\\
\label{e11}
_{11}: &&\frac{1}{A^{2}}\left[(1-kr^{2})\left(\frac{3\bar{A}^{2}}{A^{2}}+\frac{4\bar{A}}{Ar} \right)\right]+\frac{2\ddot{A}}{A^3}-\left(\frac{\dot{A}}{A^{2}}\right)^2+\frac{k}{A^2}=c_0,\\\label{e22}
_{22}: &&\frac{1}{A^{2}}\left[(1-kr^{2})\left(\frac{2\bar{\bar{A}}}{A}-\frac{\bar{A}^{2}}{A^{2}}+\frac{4\bar{A}}{Ar} \right)+\frac{2\bar{A}}{Ar}\right]+\frac{2\ddot{A}}{A^3}-\frac{\dot{A}^2}{A^{4}}+\frac{k}{A^2}=c_0,\\\label{e55}
_{55}: &&\frac{3}{A^{2}}\left[(kr^{2}-1)\left(\frac{\bar{\bar{A}}}{A}+\frac{3\bar{A}}{Ar} \right)+\frac{\bar{A}}{Ar}\right]+\frac{3\ddot{A}}{A^3} +\frac{3k}{A^2} =c_5.
\end{eqnarray}
 The role of the bulk scalar field is to provide the cosmological constant on a brane as is clear from Eqs.(\ref{a2a}-\ref{e22}). 
By imposing $c_5 = \Lambda$ one has analogous cosmological implications for suitable limits, where the warp factor has no dependence on $r$, as in the thick brane cosmology with isotropic warp factor \cite{Ahmed2013,Ahmed2014}. In an isotropic thick braneworld the condition $c_5 = 2c_0=\Lambda$ holds \cite{Ahmed2013}. Nevertheless, in this scenario 
such two constants restrict further the form of the function $A(r,t)$, when the above equations are used, by the following 
 relationship:
\begin{eqnarray}
-15\left(\frac{kr^2-1}{c_1r}\right)^2-12\frac{(1-kr^2)^{3/2}}{c_1Ar^2}+6\frac{\sqrt{1-kr^2}}{c_1Ar^2}&=&2c_0 - c_5\,.\label{opop}
\end{eqnarray}
Note that this consistency equation is trivial if the 4D scale factor is independent of the radial coordinate, as in \cite{Ahmed2013}.
Now, by computing the difference of \eqref{e11} and \eqref{e22}, one obtains the equation 
$
(1-kr^{2})\bar{R}-\frac{1}{r}=0\nonumber
$ which has solution
\begin{equation}
R(r)=\ln \frac{c_1r}{\sqrt{1-kr^{2}}}\,, 
\end{equation} where $c_1$ is a constant 
of integration. 
Moreover, the solution for Eq.~(\ref{a2a}) 
is provided by (hereon one shall notice the index $k$ in order to denote the dependence on $k=0,\pm 1$ in the following expressions):
\begin{equation}\label{A1}
A_k(t,r)=\frac{c_1}{c_1Y_k(t)+f_k(r)}\,,
\end{equation}
\noindent where $Y_k(t)$ is a constant of integration with respect to the $r$ coordinate,
\begin{eqnarray}\label{fk}
f_k(r)&\equiv&\ln \frac{\sqrt{1-kr^{2}}+1}{r}-\sqrt{1-kr^{2}}\,,
\end{eqnarray} and $A_k(t,r)$ depends on $k=0,\pm 1$. It implies hence that \begin{equation}\label{aa2}
\frac{\dot{A}_k(t,r)}{A^2_k(t,r)}=-\dot{Y}_k(t)=e^{-T(t)}\,. 
\end{equation}

We can simplify Einstein equations using (\ref{bara}), (\ref{aa0}), (\ref{aa1}), and (\ref{aa2}) to make 
 Eq.\eqref{e00} --- that corresponds to the $_{00}$ component of the Einstein equations --- to read: 
\begin{equation}\label{eqY}
\dot{Y}_k^2=-kY_k^2+w_k(r)Y_k+z_k(r)\,,
\end{equation}
where $w_k(r)$ and $z_k(r)$ are respectively given by the following expressions:
\begin{eqnarray} \nonumber
w_k(r)&=&-\frac{1}{3}\left[2(1-kr^{2})e^{-R}\left(2\bar{R}-\frac{3}{r} \right)+2\frac{e^{-R}}{r}+\frac{6kf_k(r)}{c_1}\right]\,,\\ \nonumber
z_k(r)&=&-\frac{1}{3}\left\{(1-kr^{2})e^{-R}\left[-7e^{-R}+\frac{2f_k(r)}{c_1}\left(2\bar{R}-\frac{3}{r} \right) \right]+\frac{2f_k(r)}{c_1}\frac{e^{-R}}{r}+\right.\\
&&\left.+3k\frac{f_k^2(r)}{c_1^2} -c_0\right\}\,.
\end{eqnarray}
 
The solutions for Minkowski, anti-de Sitter and de Sitter spacetimes are respectively provided by:\begin{eqnarray}\label{y1}
Y^{\pm}_{-1}(t)&=&\frac{1}{4}e^{\pm(t\mp \alpha_{-1})}\left[\left( e^{\mp(t\mp \alpha_{-1})} -w_{-1}(r) \right)^2-4z_{-1}(r) \right],\\
\label{y0}
Y^{\pm}_0(t)&=&\alpha_0 \pm \sqrt{z_0(r)}t,\\
\label{ym1}
Y^{\pm}_{1}(t)&=&\frac{1}{2}\left[w_{1}(r) \pm\sqrt{w_{1}^2(r)+4z_{1}(r)}\sin(t+|\alpha_1|) \right]\,.
\end{eqnarray}
respectively for $k=-1, 0, 1$. The constant parameters $\alpha_0, \alpha_{\pm 1}$ are constants of integration.


When $f_k(r)=0$, then $A_k(t,r)= 1/Y_k(t)$, and one has the results from \cite{Ahmed2013} for thick brane cosmology, with $c_5 = 2c_0=\Lambda$, where $\Lambda$ denotes the 4D cosmological constant.
For the case explicitly provided by Eq.~(\ref{A1}), $A_k(t,r)$ indeed does not depend on the $r$ coordinate.
Firstly, it is evident that $A_k(t,r)=0$ when $r\rightarrow\infty$, as $f_k(r)$ diverges in this case.
However, it is not properly the useful case here.
For $k=0$, $A_k(t,r)$ is independent of $r$ when $f_k(r)=0$, namely, when $r = 2/e$.
Moreover, when $k = 1$, $f_k(r)=0$ for $r = 1$, and hence $A_k(t,r)$ in Eq.~(\ref{A1}) has no dependence on the variable $r$. Finally, $A_k(t,r)$ is merely a function of the cosmic time $t$ for $k = -1$ when $r$ solves the algebraic equation $\frac{\sqrt{1+r^2}+1}{r} = \exp\left(\sqrt{1+r^2}\right)$.

Eqs.~(\ref{y1})-(\ref{ym1}) lead to the solutions \cite{Ahmed2013}
\begin{eqnarray}
{a}(t) &\sim&\begin{cases}
\sec{\!\rm h}(t+\alpha_{-1})\,, & k=-1 \quad (\Lambda<0)\\
1/(t+\alpha_0)\,, & k=0 \quad (\Lambda>0)\\
\sec(t+\alpha_1)\,,& k=+1 \quad (\Lambda>0)\end{cases}
\label{aqeel}
\end{eqnarray}
For $k = 0, 1$ the metric is singular at a finite time $t =-\alpha_a + (n + 1/2)\pi k$, ($a=0,1$), for $n\in\mathbb{Z}$ \cite{Ahmed2013}.
It further implies that in this particular case the Hubble parameter reads
\begin{eqnarray}
{H}(t) &=&\begin{cases}
\tanh(t+\alpha_{-1})\,, & k=-1 \quad (\Lambda<0)\\
1/(t+\alpha_0)\,, & k=0 \quad (\Lambda>0)\\
\tan(t+\alpha_1)\,,& k=+1 \quad (\Lambda>0)\end{cases}
\label{aqeel1}
\end{eqnarray}
for the appropriate limits above analyzed where $f_k(r)=0$.

Once the $_{00}$ component of the Einstein equations is considered, one can further analyze the $_{11}$ component. Eq.~\eqref{e11} thus reads
\begin{eqnarray}\nonumber
&&(1-kr^{2})e^{-R}\left(3e^{-R}+\frac{4}{c_1r}f_k(r) \right)+\frac{k}{c_1}f_k(r)+kY\,\\
&&+\left[(1-kr^{2})e^{-R}\frac{4}{r} +\frac{2k}{c_1}f_k(r)\right]Y-2\frac{f_k(r)}{c_1}\ddot{Y}-2Y\ddot{Y}+3{\dot{Y}^{2}}=c_0\,.\nonumber
\end{eqnarray}
It implies that 
\begin{equation}\label{yh1}
2\ddot{Y}\left( Y+\frac{f_k(r)}{c_1}\right)=3\dot{Y}^2+kY^2+u(r)Y+v(r)\,,
\end{equation}
where $u_k(r)$ and $v_k(r)$ are respectively given by:
\begin{eqnarray} \nonumber
u_k(r)&=&\frac{4}{r}(1-kr^{2})e^{-R}+\frac{2k}{c_1}f_k(r)\,,\\ \nonumber
v_k(r)&=&u_k(r)\frac{f_k(r)}{c_1}+3(1-kr^{2})e^{-2R}-\frac{k}{c_1}f_k(r) -c_0\,.
\end{eqnarray}
Moreover, the $_{22}$ and $_{33}$ components of Einstein equations are provided by Eq.~\eqref{e22}, yielding
\begin{equation}\label{yh2}
2\ddot{Y}\left( Y+\frac{f_k(r)}{c_1}\right)=3\dot{Y}^2+kY^2+m(r)Y+n(r)\,,
\end{equation}
where
\begin{align}
m(r)&=e^{-R}\left[-2(kr^2-1)\bar{R}+\frac{6}{r}-4kr\right]+2k\frac{f_k(r)}{c_1}\,,\nonumber\\
n(r)&=m(r)\frac{f_k(r)}{c_1}+3e^{-2R}(1-kr^2)-k\frac{f_k^2(r)}{c_1^2}-c_0\,.\nonumber
\end{align}
Analogously, the $_{55}$ component of Einstein equations Eq.~\eqref{e55} reads 
\begin{equation}\label{yh3}
\ddot{Y}\left( Y+\frac{f_k(r)}{c_1}\right)=2\dot{Y}^2+kY^2+g(r)Y+h(r)\,,
\end{equation}
where
\begin{align}
g(r)&=e^{-R}\left[(kr^2-1)\left(\frac{3}{r}-\bar{R}\right)+1\right]+k\frac{f_k(r)}{c_1}\,,\nonumber\\
h(r)&=g(r)\frac{f_k(r)}{c_1}-k\frac{f_k^2(r)}{c_1^2}-2\left(\frac{1-(kr^2)^2}{c_1r}\right)-\frac{c_5}{3}\,.\nonumber
\end{align}
Eqs.~(\ref{yh1}), (\ref{yh2}) and (\ref{yh3}) can be reduced to first order EDOs. By defining a new variable 
$X=\dot{Y}$, Eqs.~(\ref{yh1}) and (\ref{yh2}) can be written as
\begin{equation}\label{reduc1}
2X\frac{dX}{dY}\left(Y+\alpha\right) = 3X^2 + kY^2 + bY + c,
\end{equation}
where $\alpha = f_k(r)/c_1$, for 
Eqs.~(\ref{yh1}) and (\ref{yh2}), by identifying respectively $b$ to $u(r)$ and $m(r)$, and $c$ to $v(r)$ and $n(r)$.
Solutions 
are provided by
\begin{equation}
\dot{Y}^2=k_1 (Y+\alpha)^3-kY^2-\frac{1}{2}bY-\frac{1}{3}k\alpha^2-\frac{1}{6}\alpha b-\frac{1}{3}c,
\end{equation}
where $k_1$ is a constant of integration. Note that when $k_1=0$ the above equation has exactly the same form of Eq.~\eqref{eqY}, thus has the same kind of solutions.

Moreover Eq.~(\ref{yh2}) can be recast 
analogously as 
\[X\frac{dX}{dY}(Y+\alpha) = 2X^2 + kY^2 + g(r)Y + h(r)\,,\] and reduced to a first order EDO in a similar way, giving
\begin{equation}
\dot{Y}^2=k_1 (Y+\alpha)^4-kY^2-\frac{2}{3}(\alpha k+g(r))Y-\frac{1}{3}(\alpha^2k+\alpha g(r)+3h(r)).
\end{equation}
In general, the above obtained equations do not exhibit analytical solutions. However when $k_1=0$ the same form of the Eq.~\eqref{eqY} is again achieved.

Given the form of the above equations, the constant parameter $k_1$ fixes the initial acceleration associated to the scale factor of the universe. One can compare it to the expected data for the dynamics of the scale factor and set $k_1$ according to the initial conditions.

\begin{figure}[!ht]
\begin{minipage}{14pc}
\includegraphics[width=14pc]{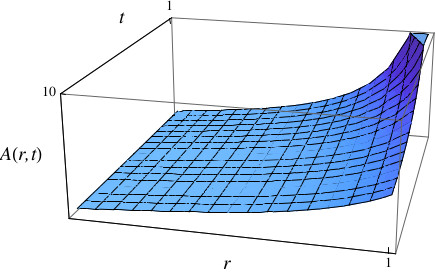}
\caption{\label{1} \footnotesize\; { Warp factor $A_{-1}(t,r)$ in (\ref{A1}), for $c_1=2$.}}
\end{minipage}\hspace{7pc}%
\begin{minipage}{14pc}
\includegraphics[width=14pc]{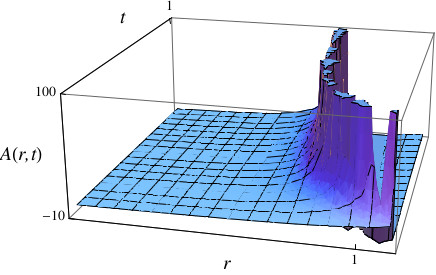}
\caption{\label{2} { \footnotesize\; Warp factor $A_0(t,r)$ in (\ref{A1}), for $c_1=2$.
}}\end{minipage}\hspace{7pc}%
\begin{minipage}{14pc}
\includegraphics[width=14pc]{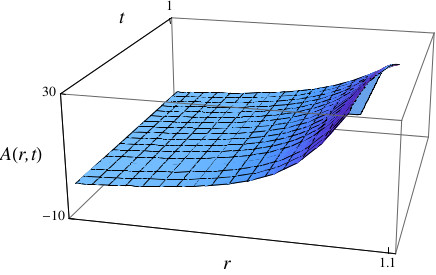}
\caption{\label{3} \footnotesize\; { Warp factor $A_1(t,r)$ in (\ref{A1}), for $c_1=2$. }}
\end{minipage}
\end{figure}
\begin{figure}[!ht]
\begin{minipage}{14pc}
\includegraphics[width=14pc]{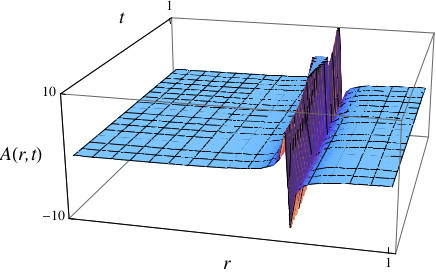}
\caption{\label{4} \footnotesize\; { Warp factor $A_{-1}(t,r)$ in (\ref{A1}), for $c_1=0.1$. }}
\end{minipage}\hspace{7pc}%
\begin{minipage}{14pc}
\includegraphics[width=14pc]{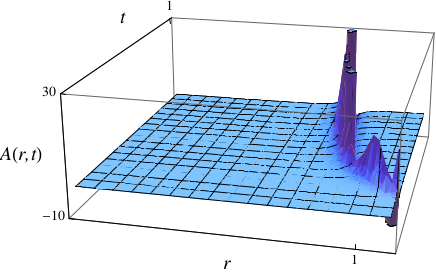}
\caption{\label{5} { \footnotesize\; Warp factor $A_0(t,r)$ in (\ref{A1}), for $c_1=0.1$. 
}}\end{minipage}\hspace{7pc}%
\begin{minipage}{14pc}
\includegraphics[width=14pc]{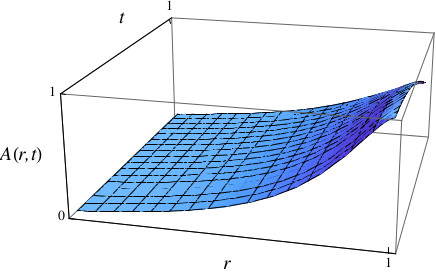}
\caption{\label{6} \footnotesize\; { Warp factor $A_1(t,r)$ in (\ref{A1}), for $c_1=0.1$. }}
\end{minipage}
\label{figuraB}
\end{figure}

In the set of Figs.~\ref{1}-\ref{3} and Figs.~\ref{4}-\ref{6} one respectively depicts the form for the warp factor $A_k(t,r)$ for $k=0,\pm1$, and the associated Hubble like parameter, calculated from the respective warp factor.

The above two sets of plots 
indicate a dependence on the 
integration parameter $c_0$, that is a multiple of the brane cosmological constant in an isotropic thick braneworld. Instead, here 
the constant $c_0$ is related to the cosmological constant $c_5 = \Lambda$
by Eq.~(\ref{opop}), and the spherical symmetry of the warp factor sets in. This explains the dependence of the $A_k(t,r)$ on the 
parameter $c_0$.

When the constant of integration $c_1$ in (\ref{opop}) is assumed to be equal to 2, Fig.~\ref{1} illustrates a monotonically increasing scale factor both radially and temporally. The larger is the radial position on the brane the steeper is the time dependence is.
Fig.~\ref{2} presents a range of singularity that attains lower values for the radial coordinate as time elapses. Fig.~\ref{3} illustrates a scale factor that increases in the range presented therein.
However such an increment is smoother as the cosmic time elapses.
When $c_1=0.1$, Fig.~\ref{4} depicts a time-independent singularity for a fixed value $r_0 \approx 0.663$ for the scale factor of a closed Universe, whereas the singularity evinced in Fig.~\ref{2} is smoother in Fig.~\ref{5}. Finally, Fig.~\ref{6} shows a similar 
profile as that one in Fig.~\ref{3}, instead the radial increment is planer.

\begin{figure}[!ht]
\begin{minipage}{14pc}
\includegraphics[width=14pc]{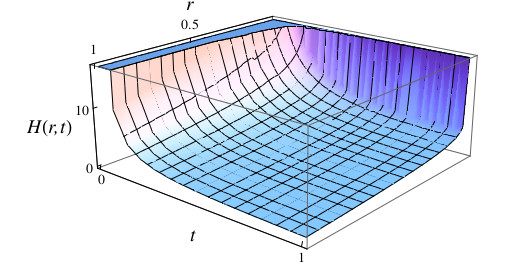}
\caption{\label{10} \footnotesize\; { Graphic of the Hubble like parameter $H_{-1}(t,r)=\dot{A}_{-1}(t,r)/A_{-1}(t,r)$ in (\ref{A1}), for $c_1=0.1$. }}
\end{minipage}\hspace{7pc}%
\begin{minipage}{14pc}
\includegraphics[width=14pc]{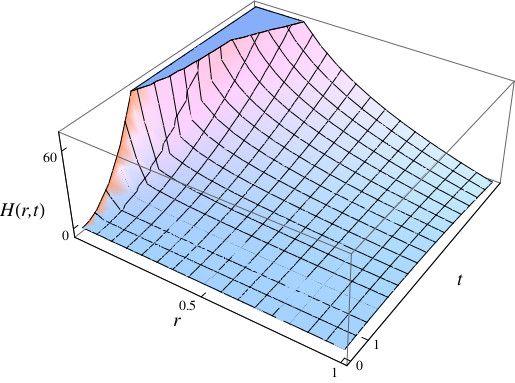}
\caption{\label{11} { \footnotesize\; Graphic of the Hubble like parameter $H_0(t,r)=\dot{A}_0(t,r)/A_0(t,r)$ in (\ref{A1}), for $c_1=0.1$. 
}}\end{minipage}\hspace{7pc}%
\begin{minipage}{14pc}
\includegraphics[width=14pc]{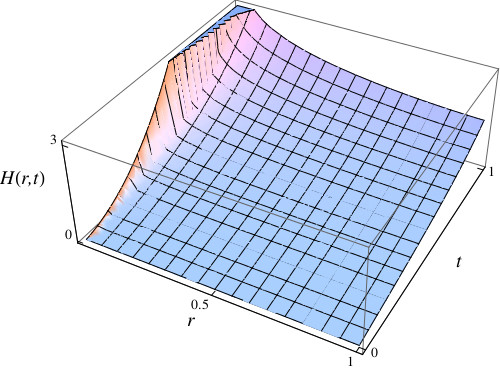}
\caption{\label{12} \footnotesize\; {Graphic of the Hubble like parameter $H_1(t,r)$ in (\ref{A1}), for $c_1=0.1$. }}
\end{minipage}
\label{figuraC}
\end{figure}

Hereupon the Hubble like parameter can also be depicted for $k=0,\pm1$. 
Their profiles are still dependent on the constant $c_1$ in (\ref{opop}), nonetheless the range of $H_k(t,r)$ changes slightly for different values of the $c_1$.
For the sake of completeness, and in order to match the results from Figs.~\ref{1}-\ref{6}, one  the Hubble like parameter for $c_1 = 0.1$ can be depicted in Figs.~\ref{10}-\ref{12}.
The above Hubble like parameters are led to the respective 
Hubble parameters  (\ref{aqeel1}), when the suitable respective limits mentioned before Eq.~(\ref{aqeel}) are taken into account. For the case $k=1$, the Hubble like parameter increases monotonically, being steepest for lower values of the radial coordinate.

The $y$-dependent part of the solutions are determined by
Eqs.~~\eqref{by}. By defining $\bar B(y)\equiv e^{b(y)}$, it can be written as
\begin{align}
3B^{\prime\prime}+\frac32\Lambda e^{-2b}&=-\phi^{\prime2}, \label{b1}\\
6B^{\prime2}-3\Lambda e^{-2b} &=\frac{1}{2}\phi^{\prime2}-V(\phi), \label{b2}\\
\phi^{\prime\prime}+4B^{\prime}\phi^\prime -\frac{dV}{d\phi}&=0. \label{b3}
\end{align}
As such equations are the same as those ones obtained in \cite{Ahmed2013}, the results for the extra dimensional profiles are likewise similar here.
When one assumes a
specific function $B(y)$, Eqs.~(\ref{b1})-(\ref{b3}) determine $\phi(y)$ and $V(\phi)$, or vice-versa 
\cite{DeWolfe:1999cp,Gremm:2000dj,Afonso:2006gi,Bazeia:2012qh}. For instance, the warp factor $B(y)=\ln[\sinh(\beta y)]$ was adopted in \cite{Ahmed2013}, where $\beta$ was assumed as a constant parameter, in order to have $A(y)\propto   e^{-|y|}$ when $y\to +\infty$, thus recovering the Randall-Sundrum model.
For  small (large) values of $y$, $|y|\lesssim \beta^{-1}$ ($|y|\gtrsim \beta^{-1}$), the kink solutions are given respectively by 
\ben
\phi_{\rm small}(y)&=&2\sqrt3\arctan[\tanh(\beta y/2)], \nonumber\\
\phi_{\rm large}(y)&=&\sqrt{-\frac{3\Lambda}2}\frac{1}{\beta}\sinh(\beta y).
\een

These results can be turned into asymmetric thick braneworld scenarios, generated after adding a constant to the superpotential associated to the scalar field (see \cite{Bazeia:2013usa, Ahmed2013}). Asymmetric branes can be generated irrespective of the potential being symmetric or asymmetric, and the sine-Gordon-type model in this context can be shown to have a stable graviton zero mode, despite the presence of an asymmetric volcano potential \cite{Bazeia:2013usa}. 
Indeed, the superpotential method described in \cite{Ahmed2013} can be further extended when one proposes the sine-Gordon-type model determined by the superpotential 
\[
W_c(\phi) = 2\sqrt{\frac{3}{2}}\sin\left(\sqrt{\frac{3}{2}}\phi\right)+c,
\]
that is obtained by the standard one, by shifting it with a constant parameter $c$ such that $|c|\leq\sqrt{6}$ \cite{Bazeia:2013usa}.
The 
solutions for the equations
\ben
\phi^\prime &=& \frac12 W_\phi,\label{foA}\\ 
A^\prime&=&-\frac13 W(\phi), \label{firstorderA}
\een
were obtained \cite{Bazeia:2013usa}:
\ben
\phi(y)&=& \sqrt{\frac{3}{2}}\arcsin(\tanh(y)),
\\
A_c(y)&=& -\ln[{\rm sech}(y)]-\frac13 c y,
\een 
where $\phi(y)$ is the standard solution of the sine-Gordon model, for $c=0$.

Finalizing, the above system could be further analyzed by including the radial dependence in the bulk scalar field that supports the radial dependence in the 4D scale factor. However, in this case, we cannot see a possible way to solve the equations analytically.

\section{Braneworld Black Holes}

The new features brought by braneworld scenarios would obviously have effects in black hole physics. Although 4D Einstein gravity allows only 4 asymptotically flat vacuum or electrovacuum black hole solutions, namely the Schwarzchild, Reissner–Nordström, Kerr and Kerr-Newman (see Sec. \ref{asyflat}), extra dimensional gravity is much less restrict. In fact, there is a wide variety of solutions depending on the brane model assumed \cite{Horowitz:2012nnc}. Such a wild environment deserves a good reason to be explored, otherwise we could still doing well with the sufficiently controlled 4D black hole solutions. Among the reasons it should be interesting to study extra dimensional black hole solutions, we may mention that \cite{Emparan:2008eg}:
\begin{itemize}
\item It could clarify physics of micro states in black hole thermodynamics. In fact, the first successful statistical counting of black-hole entropy in string theory was performed for a 5D black hole \cite{Strominger:1996sh};
\item Applications of the AdS/CFT, from QCD to condensate matter, make use of black hole physics with extra dimensions \cite{Casadio:2016zhu,Maldacena:1997re,Lin:2015kjk,Morgan:2009pn,Galow:2009kw};
\item The possible production of higher-dimensional black holes in colliders and its potential unveiling of new physics;
\item From the geometrical point of view, black hole space-times are important Lorentzian Ricci-flat manifolds.
\end{itemize}

Actually we are interested in two particular solutions of braneworld scenarios that can be seen effectively as small deviations of the the 4D Schwarzchild solution, namely the Casadio-Fabbri-Mazzacurati and the Minimal Geometric Deformation solutions. They were found from the effective field equations for warped extra dimensions. In the next chapter we are going to analyse the effect of the mentioned deviations in the non linear regime of gravitational lensing. Furthermore, the simplest extra dimensional extension of the Schwarzchild solution will also be introduced. Such solution will be used in the Chapter \ref{cap6} for modelling extra dimensional micro black hole formation.
\subsection{The Schwarzchild Extra Dimensional Solution}\label{edschwarzchild}

The simplest way to obtain an extra dimensional Schwarzchild solution is by assuming that the  extra dimensions are flat and sufficiently large. Likewise in the ADD model. By sufficiently large we mean an extra dimensional length $L$ such that $L\gg R_D$, where $R_D$ is the radius of the higher dimensional black hole horizon. This condition obviously does not hold for astrophysical black holes in the ADD model, but it could be a good approximation for black holes whose radius is close to the Planck scale. We are going to deal with those sort of black holes and its probability of formation in the Section \ref{HDHWF}.

Analogously to the 4D GR, the vacuum field equation of a $D+1$ dimensional space-time is given by
\begin{align}
R_{AB}=0.
\end{align}
A spherically symmetric solution should have the form 
\begin{align}
ds^2=-A(r)dt^2+B(r)dr^2+r^2d\Omega_{D-2}^2,
\end{align}
where $d\Omega_{D-1}^2$ is the line element of the unit $D$-sphere. The field equations then imply that \cite{Myers:1986un}
\begin{align}
A(r)=B(r)^{-1}=\left(1-\frac{C}{r^{D-2}} \right),
\end{align}
where $C$ is an integration constant related to the black hole mass by
\begin{align}
C=\frac{16\pi G M}{(D-1)\Omega_{D-1}}.
\end{align}
Here $\Omega_{D-1}$ denotes the area of a unit $D-1$-sphere and the Schwarzchild radius follows straightforward as
\begin{align}
R_D=\left(\frac{16\pi G M}{(D-1)\Omega_{D-1}}\right)^{\frac{1}{D-2}}.
\end{align}

A different family of 5D solutions could be found by assuming a metric like
\begin{align}
ds^2=g_{\mu\nu}(x^\alpha)dx^\mu dx^\nu+dy^2,
\end{align}
where the metric coefficients do not depend on the extra dimension $y$. In this case any 4D solution is also a 5D solution by construction \cite{Horowitz:2012nnc}. Solutions of this family are the so called black strings. Besides its appealing simplicity, the Schwarzchild black string were shown to have a central singularity extending
all along the extra dimension and a singular bulk horizon~\cite{cfabbri}, a configuration which moreover
suffers of the well-known Gregory-Laflamme instability~\cite{Gregory:1993vy}.

\subsection{The Casadio-Fabbri-Mazzacurati (CFM) Solutions}

We have seen in the Section \ref{brane} that solutions of the effective field equations on the brane are not uniquely determined
by the matter energy density and pressure, since gravity can propagate into the bulk
and generates a Weyl term onto the brane itself. 
Taking that into account, the CFM metrics~\cite{cfabbri,Casadio:2002uv} were found as vacuum brane solutions,
like the tidally charge metric in Ref.~\cite{Dadhich:2000am}, and contain a PPN parameter $\delta$ measured on the brane.
The case $\delta = 1$ corresponds to the exact Schwarzschild solution on the brane, and
extends into a homogeneous black string in the bulk.
Furthermore, it was observed in Ref.~\cite{cfabbri} that $\delta \approx 1$, in solar system
measurements.
More precisely, the deflection of light in the classical tests of GR provides the bound
$|\delta-1|\lesssim 0.003$~\cite{will}.
The parameter $\delta$ also measures the difference between the inertial mass
and the gravitational mass of a test body, besides affecting the perihelion shift and describing the
Nordtvedt effect~\cite{cfabbri}. 
Finally, measuring $\delta$ provides information regarding the vacuum energy of the braneworld or,
equivalently, the cosmological constant~\cite{Maartens:2010ar,cfabbri}. 
\par
By assuming a vacuum solution with a vanishing $\Lambda_4$, equivalent to fine tuning the $\Lambda_5$ on Eq. \eqref{la4la5}, the field equations \eqref{eff5d} reads
\be
R_{\mu\nu}= -{\cal E}_{\mu\nu}.
\label{0fe}
\ee
Departing from spherically symmetric systems on the brane, whose general metric can be
written as
\beq
g_{\mu\nu}\,dx^{\mu}\,dx^{\nu}
=
-A(r)\,dt^2+B(r)\,dr^2 + r^2\,d\Omega^2,
\label{cmetric}
\eeq
the CFM metrics are obtained by relaxing the condition $A(r)=B^{-1}(r)$. Such condition is valid for the Schwarzschild and
Reissner-Nordstr\"om metrics. However, in braneworld scenarios
it results in the Schwarzschild black string solution mentioned in the previous Section.
The CFM solution~I is obtained by fixing $A(r)$ equal to the Schwarzschild form, then determining
$B(r)$ from the field equations~\eqref{0fe}, whereas the CFM solution~II follows from the same procedure
but starting from a metric coefficient $A(r)$ of the Reissner-Nordstr\"om form~\cite{Dadhich:2000am}.
\subsubsection*{CFM solution I}
\label{3.1}
For the first case, the metric coefficients in Eq.~\eqref{cmetric} are given by
\beq
\!\!\!\!\!\!\!\!\!\!\!\!\!\!\!\!\!\!A_I(r)
&=&
1- \frac{2\,\GN\, M}{r},
\\
\qquad
\!\!\!\!\!\!\!\!\!\!\!\!\!\!\!\!\!\!B_I(r)
&=&
\frac{1-\frac{3\,\GN\,M}{2\,r}}
{\left(1- \frac{2\,\GN\,M}{r}\right)\left[1-\frac{\GN\,M}{2\,r}(4\delta-1)\right]}
\equiv
B(r).
\label{ar}
\eeq 
The solution~\eqref{ar} depends upon just one parameter $\delta$ and the
Minkowski vacuum is recovered for $M \rightarrow 0$ as well as the Schwarzchild one for $\delta\to 1$.
The horizon radius $r=R$ on the brane is then determined by the algebraic equation $1/B(R)=0$,
and this black hole is either hotter or colder than the Schwarzchild black hole of equal mass
$M$ depending upon the sign of $(\delta - 1)$~\cite{cfabbri}.
For example, assuming $\delta = 5/4$, one finds two solutions equal to the Schwarzchild
radius $r_\mais = {2\,\GN\,M}$.
%
%
%
\subsubsection*{CFM solution II}
\label{3.2}
The second solution for the metric coefficients reads
\begin{eqnarray}
\!\!\!\!\!\!\!\!A_{II}(r)
&=&
1- \frac{2(2\delta-1)\,\GN\,M}{r}\\
\quad
\!\!\!\!\!\!\!\! B_{II}(r)\!
 &\!=\!&\!
 \frac{1}{(2\delta\!-\!1)^2}\!\left(2(\delta\!-\!1)\!+\!\sqrt{1\!-\! \frac{\!2(2\delta\!-\!1)\,\GN\,M}{r}}\right)^2
\ .
\label{ar112} 
\end{eqnarray}
In this case the classical radius $R$ for the black hole horizon is given by
$R = r_\mais$ and $R = r_\mais\,(\delta-1/4)$. 
A comprehensive analysis of the causal structure and further features on both CFM
solutions can be found in Refs.~\cite{cfabbri,Casadio:2002uv,daRocha:2013ki}.

\subsection{The Minimal Geometric Deformation (MGD) Solution}
\label{MGD}

The minimal geometrical deformation (MGD) approach also generalizes GR solutions also  with small corrections coming, in principle, from extra dimensions. The basic idea is to consider an effective gravitational action of the form $S_G=S_{EH}+\upalpha (\mbox{corrections})$. The parameter $\upalpha$ is supposed to be small in order to have GR back when $\upalpha \to 0$. The previous action should produce a correction to the energy-momentum tensor, suchlike \cite{Ovalle:2016pwp}
\begin{align}\label{mgdcorrect}
R_{\mu \nu}-\frac{1}{2}Rg_{\mu\nu}=\kappa^2T_{\mu \nu}+\upalpha(\mbox{additional terms})_{\mu\nu}.
\end{align}
From its part, the additional term would generate a correction to the GR solution, lets say the Schwarzchild solution, thus producing a deformed Schwarzchild metric. The procedure of obtaining a minimum deformed metric that encompass corrections coming from Eq. \eqref{eff5d} and having GR as a limit case is called minimum geometrical deformation \cite{Ovalle:2016pwp}.

In braneworld scenarios, the brane self-gravity is encrypted in the brane tension $\sigma$. The brane tension was previously bounded in  the DMPR (Dadhich-Maartens-Papadopoulos-Rezania) \cite{Dadhich:2000am} and CFM solutions, by the classical tests of GR~\cite{Bohmer:2009yx}.
In the MGD procedure context, the bound $\sigma \gtrsim  5.19\times10^6 \;{\rm MeV^4}$ was obtained \cite{Casadio:2015jva},  providing a  stronger bound, contrasted to the one provided by cosmological nucleosynthesis. 
The MGD is a deformation of the Schwarzchild solution, 
which naturally led back to the Schwarzchild metric when $\sigma^{-1}\to 0$. From the effective field equations \eqref{eff5d} one argues that it suggests the inverse of the brane tension  $\sigma^{-1}\to 0$ playing the role of the parameter $\upalpha$ on the Eq. \eqref{mgdcorrect}.

In order to proceed one departs from the effective field equations \eqref{eff5d} supposing a vanishing bulk energy-momentum tensor ($\tau_{\mu\nu}=0$) and cosmological constant $\Lambda_4$ as well as that there is no exchange of energy between the bulk and the brane. 
MGD metrics  are thereafter exact solutions of Eq. \eqref{eff5d} \cite{ovalle2007}, yielding physical stellar inner   solutions \cite{Ovalle:2007bn}, having Schwarzchild outer solution  that does not jet energy into the extra dimension~\cite{darkstars}. Extensions of MGD and other 5D solutions were further obtained in various contexts ~\cite{daRocha:2012pt,Casadio:2013uma,daRocha:2013ki,Anjos:2015coa}.
\par The requirement that GR must be the low energy dominant regime at
$\sigma^{-1}\to 0$ derives  a deformed radial component of the metric, by bulk effects, yielding~\cite{covalle2}
\begin{eqnarray}
\label{edlrwssg}
B(r)
&=&
\nu(r)
+f(r)
\ ,
\end{eqnarray}\label{f}
where (hereon we denote $\frac{GM}{c^2}\mapsto M$) 
\bea\!\!\! f(r)\!&\!\!=\!\!&\!e^{-I}\!\left(\int_0^r\!\!\!\frac{e^I}{\frac{A'^{2}}{2A^2}\!+\!\frac{2}{x}}\!
\left[H(p,\rho,A)\!+\!\frac{\rho^2\!+\!3\rho\,p}{\sigma}\right]
\!dx\!+\!\zeta\right),\\
 I(r)&=&
\label{I}
\int^r_{r_0}
\frac{\!\frac{A''A}{A'^{2}}\!-\!1
\!+\!\frac{A'^{2}}{A^2}\!+\!\frac{2A'}{Ar}\!+\!\frac{1}{r^{2}}}
{\frac{A'}{2A}+\frac{2}{r}}\,dr
\ ,\\
\nu(r)
&=&
1-\strut\displaystyle\frac{2\,M^*}{r}
\ ,
\end{eqnarray} where $M^*=M$ for the outer solution ($r>R$) and $M^*=M(r)$ for the inner solution ($r\,\leq\,R$), 
where $\displaystyle M(r)\sim \!\!\int_0^r\! x^2\rho(x)\, dx$. 
 The function $H$ in Eq.~\eqref{f}, \begin{eqnarray}
\label{H}
H(p,\rho,A(r))
&\equiv&
\,p-\left[ \frac{A'}{A}\left( \frac{B'}{2B}+\frac{1}{r}\right)+(\ln B -1)r^{-2}\right.
\nonumber
\\
&&+
\left.\ln B\!\left(\!\frac{A''A}{A'^{2}}\!-\!1
\!+\!\frac{A'^{2}}{A^2}\!+\!\frac{2A'}{Ar}\right)\!
\right]\,,
\end{eqnarray} encompasses anisotropic effects of bulk gravity, the pressure, and the density. Here we are interested only on the outer solution. Towards $r>R$ the 
deformation $f(r)$ in Eq.~\eqref{edlrwssg} is minimal  for 
$f(r) \to 
f^+(r)=
\left.f(r)\right|_{p=\rho=H=0}
=
\zeta\,e^{-I}$ \cite{ovalle2007}. The outer radial component in Eq. \eqref{edlrwssg} is, accordingly
\begin{eqnarray}
\label{g11vaccum}
B_+(r)
=
{1-\frac{2\,M}{r}}
+\zeta\,e^{-I}.
\end{eqnarray}
 The parameter $\zeta$
carries the 5D correction to the vacuum, evaluated at the
star surface \cite{Casadio:2015jva}. 
For $r>R$,  the metric, following Eq.~\eqref{g11vaccum}, reads 
\begin{equation}
\label{genericext}
ds^2
=
A_+(r)\,dt^2-\frac{dr^2}{1-\frac{2M}{r}+f^+(r)}-r^2\,d\Omega^2
\ .
\end{equation}
With the MGD function $f=f^+(r)$ having the form \cite{ovalle2007,Casadio:2013uma}
\begin{equation}
\label{hhh}
f^+(r)
=    
\,\left({1-\frac{2M}{r}}\right)\left({1-\frac{3M}{2\,r}}\right)^{-1}\,
\frac{\zeta\xi}{r}
\ ,
\end{equation}
where $\xi$ is a length given by
\begin{equation}
\label{L}
\xi
\equiv
R{\left(1-\frac{3M}{2R}\right)}{\left(1-\frac{2M}{R}\right)^{-1}}
\,.
\end{equation}
In this way, the deformed outer metric reads
\begin{subequations}
\ba
\label{nu}
\!\!\!\!\!\!A(r)
&=&
1-\frac{2\,M}{r}
\ ,
\\
\!\!\!\!\!\!B(r)
&=&
\left(1-\frac{2\,M}{r}\right)^{-1}
\left[1+\left({1-\frac{3\,M}{2\,r}}\right)^{-1}\,\frac{{\zeta}\xi}{r}\right]^{-1},
\label{mu}
\ea
\end{subequations}
enclosing the vacuum solution in Ref.~\cite{germ} in the particular case
when $\zeta\,\xi={k}/{\sigma}$, $k>0$.  
The outer geometry, governed by Eqs.~\eqref{nu} and \eqref{mu}, has two horizons:
\be
r_\mais = 2\,M
\quad
{\rm and}
\quad
r_2=\frac{3M}{2} - \zeta\,\xi
\ .
\ee
However, one must have $r_2< r_\mais$, since the approximation $\zeta\sim \sigma^{-1}$ should hold in the GR limit.
It implies that the outer horizon  is the Schwarzchild one $r_\mais = 2\,M$.
Note that the specific value $\zeta=-M/2$ would produce a
single horizon $r_\mais = 2\,M$. 
In particular, $\zeta$ can be derived by considering the exact inner braneworld solution
of Ref.~\cite{ovalle2007},
\begin{eqnarray}
\label{betasigma}
\zeta(\sigma,R)
=
-\frac{C_0}{R^2\,\sigma}
\ .
\label{c0}
\end{eqnarray}
where $C_0\simeq 1.35$. {Despite being appeared for describing braneworld stars, the MGD solution is also applied as a braneworld black hole solution \cite{Casadio:2012pu}.}

In the next Chapter we shall apply and analyse the above described metric to study the modifications on the gravitational lensing 
effects, when compared to Shcwarzchild solution. In particular, we shall investigate the role of the PPN parameter and the 
 brane tension in the MGD, on the observables derived  in the strong deflection limit.


\chapter{Strong Deflection Limit in Gravitational Lensing} 

\label{cap5} 



\begin{figure}[!htb]
\centering
 \scalebox{.99}{\definecolor{wqwqwq}{rgb}{0.3764705882352941,0.3764705882352941,0.3764705882352941}
\begin{tikzpicture}[line cap=round,line join=round,>=triangle 45,x=1.0cm,y=1.0cm,scale=0.
35,every node/.style={scale=0.75}]
\draw [fill=black,fill opacity=1.0] (8.,7.) circle (0.4cm);
\draw(8.,7.) circle (0.6cm);
\draw [shift={(9.879038294607481,10.247139613384093)}] plot[domain=3.3558917962239487:4.595789826732699,variable=\t]({1.*12.119884205809411*cos(\t r)+0.*12.119884205809411*sin(\t r)},{0.*12.119884205809411*cos(\t r)+1.*12.119884205809411*sin(\t r)});
\draw (8.469069966991903,-1.7904507799358815)-- (25.40007470866168,-3.6913207639875796);
\draw (17.14423957071595,-2.7644254351842754) -- (16.904307056540667,-3.010457928533505);
\draw (17.14423957071595,-2.7644254351842754) -- (16.96483761911292,-2.4713136153899615);
\draw [->] (8.268591539736523,7.536524542571879) -- (8.601688803088534,8.774849829095649);
\draw (6.1941943975576161,9.9558506587848465) node[anchor=north west] {Photon sphere};
\draw (7.5067194706093023,5.7419191400590007) node[anchor=north west] {Black hole};
\draw (-5.7556376638378978,21.8413826784365405) node[anchor=north west] {Source};
\draw (24.506651824211561,-4.1266002387072985) node[anchor=north west] {Observer};
\draw [->] (8.,7.) -- (9.087957868776178,5.7246165988731565);
\draw (-4.085513183489083,19.295944159428302) node[star,star points=10,inner sep=0.05cm,draw] at (-4.085513183489083,20.1) {$\circ$}-- (-1.963611805943001,7.669692861623731);
\draw (-2.986681549488336,13.275262498132541) -- (-3.2914202249362283,13.434114438090393);
\draw (-2.986681549488336,13.275262498132541) -- (-2.7577047644958586,13.53152258296164);
\draw [fill=wqwqwq,fill opacity=0.3] (26.21186273567742,-3.7475833005134223) circle (0.2cm);
\end{tikzpicture}}
\caption{Typical deflection of photons' trajectory by gravitational fields. The concept of photon sphere shall be defined in the next section.}\label{fig:WDL}
\end{figure}
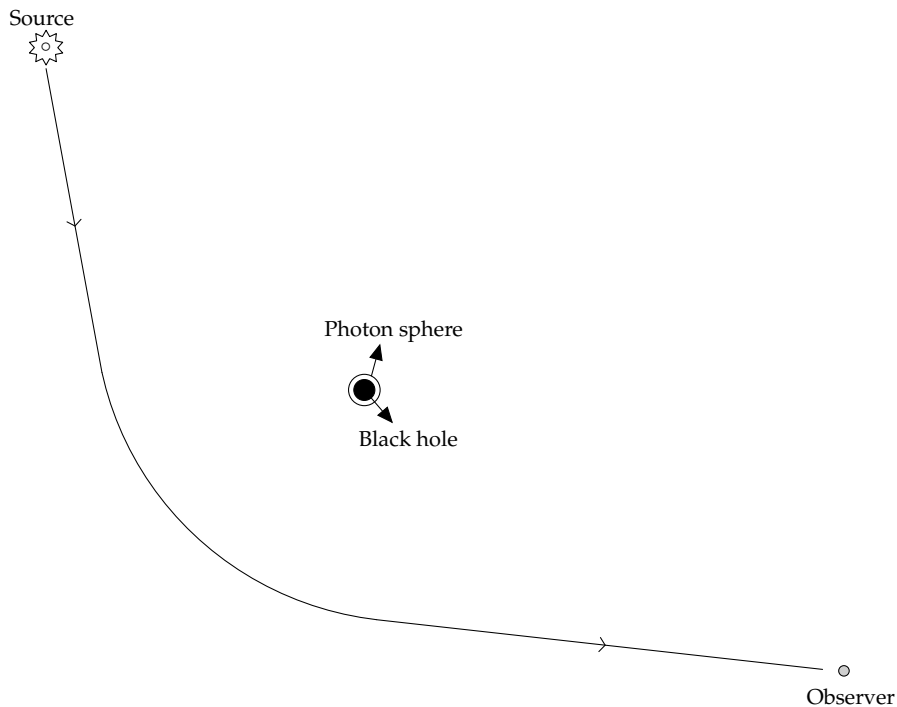

The deflection of light by gravitational fields was first observed in 1919 by Dyson, Eddington and Davidson \cite{Dyson:1920cwa}, whose modern refinements  have  become one of the experimental grounds of GR. Thereafter, the deflection of light was found to imply  lens effects, that could magnify or {}{even} create multiple images of astrophysical objects \cite{Zwicky:1937zzb}. This was a landmark for a contemporary field of research, known as gravitational lensing (GL). The works by Liebes  \cite{Liebes:1964zz} and Refsdal \cite{Refsdal:1964yk} developed, in the theory of gravitational lensing, the so called weak deflection limit (WDL), where the lens equations and the expression for the deflection angle are quite simplified, hence allowing one to solve them exactly. The predictions of the theory, in this regime, have been thoroughly supported by experiments {}{and observations}. It is worth mentioning, for example,  the observation of twin quasars separated by arc-seconds ({}{arcsec}), at same redshifts and magnitudes \cite{Walsh:1979nx}, images of distorted galaxies inside another galaxy \cite{paczynski1987giant}, among others \cite{deXivry:2009ci}.  Fig. \ref{fig:WDL} shows the kind of deflection appropriately described by the WDL.

With all this success in the weak field limit, the question about what happens in the strong field regime was driven by the possibility of existence of galactic supermassive black holes at the centre of our galaxy. The subject was brought back by Virbhadra and Ellis \cite{Virbhadra:1999nm}, who theoretically investigated the strong field region of a Schwarzschild type lens. They found that, in the strong field regime, the theory predicts a large number of images of an observed  object -- theoretically, an infinite sequence of images, with an adherent point. The result contrasts with a pair of images, or an Einstein ring, predicted in the weak field limit regime. Following the work of Virbhadra and Ellis, Bozza \cite{Bozza:2001xd,Bozza:2002zj} has found an interesting simplification for the lens equation beyond the WDL, finding the expression for observable quantities in the so-called strong deflection limit  (SDL) regime. Bozza proved  that, when the angle between the source and the lens tends to zero, the deflection angle diverges logarithmically. Furthermore, it can be integrated up to first order, from where the GL observables can be derived.

\begin{figure}[!htb]
\captionsetup[subfigure]{font=footnotesize}
\centering
{%
 \scalebox{.85}{\definecolor{wqwqwq}{rgb}{0.3764705882352941,0.3764705882352941,0.3764705882352941}
\begin{tikzpicture}[line cap=round,line join=round,>=triangle 45,x=1.0cm,y=1.0cm,scale=0.
35,every node/.style={scale=0.75}]
\draw [fill=black,fill opacity=1.0] (8.,7.) circle (0.4cm);
\draw(8.,7.) circle (0.6cm);
\draw[fill=wqwqwq,fill opacity=0.3] (23.680048592014483,-3.5868331961538704) circle (0.2cm);
\draw [shift={(8.31228256387508,7.150787562049857)}] plot[domain=-2.5801642307375414:2.6524219317630933,variable=\t]({1.*3.327943936068343*cos(\t r)+0.*3.327943936068343*sin(\t r)},{0.*3.327943936068343*cos(\t r)+1.*3.327943936068343*sin(\t r)});
\draw [shift={(10.023774539756102,5.957745144069343)}] plot[domain=2.6063550550884744:4.497450563883,variable=\t]({1.*5.405054664701047*cos(\t r)+0.*5.405054664701047*sin(\t r)},{0.*5.405054664701047*cos(\t r)+1.*5.405054664701047*sin(\t r)});
\draw (8.87094522789078,0.6770633219832312)-- (22.93657935935155,-3.422064339185338);
\draw (16.106320383955197,-1.4315317234984004) -- (15.827865017324582,-1.6329323390305195);
\draw (16.106320383955197,-1.4315317234984004) -- (15.979659569917754,-1.1120686781715872);
\draw (-3.145125072985704,19.082950271151905) node[star,star points=10,inner sep=0.05cm,draw] at (-3.75513183489083,19.8) {$\circ$}-- (5.495193036340197,5.379003874500119);
\draw (1.2875603584493296,12.052505005480523) -- (0.9455698950901877,12.086300302690479);
\draw (1.2875603584493296,12.052505005480523) -- (1.4044980682643053,12.375653842961547);
\end{tikzpicture}}}%
\caption{Sketch of the non-linear lensing deflection effect present in the SDL regime. }\label{fig:SDL}
\end{figure}
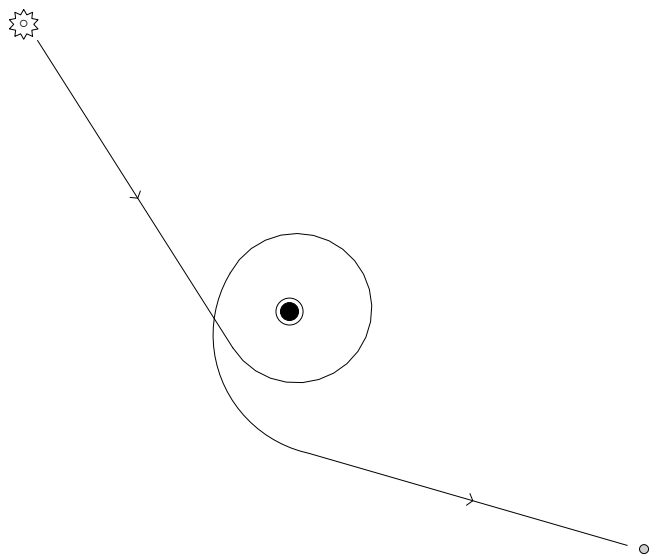

 In a series of papers,  Keeton and Petters studied a general framework for the weak field limit, comprising the case of a static and spherically symmetric solution \cite{Keeton:2005jd}, the post-Newtonian metrics \cite{Keeton:2006sa} and brane-world gravity \cite{Keeton:2006di}. Whisker considered the gravitational lensing at the strong deflection limit \cite{Whisker:2004gq} as a way to seek for signatures of solutions of five-dimensional (5D) brane-world gravity. For some 5D brane-world solutions, the differences in the observables were found to be very small from the 4D Schwarzschild  case. Thereafter,  Bozza revised the theoretical and observational aspects of gravitational lensing produced by black holes \cite{Bozza:2009yw}. {}{In a} recent paper, the SDL approach was applied to study Galileon black holes \cite{Zhao:2016kft}.

 In this chapter, we apply the SDL to obtain the GL observables for two remarkable black hole solutions beyond GR, namely the Casadio-Fabbri-Mazzacurati   and the Minimal Geometric Deformation, both introduced in the previous chapter. The results support the fact that an increasing precision of some astronomical observation could allow us to fully apply the SDL approach. Not only to distinguish black hole flavours, but also to test deviation of GR predicted by alternative theories of gravity.

\section{Exact Solutions, the Weak and the Strong Deflection Limit}

Whenever bending of light by gravitational fields is detected in
astronomical observations, it is generally very weak.  This leads to the conclusion that the first order
Einstein formula for the deflection angle is sufficient to explain
all observed phenomenology
. On the other hand, in the neighbourhood
of compact astronomical objects, electromagnetic radiation 
travels through very strong gravitational fields. In such extreme
cases, the calculation of the deflection of photons needs to be
pushed beyond the first order approximation on which the Einstein
formula relies. The SDL was proposed in order to contemplate lensing effectes generated by those compact objects. Now we are going to discuss some basic features and applicabilities of both approaches, namely the WDL and SDL.

Consider a ray of light that travels from a source \textit{S} to an observer \textit{O}. If there exist local matter inhomogeneities in between them, the
gravitational field along its trajectory will continuously deflect it. Such deflection is similar to the light deflection by lens, hence the phenomena became known as gravitational lensing.

The general setup of a gravitational lens is depicted in Fig. 1. The light of a source \textit{S} is deflected by the gravitational field of the lens \textit{L}. The image of \textit{S} appears to the observer \textit{O} in a position characterised by the angle $\theta$, rather than the angle $\beta$. The deflection is then given by the angle $\alpha$. A basic assumption here is that the source and the observer are far from the lens, in an asymptotic  flat region. Thus the light ray is assumed to be coming in a straight line from infinity. As it approaches the lens it is deflected by an angle $\alpha$, finally going to infinity again in a straight line. In such setup, the deflection angle can be obtained by integrating the null geodesics.

\begin{figure}[!htb]
\centering
\begin{tikzpicture}[line cap=round,line join=round,>=triangle 45,x=1.0cm,y=1.0cm]
\draw [shift={(-4.22,0.24)},fill=black,fill opacity=0.1] (0,0) -- (81.49812058851853:1.8) arc (81.49812058851853:90.:1.8) -- cycle;
\draw [shift={(-4.22,0.24)},fill=black,fill opacity=0.1] (0,0) -- (59.01453229825211:1.1) arc (59.01453229825211:90.:1.1) -- cycle;
\draw (-5.96,8.)-- (-5.96,0.24);
\draw [dash pattern=on 5pt off 5pt] (-4.22,8.)-- (-4.22,0.24);
\draw [dash pattern=on 5pt off 5pt] (-3.06,8.)-- (-4.22,0.24);
\draw [line width=1pt] (-3.06,8.) node[star,star points=10,inner sep=0.05cm,draw,fill=white]  {$\circ$}-- (-2.3,4.66);
\draw [line width=1pt] (-2.6533751848318867,6.212990943866453) -- (-2.8262613201669344,6.29671898103986);
\draw [line width=1pt] (-2.6533751848318867,6.212990943866453) -- (-2.5337386798330637,6.3632810189601425);
\draw [shift={(-4.22,4.11)},line width=1pt]  plot[domain=-0.4861157375866316:0.2647869679917831,variable=\t]({1.*1.9906029237394383*cos(\t r)+0.*1.9906029237394383*sin(\t r)},{0.*1.9906029237394383*cos(\t r)+1.*1.9906029237394383*sin(\t r)});
\draw [dash pattern=on 1pt off 2pt on 5pt off 4pt] (0.44,8.) node[line width=1pt, dash pattern=,star,star points=10,inner sep=0.05cm,draw,fill=white]  {$\circ$} -- (-4.22,0.24);
\draw [line width=1pt] (-2.46,3.18)-- (-4.22,0.24);
\draw [line width=1pt] (-3.401636457618115,1.6070390992061045) -- (-3.46870112599237,1.7870455720226432);
\draw [line width=1pt] (-3.401636457618115,1.6070390992061045) -- (-3.2112988740076323,1.632954427977357);
\draw (-5.62,8.)-- (-5.62,4.25);
\draw (-5.62,3.97)-- (-5.62,0.24);
\draw [dash pattern=on 5pt off 5pt] (-4.22,8.)-- (0.44,8.);
\draw [dotted] (-4.22,4.11)-- (-2.5119574482627307,3.08429403465262);
\draw [dotted] (-4.22,4.11)-- (-2.2310228866342507,4.190436574437585);
\draw [shift={(-2.0974708176905086,3.774512114747136)}] plot[domain=1.0299978951290354:1.7956491073448766,variable=\t]({1.*1.0323811169938255*cos(\t r)+0.*1.0323811169938255*sin(\t r)},{0.*1.0323811169938255*cos(\t r)+1.*1.0323811169938255*sin(\t r)});
\begin{scriptsize}
\draw [fill=black,shift={(-5.96,8.)}] (0,0) ++(0 pt,3.75pt) -- ++(3.2475952641916446pt,-5.625pt)--++(-6.495190528383289pt,0 pt) -- ++(3.2475952641916446pt,5.625pt);
\draw [fill=black,shift={(-5.96,0.24)},rotate=180] (0,0) ++(0 pt,3.75pt) -- ++(3.2475952641916446pt,-5.625pt)--++(-6.495190528383289pt,0 pt) -- ++(3.2475952641916446pt,5.625pt);
\draw[color=black] (-6.58,4.42) node {{\large $D_{os}$}};
\draw [fill=black,shift={(-5.62,8.)}] (0,0) ++(0 pt,3.75pt) -- ++(3.2475952641916446pt,-5.625pt)--++(-6.495190528383289pt,0 pt) -- ++(3.2475952641916446pt,5.625pt);
\draw [fill=black,shift={(-5.62,0.24)},rotate=180] (0,0) ++(0 pt,3.75pt) -- ++(3.2475952641916446pt,-5.625pt)--++(-6.495190528383289pt,0 pt) -- ++(3.2475952641916446pt,5.625pt);
\draw [fill=black] (-4.22,0.24) ++(-1.5pt,0 pt) -- ++(1.5pt,1.5pt)--++(1.5pt,-1.5pt)--++(-1.5pt,-1.5pt)--++(-1.5pt,1.5pt);
\draw[color=black] (-3.9,0.09) node {{\large $O$}};
\draw [fill=black] (-4.22,4.11) circle (4.5pt);
\draw[color=black] (-4.6,4.4) node {{\large $L$}};
\draw [fill=black] (-3.06,8.) ++(-3.0pt,0 pt) -- ++(3.0pt,3.0pt)--++(3.0pt,-3.0pt)--++(-3.0pt,-3.0pt)--++(-3.0pt,3.0pt);
\draw[color=black] (-2.87,8.5) node {{\large $S$}};
\draw [fill=black] (0.44,8.) ++(-3.0pt,0 pt) -- ++(3.0pt,3.0pt)--++(3.0pt,-3.0pt)--++(-3.0pt,-3.0pt)--++(-3.0pt,3.0pt);
\draw[color=black] (0.6,8.5) node {{\large $I$}};
\draw [fill=black,shift={(-5.62,4.25)},rotate=180] (0,0) ++(0 pt,3.75pt) -- ++(3.2475952641916446pt,-5.625pt)--++(-6.495190528383289pt,0 pt) -- ++(3.2475952641916446pt,5.625pt);
\draw [fill=black,shift={(-5.62,3.97)}] (0,0) ++(0 pt,3.75pt) -- ++(3.2475952641916446pt,-5.625pt)--++(-6.495190528383289pt,0 pt) -- ++(3.2475952641916446pt,5.625pt);
\draw[color=black] (-5.2,6.22) node {{\large $D_{ls}$}};
\draw[color=black] (-5.2,2.32) node {{\large $D_{ol}$}};
\draw[color=black] (-3.2,3.24) node {{\large $u$}};
\draw[color=black] (-4.05,2.32) node {{\large $\beta$}};
\draw[color=black] (-3.75,1.5) node {{\large $\theta$}};
\draw[color=black] (-2.94,4.4) node {{\large $r_0$}};
\draw[color=black] (-1.9,5.0) node {{\large $\alpha$}};
\end{scriptsize}
\end{tikzpicture}
\caption{Gravitational lensing setup, where $r_{0}$ is  the closest approximation of the light rays to the lens and $u$ denotes the impact parameter.}\label{fig:lsetup}
\end{figure}
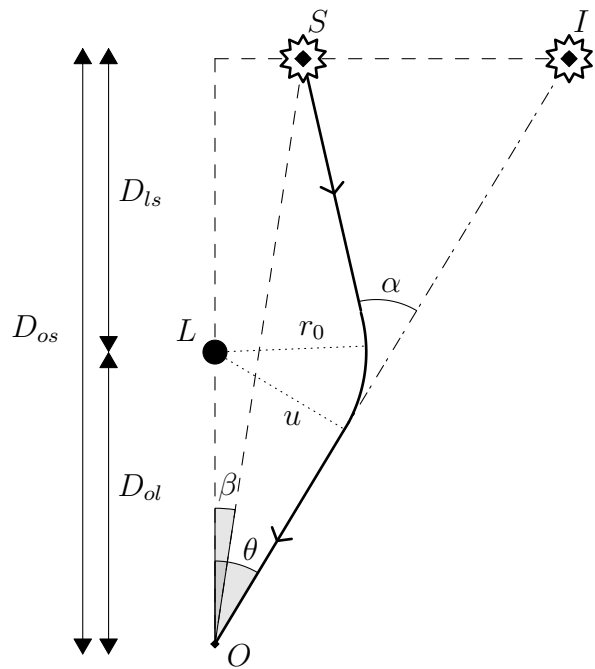

Here we are interested in spherically symmetric solutions of the Einstein Equations, whose general form, adopting the $(+,-,-,-)$\footnote{Such signature shall be adopted up to the end of this chapter.} signature, is given by
\begin{equation}
ds^2=A(r)dt^2-B(r)dr^2-r^2(d\vartheta^2+\sin^2\vartheta d\phi^2).
\label{Sph Metric}
\end{equation}
According to the above setup, let us suppose that a photon
comes from an infinite distance, approaches the black hole at a minimum
distance $r_0$ and goes away to infinity. The deflection angle,
defined as the angle between the asymptotic incoming and outgoing
trajectories, is then twice the deflection between the trajectory at $r_0$ and infinity. It can be derived from the geodesics equations (see \cite{weinberg1972gravitation} for a detailed derivation)
\begin{align}\nonumber
\alpha &=2\int\limits_{r_0}^{\infty}
\frac{u}{r}\sqrt{\frac{B(r)}{r^2/A(r)-u^2}}\d r-\pi\\
&\equiv {I}(r_0)-\pi
\label{alpha Sch}
\end{align}
where $u$ is the impact parameter, which gives a mesure of the distance between
the black hole and each of the asymptotic photon trajectories \cite{weinberg1972gravitation}. It is
related to the closest approach distance $r_0$ by
\begin{equation}
u^2=\frac{r_0^2}{A(r_0)}. \label{u rm}
\end{equation}
The impact parameter is a constant of motion and, in the case of massive particles, is proportional to the angular momentum per unit mass. It is also related to the angle $\theta$, as can be seen from  Fig. \ref{fig:lsetup}, by 
\begin{align}\label{utheta}
\sin \theta=\frac{u}{D_{ol}}.
\end{align}
There is no analytical solution to the equation \eqref{alpha Sch} in terms of standard functions. However, an approximation has been used with great success since the early times of general relativity: the  weak deflect limit. In this case the light is assumed to be propagating far from the lens ($r\geq r_0 \gg 2M$), resulting in a small gravitational field, and consequently a small deflection angle. The resulting deflection angle, for the Schwarzschild metric, is given by\footnote{From here up to the end of this section, except for the photon sphere equation, we are going to deal only with the Schwarzschild solution.} (see \cite{Carroll:2004st} for a detailed derivation)
\begin{equation}
\alpha_{WDL} = 4M/u. \label{alpha WDL Sch}
\end{equation}

\begin{figure}[!htb]
\center
\includegraphics[scale=.8]{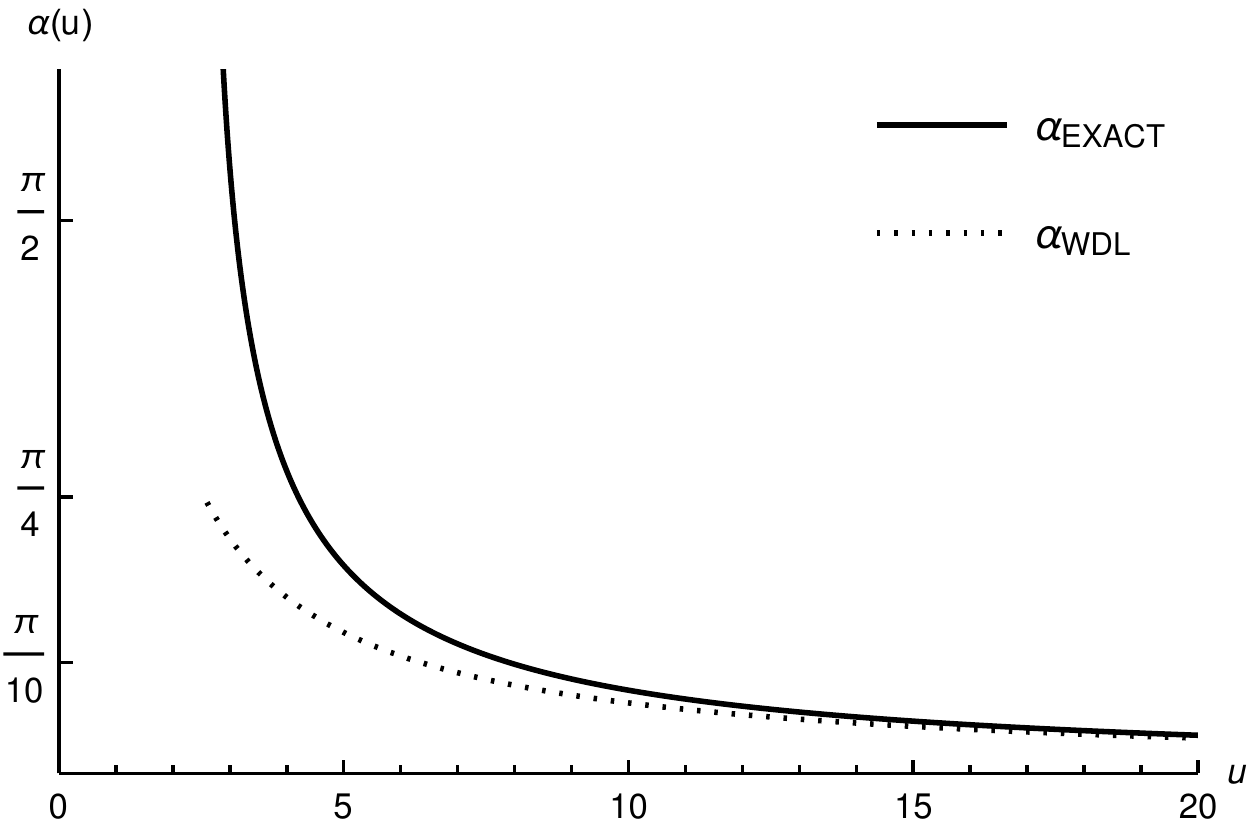}
\caption{Comparison between the deflection angle from the exact solution and the one given by the WDL, both using the Schwarzsschild metric.} \label{Fig alphaw}
\end{figure}

An exact solution was found by Darwin \cite{Darwin180,Darwin39,Frolov2011}, in terms of elliptic integrals. The solution reads
\begin{equation}
\alpha=-\pi+4\sqrt{r_0/s} F(\varphi,m), \label{alpha Darwin}
\end{equation}
where $F(\varphi,m)=\int\limits_0^\varphi
\frac{d\vartheta}{\sqrt{1-m \sin^2\vartheta}}$ is the elliptic integral of the first
kind and
\begin{eqnarray}
&&s=\sqrt{(r_0-2M)(r_0+6M)} \\
&& m=(s-r_0+6M)/2s \\
&& \varphi=\arcsin \sqrt{2s/(3r_0-6M+s)}.
\end{eqnarray}
In order to compare the above results, we have to express both in terms of the same variable, as the first is a function of $u$ and the second of $r_0$. We choose to express both in terms of $u$, as it is a constant of motion and hence coordinate independent. However let us first, as a matter of simplification and being able to numerically solve the elliptic integral, express the parameters of the deflection angles in units of Schwarzschild radius,

\begin{equation}
\alpha_{WDL} = 2/u, \label{alpha WDL Sch}
\end{equation} 
 \begin{equation}
\alpha=-\pi+4\sqrt{r_0/s} F(\varphi,m), \label{alpha Darwin}
\end{equation}
\begin{eqnarray}
&&s=\sqrt{(r_0-1)(r_0+3)} \\
&& m=(s-r_0+3)/2s \\
&& \varphi=\arcsin \sqrt{2s/(3r_0-3+s)}.
\end{eqnarray}
The Eq. (\ref{alpha Darwin}) can be expressed as a function of $u$ by using the Eq. \eqref{u rm}. Thus we have $r_0(u)$ as the real solution to the equation\footnote{Already in units of Schwarzschild radius.} $r_0^3-u^2r_0+u^2=0$.  Now we are able to compare the deflection angle in both approaches. The results are shown in Fig. \ref{Fig alphaw}. There we can see that the WDL gives a very accurate result for large impact parameters, which means large minimum distance between the photons' trajectory and the lens. It agrees, as expected, with the basic assumptions of the WDL.

Despite the good accuracy of the WDL for large $u$, it does not catch the divergence of the deflection angle when $u\to 3\sqrt{3}M\equiv u_m$. Such phenomena emerge as a result of the full non-linear regime of GR. Large deflection angle means that the light ray is deflected by $\alpha \in [2n\pi,2(n+1)\pi]$, performing $n$ complete loops around
the black hole before leaving it.  The images originated from those photons are known as relativistic images \cite{Virbhadra:1999nm}.  Photons with impact parameters smaller than
$u_m$ are simply captured by the black hole's gravitational field and fall into the
horizon. A light ray with impact
parameter exactly equal to $u_m$, on the other hand, would perform an infinite
number of loops as $r\to r_m$ indefinitely.  The length $r_m$ is also called the
radius of the photon sphere, a null surface that, on the Schwarzschild space-time, is characterized by the the statements \cite{Virbhadra:1999nm,Claudel:2000yi}:
\begin{itemize}
\item Any future endless null geodesic starting at some point with $r > r_m$ and initially directed outwards, will continue outwards and escape to
infinity;
\item Any future endless null geodesic starting at some point with $r <r_m$
and initially directed inwards, will continue inwards and fall into the black
hole;
\item Any null geodesic starting at some point of the photon sphere, and initially tangent to it, will remain on the photon sphere.
\end{itemize}
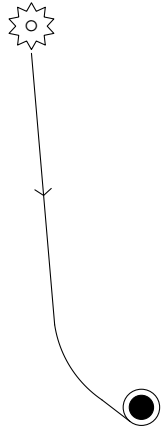
\begin{figure}[!htb]
\centering
\definecolor{wqwqwq}{rgb}{0.3764705882352941,0.3764705882352941,0.3764705882352941}
\begin{tikzpicture}[line cap=round,line join=round,>=triangle 45,x=1.0cm,y=1.0cm, scale=0.
4]
\draw [fill=black,fill opacity=1.0] (8.067194706093023,6.665004709376533) circle (0.4cm);
\draw(8.067194706093023,6.665004709376533) circle (0.6cm);
\draw (6.769394357352941,6.911197528194522)-- (7.699813427695275,6.190630388525545);
\draw [shift={(8.85655372676281,9.946679650071836)}] plot[domain=3.2911596890429307:4.1100422636407785,variable=\t]({1.*3.6838004758065006*cos(\t r)+0.*3.6838004758065006*sin(\t r)},{0.*3.6838004758065006*cos(\t r)+1.*3.6838004758065006*sin(\t r)});
\draw (4.450317358003112,18.399762327623808) node[star,star points=10,inner sep=0.05cm,draw] at (4.45,19.3) {$\circ$} -- (5.213880353710982,9.397756483488912);
\draw (4.849930830905812,13.688529804981451) -- (4.5618036551178776,13.875832580493663);
\draw (4.849930830905812,13.688529804981451) -- (5.102394056596215,13.92168623061906);
\end{tikzpicture}
\caption{Trajectory of photons tangentially reaching the photon sphere and captured in a circular orbit.}\label{fig:SDL3}
\end{figure}
In spherically symmetric space-times, the photon sphere can also be defined as the boundary surface whose radius satisfies the photon sphere equation \cite{Bozza:2002zj}. See \cite{Claudel:2000yi} for general definitions and rigorous results concerning it. The photon sphere can be found by combining the above last statement (Fig. \ref{fig:SDL3}) with the metric and the radial geodesic equations. At first, by isotropy, we consider the orbit of the photons to be confined to the equatorial plane ($\vartheta=\pi/2$). As we are interested in circular orbits, the radial derivatives must be null. Thus, from the metric \eqref{Sph Metric} it follows
\begin{align}
\left(\frac{d\phi}{dt}\right)^2=\frac{A(r)}{r^2}.
\end{align}
Likewise, from the radial geodesic equation it follows
\begin{align}
\left(\frac{d\phi}{dt}\right)^2=\frac{A'(r)}{2r}.
\end{align}
Combining those equations we finally find the photon sphere equation
\begin{align}\label{photon_sphere}
\frac{A(r)}{r}=\frac{A'(r)}{2}.
\end{align}

\begin{figure}[!htb]
\centering
\includegraphics[scale=.8]{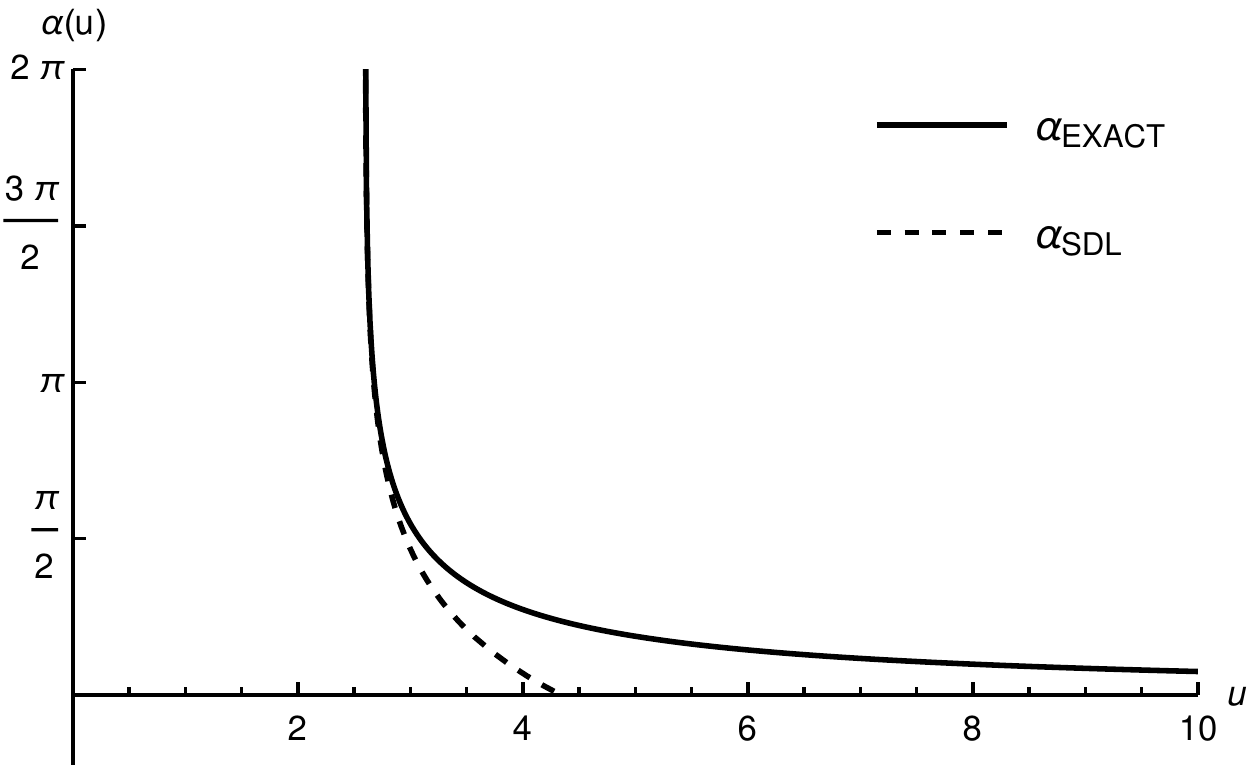}
\caption{Comparison between the deflection angle from the exact solution and the one given by the SDL. For $\alpha \approx 2\pi$ the estimated error about 0.06\%  \cite{Bozza:2002zj}}
\label{Fig_pas}
\end{figure}

It has been shown that the divergence of the  deflection angle, as the closest approximation $r_{0}$ {}{tends} to the photon sphere, is logarithmic \cite{Bozza:2002zj}. Doing the appropriate expansion around $r_m$, as we are going to discuss in the next section, the expression for the deflection angle in the SDL reads \cite{Bozza:2002zj} 
\begin{equation}
\alpha_{SDL}=-\log (u/u_m-1)+\log\left[216(7-4\sqrt{3})\right]-\pi, \label{alpha SDL Sch}
\end{equation}
which is compared to the exact result in Fig.
\ref{Fig_pas}.

\begin{figure}[!htb]
\centering
\includegraphics[scale=.8]{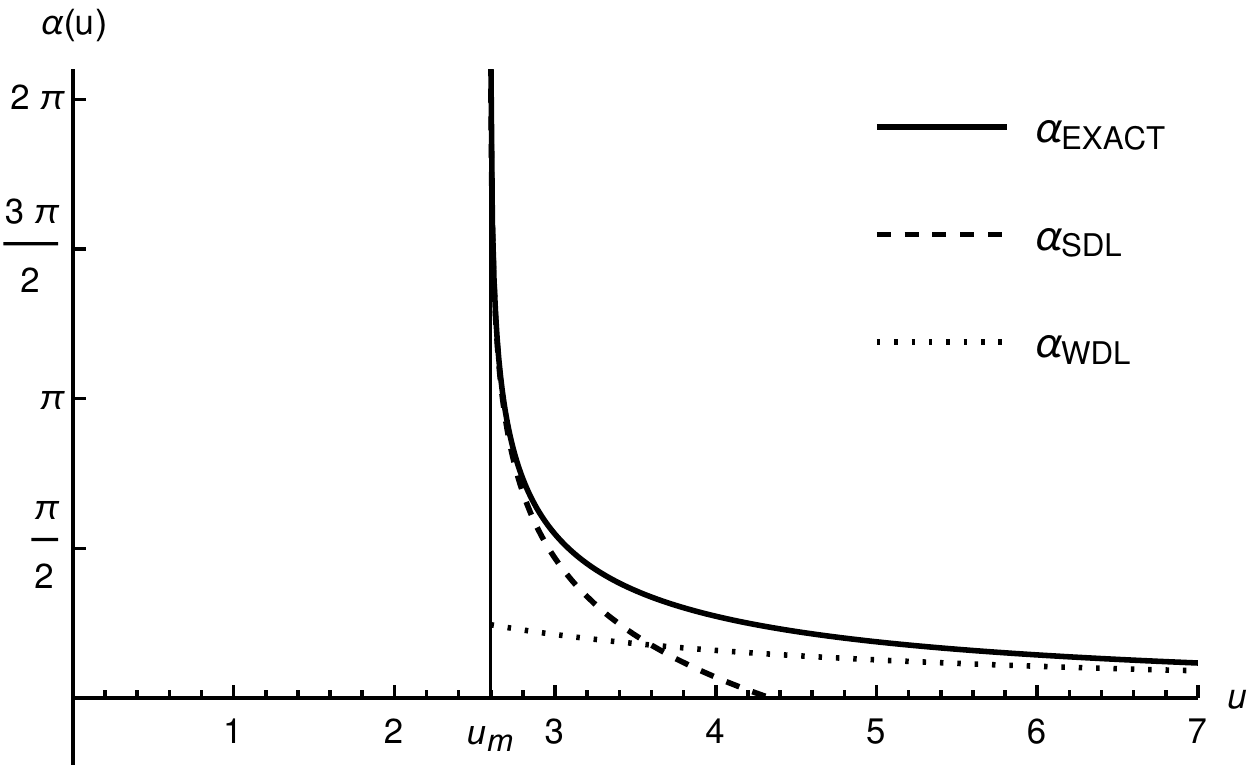}
\caption{Comparison between the deflection angle from the exact solution and the one given by both the WDL and SDL.}
\label{Fig_pasw}
\end{figure}

Now it is clear that both the WDL and SDL approximations have good accuracy on their range of applicability. The equations of the WDL are simple and effective for light rays travelling far from the lens. Their predictions have been successfully tested for decades. The $n$-loop images, although, are missed in the WDL.  However, as it keeps the non-linearity of GR, those images can be
well-described using the SDL approximation.  Fig. \ref{Fig_pasw} compares the deflection angle predicted by the three regimes in a range distance corresponding to $3M<r_0\lesssim 12.87M$.

The WDL and SDL approximation are both great theoretical achievements of gravitational lensing.  The later, however, has not been fully explored due to observational limitations. Its predictions are still beyond the current precision.

In the next section we are going to derive the SDL coefficients for the metric \eqref{Sph Metric}, following the seminal Bozza paper \cite{Bozza:2002zj}. 

\section{The SDL coefficients for Spherically Symmetric Space-Times}

Here we are interested in the full non-linear regime of GR, that translates in considering the light trajectory in the neighbourhood of compact objects like black holes. The basic equations that we are going to use in this regime were introduced by Bozza \textit{et al} in the references \cite{Bozza:2001xd,Bozza:2002zj}. Their idea was to split the integral of the equation \eqref{alpha Sch} into a singular and a regular part. Doing so, they were able to show that the singular part has a logarithmic divergence, which could be expanded and appropriately approximated. The regular part, on the other hand, could be calculated numerically and added to the final result. 

Let us establish a simplified notation: in this section we are going to denote $A(r_0)\equiv A_0$ as well as $A'(r_0)\equiv A'_0$ and $A''(r_0)\equiv A''_0$. The equations are derived as follows. By doing the change of variables
\begin{eqnarray}\label{rtoz}
&& z= \frac{A(r)-A(r_0)}{1-A(r_0)}\equiv  \frac{A(r)-A_0}{1-A_0},
\end{eqnarray}
the integral (\ref{alpha Sch})
becomes
\begin{align}
I(r_0)=\int\limits_0^1 R(z,r_0) f(z,r_0) dz \label{I z},
\end{align}
with
\begin{align}
 R(z,r_0)&=2\frac{\sqrt{B(r(z)) A(r(z))}}{r(z)^2 A'(r(z)) }\left( 1-A_0 \right)
r_0, \label{R} 
\end{align}
and
\begin{align}
 f(z,r_0)&=\frac{1}{\sqrt{A_0- \left[ \left(1-A_0 \right) z+ A_0
\right]\frac{r_0^2}{r(z)^2}}}.
\end{align}
The metric coefficients, initially depending on $r$, are made $z$ dependent by expressing $r$ as a function of $z$. It is done by inverting the equation \eqref{rtoz}, with $r(z)=A^{-1} \left[\left(1-A(r_0) \right) z+ A(r_0) \right]$. In the Eq. \eqref{I z} it was assumed that the space-time is asymptotically flat.

As is straightforward to see, the divergence is carried by $f(z,r_0)$,  when $z \rightarrow 0$. Its analysis is made by expanding the argument of the square root in $f(z,x_0)$, up to the second order in
$z$. Performing the expansion gives
\begin{align}
f(z,r_0)&=f_0(z,r_0)+\mathcal{O}(z^{-3/2})
\end{align}
where
\begin{eqnarray}
f_0(z,x_0)= \frac{1}{\sqrt{\kappa z +\eta
z^2}}
\end{eqnarray}
and
\begin{eqnarray}
& \kappa=\kappa(r_0) & =\frac{1-A_0}{r_0 A'_0}  \left(2 A_0-r_0 A'_0
\right) \label{alpha} \\%
& \eta= \eta(r_0) & = \frac{\left( 1-A_0 \right)^2}{r_0^2 {A'_0}^3} \left[
 2r_0 {A'_0}^2  -3A_0 A'_0 -r_0A_0
A''_0 \right]. \label{beta}
\end{eqnarray}
Here the identities $(A^{-1})'(A_0)=1/A_0'$ and $(A^{-1})''(A_0)=-A_0''/{A'_0}^3$ have been used in order to get the above results. Using the approximation $f(z,r_0) \approx f_0(z,r_0)$, the divergence in \eqref{I z} occurs when $\kappa\to 0$. From the Eq. \eqref{alpha} we see that it happens when $2 A_0-r_0 A'_0=0$, which is equivalent to the Eq. \eqref{photon_sphere}, for the radius $r_m$ of the photon sphere. It simply means that, as expected, the deflection angle diverges when the closest approximation coincides with the photon sphere. The key point now is to rewrite the Eq. \eqref{I z} as
\begin{align}\nonumber
I(r_0)&=\int\limits_0^1 R(z,r_0) f(z,r_0) dz\\ \nonumber
&=\int\limits_0^1 \left\{R(z,r_0) f(z,r_0)-\left[R(z,r_0)f_0(z,r_0)-R(z,r_0)f_0(z,r_0) \right]_{z\to 0, r_0\to r_m}\right\}dz\\\nonumber
&=\int\limits_0^1 \left[R(z,r_0) f(z,r_0)-R(0,r_m) f_0(z,r_0)\right]dz+\int\limits_0^1 R(0,r_m) f_0(z,r_0)dz\\\nonumber
&\equiv\int\limits_0^1 g(z,r_0)dz+\int\limits_0^1 R(0,r_m) f_0(z,r_0)dz\\
&\equiv I_R(r_0)+I_D(r_0).
\end{align}
with\footnote{We have used the Eq. \eqref{photon_sphere}.} 
\begin{align}
R(0,r_m)=\sqrt{\frac{B(r_m)}{A(r_m)}}[1-A(r_m)].
\end{align}
Then the integral $I(r_0)$ splits into a regular part $ I_R(r_0)$ and a divergent part $ I_D(r_0)$. As mentioned before, the regular part could be numerically integrated  and added to the final result. It will be denoted as $b_R$. The divergent part can be expanded at the photo sphere neighbourhood, yielding \cite{Bozza:2002zj}

\begin{equation}
I_D(r_0)=-a \log
\left(\frac{r_0}{r_m}-1 \right)+b_D + \mathcal{O}(r_0-r_m), \label{ID}
\end{equation}
and the coefficients
\begin{eqnarray}
&& a=\left.\frac{R(0,r_m)}{\sqrt{\eta}}\right|_{r_0=r_m} \label{a} \\ %
&& b_D=\left.\frac{R(0,r_m)}{\sqrt{\eta}} \log \frac{2(1-A_0)}{A'_0
r_0}\right|_{r_0=r_m} \label{bD}.
\end{eqnarray}
Up to first order in $(r_0-r_m)$, the deflection angle can then be expressed as
\begin{align}\label{alphar}
\alpha(r_0)\approx -a \log
\left(\frac{r_0}{r_m}-1 \right)+b_D +b_R-\pi.
\end{align}
This is the approximation of the SDL for $\alpha$, depending on the minimum distance $r_0$.

\subsection{SDL Coefficients for $\alpha(\theta)$}

The distance $r_0$ is not an observable quantity, in this sense seems reasonable that the angle $\theta$ is more suitable to be the dependent variable of the function $\alpha$. The impact parameter is connected to both quantities by the Eqs. \eqref{u rm} and \eqref{utheta}, with the latter being approximated to $u=\theta D_{ol}$ for small $\theta$. Those facts can thus be used to convert $\alpha(r_0)$ into $\alpha(\theta)$. In fact, from the Eq. \eqref{u rm}, the impact parameter on the photon sphere is
\begin{equation}
u_m=\left.\frac{r_0}{\sqrt{{A_0}}}\right|_{r_0=r_m}\equiv \frac{r_m}{\sqrt{{A_m}}}. \label{um}
\end{equation}
The quantity $u-u_m$ can be expanded around $r_0=r_m$ as
\begin{eqnarray}\label{uum}
&& u-u_m=\chi \left(r_0-r_m \right)^2,
\end{eqnarray}
with $\chi$ given by
\begin{eqnarray}
&& \chi=\frac{2 A_m -r_m^2 A''_m}{4 r_m\sqrt{A_m^3}}.
\end{eqnarray}
Rearranging the Eq. \eqref{uum} reads
\begin{align}
\sqrt{\frac{u_m}{\chi r_m^2}}\sqrt{\frac{u}{u_m}-1}=\frac{r_0}{r_m}-1.
\end{align}
Inserting it into the Eq. \eqref{alphar} we finally have
\begin{align}\label{alphat}\nonumber
\alpha(\theta)&\approx  -a \log
\left(\sqrt{\frac{u_m}{\chi r_m^2}}\sqrt{\frac{u}{u_m}-1} \right)+b_D +b_R-\pi\\\nonumber
&= -\frac{a}{2} \log
\left({\frac{u}{u_m}-1} \right)+\frac{a}{2} \log
\left({\frac{\chi r_m^2}{u_m}} \right)+b_D +b_R-\pi\\
&\equiv -\bar{a} \log
\left({\frac{\theta D_{ol}}{u_m}-1} \right)+\bar{b},\\
\end{align}
where the SDL coefficients $\bar{a}$ and $\bar{b}$ are given by,
\begin{eqnarray}
&& \overline{a}=\frac{a}{2}= \frac{R(0,r_m)}{2\sqrt{\eta_m}} \label{oa}%
\end{eqnarray}
and
\begin{eqnarray}
&& \overline{b}=\frac{a}{2} \log
\left({\frac{\chi r_m^2}{u_m}} \right)+b_D +b_R-\pi.\label{ob}
\end{eqnarray}
Those coefficients play a prominent role on the observables of the SDL regime. We shall better discuss how it allows the derivation of observable quantities for the gravitational lensing in Sect. \ref{obs}.  In the next section we are going to obtain the coefficients for the CFM and MGD solutions.

\section{Deflection and Coefficients of the MGD and CFM Solutions}

Before calculating the SDL coefficients of the CFM and MGD solutions we are going to, as a matter of comparison, calculate the same coefficients of the Schwarzschild solution. Its strong deflection limit gravitational
lensing was studied in Ref. \cite{Bozza:2001xd}. Those coefficients will also be used on the comparison between observables that we are going to do in the next section.

As it is well known, the Schwarzschild solution, in Schwarzschild units\footnote{Up to the end of this chapter, the lenghs shall be given as multiples of the Schwarzschild radius $r_s$, such that $r_s=1$.}, is given by
\begin{align}
ds^2=\left(1-\frac{1}{r}\right)dt^2-\left(1-\frac{1}{r}\right)^{-1}dr^2-r^2(d\vartheta^2+\sin^2\vartheta d\phi^2)
\end{align}
The integrands of the regular part $I_R(r_0)$, namely $R(z,x_0)$ and $f(z,x_0)$, read
\begin{eqnarray}
&&R(z,x_0)=2 \\%
&&f(z,x_0)=\frac{1}{\sqrt{\left( 2r_0-{3} \right) \frac{z}{r_0}+ \left(
{3} -r_0 \right) \frac{z^2}{r_0}-\frac{z^3}{r_0}}}.
\end{eqnarray}
From Eqs. (\ref{alpha}) and (\ref{beta}) we find $\kappa$ and $\eta$, the expansions coefficients of $f(z,x_0)$
\begin{eqnarray}
&& \kappa= 2-\frac{3}{r_0} \label{alpha sch}\\%
&& \eta= \frac{3}{r_0} -1. \label{beta sch}
\end{eqnarray}
As its solution coincides with the one for the photon sphere equation, the equation $\kappa=0$ gives us the radius of the photon sphere
\begin{equation}
r_m=\frac{3}{2},
\end{equation}
following
\begin{equation}
\eta_m=1.
\end{equation}
For the regular integral $I_R(r_0)$, there is no necessity for numerical integration in this case, as it can be integrated analytically, yielding
\begin{equation}
b_R=2 \log \left[ 6 \left( 2-\sqrt{3} \right) \right].
\label{bR Sch}
\end{equation}

From Eqs. (\ref{oa}), (\ref{ob}) and \eqref{um} follow the coefficients
$\overline{a}$, $\overline{b}$ and $u_m$ of the deflection angle
\begin{eqnarray}
&& \overline{a}=1 \\ %
&& \overline{b}=-\pi+b_R+\log 6\approx-0.4002 \\%
&& u_m=\frac{3 \sqrt{3}}{2}\approx 2.5981.
\end{eqnarray}
Then the Schwarzschild deflection angle, in the strong field
limit, is then
\begin{equation}
\alpha(\theta)=-\log \left(\frac{2\theta D_{ol}}{3\sqrt{3}} -1
\right) + \log \left[ 216 \left( 7-4\sqrt{3} \right) \right]-\pi.
\end{equation}

\subsection{MGD}

Here we are going to deal with the MGD solution introduced in the previous chapter. Its metric coefficients are
\begin{eqnarray}
&&A(r)= 1- \frac{1}{r}\\ %
&&B(r)=\left(1-\frac{1}{r}\right)^{-1}
\left[1+\left({1-\frac{3}{4\,r}}\right)^{-1}\,\frac{{\zeta}\upxi}{r}\right]^{-1},\label{BMGD}\\ 
&&C(r)= r^2 .
\end{eqnarray}

From the regular part of $I_R(r_0)$ we find
\begin{eqnarray}
&&R(z,x_0)=2\sqrt{\frac{ (3 {r_0}+4 z-4){r_0}}{4 \zeta\upxi (z-1)^2+{r_0} (3 {r_0}+4 z-4)}} \\%
&&f(z,x_0)=\frac{1}{\sqrt{\left( 2r_0-{3} \right) \frac{z}{r_0}+ \left(
{3} -r_0 \right) \frac{z^2}{r_0}-\frac{z^3}{r_0}}}.
\end{eqnarray}
The function $f(z,x_0)$ is identical to one of the Schwarzschild case, thus the same happens for $\kappa$ and $\eta$ 
\begin{eqnarray}
&& \kappa= 2-\frac{3}{r_0}\\%
&& \eta= \frac{3}{r_0} -1, 
\end{eqnarray}
as well as for $r_m=3/2$ and $\eta_m=1$. %
The integral $I_R(r_0)$ is not solved analytically, we shall simply represent it as $b_R$.

\begin{figure}[!htb]
\centering
\includegraphics[scale=.67]{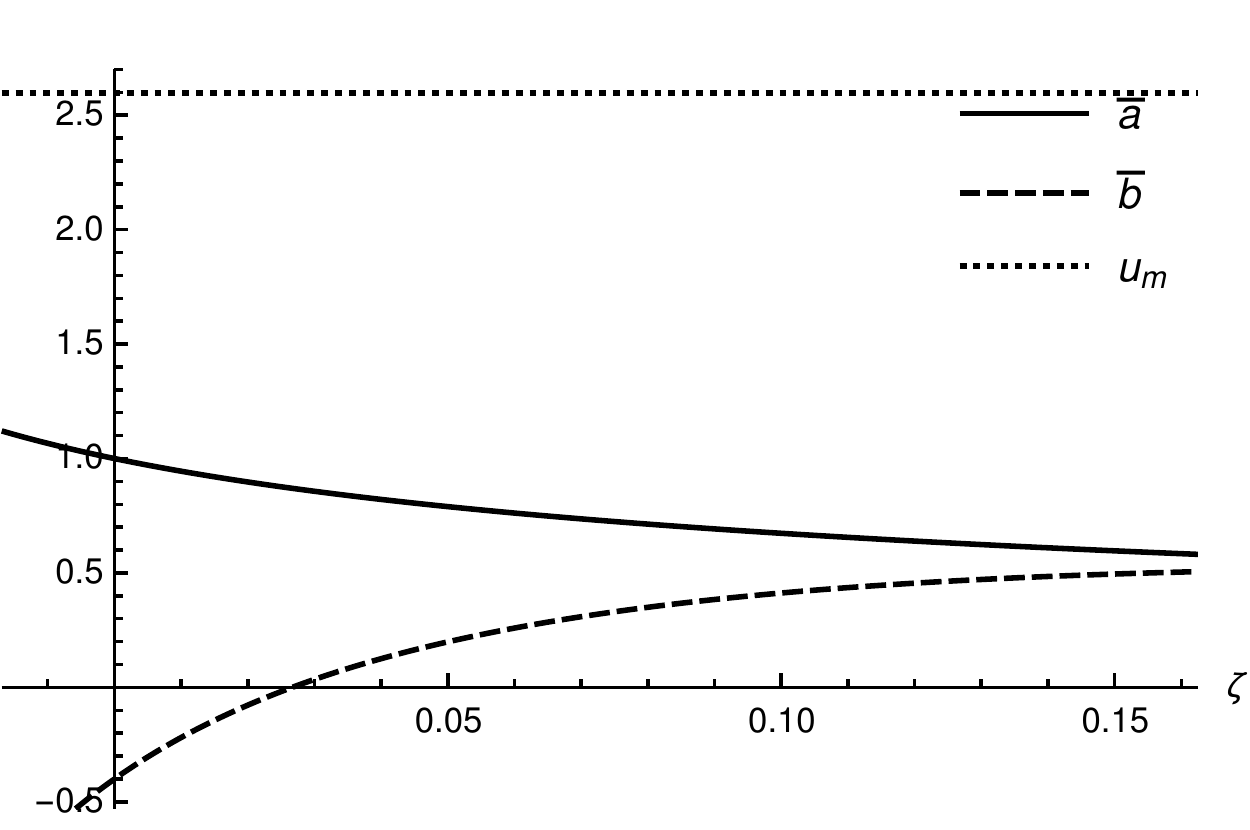}
\caption{SDL coefficients for the MGD solution as functions of the parameter $\zeta$ in Eq. (\ref{BMGD}).}
\label{fig_coeff_mgd}
\end{figure}

We use again Eqs. (\ref{oa}), (\ref{ob}) and \eqref{um} to find the coefficients
\begin{eqnarray}
&& \overline{a}=\sqrt{\frac{3}{16 \zeta \upxi+3}}, \\ %
&& \overline{b}=-\pi+b_R+\sqrt{\frac{3}{16 \zeta \upxi+3}}\log 6, \\%
&& u_m=\frac{3 \sqrt{3}}{2}\approx 2.5981.
\end{eqnarray}
Using the above coefficients we find the MGD deflection angle, 
\begin{equation}
\alpha(\theta)=-\sqrt{\frac{3}{16 \zeta \upxi+3}}\log \left(\frac{2\theta D_{ol}}{3\sqrt{3}} -1
\right) +b_R+\sqrt{\frac{3}{16 \zeta \upxi+3}}\log 6-\pi.
\end{equation}
Fig.  \ref{fig_coeff_mgd} shows the  behaviour of the SDL coefficients, upon varying the parameter $\zeta$ of the MGD solution. We assumed the value of 1.437$R_S$ for the parameter $\upxi$, according to the reason mentioned in the next section.


\subsection{CFM I \& II} 

Now the coefficients for both CFM solutions will be calculated. We start by the CFM I:
\begin{eqnarray}
&&A_I(r)= 1- \frac{1}{r}\\ %
&&B_I(r)=\frac{1-\frac{3}{4\,r}}
{\left(1- \frac{1}{r}\right)\left[1-\frac{1}{4\,r}(4\delta-1)\right]}\\
&&C_I(r)= r^2 .
\end{eqnarray}
As in the previous cases, the integrands of $I_R(r_0)$ are
\begin{eqnarray}
&&R(z,x_0)={2  \sqrt{\frac{4 {r_0}+3 z-3}{4 \delta (z-1)+4 {r_0}-z+1}}} \\%
&&f(z,x_0)=\frac{1}{\sqrt{\left( 2r_0-{3} \right) \frac{z}{r_0}+ \left(
{3} -r_0 \right) \frac{z^2}{r_0}-\frac{z^3}{r_0}}}.
\end{eqnarray}
Again $f(z,x_0), \kappa,\eta, r_m$ and $\eta_m$ are all identical to the Schwarzschild  case. The integral $I_R(r_0)$ is denoted by $b_R(r_0,\delta)$. Finally the coefficients of the strong deflection limit are
\begin{eqnarray}
& \overline{a} &=\sqrt{\frac{3}{7-4\delta}}, \\%
& \overline{b} &= -\pi +b_R(r_0,\delta)+\sqrt{\frac{3}{7-4\delta}}\log 6, \\%
& u_m &=\frac{3 \sqrt{3}}{2}.
\end{eqnarray}
Leading to the deflection angle
\begin{equation}
\alpha(\theta)=-\sqrt{\frac{3}{7-4\delta}}\log \left(\frac{2\theta D_{ol}}{3\sqrt{3}} -1
\right) + \sqrt{\frac{3}{7-4\delta}}\log 6  +b_R(r_0,\delta) -\pi.
\end{equation}

For the CFM II, the metric coefficients are
\begin{align}
A_{II}(r)
&=1- \frac{2\delta-1}{r}\\
 B_{II}(r)&=
 \frac{1}{(2\delta\!-\!1)^2}\!\left(2(\delta\!-\!1)\!+\!\sqrt{1\!-\! \frac{2\delta\!-\!1}{r}}\right)^2\\
C_{II}(r)&= r^2.
\end{align}

The integrands of $I_R(r_0)$ read
\begin{eqnarray}\nonumber
&&R(z,x_0)=2\left[\sqrt{\frac{2 \delta (z-1)+r_0-z+1}{r_0}}+2 \delta-2\right]\left[{\frac{ (2 \delta (z-1)+r_0-z+1)}{(1-2 \delta)^2 r_0}}\right]^{1/2} \\%
&&f(z,x_0)=\left[\frac{r_0}{z \left[ ((3-z) z-3)2 \delta-(r_0+3) z+2 r_0+z^2+3\right]}\right]^{1/2},
\end{eqnarray}
with the coefficients $\kappa$ and $\eta$ 
\begin{eqnarray}
&& \kappa= \frac{2 r_0+3-6 \delta}{r_0}\\%
&& \eta= -\frac{r_0+3-6 \delta}{r_0}. 
\end{eqnarray}
For this case we have $r_m=(6\delta-3)/2$ and $\eta_m=1$. Denoting again $I_R(r_0)$  as $b_R(r_0,\delta)$ we can find the SDL coefficients
\begin{eqnarray}
& \overline{a} &=\left[\frac{6 \delta+\sqrt{3} \sqrt{5-4 \delta}-6}{3(1-2 \delta)}\right]\sqrt{\frac{5-4 \delta}{12 \delta-9}}, \\%
& \overline{b} &= -\pi +b_R(r_0,\delta)+\overline{a}\log \frac{6 (4 \delta-3)}{5-4 \delta}, \\%
& u_m &=\frac{3\sqrt{3}}{2}  \sqrt{\frac{1}{5-4 \delta}},
\end{eqnarray}
and the deflection angle
\begin{equation}
\alpha(\theta)=- \overline{a}\log \left(\frac{2\theta D_{ol}}{3\sqrt{3}} \sqrt{{5-4 \delta}} -1
\right) + \overline{a}\log \frac{6 (4 \delta-3)}{5-4 \delta}  +b_R(r_0,\delta) -\pi.
\end{equation}

\begin{figure}[!htb]
\centering
\includegraphics[scale=.67]{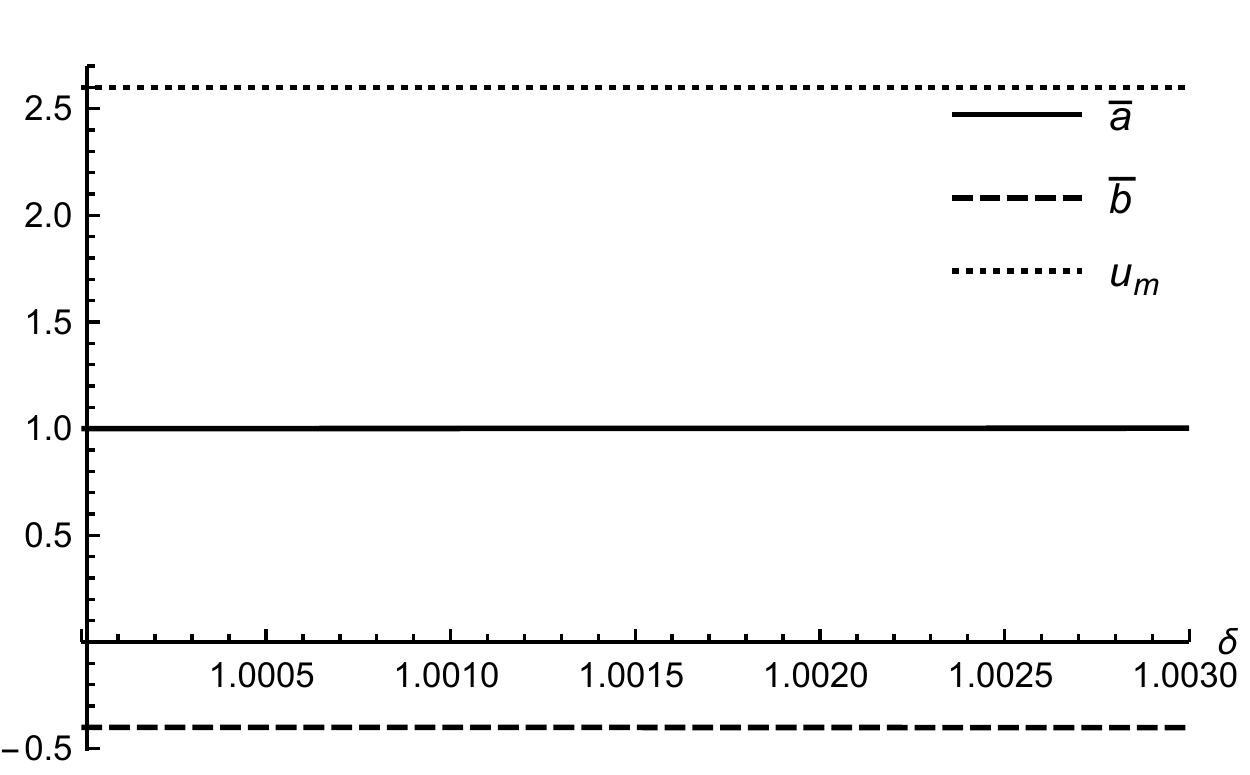}
\caption{SDL coefficients for the CFM I solution, as functions of the PPN parameter $\delta$.}
\label{fig_coeff_CFM I}
\end{figure}

\begin{figure}[!htb]
\centering
\includegraphics[scale=.67]{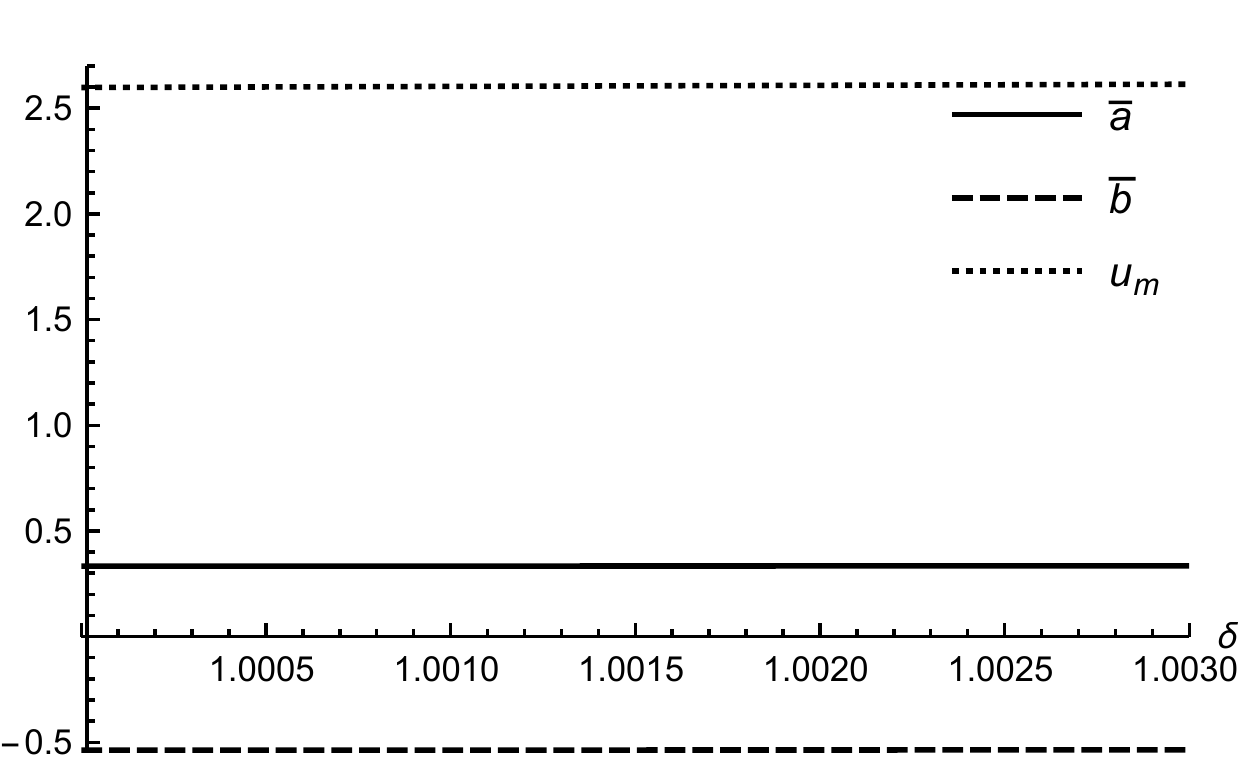}
\caption{SDL coefficients for the CFM II solution, as functions of the PPN parameter $\delta$.}
\label{fig_coeff_CFM II}
\end{figure}

 Figs. \ref{fig_coeff_CFM I} and \ref{fig_coeff_CFM II} show that, for the CFM I and CFM II solutions, the coefficients do not have an appreciable variation in the allowed range of the parameter $\delta$. It also can be seen in Table \ref{tabcfm1}. It is worth mentioning that the lines in Figs. \ref{fig_coeff_CFM I} and \ref{fig_coeff_CFM II} are not really straight, but just 
a resolutional consequence of the tiny range determined by the (currently observed) PPN parameter $\delta$. Furthermore, Figs. \ref{fig_coeff_mgd}, \ref{fig_coeff_CFM I} and \ref{fig_coeff_CFM II} show that the SDL coefficients are smooth in the allowed range.

\section[Observables in the {strong deflection limit}]{Observables in the {strong deflection limit}\protect\footnote{We have published the results of this section in Ref. \cite{Cavalcanti:2016mbe}.}}\label{obs}

One of the important aspects of the SDL approach is the potential to identify different black hole solutions. However, the applicability of the approach is hugely increased by the possibility of testing extensions of GR. In fact, any testable features of  theories beyond GR could provide us important hints on the nature of gravity, in such regimes. Our aim, in this section, is to identify the deviation of the MGD and CFM solutions from the classical Schwarzschild black hole. It is accomplished by analysing the observables of the SDL found in Ref. \cite{Bozza:2002zj} as well as the time delay of relativistic images \cite{Bozza:2003cp} and, subsequently, comparing them with the standard Schwarzschild one.

Some observable features of the SDL regime can be calculated using only the photon sphere impact parameter $u_{m}$ and the expansion coefficients $\bar{a}$ and $\bar{b}$. The first observable introduced in \cite{Bozza:2002zj} is the angular position $\theta_\infty$ of the accumulating relativistic images, which tends to the angular position of the photon sphere. The second, denoted by $s$, is related to the distance of relativistic images. It is defined as the angular distance between the largest and smallest orbit of the light rays winding around the black hole. The third one is the magnification of the images after the lensing effect. 

Despite the angle $\alpha$ between the lens and the image being large in the SDL, the effects are more prominent when the source, the lens, and the  observer are highly aligned ($\beta \approx 0$ in Fig. \ref{fig:lsetup}) \cite{Virbhadra:1999nm}. It means that the deflection angle $\alpha$ should be a small deviation of $2n\pi$, that is $\alpha(\theta)=2n\pi+\Delta \alpha_n$, with $\Delta \alpha_n$ small. The next step is to find the angles $\theta_n^0$ such that $\alpha(\theta_n^0)=2n\pi$, which represents a null deviation $\Delta \alpha_n$, and expand $\alpha(\theta)$ around $\theta_n^0$. Hence, the lens equation \eqref{alphat} straightforwardly yields  
\begin{equation}
\theta_n^0=\frac{u_m}{D_{ol}}\left[1+\exp\left(\frac{\bar{b}-2n\pi}{\bar{a}}\right)\right]\,,
\end{equation}
resulting in
\begin{equation}
\Delta\alpha_n=\frac{\bar{a}D_{ol}}{u_m \exp\left(\frac{\bar{b}-2n\pi}{\bar{a}}\right) }(\theta_n^0-\theta).
\end{equation}
It allows one to define the two following observables,\vspace*{-0.2cm}
\begin{align}
\theta_\infty\equiv&\lim_{n\rightarrow \infty}\theta_n^0=\frac{u_m}{D_{ol}},\\
s\equiv &\; \theta_1-\theta_\infty=\theta_\infty \exp\left(\frac{\bar{b}-2\pi}{\bar{a}}\right).
\end{align}\vspace*{-0.03cm}
\!\!The magnification resulting from the lensing effect, defined as the ratio of the flux of the image to the flux of the non lensed source \cite{Virbhadra:1999nm}, is the third observable quantity in the SDL regime. Given the flux\vspace{-0.3cm}
\begin{equation}\vspace*{-0.2cm}
\mu=\frac{\sin \theta}{\sin \beta}\left(\frac{d \beta}{d \theta}\right)^{-1},
\end{equation}
the ratio of flux of the images to the flux of the source, expressed as function of the SDL coefficients, reads \cite{Bozza:2002zj,Bozza:2009yw}\vspace*{-0.6cm}
\begin{equation}
k=\frac{\mu_1}{\sum_{n=2}^\infty \mu_n}=\exp\left(\frac{2\pi}{\bar{a}}\right).\end{equation}

Another useful information that can be extracted from the SDL regime and its coefficients is the time delay of relativistic images. A simple formula was derived by Bozza and Mancini in \cite{Bozza:2003cp}, relating the delay between the $n$-loop image and the $m$-loop image. According to their result, when the  lens and the observer are nearly aligned and the black hole has spherical symmetry,  the time delay is given by
\begin{equation}
\Delta T_{n,m}=T_{n,m}^0+T_{n,m}^1\,,\end{equation}
where
\begin{equation}
\Delta T_{n,m}^0 = 2\pi(n-m)u_m\,,\end{equation}
and the first correction  is given by
\begin{align}\nonumber
\Delta T_{n,m}^1=&2 \sqrt{\frac{B_m}{A_m}} \sqrt{\frac{u_m}{\chi}} \exp\left(\frac{\bar{b}}{2\bar{a}}\right)\left[\exp\left(-\frac{\pi  m}{\bar{a}}\right)-\exp\left(-\frac{\pi  n}{\bar{a}}\right)\right].
\end{align}
The first correction $T_{2,1}^1$, in the Schwarzschild case, was shown to contribute with only 1.4\% {}{to} the total time delay \cite{Bozza:2003cp}. 
 The cases analysed in the present chapter lead to a correction of order of $0.2-2$\%, 1.5\% and 0.0005\%, for the MGD, CFM I, and CFM II {}{metrics}, respectively.

{}{Having} introduced all the observable quantities, now we can use the known data from Sagittarius A$^*$, the galactic supermassive black hole at the center of our galaxy, to compare SDL lensing results of the Schwarzschild, MGD, and CFM solution. The observables and the SDL coefficients for MGD and CFM solutions are displayed in  Tables \ref{tabmgd1}, \ref{tabmgd2}, \ref{tabcfm1} and \ref{tabcfm2}. In both cases we calculated the results considering the allowed range of the parameter $\zeta$ for the MGD metric and the $\delta$ for the CFM metrics. We have used the recent results for the mass and distance of Sagittarius A$^*$, updating also the results for the Schwarzschild black hole. According to \cite{SgtA}, they are given by $M=4.02\times 10^6M_\odot$ and $D_{ol}=7.86$ kpc respectively. In what follows, the value $R=1.437 R_S$, regarding the MGD solutions, shall be taken into account in $\upxi$, corresponding to a compact object surface, by considering  the current value for the cosmological constant in the matching conditions at the {}{compact object} surface. In addition, in the tables below we denote $k_m=2.5\log k$ as the magnitude; $u_m/R_S$ is the normalised minimum impact parameter; $\Delta T_{2,1}$ denotes the time delay between the 1-loop and 2-loop relativistic images given in Schwarzschild time ($2GM/c^3 \approx 39,56s$) and $\Delta T_{2,1}(\mbox{min})$ is the same time delay, given in minutes. {It should be noticed that $\theta_\infty$ and $s$ are converted from radians to arcsec.}

\begin{table}
\centering
\begin{tabular}{l c c c c}
\toprule
 &{Schwarzschild}& \multicolumn{3}{c}{MGD} \\ 
 &  ($\zeta\equiv0$)&$\zeta= -1.69\times 10^{-2}$ & 1.31$\times 10^{-2}$ & 4.31$\times 10^{-2}$ \\
\toprule
$\theta_\infty$ $(\mu{\rm arcsec})$ & 20.21 & 26.21 &26.21 & 26.21 \\
$s$ $(\mu{\rm arcsec})$& 0.0328 & 0.0291 &0.0452 & 0.0136  \\ 
$k_m$ & 6.82 & 6.09 & 7.34 & 8.41  \\
$u_m/R_S$ & 2.6 & 2.6 & 2.6 &2.6  \\ 
$\bar{a}$ & 1 & 1.12 & 0.93 & 0.81  \\ 
$\bar{b}$ & $-$0.4002 & $-$0.8459 & $-$0.1715 & 0.1499  \\  
$\Delta T_{2,1}$ &16.573&16.646 &16.529&16.457 \\ 
$\Delta T_{2,1}(\mbox{min})$&10.93 &10.97&10.90 &10.85 \\ 
\bottomrule
\end{tabular}
\caption{Observables in the strong deflection limit for the MGD solution. }
\label{tabmgd1} 
\end{table}

\begin{table}
\centering
\begin{tabular}{lcccc}
\toprule
 & Schwarzschild& \multicolumn{3}{c}{MGD} \\ 
 &  ($\zeta\equiv0$)& 7.31$\times 10^{-2}$ & 10.31$\times 10^{-2}$ & 13.31$\times 10^{-2}$ \\
\toprule
$\theta_\infty$ $(\mu{\rm arcsec})$ & 20.21  &26.21 & 26.21 & 26.21\\
$s$ $(\mu{\rm arcsec})$& 0.0328  & 0.0074 & 0.0040 & 0.0022 \\
$k_m$ & 6.82  & 9.36 & 10.22 & 11.00 \\ 
$u_m/R_S$ & 2.6  &2.6 & 2.6 & 2.6 \\
$\bar{a}$ & 1  & 0.73 & 0.67 & 0.62 \\
$\bar{b}$ & $-$0.4002  & 0.3225 & 0.4203 & 0.4759 \\ 
$\Delta T_{2,1}$ &16.573&16.413&16.385&16.366 \\
$\Delta T_{2,1}(\mbox{min})$&10.93 &10.82&10.80&10.79 \\ 
\bottomrule
\end{tabular}
\caption{Observables in the strong deflection limit for the MGD solution.  }
\label{tabmgd2}
\end{table}

\begin{table}
\centering
\begin{tabular}{lcccc}
\toprule
 & Schwarzschild &   \multicolumn{3}{c}{CFM I} \\ 
 & ($\delta\equiv1$) &  $\delta=1.001$ &  $1.002$ &  $1.003$   \\
\toprule
$\theta_\infty$ $(\mu{\rm arcsec})$& 26.21 & 26.21 & 26.21 & 26.21\\
$s$ $(\mu{\rm arcsec})$& 0.0328 & 0.0329 &0.0331 & 0.0332  \\
$k_m$ & 6.82 & 6.817 & 6.813 & 6.808    \\
$u_m/R_S$ & 2.6 & 2.6 & 2.6 & 2.6 \\
$\bar{a}$ & 1 & 1.0007 & 1.0013& 1.0020    \\
$\bar{b}$ & $-0.4002$ & $-$0.4007 & $-$0.4011 & $-$0.4016   \\
$\Delta T_{2,1}$ &16.573 &16.574&16.574&16.575  \\
$\Delta T_{2,1}(\mbox{min})$ &10.93 &10.93&10.93&10.93 \\
\bottomrule
\end{tabular}
\caption{Observables in the strong deflection limit for the CFM I solution.}
\label{tabcfm1} 
\end{table}

\begin{table}
\centering
\begin{tabular}{lcccc}
\toprule
 & Schwarzschild &   \multicolumn{3}{c}{CFM II} \\ 
 & ($\delta\equiv1$) &  $\delta=1.001$ &  $1.002$ &  $1.003$ \\
\toprule
$\theta_\infty$ $(\mu{\rm arcsec})$& 26.21 &   26.26 & 26.31 & 26.36  \\
$s$ $(\mu{\rm arcsec})$& 0.0328   & $\sim 10^{-8}$ & $\sim 10^{-8}$ & $\sim 10^{-8}$ \\
$k_m$ & 6.82   & 20.436 & 20.410 & 20.377  \\
$u_m/R_S$ & 2.6 & 2.6 & 2.61& 2.61 \\
$\bar{a}$ & 1 & 0.3338 & 0.3348 & 0.3348  \\
$\bar{b}$ & $-0.4002$  & $-$0.5368 & $-$0.5364 & $-$0.5361  \\
$\Delta T_{2,1}$ &16.573 &16.357&16.390&16.422  \\
$\Delta T_{2,1}(\mbox{min})$ &10.93 &10.78&10.81&10.83  \\
\bottomrule
\end{tabular}
\caption{Observables in the strong deflection limit for  the CFM II solution.}
\label{tabcfm2}  
\end{table}



 The effects of gravitational lensing in the strong field regime are concretely noticeable   wherein the thorough capture of the photon by the black hole is regarded. 
Regarding both the CFM solutions, and the MGD of GR, we calculated the observables of the gravitation lensing of a background source in the strong field regime. The  angular difference  $s$ from the outmost image and the adherent point formed by the other images; the parameter $k_m$ that reveals the image magnification; the normalised minimum impact parameter $u_m/R_S$ were studied, unravelling a possible observable signature 
for the CFM I, the CFM II, and the MGD solutions. Moreover, the delay between relativistic images is significant for MGD and CFMII solutions, being a potentially key information on the characterisation of the solution that may model the supermassive  black hole in the centre of our galaxy. 
The typical signature of CFM I, CFM II, and MGD solutions are, mainly, observable by the parameters $k_m$ and $\theta_\infty$, as well as the time delay. It is worth  mentioning that the CFM II solution is hugely different, when compared to the Schwarzschild, CFM I, and MGD solutions, in what regards the parameter the magnitude $k_m$. The signatures of the MGD, on the other hand, could be evinced by the combinations of the parameters $k_m$ and $s$. However, even the difference on the values of $s$ for the MGD and Schwarzschild being potentially higher than one order of magnitude, it is still a little beyond resolution ESA satellite mission, which can reach 7 $\mu$arcsec \cite{Jordan:2008ky}. Thus, from the observational perspective, the astrophysical data should be better improved in order to apply the thorough potential of the gravitational lensing in the strong field regime.


\chapter{Quantum Black Holes and the Horizon Wave Functions Formalism} 

\label{cap6} 


\par
Physics laws are well known to hold on their respective appropriate scale. It makes physicists to deal with a wide range of fields, from quantum mechanics to cosmology, and scale dependent phenomena occurring in between them. In the last century, reaching higher energies has allowed us to look closer to the structure of matter.  It raises the question: assuming higher energies are achieved, could we always study smaller structures?  Or there is a limit beyond which we cannot go? Rather this limit exists or not, any answer to these questions must take quantum mechanics into account. However, such a high energy in small volumes should change the space-time itself, thus one also cannot neglect the curvature of the background. Therefore gravity plays a prominent hole in such an extreme scenario. 

In this chapter we are going to discuss a recent proposal to study phenomenologically gravity and quantum mechanics, the so called horizon wave function. Besides its simplicity interesting results can be extracted from such approach. One of then is a new derivation of the generalized uncertainty principle. The horizon wave function is also applied in extra dimensional scenarios.
\section{Minimum Length and GUP}

In recent years efforts have been made in order to, at least theoretically, find a reasonable answer to the questions placed at the beginning of the chapter. The interesting fact is that minimum lengths are predicted from different approaches to quantum gravity, as string theory \cite{Gross:1987ar,Konishi:1989wk,Amati:1988tn}, loop quantum gravity \cite{Rovelli:1994ge}, quantum black holes \cite{Scardigli:1999jh,Maggiore:1993rv}, among others (see \cite{Hossenfelder:2012jw,Casadio2015b,Sprenger:2012uc,Tawfik:2015rva} for recent reviews ). Some of those efforts have led to the so called \textit{Generalized Uncertainty Principle} (GUP), from where a minimum length raises naturally. 

Let us start with an heuristic argument for introducing the idea of a minimum length scale. From optical microscopes, we know that for probing structures of a certain length $\Delta x$ we have to use photons whose wavelength is smaller than $\Delta x$, demanding  high energetic photons for tiny $\Delta x$.  On the other hand, when gravity is taken into account, there exists the possibility that the amount of energy necessary for probing the structure is high enough to form a horizon. In this case a black holes would be formed and the observation would not be completed. We can thus associate a minimum length to the length given by the inverse of the minimum energy necessary to form a black hole in the previous thought experiment. A general criteria for deciding if a highly dense region would or not become a black hole is the Kip Thorne's hoop conjecture \cite{Thorne:1972ji}. It states that a black hole is formed  if an amount of energy $\omega$ is compacted at any time into a region whose circumference in every direction is $R \leq 4\pi G \omega$. The hoop conjecture has not been proved yet, however there are analytical and numerical results that corroborates it \cite{Eardley:2002re,Hsu:2002bd}. 

Now assume that we have a particle of energy $\omega$. Its extension $R$ has to be larger than the Compton wavelength associated to the energy, so $R\geq 1/\omega$. Thus a larger energy implies better focus on the particle. However, according to the hoop conjecture a black hole with radius $2G\omega$ is formed if the extension drops below $4GE\omega$. The minimal length expected to be achievable before the black hole formation is the Planck length $\ell_p=\sqrt{\hbar G/c^3} \sim 10^{-35}$m, which is way above the current experimental capacity. As a result, the interesting fact is that the Planck length is not only the smallest scale in particle physics, but also the smallest admissible size of a black hole.

In quantum mechanics one of the key features of the theory is the Heisenberg Uncertainty Principle (HUP), which states that
\begin{align}
\Delta x \Delta p & \gtrsim \hbar.
\end{align}
The HUP means that we cannot know precisely the position and momentum of the particle simultaneously. Any increment of precision in the position results in a decrement precision in the momentum and \textit{vice versa}.

Suppose now that, for some reason, when we approach the quantum gravity scale the HUP changes to a deformed version \cite{Sprenger:2012uc}, as
\begin{align}\label{GUP0}
\Delta x \Delta p & \gtrsim \hbar\left(1+\alpha(\Delta p)^2\right)
\end{align}
or equivalently
\begin{align}
\Delta x & \gtrsim \hbar\left(\frac{1}{\Delta p}+\alpha\Delta p\right).
\end{align}
In this case, the above interpretation concerning the precision of position and momentum is no longer valid. In fact, for increasing momentum the system could reach a state where not even the inequality \eqref{GUP0} is still valid. Analogously to the heuristic discussion above, this point characterizes a scale, that is, the minimum scale length $\Delta x$ such that the inequality \eqref{GUP0} remains valid.

The apparent arbitrary deformation on the HUP is not as arbitrary as one could think. Actually that is the expression for the GUP emerging from different approaches to quantum gravity phenomenology \cite{Casadio2015b,Tawfik:2015rva} and, as we are going to see in this chapter, from the horizon have function formalism \cite{Casadio:2013aua}.

From the many different derivations of the GUP, the one obtained by Maggiore \cite{Maggiore:1993rv} figures among the most famous. He derives the GUP from a thought experiment of measuring the area of the apparent horizon of a black hole in the quantum gravity regime. The experiment consists in observing photons scattered by a black hole and use them to reconstruct the apparent horizon. Recording photons scattered due to Hawking radiation, in principle we would be able to obtain an "image" of the black hole and capturing photons from different angles locate its center. We thus would be able to measure the horizon radius. The measuring process is subject to two different source or errors. The first one comes from the limitations of the HUP as described above, where the observer plays the role of a microscope and
\begin{align}
\Delta x ^{(1)} \sim \frac{\lambda}{\sin \theta},
\end{align} 
where $\theta$ is the scattering angle. The second source of error comes from the effect of the Hawking radiation itself, according to the horizon radius  decreases from $2(M+\Delta M)$ to $2M$. Here  $\Delta M =h/\lambda$ is the energy of the emitted photon. The corresponding error is intrinsic to the measurement and it value is
\begin{align}
\Delta x ^{(2)} \sim {2G}\Delta M,
\end{align}
or equivalently
\begin{align}
\Delta x ^{(2)} \sim {2G}\frac{h}{\lambda}.
\end{align}
Using the inequality $\frac{\lambda}{\sin \theta}\geq \lambda$ and combining linearly the errors $\Delta x ^{(1)}$ and $\Delta x ^{(2)}$ one obtains
\begin{align}
\Delta x \gtrsim \lambda+\alpha {2G}\frac{h}{\lambda}\sim \frac{h}{\Delta p}+\alpha{2G}\Delta p,
\end{align}
which coincides with the GUP introduced previously.

A strength of the  Maggiore's above derivation is that the only assumption made is, besides the linear combination of the errors, the existence of the Hawking radiation. Making no reference to results coming from any quantum gravity candidate, the derivation could thus be called model-independent.  The price paid is that, in this derivation, there is no prediction for the value of the parameter $\alpha$. 

Once we have introduced the idea behind the GUP and the minimum length, in the next section we are going to discuss the Horizon Wave Function (HWF) formalism and rederive the GUP. The HWF has been shown as a fruitful route in quantum black hole phenomenology and beyond \cite{Casadio:2014vja,Casadio:2015qaq}.

\section{The Horizon Wave Functions Formalism}

The horizon wave function formalism raises as an effective approach to give us hints on what would be expected from black hole physics near the Planck scale. The main idea is to push basic features of quantum mechanics and gravity beyond our present experimental limits. In doing so one faces the  conceptual challenge of describing simultaneously classical and quantum objects, such as horizons and particles, consistently, in the same approach. The horizon wave function formalism, also called horizon quantum mechanics, proposes an effective merge of the mentioned classical and quantum objects by associating a wave function to the particle horizon. Such association enables the use of the quantum mechanics machinery in order to distinguish particles and black holes as well as to derive the GUP. Now we are able to introduce the HWF in the standard way.

As discussed in the Chapter \ref{cap0}
, horizons of black holes are described in general relativity as trapping surfaces, their location are determined by
\be
g^{ij}\nabla_i r \nabla_j r
\,=\,
0
\ ,
\ee
where $\nabla_i r$ is orthogonal to surfaces of constant area
$\mathcal{A} = 4\pi r^2$. As it is well known, from spherical symmetry and Einstein field equations follows
\be
g^{rr}
\,=\,
1-\frac{2\lp(m/m_p)}{r}
\ ,
\ee
where $m=m(r,t)$ is the Misner-Sharp mass
\be\label{msmass}
m(r,t)
\,=\,
4\pi \int_0^r \rho(\bar{r},t)\bar{r}^2 \, \d \bar{r}
\ ,
\ee
and $\rho=\rho(r,t)$ denotes the matter density.
A trapping surface then exists if there are values of $r$ 
and $t$ such that the gravitational radius
$
\Rh
=
2\,\lp \, {m}/{m_p}
\,
$
satisfies
\be
\Rh(r,t)
\,\geq\,
r
\ .
\ee
\par
Considering a spin-less point-particle of mass $m$, it follows from the HUP an uncertainty in the particle spatial localisation of the order of the Compton
scale $\lambda_m \simeq \hbar/m=\lp\,m_p/m$. Arguing that quantum mechanics gives a more precise description of physics, $\Rh$ makes sense only if it is larger than the Compton wave length associated to the same mass, namely $\Rh
\,\gtrsim\,
\lambda_m$, thus
\be
\,\lp \, {m}/{m_p}
\,\gtrsim\,
\lp\,m_p/m
\quad\Longrightarrow\quad
m
\,\gtrsim\,
m_p
\ .
\ee
It suggests the Planck mass as the minimum mass such that the Schwarzchild radius could be defined. From quantum mechanics, the spectral decomposition of a spherically symmetric matter distribution is given by the
wave-function
\be
\Ket{\psi_{\rm S}}
=
\sum_{E} C(E) \Ket{\psi_{E}}
\ ,
\ee
with the usual eigenfunction equation
\be
\hat{H} \Ket{\psi_{E}}
=
E\Ket{\psi_{E}}
\ ,
\ee
regardless of the specific form of the actual Hamiltonian operator  $\hat{H}$. 
Using the energy spectrum and inverting the expression of the Schwarzschild radius we have
\begin{align}
E=m_p\frac{\rh}{2\ell_p}.
\end{align}
Putting it back into the wave function one can define the (unnormalised) HWF as
\begin{align}
\psi_H(\rh)=C\left(m_p\frac{\rh}{2\ell_p}\right)
\end{align}
whose normalisation is fixed, as usual, by the inner product
\begin{align}
\langle {\psi_H}\Ket{\,\phi_H}=4\pi\int_0^\infty\psi^*_H(\rh)\phi_H(\rh)\rh^2\d \rh.
\end{align}
Note though that here, the classical radius $R_H$ is thus replaced by the expectation value of the operator $\hat{R}_H$. From the uncertainty on the expectation value follows that the radius will necessarily be ``fuzzy”, like the position of the source itself.%

One thing one have to know in order to establish a criterion for deciding rather a mass distribution do form or not a black hole is if it lies inside its own horizon of radius $r = \rh$. From quantum mechanics one finds that it is given by the product
\begin{align}\label{ppbh}
\mathcal{P}_<(r<\rh)=P_S(r<\rh)\mathcal{P}_H(\rh)
\end{align}
where the first term,
\begin{align}\label{ppsph}
P_S(r<\rh)=4\pi \int_0^{\rh}|\psi_S(r)|^2r^2\d r
\end{align}
is the probability that the particle resides inside the sphere of radius $r = \rh$, while the second term,
\begin{align}\label{ppgrad}
\mathcal{P}_H(\rh)=4\pi \rh^2|\psi_H(\rh)|^2
\end{align}
is the probability density that the value of the gravitational radius is $\rh$. Finally, the probability that the particle described by the wave-function $\psi_S$ is a BH will then be given by the integral of \eqref{ppbh} over all possible values of the horizon radius $\rh$, namely
\begin{align}
P_{BH}=\int_0^\infty \mathcal{P}_<(r<\rh) \d \rh
\end{align}
which is the main outcome of the HWF formalism.

\subsection{Gaussian Sources}

The previous construction could be made explicit by applying a model for the wave function. For simplicity and practical reasons, one usually chooses a spherically symmetric Gaussian wave-function describing a massive particle at rest, such as

\be
\psi_{\rm S}(r)
=
\frac{e^{-\frac{r^2}{2\,\ell^2}}}{(\ell\, \sqrt{\pi})^{3/2}}.
\ 
\label{Gauss}
\ee
The corresponding function in momentum space is thus given by
\begin{align}\nonumber
\tilde{\psi}_{\rm S}(p)
&=4\pi\int_0^\infty\frac{\sin(rp)}{\sqrt{8\pi^3}rp}\frac{e^{-\frac{r^2}{2\,\ell^2}}}{(\ell\, \sqrt{\pi})^{3/2}}r^2\d r\\
&=\frac{e^{-\frac{p^2}{2\,\Delta^2}}}{(\Delta\, \sqrt{\pi})^{3/2}}
\ ,
\label{momGauss}
\end{align}
where $\Delta=m_p \,\ell_p/\ell$ is the spread of the wave-packet in momentum space whose width $\ell$ should be diminished by the Compton length of the particle,
\be\label{compt}
\ell
\geq
\lambda_m
\sim
\frac{m_p\,\ell_p}{m}
\ .
\ee

Now the relativistic mass-shell relation in flat
space~\cite{Casadio:2013tma},
\be
p^2
=
E^2-m^2
\ ,
\ee
is assumed and the energy $E$ of the particle is expressed in terms of the related horizon radius $\rh=\Rh(E)$, following 
\begin{align}
E^2&=\frac{m_p^2\rh^2}{4\ell_p^2}.
\end{align}
Thus, from \eqref{momGauss} one find the HWF
\begin{align}\nonumber
\psi_{\rm H} (\rh)&=\mathcal{N}_{\rm H}\Theta(\rh-\Rh) \, e^{-\frac{m_p^2\rh^2}{8\Delta \ell_p^2}}\\
&=\mathcal{N}_{\rm H}\,
\Theta(\rh-\Rh)
\,  e^{-\frac{\ell^2\rh^2}{8\ell_p^4}}.
\label{nnormHWF1}
\end{align}
The Heaviside step function $\Theta$ appears above due to the imposition $E \geq m$. 
The normalisation factor $\mathcal{N}_{\rm H}$ is fixed according to
\begin{align}\nonumber
\mathcal{N}_{\rm H}^{-2} &=4\pi
\int_0^\infty |\psi_{\rm H}(\rh)|^2 \, \rh^{2} \, \d \rh \\ \nonumber
&=4\pi \,
\int_{0}^\infty 
\Theta(\rh-\Rh)
\,  e^{-\frac{\ell^2\rh^2}{4\ell_p^4}}
\rh^{2} \, \d \rh
\notag\\
&=4\pi \,
\int_{\Rh}^\infty 
\,  e^{-\frac{\ell^2\rh^2}{4\ell_p^4}}
\rh^{2} \, \d \rh.
\end{align}
Making $\rho=\frac{\ell^2 \rh^2}{4\ell^4}$,
\begin{align}\label{N2}\nonumber
\mathcal{N}_{\rm H}^{-2} &=\frac{16\pi\ell_p^6}{\ell^3}
\int_{\frac{m^2}{\Delta^2}}^\infty \rho^{3/2-1}e^\rho \, \d \rho \\ \nonumber
 &=\frac{16\pi\ell_p^6}{\ell^3}
\Gamma\left(\frac{3}{2},\frac{m^2}{\Delta^2}\right)\\ 
 &=\left[\frac{1}{4\ell_p^3}\sqrt{\frac{\ell^3}{\pi
\Gamma\left(\frac{3}{2},\frac{m^2}{\Delta^2}\right)}}\right]^{-2},
\end{align}
where $\Gamma(s,x)$ is the upper incomplete Euler-Gamma function, defined by
\be
\Gamma(s,x)
=
\int_x^\infty t^{s-1} \, e^{-t} \, \d t.
\ee
From Eqs. \eqref{nnormHWF1} and \eqref{N2} the normalized HWF follows straightforward
\begin{align}\label{hwfnorm}\nonumber
\psi_{\rm H} (\rh)&=\frac{1}{4\ell_p^3}\sqrt{\frac{\ell^3}{
\pi\Gamma\left(\frac{3}{2},\frac{m^2}{\Delta^2}\right)}}\,
\Theta(\rh-\Rh)
\,  e^{-\frac{\ell^2\rh^2}{8\ell_p^4}}.
\end{align}
The expression above has two parameters non fixed \textit{a priori}, the particle mass $m$ and the Gaussian width $\ell$, resulting that the probability $P_{BH} = P_{BH}(\ell,m)$ will also depend on both. 

According to the previous section, the first thing we have to calculate in order to find the probability of black hole formation is the probability that the particle resides inside a sphere of radius $r=\rh$. From Eqs. \eqref{ppsph} and \eqref{Gauss} one obtains
\begin{align}\nonumber
P_S(r<\rh)&=4\pi\int_0^{\rh}|\psi_S(r)|^2r^2\d r\\\nonumber
&=\frac{4\pi}{\ell^3\pi^{3/2}}\int_0^{\rh} e^{-\frac{r^2}{\ell^2}}r^2\d r\\\nonumber
&=\frac{4}{\ell^3\pi^{1/2}}\frac{\ell^3}{2}\int_0^{\frac{\rh^2}{\ell^2}} y^{\frac{3}{2}-1}e^{y}\d y\\
&=\frac{2}{\sqrt{\pi}}\gamma\left(\frac{3}{2},\frac{\rh^2}{\ell^2}\right)
\end{align}
where we took $y={r^2}/{\ell^2}$ and $\gamma(s,x) = \Gamma(s) - \Gamma(s,x)$ is the lower incomplete Gamma function. From the Eqs. \eqref{ppgrad} and \eqref{hwfnorm} yields
\begin{align}\nonumber
\mathcal{P}_H(\rh)&=4\pi \rh^2|\psi_H(\rh)|^2\\
&=\frac{1}{4\ell_p^6}{\frac{\ell^3}{
\Gamma\left(\frac{3}{2},\frac{m^2}{\Delta^2}\right)}}\,
\Theta(\rh-\Rh)
\,  e^{-\frac{\ell^2\rh^2}{4\ell_p^4}}\rh^2.
\end{align}
Combining the previous results one finds the probability density that the particle resides in its own gravitational radius
\begin{align}\nonumber
\mathcal{P}_<(r<\rh)&=P_S(r<\rh)\mathcal{P}_H(\rh)\\
&=\frac{\ell^3}{2\sqrt{\pi}\ell_p^6}{\frac{\gamma\left(\frac{3}{2},\frac{\rh^2}{\ell^2}\right)}{
\Gamma\left(\frac{3}{2},\frac{m^2}{\Delta^2}\right)}}\,
\Theta(\rh-\Rh)
\,  e^{-\frac{\ell^2\rh^2}{4\ell_p^4}}\rh^2.
\end{align}
The probability that the particle described by the Gaussian is a black hole is finally given by
\begin{align}\label{pbh3d}
P_{BH}(\ell,m)=\frac{\ell^3}{2\sqrt{\pi}\ell_p^6}{\frac{1}{
\Gamma\left(\frac{3}{2},\frac{m^2}{\Delta^2}\right)}}\int_{\Rh}^\infty\,
\gamma\left(\frac{3}{2},\frac{\rh^2}{\ell^2}\right)
\,  e^{-\frac{\ell^2\rh^2}{4\ell_p^4}}\rh^2\d \rh,
\end{align}
which we calculate numerically. Before performing the calculations it is convenient  to express the the probability in terms of dimensionless variables, with lengths in unities of the Planck length ($\ell/\ell_p\to \ell$,  $\rh/\ell_p\to\rh$ and $\Rh/\ell_p \to \Rh$) and the particle mass in units of the Planck mass ($m/m_p\to m$). The probability above thus turns into
\begin{align}\label{pbh3dlm}
P_{BH}(\ell,m)=\frac{\ell^3}{2\sqrt{\pi}}{\frac{1}{
\Gamma\left(\frac{3}{2},m^2\ell^2\right)}}\int_{2m}^\infty\,
\gamma\left(\frac{3}{2},\frac{\rh^2}{\ell^2}\right)
\,  e^{-\frac{\ell^2\rh^2}{4}}\rh^2\d \rh.
\end{align}

If we assume the Gaussian width having the same order of the particle Compton length we could set $\ell \sim m^{-1}$ on the Eq. \eqref{pbh3dlm} and find the probability depending on only one of $\ell$ or $m$. On the other hand, departing again from the Eq. \eqref{compt} we may set values for $m$ in terms of the Planck mass and find the probability in this scenario. Applying  $\ell \sim m^{-1}$ yields	
\begin{align}\
P_{BH}(\ell)=\frac{\ell^3}{2\sqrt{\pi}}{\frac{1}{
\Gamma\left(\frac{3}{2},1\right)}}\int_{2\ell^{-1}}^\infty\,
\gamma\left(\frac{3}{2},\frac{\rh^2}{\ell^2}\right)
\,  e^{-\frac{\ell^2\rh^2}{4}}\rh^2\d \rh.
\end{align}
or
\begin{align}
P_{BH}(m)=\frac{1}{2m^3\sqrt{\pi}}{\frac{1}{
\Gamma\left(\frac{3}{2},1\right)}}\int_{2m}^\infty\,
\gamma\left(\frac{3}{2},m^2{\rh^2}\right)
\,  e^{-\frac{\rh^2}{4m^2}}\rh^2\d \rh.
\end{align}
The resulting probabilities are shown in the figures  \ref{pbh3d1} and \ref{pbh3d2} below.

\begin{figure}[h!]
\centering
\raisebox{3.75cm}{\footnotesize $P_{BH}$}
\includegraphics[width=5\textwidth/12]{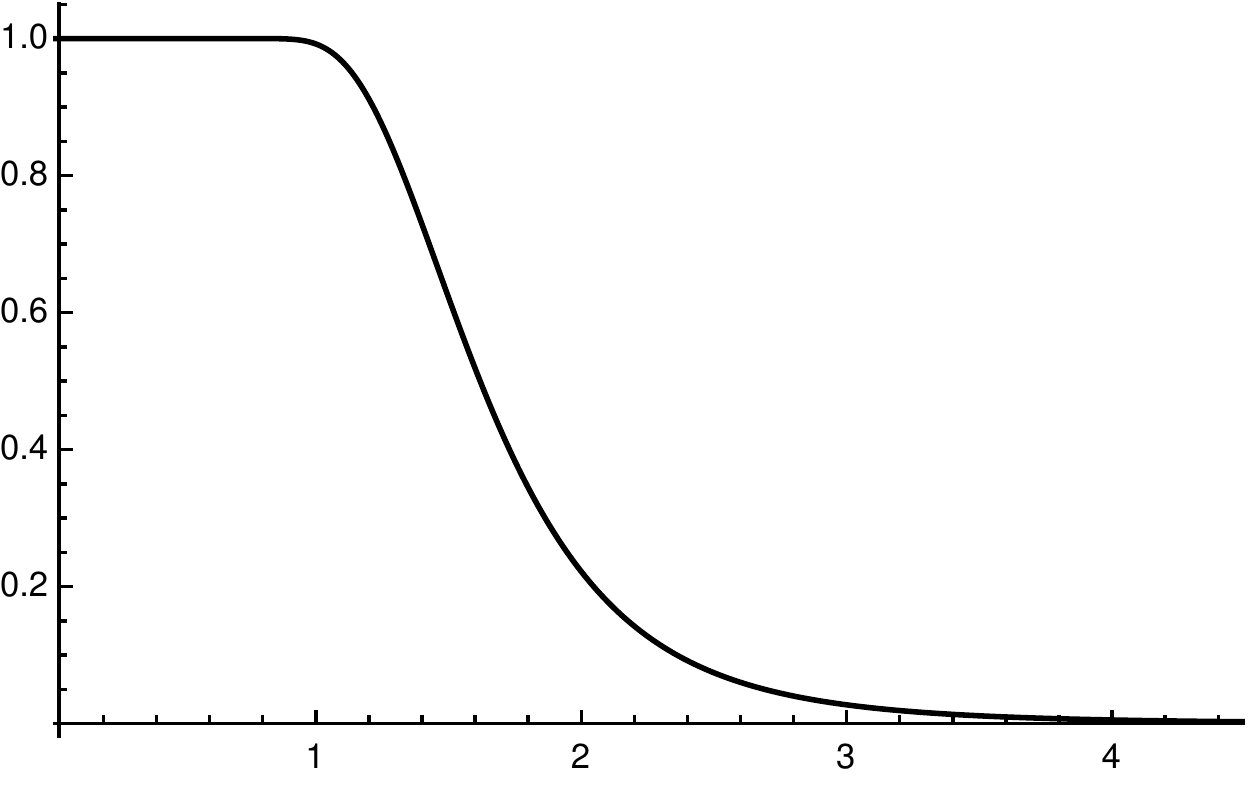}
$\quad$
\includegraphics[width=5\textwidth/12]{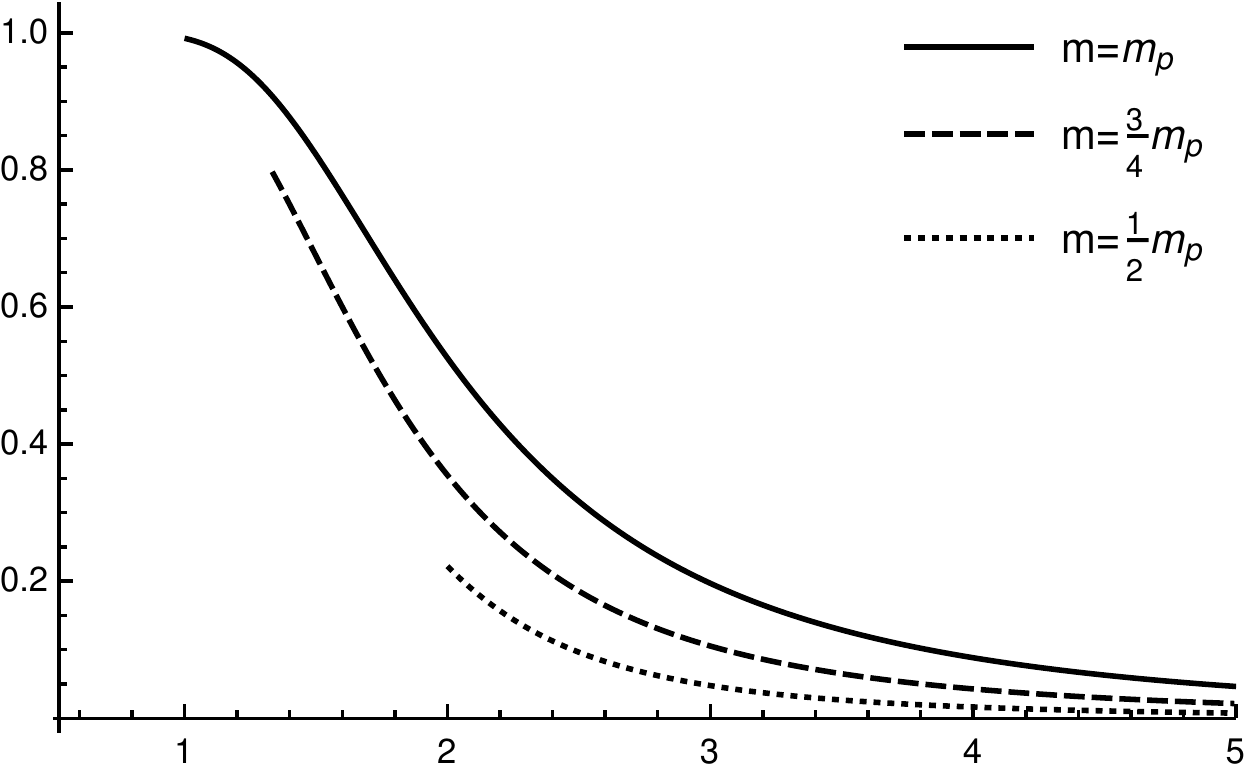}
{\footnotesize $\ell/\ell_D$}
\\
\begin{footnotesize}
$\ell \sim m^{-1}$
\hspace{6cm}
$\ell > m^{-1}$
\end{footnotesize}
\caption{The plot on the left hand side shows the probability of a "particle" being a black hole depending on the Gaussian width, assuming $\ell \sim m^{-1}$. On the right  hand side we see the same probability for different feactions of the planck mass  $m=m_p$ (solid), $m=3m_p/4$ (dashed) and $m=m_p/2$ (dotted).}
\label{pbh3d1}
\end{figure}

\begin{figure}[h!]
\centering
\raisebox{5.75cm}{\small $P_{BH}$}
\includegraphics[width=5\textwidth/8]{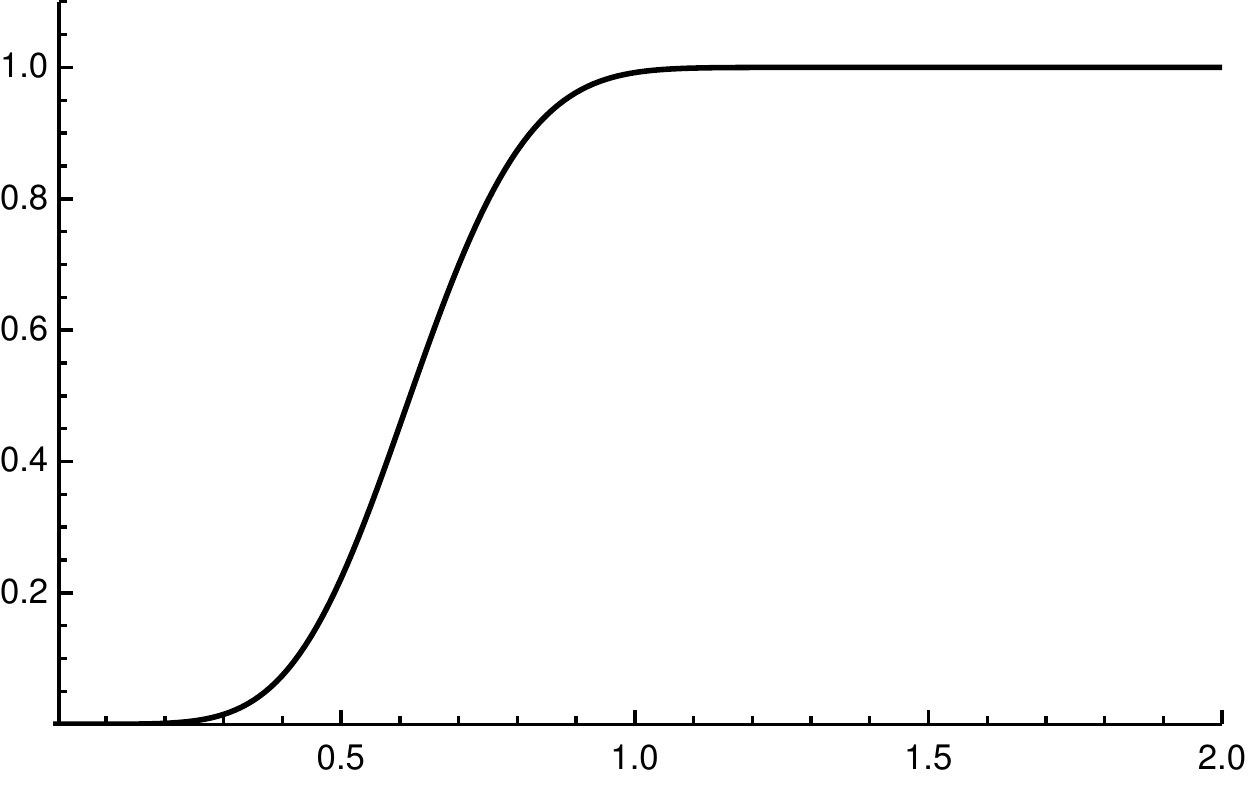}
{\small $m/m_p$}
\caption{Probability of a ``particle" being a black hole depending on the particle's mass and assuming $\ell \sim m^{-1}$.
\label{pbh3d2}}
\end{figure}

\subsection{GUP Revisited}

As the horizon wave function formalism applies the standard wave function description for particles, a question that raises naturally is rather and how it affects the HUP. As might be expected it results in a GUP,  like the one discussed at the beginning of the chapter. In quantum mechanics one derives the HUP by calculating the uncertainty associated to the wave function. Here, the starting point is the same, from the Gaussian wave-function \eqref{Gauss} follows the uncertainty in the particle size, given by
\begin{align}\label{dr}\nonumber
\Delta {r}_{0}^2&=\langle{r}^2\rangle-\langle {r}\rangle^2\\\nonumber
&4\pi\int_0^\infty|\psi_S(r)|^2r^4\d r-\left(4\pi\int_0^\infty|\psi_S(r)|^2r^3\d r\right)^2\\
&\frac{3}{2}\ell^2-\frac{4}{\pi}\ell^2=\frac{3\pi-8}{2\pi}\ell^2.
\end{align}
One might find the uncertainty in the horizon radius in analogous way
\begin{align}\label{drh}\nonumber
\Delta {r}_{\rm H}^2&=\langle{r}_{\rm H}^2\rangle-\langle {r}_{\rm H}\rangle^2\\\nonumber
&=4\pi\int_0^\infty|\psi_H(\rh)|^2\rh^4\d \rh-\left(4\pi\int_0^\infty|\psi_H(\rh)|^2\rh^3\d \rh\right)^2\\\nonumber
&=\frac{4\ell_p^4}{\ell^2}\frac{\Gamma\left(\frac{5}{2},1\right)}{\Gamma\left(\frac{3}{2},1\right)}-\left(\frac{2\ell_p^2}{\ell}\frac{\Gamma\left(2,1\right)}{\Gamma\left(\frac{3}{2},1\right)}\right)^2\\\nonumber
&=\frac{4\ell_p^4}{\ell^2}\left[\frac{\Gamma\left(\frac{5}{2},1\right)}{\Gamma\left(\frac{3}{2},1\right)}-\left(\frac{\Gamma\left(2,1\right)}{\Gamma\left(\frac{3}{2},1\right)}\right)^2\right]\\
&=\frac{4\ell_p^4}{\ell^2}\left[\frac{E_{-\frac{3}{2}}\left(1\right)}{E_{-\frac{1}{2}}\left(1\right)}-\left(\frac{E_{-1}\left(1\right)}{E_{-\frac{1}{2}}\left(1\right)}\right)^2\right]
\end{align}
where we have used the generalised exponential integral $E_{k}(x)$, defined by
\begin{align}
E_{k}(x)=\int_1^\infty \frac{e^{-xt}}{t^k}\d t
\end{align}
and related to the incomplete gamma function by the expression
\begin{align}
\Gamma(s,x)&=x^sE_{1-s}(x).
\end{align}
Note that $\Delta r_0$ and $\Delta \rh$ have difference dependence on $\ell$.  Such a dependence results in a vanishing $\Delta \rh$ for $\ell \gg \ell_p$, when the Heisenberg uncertainty is recovered, as expected. Furthermore, from Eq. \eqref{drh} we see that $\Delta r_{\rm H}\sim \ell^{-1}$, and from $\ell\sim m^{-1}$ follows $\Delta r_{\rm H}\sim m$. It means that the uncertainty on the horizon radius has the same order of magnitude of the horizon radius itself. That is an intriguing feature of quantum black holes derived from HWF that obviously is not expected for astrophysical black holes.
 
\begin{figure}[h!]
\centering
\raisebox{6.31cm}{\Large $\frac{\Delta r}{\ell_p}$}
\includegraphics[width=6\textwidth/8]{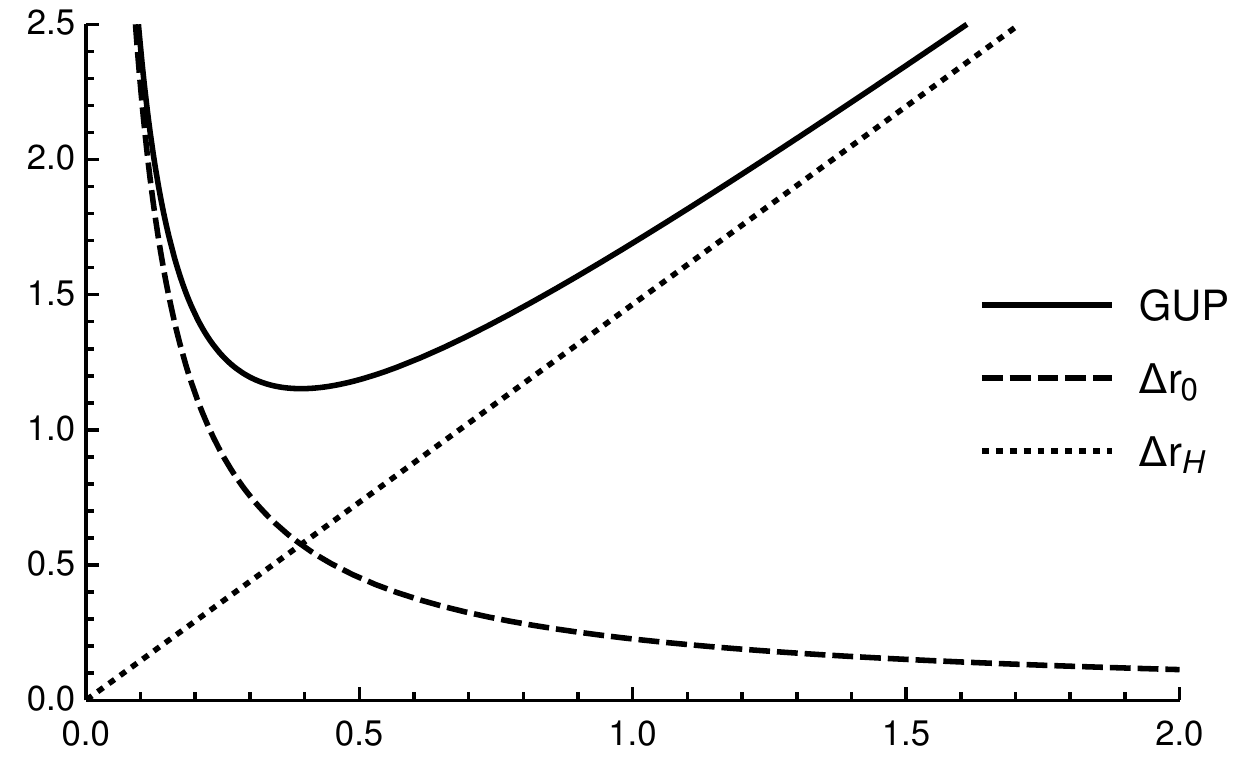}
{\Large $\frac{\Delta p}{m_p}$}
\caption{GUP profile emerged from the horizon wave function formalism for $\alpha=1$.
\label{gup3dp}}
\end{figure}

The total minimum radial uncertainty could now be taken as a linear combination of the quantities calculated above, given
\begin{align}\label{gup3d0}\nonumber
\Delta r&=\Delta r_{0}+\alpha\Delta \rh\\
&=\Delta^{1/2}_{0}\ell+\alpha\Delta^{1/2}_{0}\Delta_{\rm H}\frac{\ell_p^2}{\ell}
\end{align}
where $\Delta_0=\frac{3\pi-8}{2\pi}$ and
\begin{align}
\Delta^2_{\rm H}&=\frac{4}{\Delta_0}\left[\frac{E_{-\frac{3}{2}}\left(1\right)}{E_{-\frac{1}{2}}\left(1\right)}-\left(\frac{E_{-1}\left(1\right)}{E_{-\frac{1}{2}}\left(1\right)}\right)^2\right].
\end{align}
For the uncertainty in momentum we have
\begin{align}\nonumber
\Delta {p}^2&=\langle{p}^2\rangle-\langle {p}\rangle^2\\\nonumber
&4\pi\int_0^\infty|\tilde{\psi}_S(p)|^2p^4\d p-\left(4\pi\int_0^\infty|\tilde{\psi}_S(p)|^2p^3\d p\right)^2\\\nonumber
&=\frac{3}{2}\Delta^2-\frac{4}{\pi}\Delta^2=\frac{3\pi-8}{2\pi}\Delta^2\\
&=\Delta_{0}\frac{m_p^2\ell_p^2}{\ell^2}.
\end{align}
Note that the momentum uncertainty and the width $\ell$ are related by the expression above as $\Delta p/m_p\ell_p=\Delta^{1/2}_0/\ell$. Using this relation in the Eq. \eqref{gup3d0} results in
\begin{align}
\frac{\Delta r}{\ell_p}=\Delta_0\frac{m_p}{\Delta p}+\alpha\Delta_{\rm H}\frac{\Delta p}{ m_p},
\end{align}
which has the same form of the GUP discussed at the beginning of the chapter. The Fig. \ref{gup3dp} shows the behavior of the GUP as function of the momentum uncertainty, taking $\alpha=1$. There we can clearly see a minimum $\Delta r$ placed between $\ell_p$ and $1.5\ell_p$. If fact, one finds $\displaystyle \Delta p=\left(\frac{\Delta_0}{\alpha \Delta_{\rm H}}\right)^{1/2}m_p$ associated with the minimum length, corresponding to $\Delta r=2\sqrt{\alpha \Delta_0 \Delta_{\rm H}}\approx 1.15\sqrt{\alpha}\ell_p \equiv \ell^*$.

\begin{figure}[h!]
\begin{minipage}{\textwidth/2}
\raisebox{4.31cm}{ $\frac{\ell^*}{\ell_p}$}
\includegraphics[width=14\textwidth/15]{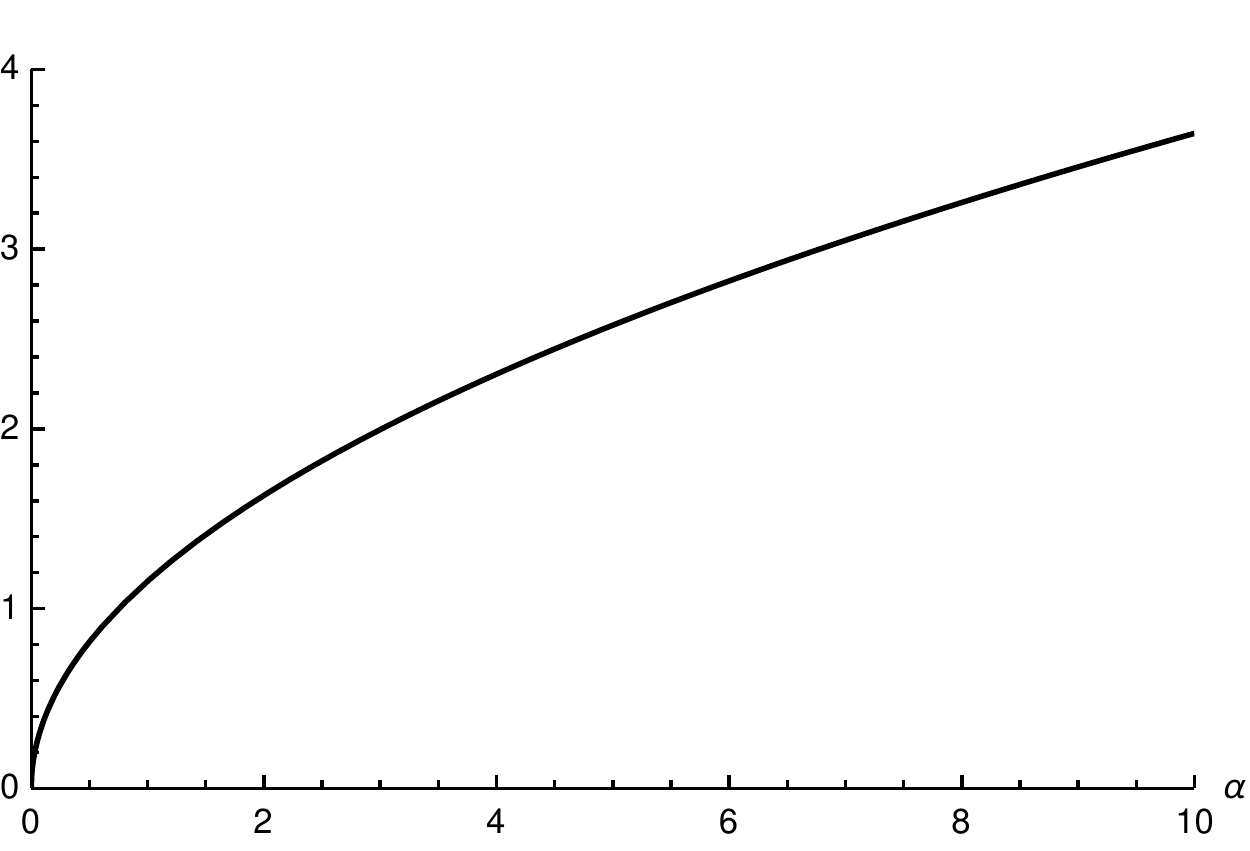}
%
\caption{Minimum length $\ell^*$ for a range of $\alpha$.
\label{minL3d}}
\end{minipage}
\begin{minipage}{\textwidth/2}
\raisebox{4.31cm}{ $\frac{m^*}{m_p}$}
\includegraphics[width=14\textwidth/15]{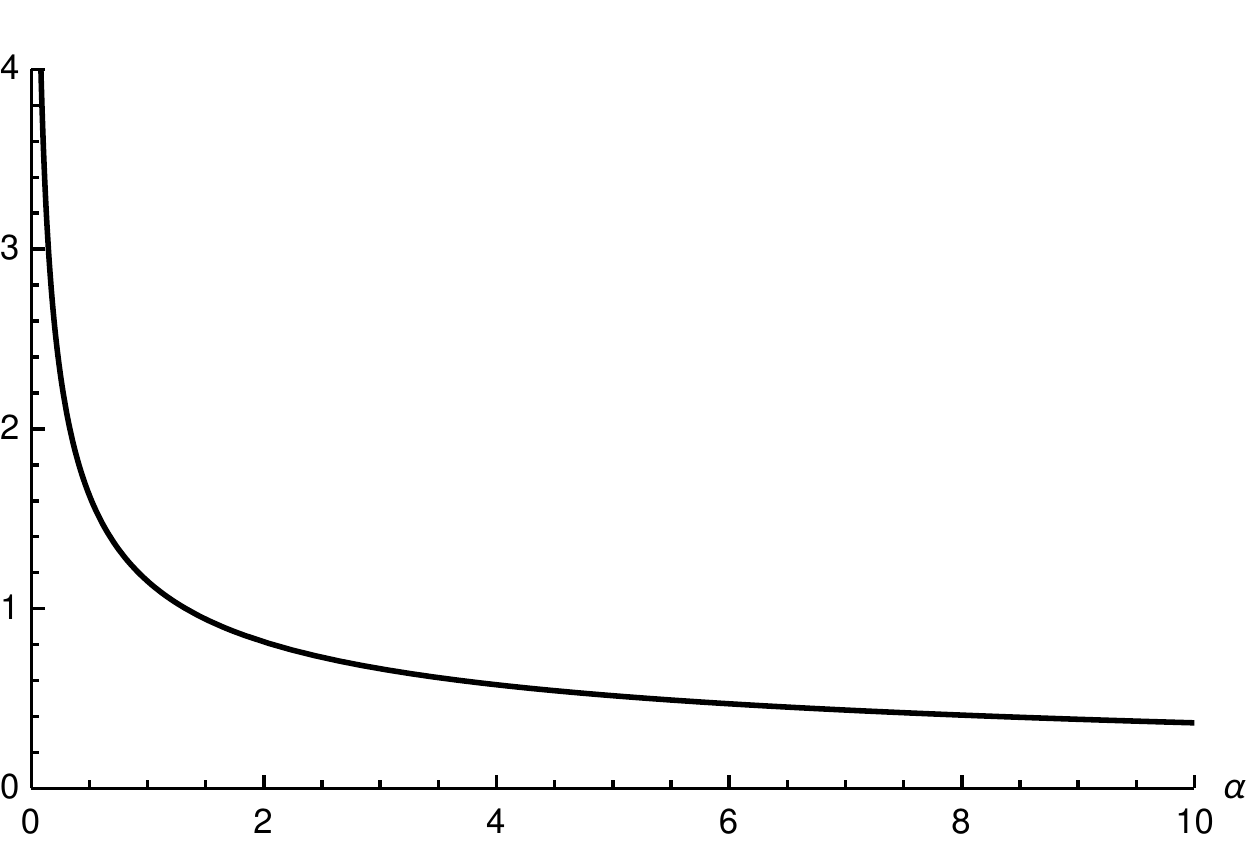}
%
\caption{Minimum mass $m^*$ for a range of $\alpha$.
\label{minM3d}}
\end{minipage}
\end{figure}

Rather than assuming $\alpha=1$, which according to the above results give us a minimum length very close to the Planck length, Figures \ref{minL3d} and \ref{minM3d} show the minimum length and minimum mass associated to the GUP for a range of $\alpha$. It is straightforward to see from the GUP expression that larger $\alpha$ means large correction for the quantum mechanics uncertainty. As a consequence we have large minimum length and small minimum mass, as can be seen in the Figures \ref{minL3d} and \ref{minM3d}. Note that not only the Planck length, but also the Planck mass emerges as a minimum scale when $\alpha\approx 1$.

\section[HWF and Extra Dimensions]{HWF and Extra Dimensions\protect\footnote{We have published the results of this section in Ref. \cite{Casadio:2015jha}.}}\label{HDHWF}

In this section we are going to study the effects of the horizon wave function formalism when the existence of extra dimensions is assumed. Besides some awkward expression on the way, the  approach in this session is pretty similar to the previous one, while some new interesting results are found.   

First of all we recall that given any spherically symmetric function $f=f(r)$ in $D$ spatial dimensions,
the corresponding function in momentum space is given by
\begin{align}\nonumber
\tilde{f}(p)&=\int \d \vec{x} \,
\frac{e^{-i \, \vec{p}\cdot \vec{x} / \hbar}}{(2\pi\hbar)^{D/2}}
\, f(r)\\\nonumber
&=\int_0^\infty \d r
\,r^{D-1}\, 
f(r)
\left[ \frac{\Omega_{D-2}}{(2\pi\hbar)^{D/2}}\int_0^\pi \d\theta\,
e^{-i\,p\,r\cos\theta /\hbar}\,(\sin \theta)^{D-2} \right]\\\nonumber
&=\int_0^\infty \d r 
\,r^{D-1}\, 
f(r)
\left[ \frac{(p\,r)^{\frac{2-D}{2}}}{\hbar} \,J_{\frac{D-2}{2}}\left(\frac{r\, p}{\hbar}\right)\right]\\
&= \frac{p^{\frac{2-D}{2}}}{\hbar}\int_0^\infty \d r \, r^{D/2}\,  J_{\frac{D-2}{2}}\left(\frac{r p}{\hbar}\right)f(r)
\end{align}
where $\Omega_{D-2}$ is the volume of the $D-1$ dimensional unity sphere and the normalised radial modes are given by the Bessel functions
\be
J_{\frac{D-2}{2}}\left(\frac{r p}{\hbar}\right)
=
\frac{\Omega_{D-2}}{(2\pi)^{D/2}} \, \left(\frac{r \,p}{\hbar}\right)^{\frac{D-2}{2}}
\int_0^\pi \d\theta\, e^{-i\,p\,r\,\cos\theta /\hbar}\,(\sin \theta)^{D-2}
\ ,
\ee
Accordingly, the inverse transform is given by
\be
{f}(r)
=
\frac{r^{\frac{2-D}{2}}}{\hbar}
\int_0^\infty \d p
\,
p^{D/2}\,
J_{\frac{D-2}{2}}\left(\frac{r\, p}{\hbar}\right)
\tilde{f}(p)
\ .
\ee
It is necessary in order to find the extra dimensional Gaussian wave function in momentum space. 
%
%
\label{GaussModes}

Considering again a localized massive particle described by the
Gaussian wave-function in $D$ spatial dimensions
\be
\psi_{\rm S}(r)
=
\frac{e^{-\frac{r^2}{2\,\ell^2}}}{(\ell\, \sqrt{\pi})^{D/2}}
\ ,
\label{Gauss}
\ee
and the corresponding function in momentum space is thus given by
\be
\tilde{\psi}_{\rm S}(p)
=
\frac{e^{-\frac{p^2}{2\,\Delta^2}}}{(\Delta\, \sqrt{\pi})^{D/2}}
\ ,
\label{momGaussd}
\ee
where $\Delta=m_D \,\ell_D/\ell$ is the spread of the wave-packet in momenta space, $m_D$ and $\ell_D$ are the fundamental mass and length scales for $D+1$ dimensional space-times. Analogously to the previous $D=3$ case, it follows that $\Delta\leq m$, which implies
\be
\ell
\geq
\lambda_m
\equiv
\frac{\hbar}{m}
=
\frac{m_D\,\ell_D}{m}
\ .
\ee
We have seen in the Section \ref{edschwarzchild} that the Schwarzschild metric, as a solution of the vacuum Einstein equations,
generalises in $(1+D)$-dimensional space-time as
\be
ds^2
=
-\left(1-\frac{\RSD}{r^{D-2}}\right) \, \d t^2
+\left(1-\frac{\RSD}{r^{D-2}}\right)^{-1} \, \d r^2
+r^{D-1} \, \d\Omega_{D-1}
\ ,
\ee
where the classical horizon radius is given by
\be
\RSD
=
\left(\frac{16\pi G_D m}{(D-1)\Omega_{D-1}}\right)^{\frac{1}{D-2}} 
\label{SchwD}
\ee
As expected, if $D=3$ we have the standard result of the previous session $R_{3}=\Rh$.
\par
As in $D=3$, we assume the relativistic mass-shell relation in flat
space~\cite{Casadio:2013tma},
\be
p^2
=
E^2-m^2
\ ,
\ee
and define the HWF expressing the energy $E$ of the particle
in terms of the related horizon radius~\eqref{SchwD}, $\rh=R_D(E)$.
From Eq.~\eqref{momGaussd}, we then get
\begin{align}\nonumber
\psi_{\rm H} (\rh)&=\mathcal{N}_{\rm H}\,
\Theta(\rh-R_D)
\, \exp\left\{ -
\frac{1}{2}\left(\frac{D-2}{2\,\gd\, \Delta}\right)^2
\, \left[ \rh^{2\,(D-2)} - R_D^{2\,(D-2)}
\right] \right\}\\
&=\Theta(\rh-\RS) \, \mathcal{N}_{\rm H} \, e^{m^2/2\Delta^2}
\, \exp\left\{ -
\frac{1}{2}\left(\frac{D-2}{2\gd \Delta}\right)^2
\, \rh^{2(D-2)} \right\},
\label{nnormHWF}
\end{align}

with normalisation $\mathcal{N}_{\rm H}$ 
\begin{align}\nonumber
\mathcal{N}_{\rm H}^{-2} e^{-m^2/\Delta^2}&={\Omega_{D-1}} \,
\int_0^\infty |\psi_{\rm H}(\rh)|^2 \, \rh^{D-1} \, \d \rh \\ \nonumber
&=\frac{2\,\pi^{D/2}}{\Gamma(D/2)} \,
\int_{R_D}^\infty 
\exp\left\{-\left(\frac{D-2}{2\gd \Delta}\right)^2 \, \rh^{2(D-2)} \right\}
\rh^{D-1} \, \d \rh
\notag\\
&=\frac{\pi^{D/2}}{(D-2)\Gamma(D/2)} \, \left(\frac{2\gd \Delta}{D-2}\right)^{\frac{D}{D-2}}
\int_{m^2/\Delta^2}^\infty e^{-\rho_{\rm H}}
\, \rho_{\rm H}^{\frac{D}{2D-4}-1} \, \d \rho_{\rm H}
\notag\\
&=\frac{\pi^{D/2}}{D-2} \, \left(\frac{2\gd \Delta}{D-2}\right)^{\frac{D}{D-2}}
\, \frac{\Gamma\left(\frac{D}{2D-4},1\right)}{\Gamma\left(\frac{D}{2}\right)},
\end{align}
and $\rho_{\rm H}$ was defined as
\be
\rho_{\rm H}
\,=\,
\left(\frac{D-2}{2\gd \Delta}\right)^2 \rh^{2(D-2)}
\ .
\label{rhoH}
\ee
The coefficient $\mathcal{N}_{\rm H}$ is thus given by
\be
\mathcal{N}_{\rm H} \, e^{m^2/2\Delta^2}
\,=\,
\left\{
\frac{D-2}{\ell_D^{\,D}\,\pi^{D/2}} \, \left[\frac{(D-2)\,m_D}{2\,\Delta}\right]^{\frac{D}{D-2}}
\, \frac{\Gamma\left(\frac{D}{2}\right)}
{\Gamma\left(\frac{D}{2D-4},\frac{m^2}{\Delta^2}\right)}
\right\}^{1/2}
\ ,
\ee

thus, combining with the Eq.~\eqref{nnormHWF} follows
\begin{align}\label{DHWF}
\psi_{\rm H}(\rh)=&\left\{
\frac{D-2}{\ell_D^{\,D} \, \pi^{D/2}}
\left[\frac{(D-2)\,m_D}{2\,\Delta}\right]^{\frac{D}{D-2}}
\, \frac{\Gamma\left(\frac{D}{2}\right)}
{\Gamma\left(\frac{D}{2D-4},\frac{m^2}{\Delta^2}\right)}
\right\}^{1/2}
\nonumber\\
&\times
\Theta(\rh-R_D)\, \exp\left\{ -
\frac{(D-2)^2}{8} \, \frac{m_D^2}{\Delta^2}
\, \left(\frac{\rh}{\ell_D}\right)^{2(D-2)} \right\}
\end{align}

\subsection{Black Hole Probability}
\label{Probability}
\begin{figure}[t!]
\centering
\raisebox{3.75cm}{ {\footnotesize $P_{\rm BH}$}}
\includegraphics[width=5\textwidth/12]{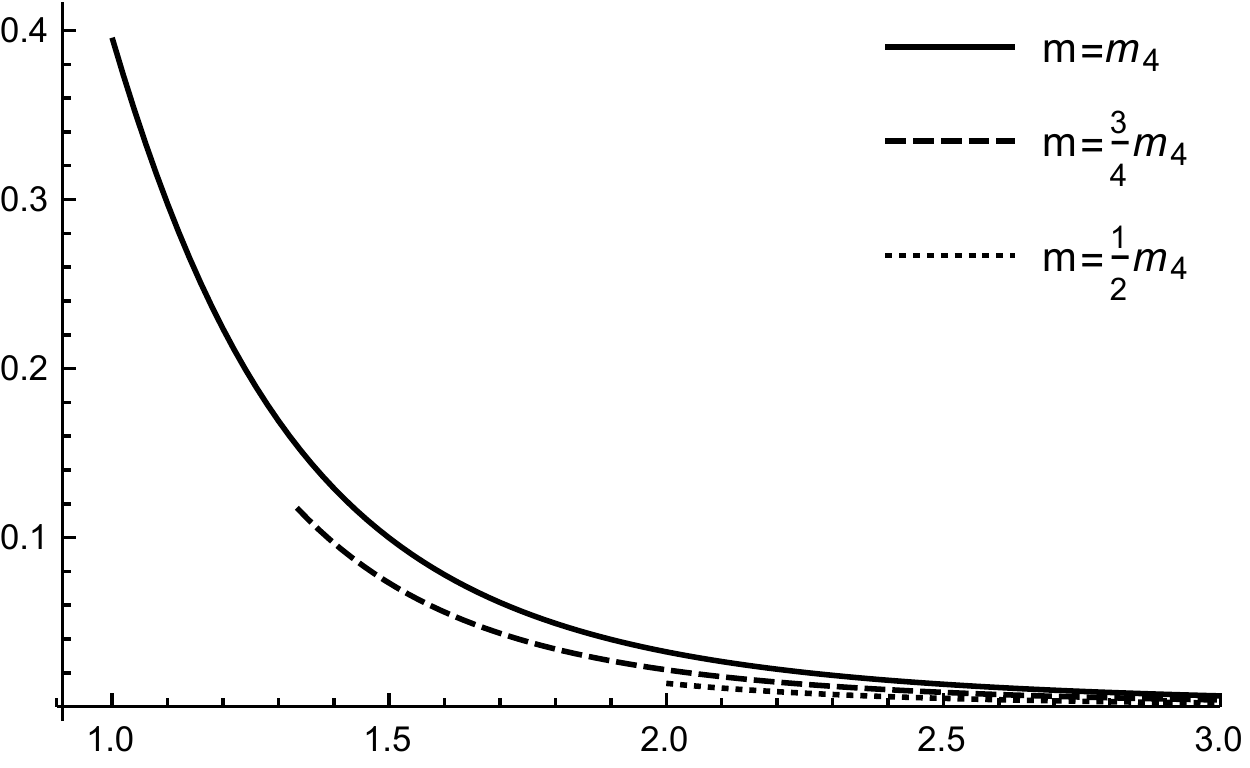}
$\quad$
\includegraphics[width=5\textwidth/12]{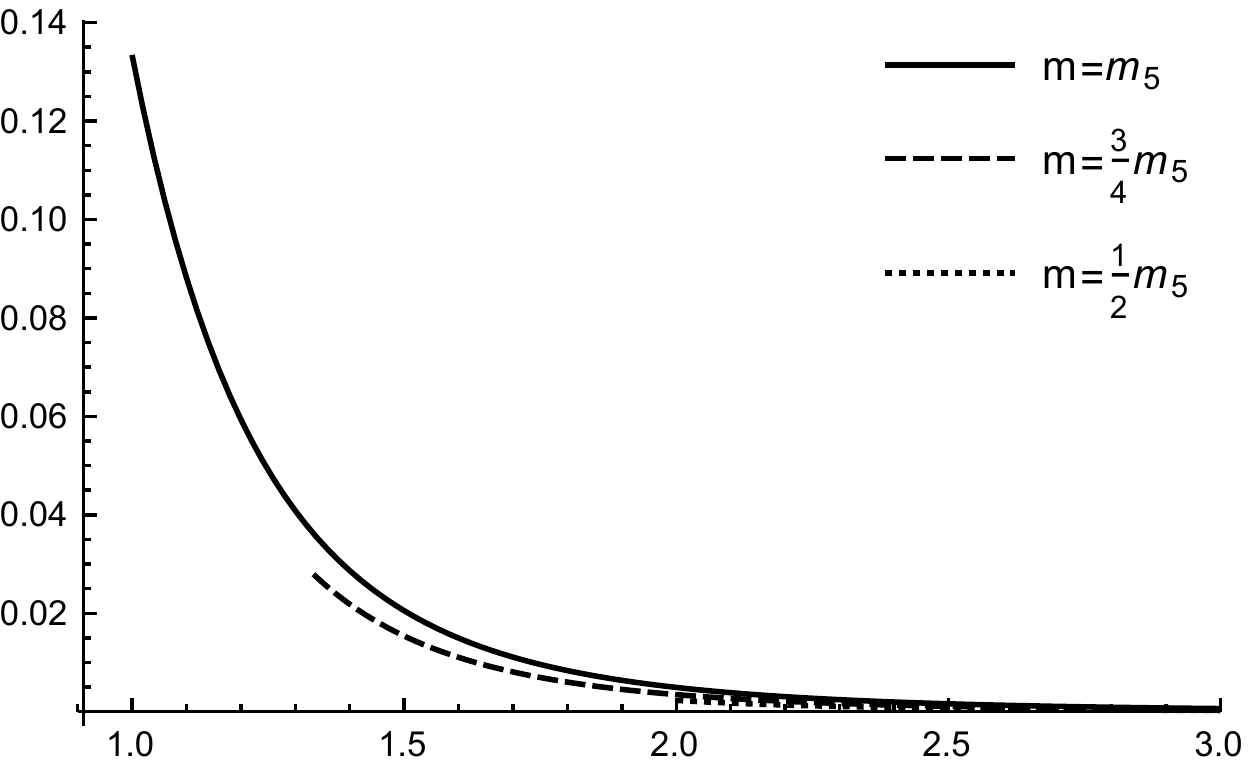}
{$\frac{\ell}{\ell_D}$}
\\
$D=4$
\hspace{6cm}
$D=5$
\\
\raisebox{3.75cm}{ {\footnotesize $P_{\rm BH}$}} 
\includegraphics[width=5\textwidth/12]{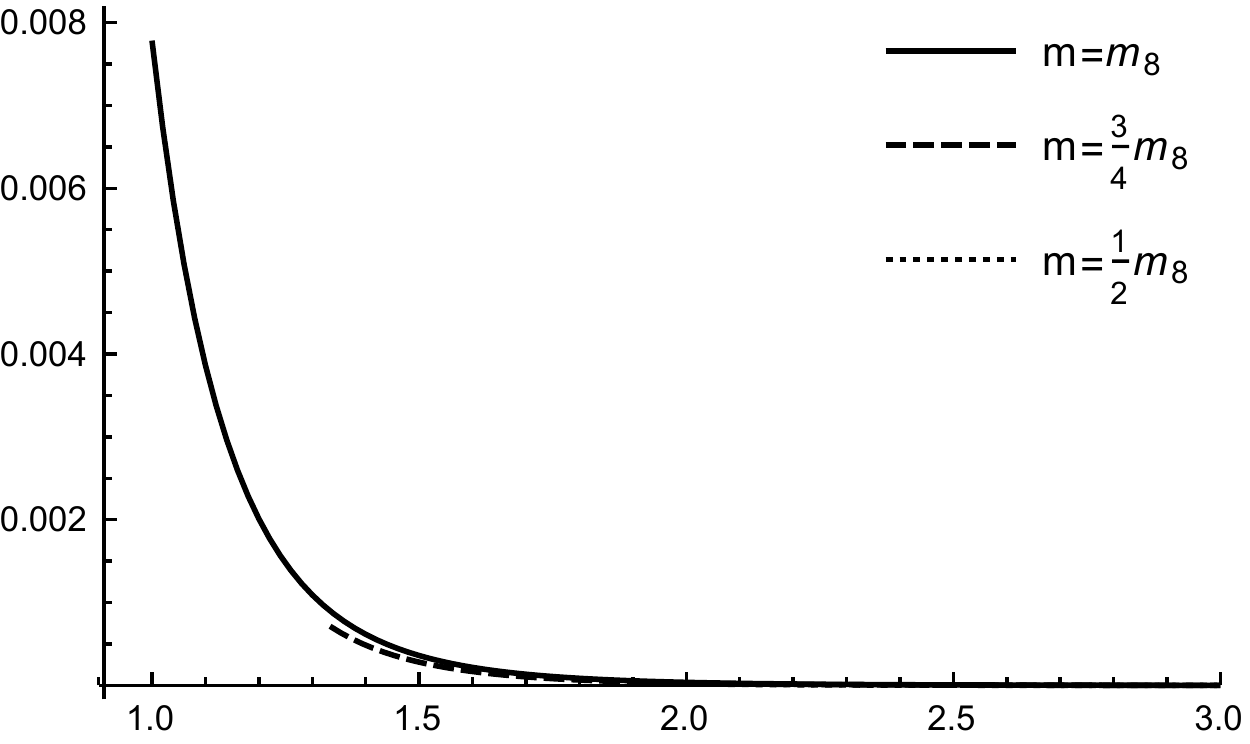}
$\quad$
\includegraphics[width=5\textwidth/12]{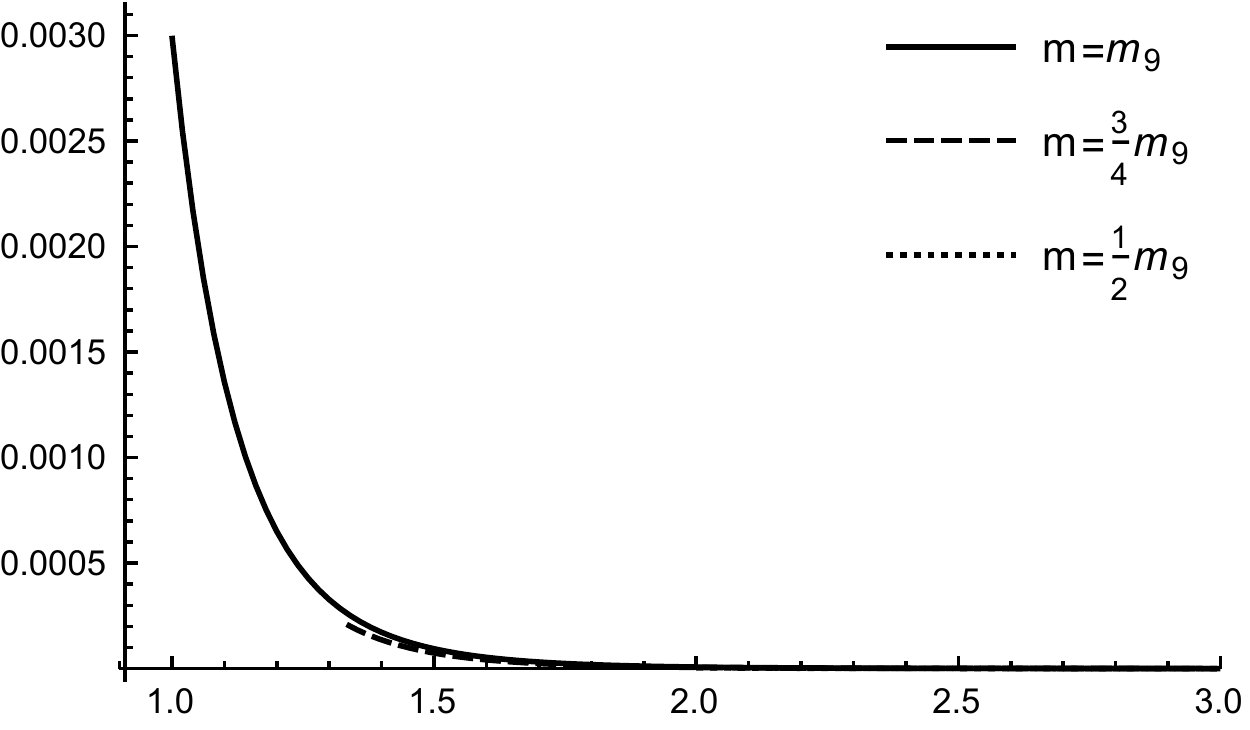}
{$\frac{\ell}{\ell_D}$}
\\
$D=8$
\hspace{6cm}
$D=9$
\caption{Probability $P_{\rm BH}(\ell,m)$ of a particle, described by a 
Gaussian with $\ell\ge\lambda_m$, to be a black hole for $m=m_D$
(solid line), $m=3\,m_D/4$ (dashed line) and $m=m_D/2$ (dotted line).
From left to right, the spatial dimensions are $D=4$ and 5 on the first line
and $D=8$ and 9 on the second line
(note the different scales on the vertical axes). 
\label{prob1}}
\end{figure}

The probability density that a particle
lies inside its own gravitational radius is defined in the same way of the previous section,
\be
\mathcal{P}_<(r<\rh)
\,=\,
P_{\rm S}(r<\rh) \, \mathcal{P}_{\rm H}(\rh)
\ ,
\label{PInside}
\ee
where the probability that the particle is inside a $D$-ball of radius $\rh$ is
\be
P_{\rm S}(r<\rh)
=
\Omega_{D-1}\int_0^{\rh} |\psi_{\rm S}(r)|^2 \, r^{D-1} \, \d r
\ ,
\ee
and the probability density that the radius of the horizon equals $\rh$ is
\be
\mathcal{P}_{\rm H}(\rh)
=
\Omega_{D-1} \, r^{D-1} \, |\psi_{\rm H}(\rh)|^2
\ee
Again, integrating~\eqref{PInside} over all the possible values of the horizon radius $\rh$, 
\be
P_{\rm BH}
\,=\,
\int_0^\infty
\mathcal{P}_<(r<\rh) \, \d \rh
\label{PBH}
\ee
gives the probability that the particle is a black hole in a $D+1$ dimensional space-time.
%
%
%
%
Applying the $D+1$ dimensional Gaussian wave function introduced in the Eq. \eqref{Gauss} we find
\be
P_{\rm S}(r<\rh)
=
\frac{\gamma\left(\frac{D}{2},\frac{\rh^2}{\ell^2}\right)}{\Gamma\left(\frac{D}{2}\right)}
\ .
\ee
Properties of the $\gamma$ ensure that $P_{\rm S}=1$ if $\rh\to\infty$, 
while $P_{\rm S}=0$ if $\rh=0$ as expected.
Then,
\begin{align}
\mathcal{P}_{\rm H}(\rh)&=\frac{2}{\ell_D^{\,D}}\, \left(\frac{(D-2) \, m_D}{2 \, \Delta}\right)^{\frac{D}{D-2}}
\frac{D-2}{\Gamma\left(\frac{D}{2\,D-4},\frac{m^2}{\Delta^2}\right)}
\nonumber\\
&\times
\Theta(\rh-R_D) \, \exp\left\{ -
\left[\frac{(D-2)\, m_D}{2 \, \Delta}\right]^2
\, \left(\frac{\rh}{\ell_D}\right)^{2(D-2)} \right\}
\, \rh^{D-1}
\end{align}
and Eq.~\eqref{PBH} finally becomes
\begin{align}\label{PBHexplicit}
P_{\rm BH}(\ell,m))&=\frac{2}{\ell_D^{\,D}}\left(\frac{(D-2)\, m_D}{2\,\Delta}\right)^{\frac{D}{D-2}}
\frac{D-2}{\Gamma\left(\frac{D}{2D-4},\frac{m^2}{\Delta^2}\right) \Gamma\left(\frac{D}{2}\right)}
\notag\\
&\times
\int_{R_D}^\infty
\gamma\left(\frac{D}{2},\frac{\rh^2}{\ell^2}\right)
 \exp\left\{ -
\left[\frac{(D-2)\, m_D}{2 \, \Delta}\right]^2
\, \left(\frac{\rh}{\ell_D}\right)^{2(D-2)} \right\}
\, \rh^{D-1} \, \d \rh
\end{align}
which yields the probability for a particle to be a black hole depending
on the Gaussian width $\ell$, mass $m$ and spatial dimension $D$.
Since the above integral cannot be analytically performed, we show
the numerical dependence on $\ell\gtrsim \lambda_m$
 of the above probability
for different masses and spatial dimensions in Fig.~\ref{prob1}.

Here one faces the first interesting results of extra dimensional HWF. The probability $P_{\rm BH}$
at given $m$ decreases significantly for increasing $D$, and for large
values of $D$ even a particle of mass $m\simeq m_D$ is most likely not
a black hole.
This result should have a strong impact on predictions of black hole
production in particle collisions.
For example, one could approximate the effective production cross-section
as $\sigma(E)\sim P_{\rm BH}(E)\,\sigma_{\rm BH}(E)$, where
$\sigma_{\rm BH}\sim 4\,\pi\,E^2$ is the usual black disk expression for a
collision with centre-of-mass energy  $E$.
Since $P_{\rm BH}$ can be very small, $\sigma(E)\ll \sigma_{\rm BH}(E)$
for $D>4$ and one in general expects much less black holes can be produced
than standard estimates~\cite{Dimopoulos:2001hw}. 
\par

\subsection{GUP From Higher Dimensional HWF}
\label{secGUP}
\begin{figure}[h!]
\centering
\raisebox{3.75cm}{\footnotesize $\frac{\expec{\hat{r}_{\rm H}}}{R_D}$}
\includegraphics[width=5\textwidth/12]{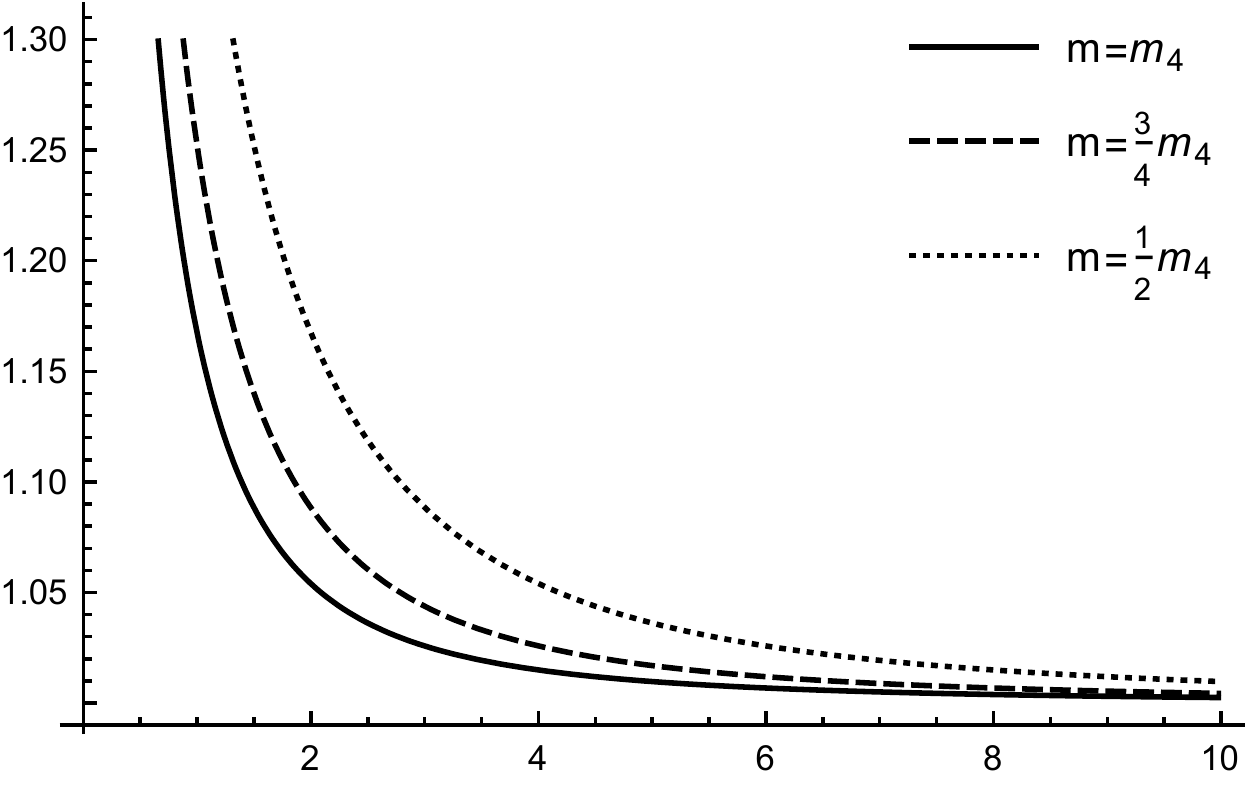}
$\quad$
\includegraphics[width=5\textwidth/12]{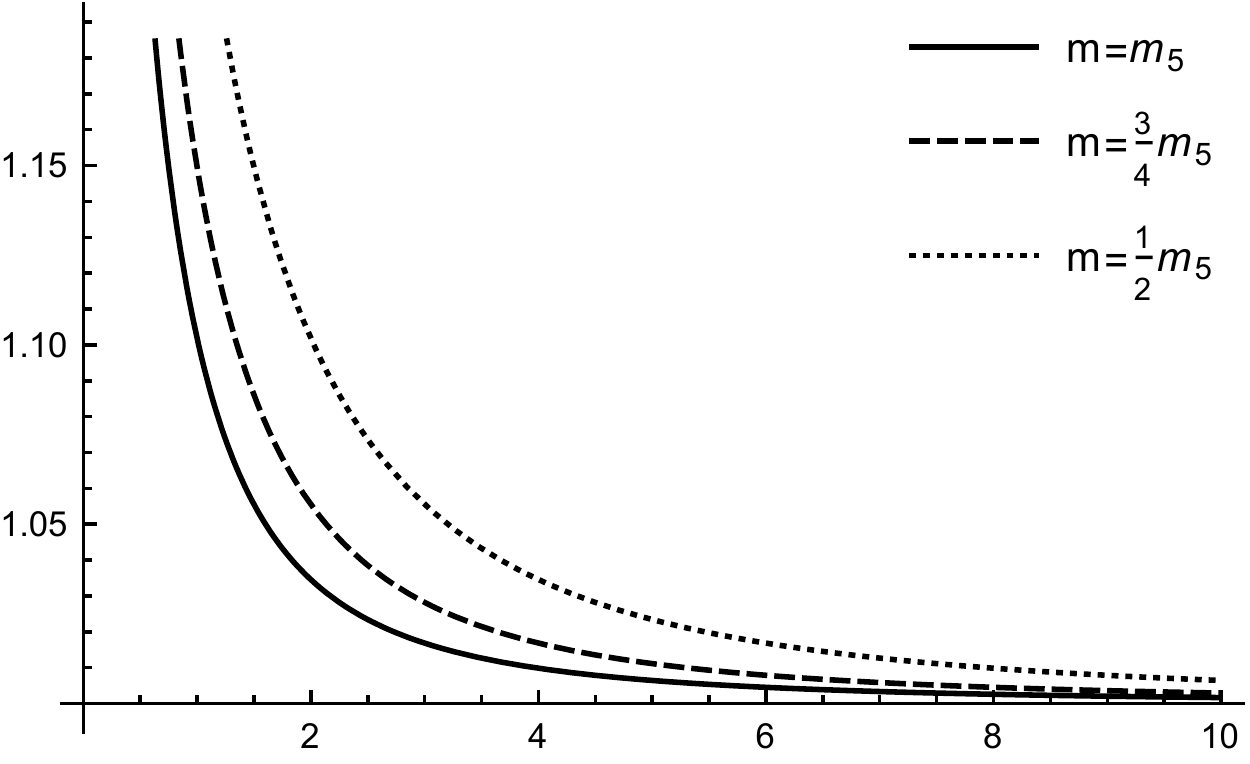}
{\footnotesize $\ell/\ell_D$}
\\
\raisebox{3.75cm}{\footnotesize $\frac{\Delta{r}_{\rm H}}{R_D}$}
\includegraphics[width=5\textwidth/12]{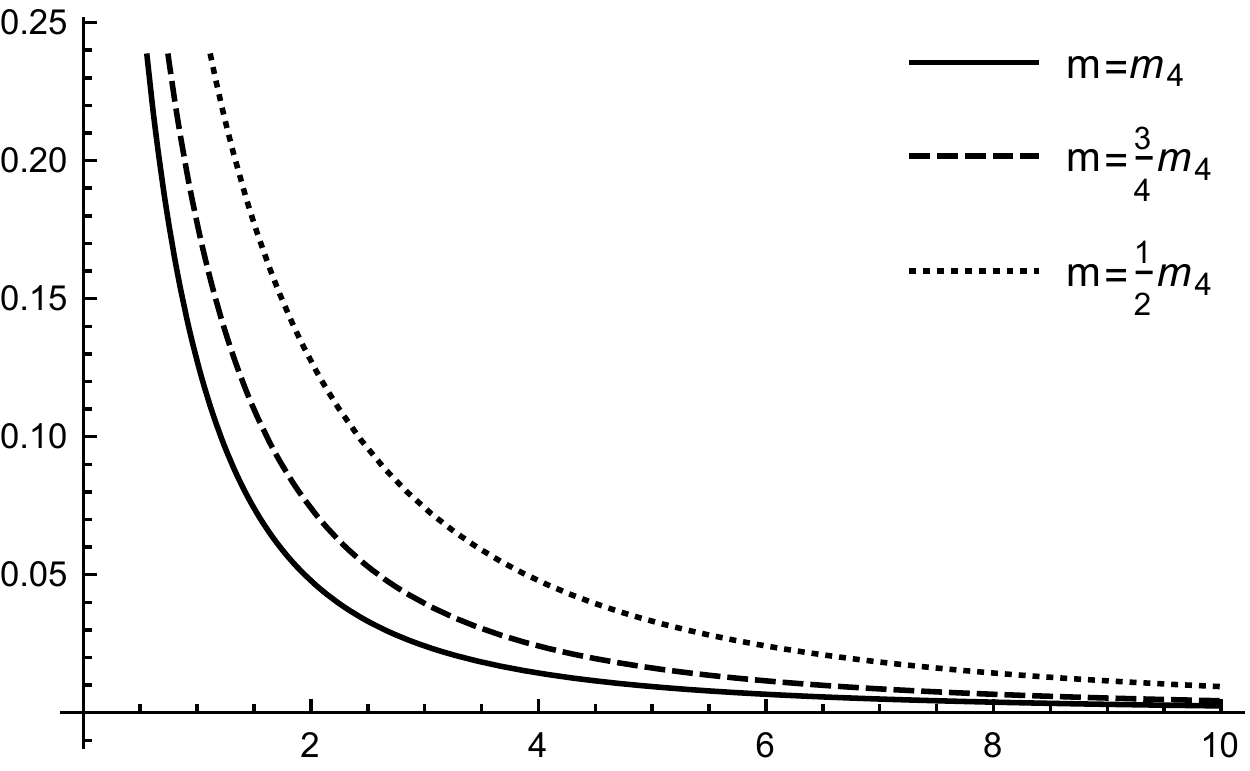}
$\quad$
\includegraphics[width=5\textwidth/12]{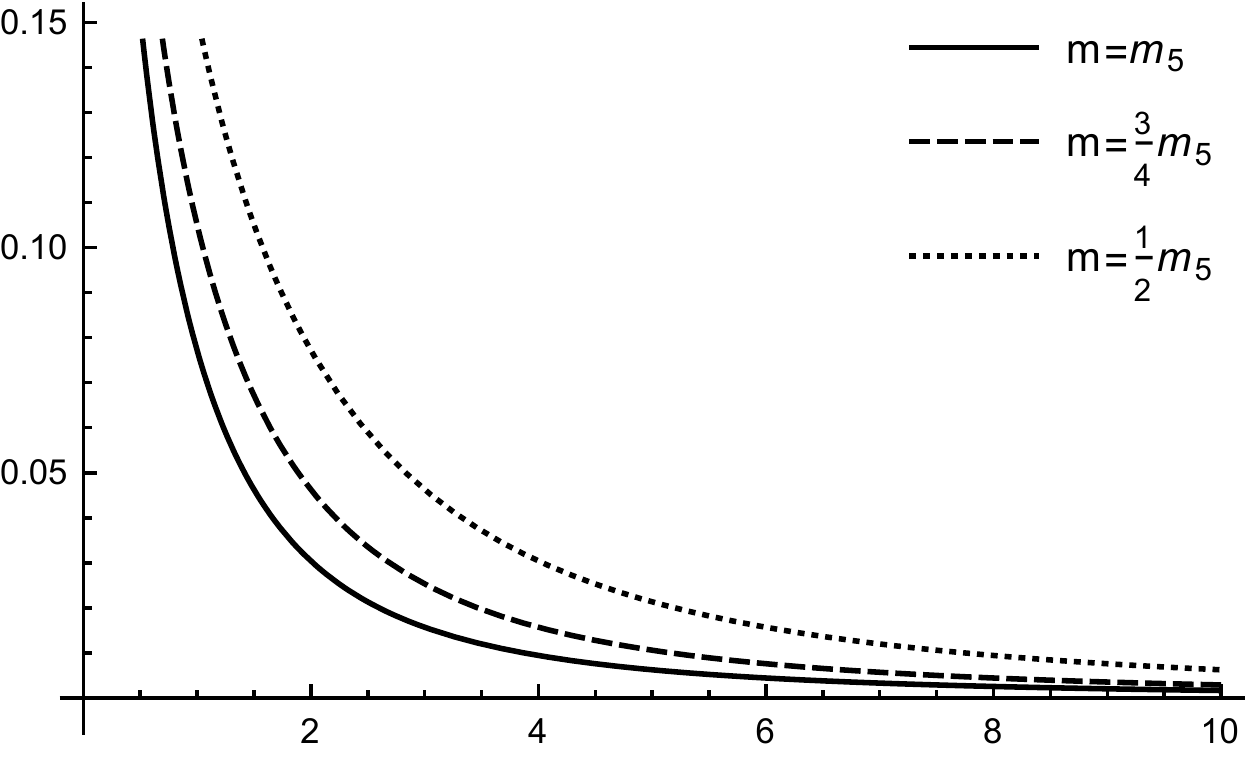}
{\footnotesize $\ell/\ell_D$}
\\
$D=4$
\hspace{6cm}
$D=5$
\caption{Plots of $\expec{\hat{r}_{\rm H}}$ (upper panel) and $\Delta \rh$ (lower panels)
as functions of $\ell/\ell_D$, for $D=4$ (left) and $D=5$ (right), with $m=m_D$
(solid line), $m=\frac{3}{4}\,m_D$ (dashed line) and $m=\frac{1}{2}\,m_D$ (dotted line).
\label{expecD}}
\end{figure}

Deriving an extra dimensional GUP requires, as in the 3D case, the expectation values of some quantities. Analogous to the 3D case we have
\be
\expec{\hat{r}_{\rm H}}
=
\frac{\Gamma\left(\frac{D+1}{2D-4},\frac{m^2}{\Delta^2}\right)}
{\Gamma\left(\frac{D}{2D-4},\frac{m^2}{\Delta^2}\right)}
\, \left(\frac{\Delta}{m}\right)^{\frac{1}{D-2}} \, R_D
\ .
\ee
or
\be
\expec{\hat{r}_{\rm H}}
=
\frac{\E_{\frac{D-5}{2D-4}}(\frac{m^2}{\Delta^2})}
{\E_{\frac{D-4}{2D-4}}(\frac{m^2}{\Delta^2})} \, R_D
\ .
\label{expecRhD}
\ee
We likewise obtain
\be
\expec{\hat{r}_{\rm H}^2}
=
\frac{\E_{\frac{D-6}{2D-4}}(\frac{m^2}{\Delta^2})}
{\E_{\frac{D-4}{2D-4}}(\frac{m^2}{\Delta^2})} \, R_D^2
\ ,
\ee
and estimate the relative uncertainty in the horizon as
\be
\Delta \rh
=
\sqrt{\expec{\hat{r}_{\rm H}^2}-\expec{\hat{r}_{\rm H}}^2}
=
\sqrt{
\frac{\E_{\frac{D-6}{2D-4}}(\frac{m^2}{\Delta^2})}
{\E_{\frac{D-4}{2D-4}}(\frac{m^2}{\Delta^2})} -
\left(\frac{\E_{\frac{D-5}{2D-4}}(\frac{m^2}{\Delta^2})}
{\E_{\frac{D-4}{2D-4}}(\frac{m^2}{\Delta^2})}\right)^2 } \, R_D
\label{deltarhD}
\ .
\ee
Since
\be
\frac{m}{\Delta}\,=\,
\frac{\ell \, m}{\ell_D \, m_D}
\,\varpropto\,
\frac{\ell}{\ell_D}
\ ,
\ee
it is also possible to see that, for $\ell\gg\ell_D$, we recover the expected classical
results
\be
\expec{\hat{r}_{\rm H}}
\simeq
R_D
\ ,
\qquad
\Delta\rh
\simeq
0
\ .
\ee
Fig.~\ref{expecD} shows the plots of~\eqref{expecRhD} and~\eqref{deltarhD} for $D>3$
as functions of $\ell/\ell_D$.
%
%
%
%
%
%
%
%
\par
Again the GUP follows by linearly combining the usual Heisenberg
uncertainty with the uncertainty in the horizon size, 
\be
\Delta r
=
\Delta r_{0} + \alpha \,\Delta \rh
\ .
\ee
We can compute the Heisenberg part starting from the state~\eqref{Gauss},
that is 
\begin{align}
\expec{\hat{r}^n}&=
\Omega_{D-1} \int_0^\infty \tilde{\psi}^*(r) \, \hat{r}^n \, \tilde{\psi}(r) \, r^{D-1} \, \d r
\notag
\\\notag
&=
\frac{2\Delta^{-D}}{\Gamma\left(\frac{D}{2}\right)} 
\int_0^\infty e^{-p^2/\Delta^2} p^{D+n-1} \, \d p\\
&=
\frac{\Gamma\left(\frac{D+n}{2}\right)}{\Gamma\left(\frac{D}{2}\right)} \, \ell^n
\ .
\end{align}

\begin{figure}[h!]
\centering
\raisebox{3.75cm}{\footnotesize $\frac{\Delta r}{\ell_D}$}
\includegraphics[width=5\textwidth/12]{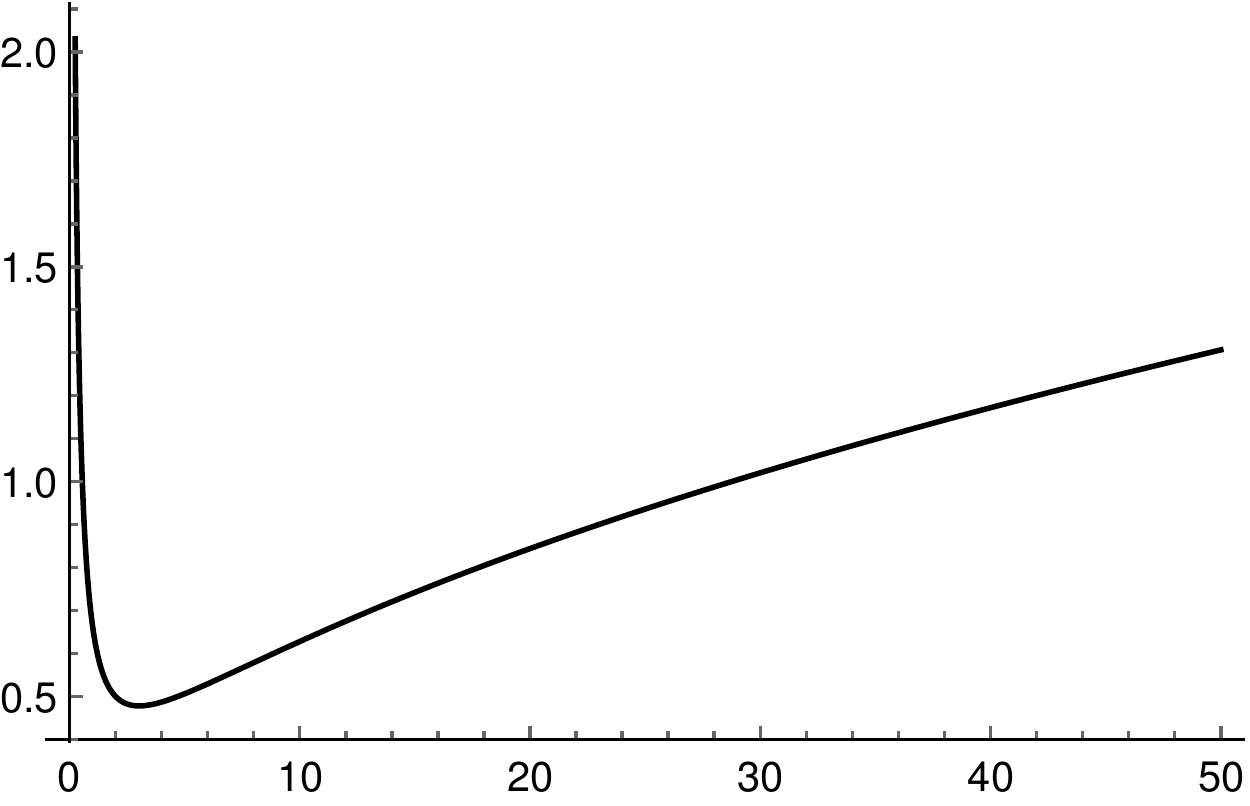}
$\quad$
\includegraphics[width=5\textwidth/12]{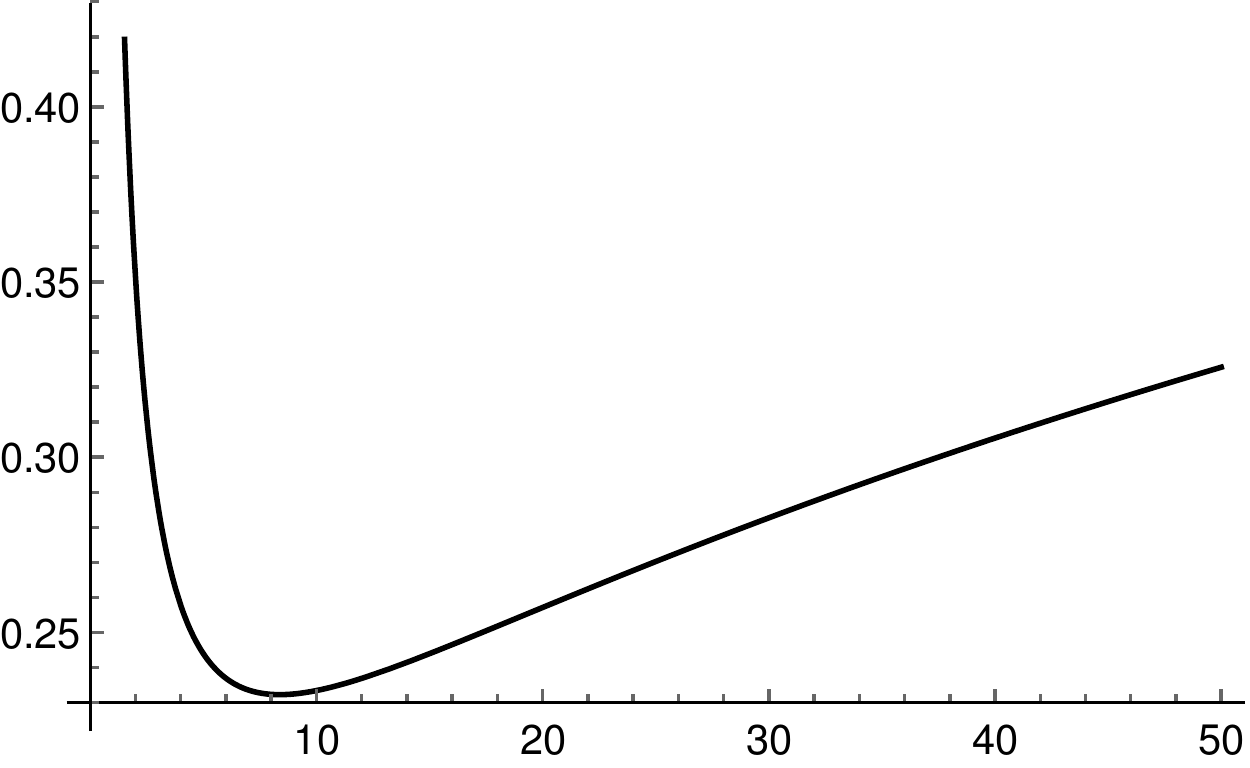}
$\frac{\Delta p}{m_D}$
\\
$D=4$
\hspace{6cm}
$D=5$
\caption{Plots of $\Delta r/\ell_D$ as function of $\Delta p/m_D$ for $D=4$ and $5$
 and $\alpha=1$.
\label{delta-r}}
\end{figure}

Using $\Gamma(z+1)=z\,\Gamma(z)$, this yields
\be
\expec{\hat{r}}
=
\frac{\Gamma\left(\frac{D+1}{2}\right)}{\Gamma\left(\frac{D}{2}\right)} \, \Delta
=
\frac{2^{1-D}\sqrt{\pi}\, (D-1)!}{\Gamma\left(\frac{D}{2}\right)^2} \, \ell
\ ,
\ee
and
\be
\expec{\hat{r}^2}
=
\frac{\Gamma\left(\frac{D}{2}+1\right)}{\Gamma\left(\frac{D}{2}\right)} \, \Delta^2
=
\frac{D}{2} \, \ell^2
\ ,
\ee
so that
\be
\Delta r_{0}
=
\sqrt{\frac{D}{2}
-\left(\frac{2^{1-D}\sqrt{\pi}}{\Gamma\left(\frac{D}{2}\right)^2} \, (D-1)!\right)^2} \, \ell
\equiv
A_D \, \ell
\label{Drell}
\ .
\ee 
Using instead the state~\eqref{momGauss} in momentum space, the same procedure
yields
\be
\Delta p
=
A_D \, \Delta
=
A_D \, \frac{m_D \, \ell_D}{\ell}
\label{Deltap}
\ .
\ee
Expressing $\ell$ and $\Delta$ from the above equation as functions
of $\Delta p$, we find
\be
\frac{\Delta r}{\ell_D}
=
A_D^2\,\frac{m_D}{\Delta p}+
\alpha \,
\sqrt{
\frac{\E_{\frac{D-6}{2D-4}}\left(\frac{A_D^2 m^2}{(\Delta p)^2}\right)}
{\E_{\frac{D-4}{2D-4}}\left(\frac{A_D^2 m^2}{(\Delta p)^2}\right)} -
\left[\frac{\E_{\frac{D-5}{2D-4}}\left(\frac{A_D^2 m^2}{(\Delta p)^2}\right)}
{\E_{\frac{D-4}{2D-4}}\left(\frac{A_D^2 m^2}{(\Delta p)^2}\right)}\right]^2 } 
\, \left(\frac{2}{|D-2|}\, \frac{m}{m_D}\right)^\frac{1}{D-2} 
\ .
\label{GUP}
\ee
Assuming the width of the wave-packet 
equal to the Compton length follows
\be
m
=
\Delta
=
\frac{\Delta p}{A_D}
\ee
and the GUP finally reads
\begin{align}
\frac{\Delta r}{\ell_D}
&=
A_D^2\frac{m_D}{\Delta p}+
\alpha \,
\sqrt{
\frac{\E_{\frac{D-6}{2D-4}}(1)}{\E_{\frac{D-4}{2D-4}}(1)} -
\left(\frac{\E_{\frac{D-5}{2D-4}}(1)}{\E_{\frac{D-4}{2D-4}}(1)}\right)^2 } 
\left(\frac{2}{|D-2|}\, \frac{\Delta p}{A_D m_D}\right)^\frac{1}{D-2} 
\nonumber
\\
&=
\frac{C_{\rm QM}}{\Delta p}
+C_{\rm H}\,\Delta p^{\frac{1}{D-2}}
\ ,
\label{GUPCompton}
\end{align}
where $C_{\rm QM}$ and $C_{\rm H}$ are constants (independent of $\Delta p$).
Fig.~\ref{delta-r} shows $\Delta r$ for different spatial dimensions,
setting $\alpha=1$.

\begin{figure}[h!]
\centering
\raisebox{3.75cm}{ \footnotesize${\frac{m^*_D}{m_D}}$}
\includegraphics[width=5\textwidth/12]{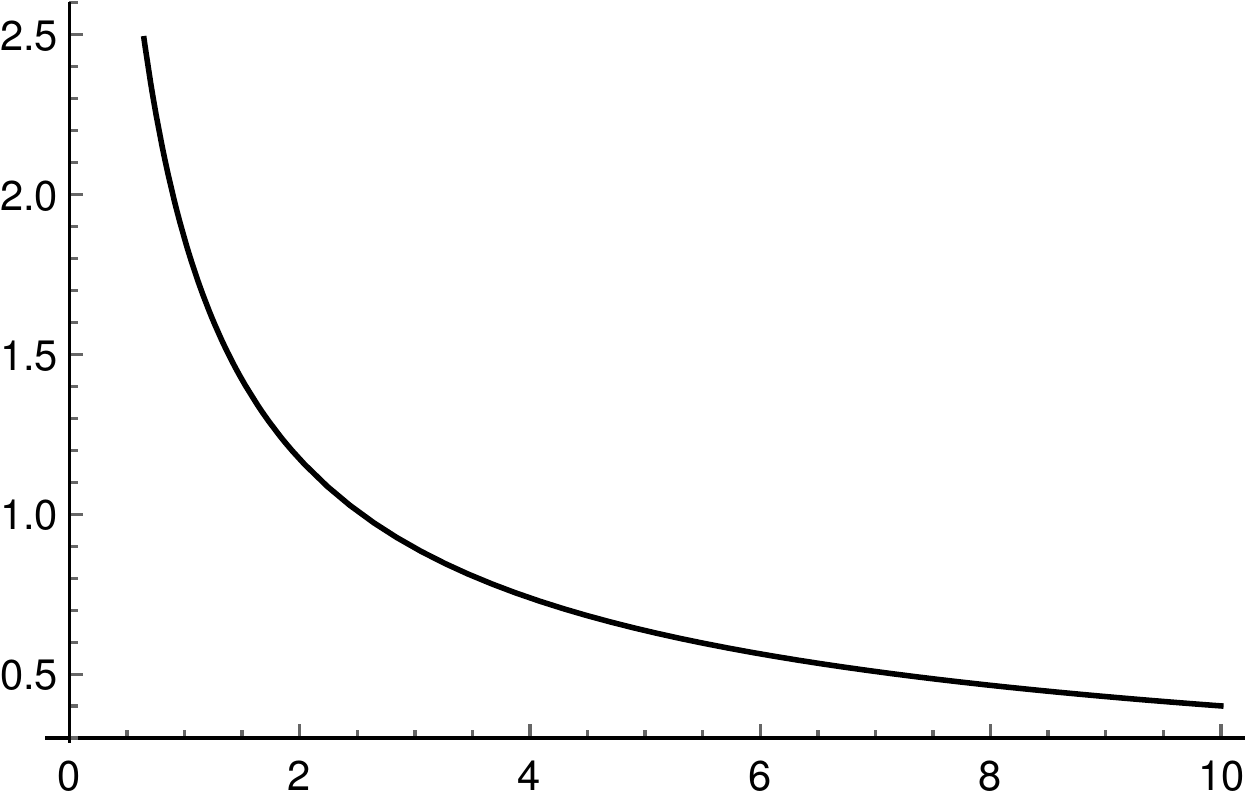}
$\quad$
\includegraphics[width=5\textwidth/12]{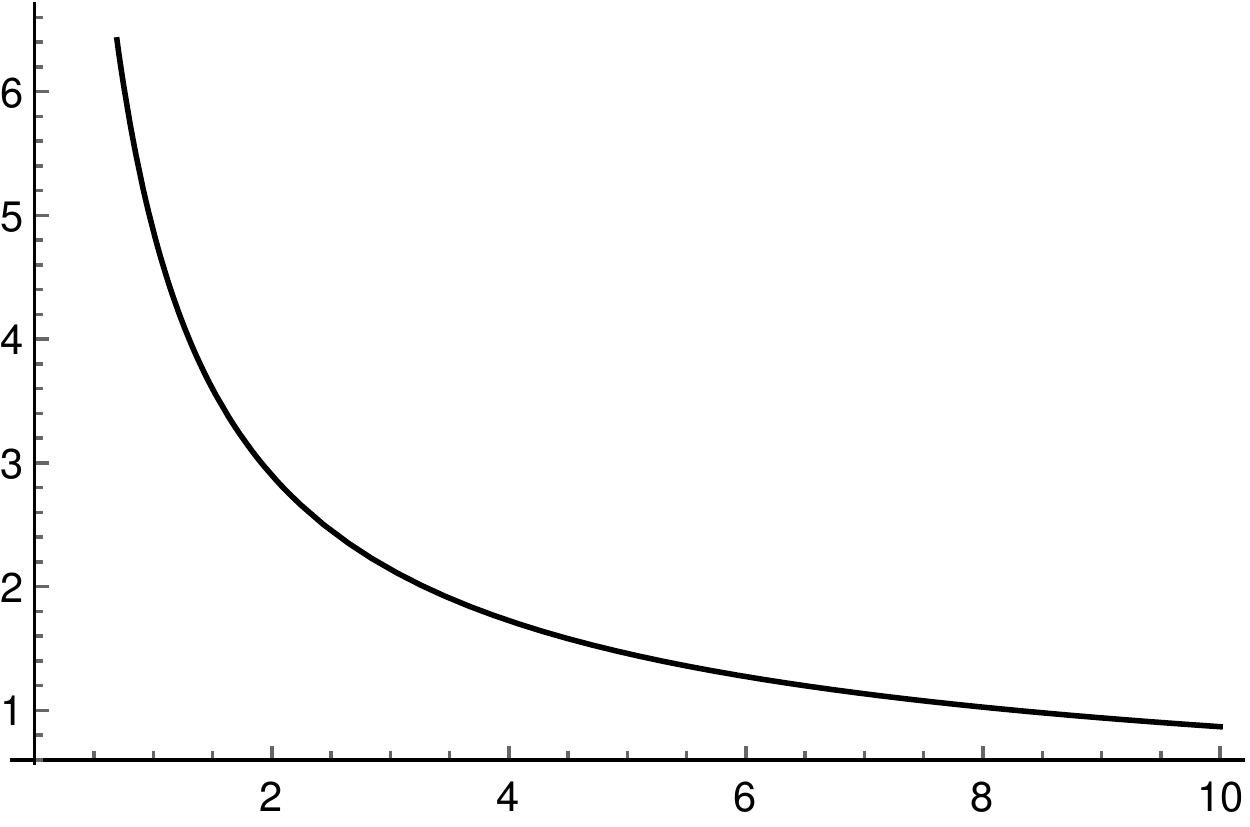}
$\alpha$
\\
$D=4$
\hspace{4cm}
$D=5$
\caption{Minimum scale $m^*_D$ as function of the parameter $\alpha$
for $D=4$ and $5$.
\label{scM}}
\end{figure}

\begin{figure}[h!]
\centering
\raisebox{3.3cm}{\tiny ${\frac{\ell^*_D}{\ell_D}}$}
\includegraphics[width=5\textwidth/12]{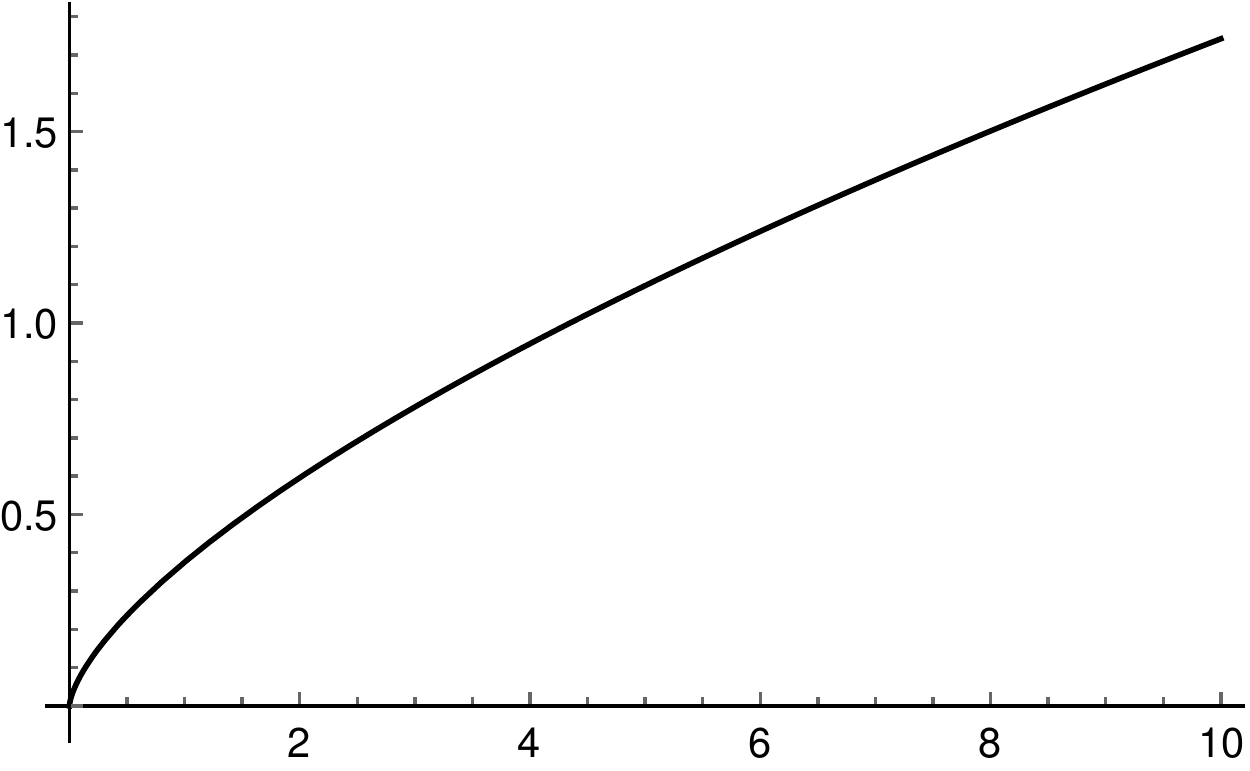}
$\quad$
\includegraphics[width=5\textwidth/12]{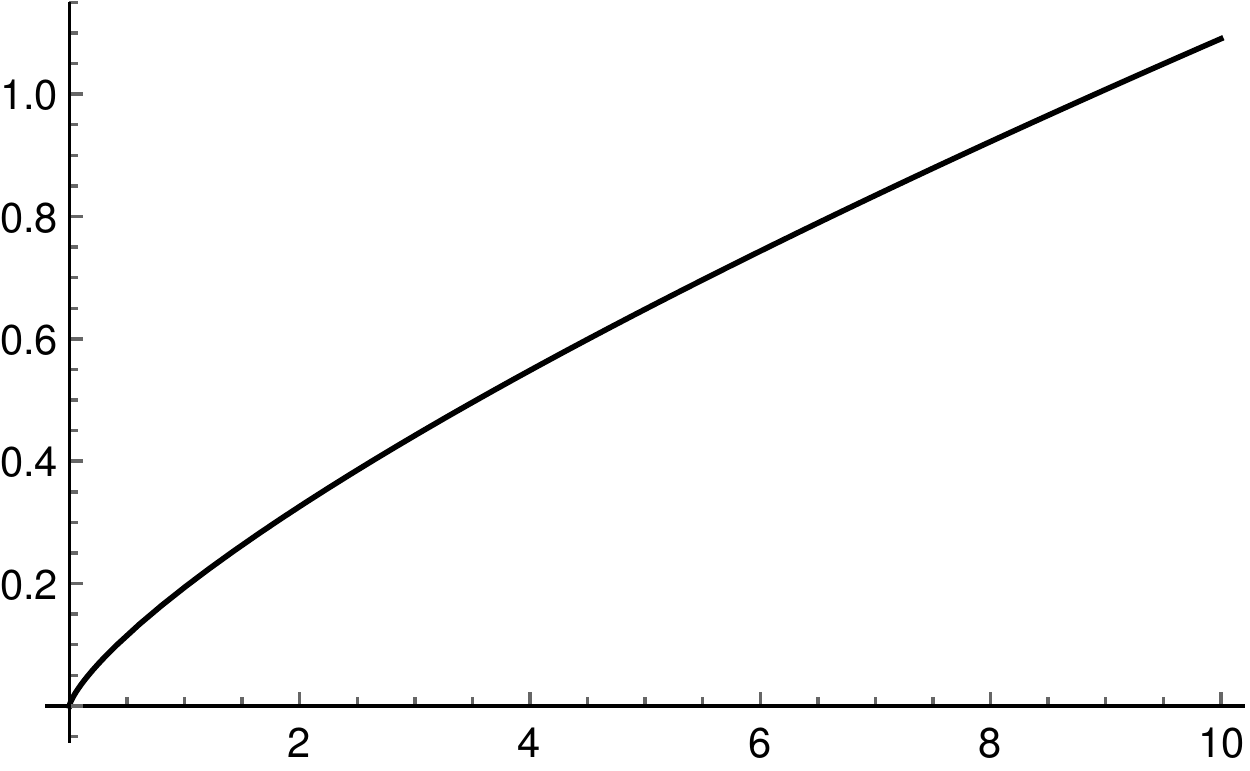}
$\alpha$
\\
$D=4$
\hspace{4cm}
$D=5$
\caption{Minimum scale $\ell^*_D$ as function of the parameter $\alpha$
for $D=4$ and $5$.
\label{scL}}
\end{figure}

In higher-dimensional cases we obtain the same qualitative behaviour of the previous section for the GUP, 
with a minimum length uncertainty $\ell^*_D$
\be
\ell^*_D
=
\ell_D\left(\frac{D-1}{D-2}\right) \, (2A_D)^{\frac{1}{D-1}}
\left(\alpha\sqrt{\frac{\E_{\frac{D-6}{2D-4}}(1)}{\E_{\frac{D-4}{2D-4}}(1)} -
\left(\frac{\E_{\frac{D-5}{2D-4}}(1)}
{\E_{\frac{D-4}{2D-4}}(1)}\right)^2 }\right)^{\frac{D-2}{D-1}}
\label{MinL}
\ee
corresponding to an energy scale $m^*_D$, satisfying
\be
m^*_D
=
m_D \, \frac{(D-2)}{2^{\frac{1}{D-1}}}\,\left[
\alpha\sqrt{\frac{\E_{\frac{D-6}{2D-4}}(1)}{\E_{\frac{D-4}{2D-4}}(1)} -
\left(\frac{\E_{\frac{D-5}{2D-4}}(1)}
{\E_{\frac{D-4}{2D-4}}(1)}\right)^2 }\right]^{\frac{2-D}{D-1}} \, A_D^{\frac{2D-3}{D-1}}
\ .
\label{MinM}
\ee
The impact of $\alpha$ on this minimum length is then shown in Fig.~\ref{scM},
where we plot the scale $m^*_D$ corresponding to the minimum $\ell^*_D$
as a function of this parameter, and in Fig.~\ref{scL}, where
we plot directly $\ell^*_D$.
A qualitative difference of the 3D case is that for all values of $D$ considered here, assuming $m^*_D\simeq m_D$ favours large
values of $\alpha$, whereas requiring $\ell^*_D\simeq \ell_D$ would favour small
values of $\alpha$.

\par

The higher-dimensional cases $D>3$ look qualitatively very similar
to the standard $(3+1)$-scenario, with a probability $P_{\rm BH}$ of similar shape,
and a related GUP leading to the existence of a minimum length scale.
However, one of the main results is that the probability $P_{\rm BH}$
for fixed mass decreases for increasing $D>3$.
In fact, for $m\simeq m_D$, one has $P_{\rm BH}\simeq 0.14$ for $D=5$,
which further decreases to $P_{\rm BH}\simeq 3\times 10^{-3}$ for $D=9$.
This implies that, although the fundamental scale $m_D$ could be smaller
for larger $D$, one must still reach an energy scale significantly larger than $m_D$
in order to produce a black hole.
It is clear that this should have a strong impact on the estimates of the number of 
black holes produced in colliders which are based on models with extra spatial dimensions,
and, conversely, on the bounds on extra-dimensional parameters obtained from the lack of
observation of these objects. %
{In fact, in Ref. \cite{Arsene:2016kvf} the black hole formation at the LHC in the present scenario was simulated by applying the package BLACKMAX to perform numerical simulations. They found that the ratio of black hole formation always decrease for an increasing number of extra dimensions, as well as that the HWF approach predicts black hole production with masses smaller than the expected in the standard scenario. It corroborates the results of the present section.}
 

\chapter{Conclusion} 

\label{cap7} 

%
In this thesis we have discussed different topics, all connected by black hole physics beyond general relativity. What makes it interesting is the fact that black holes can not be completely explained by classical GR. Nobody neither know what exactly is a space-time singularity nor understand the physical phenomena expected in such scale. For that reason black holes could effectively provide information about the merge of gravity and quantum mechanics. We have dealt with some ideas coming from quantum gravity proponents, trying to give a small contribution to the problem of finding consequences of such ideas in a tangible scale.  We have calculated the Hawking radiation associated to an effective black hole solution coming from string theory by emitting a fermion beyond the standard model. We found that, even having different dynamical equations, the associated temperature is the same of the regular Dirac fermions. Our next result was to find and exact FRW-like solution of the Einstein equation in a model with a thick brane generated by a scalar field. The solutions are pretty complicated and deserve further analysis.  

Our next two results were, in principle, closer to experimental and observational data. We have used the strong deflection limit approach to calculate observable quantities of black holes found under the extra dimensional paradigm. Those observables deviate from the ones found for the Schwarzschild case and could, again in principle,  be used to rule out black hole solutions. The last chapter deals with possible black holes formed in high energy collisions. For that the horizon quantum mechanics formalism was employed . Our result, in a collaboration developed during the time I spent at the Bologna university, consists of extending the formalism to an extra dimensional model. We found that the probability of such black hole formation is suppressed when extra dimensions are added to the model.

As perspectives for future works we could mention that  the extended MGD approach \cite{Casadio:2015gea} may be further applied to provide observational bounds on its parameters. 
Indeed, Ref. \cite{Casadio:2015gea} proposed 
an extension of the MGD, by assuming 
the more general temporal metric component \begin{eqnarray}
A(r)
=\left(1-\frac{2\,M}{r}\right)^{1+k},\end{eqnarray} instead of the standard MGD setup, given by Eq.  (\ref{nu}). 
Clearly the standard MGD metric (Eqs. \ref{mu} and \ref{nu}) is recovered when $k=0$. 
We want to apply the extended MGD in the context of our paper, Ref. \cite{Casadio:2016zhu}, regarding the ratio $\eta/s$ for predicting the new parameters.

A new perspective for black hole physics have been studied recently, by considering
black holes as Bose-Einstein condensates (BEC) of a large number $N$ of weakly
interacting gravitons close to a critical
point~\cite{Dvali:2011aa,Dvali:2012gb,Dvali:2012rt}.
This paradigm has the merit to  directly interconnect black hole physics
to the study of critical phenomena, where quantum effects are
relevant at critical points, even for a macroscopic number $N$ of
particles~\cite{Flassig:2012re}. 
Black hole features that cannot be recovered in a standard semi-classical approach  may then be encoded by the quantum state of the critical BEC~\cite{Casadio:2015lis,Casadio:2015jha},
with the semi-classical regime obtained as a particular limit for $N\to\infty$. 
We are going to use such formalism and the extended MDG to generalise the result of Ref. \cite{Casadio:2016aum}.

Regarding the horizon quantum mechanics, we are going to find consequences of applying it to the problem of abundance of primordial black holes \cite{Carr:2009jm,Carr:2016drx}. It could also result in constraints on inflationary models \cite{Green:1997sz}. The results related to HQM could also be extended to non Gaussian distributions or different models of extra dimensions.

The discussion above supports the importance of black hole physics as a tool to investigate gravity beyond GR. We hope that consequences could be tested experimentally in the near future.

\appendix 

%

\chapter{List of Published Papers} 

\label{AppendixA} 
\begin{enumerate}

\item %
  R.~da Rocha and R.~T.~Cavalcanti.\\
  \newblock {Flag-dipole and flagpole spinors fluid flows in Kerr spacetimes}.\\
  \newblock {\em Phys.\ Atom.\ Nucl.,}  80:329, 2017.

\item 
Roberto Casadio, Rogerio~T. Cavalcanti, Andrea Giugno and Jonas Mureika.\\
\newblock {Horizon of quantum black holes in various dimensions}.\\
\newblock {\em Phys. Lett. B}, 760:36--44, 2016.

\item 
R.~T. Cavalcanti, A.~Goncalves da~Silva and Roldao da~Rocha.\\
\newblock {Strong Deflection Limit Lensing Effects in the Minimal Geometric Deformation and Casadio-Fabbri-Mazzacurati Solutions}.\\
\newblock {\em Class. and Quantum Grav.}, 33:215007, 2016.

\item 
Roberto Casadio, Rogerio~T. Cavalcanti and Roldao da~Rocha.\\
\newblock {Fluid/gravity correspondence and the CFM brane-world solutions}.\\
\newblock {\em Eur. Phys. J. C}, 76:556, 2016.

\item 
R.~T. Cavalcanti and Roldao da~Rocha.\\
\newblock {Dark Spinors Hawking Radiation in String Theory Black Holes}.\\
\newblock {\em Adv. High Energy Phys.}, 2016:4681902, 2016.

\item 
R.T. Cavalcanti, R.~da~Rocha and J.M. {Hoff da Silva}.\\
\newblock Could elko spinor fields induce vsr symmetry in the DKP (meson)
  algebra?\\
\newblock {\em Adv. Appl. Clifford Algebr.}, pages 1--11, 2015. (online)\\
\newblock {\em Adv. Appl. Clifford Algebr.}, 27(1):267, 2017.

\item 
A.~E. Bernardini, R.~T. Cavalcanti and Roldao da~Rocha.\\
\newblock {Spherically Symmetric Thick Branes Cosmological Evolution}.\\
\newblock {\em Gen. Rel. Grav.}, 47(1):1840, 2015.

\item 
R.~T. Cavalcanti.\\
\newblock {Classification of Singular Spinor Fields and Other Mass Dimension
  One Fermions}.\\
\newblock {\em Int. J. Mod. Phys. D}, 23(14):1444002, 2014.

\item 
R.~T. Cavalcanti, J.~M. Hoff~da Silva and Roldao da~Rocha.\\
\newblock {VSR symmetries in the DKP algebra: the interplay between Dirac and
  Elko spinor fields}.\\
\newblock {\em Eur. Phys. J. Plus}, 129(11):246, 2014.

\item 
  R.~da Rocha, L.~Fabbri, J.~M.~H. da Silva, R.~T.~Cavalcanti and J.~A.~Silva-Neto.\\
 \newblock{Flag-Dipole Spinor Fields in ESK Gravities}.\\
 \newblock{ \em J.\ Math.\ Phys. },  54:102505, 2013.

\end{enumerate}


\chapter{Explicit Calculations of the Section \ref{thickfrw}} 

\label{AppendixB} 

This appendix is devoted to present, explicitly, some calculations of the Section \ref{thickfrw}. Here the same notation of the mentioned section will be used, in special, for derivatives
\begin{equation}
\frac{\partial f}{\partial t} \equiv \dot{f}, \qquad \frac{\partial f}{\partial r} \equiv \bar{f}, \qquad \frac{\partial f}{\partial y} \equiv {f'}.
\end{equation}
First of all we have to calculate the energy-momentum tensor
\begin{equation}
T_{MN}=\nabla_M \phi \nabla_N \phi-g_{MN}\left(\frac{1}{2}g^{AB}\nabla_A \phi \nabla_B \phi + V(\phi) \right), \qquad \phi \equiv \phi(t,y),
\end{equation}
following
\begin{eqnarray}
\frac{1}{2} g^{AB}\nabla_A \phi \nabla_B \phi&=&\frac{1}{2}g^{AB}\partial_A \phi \partial_B \phi \\
&=&\frac{1}{2}g^{00}\dot{\phi}^{2}+\frac{1}{2}g^{55}\phi'^{2}\\
&=&-\frac{1}{2a^{2}}\dot{\phi}^{2}+\frac{1}{2}\phi'^{2}
\end{eqnarray}

\begin{equation}
\Rightarrow T_{MN}=\nabla_M \phi \nabla_N \phi-g_{MN}\left(-\frac{1}{2a^{2}}\dot{\phi}^{2}+\frac{1}{2}\phi'^{2} + V(\phi) \right).
\end{equation}

Explicitly, the components read
\begin{eqnarray}
T_{00}&=&\dot{\phi}^{2}+ a^{2}\left(-\frac{1}{2a^{2}}\dot{\phi}^{2}+\frac{1}{2}\phi'^{2} + V(\phi) \right) \\
&=&\frac{1}{2}\dot{\phi}^{2}+a^{2}\left(\frac{1}{2}\phi'^{2} + V(\phi) \right)\\
&=&a^{2}\left(\frac{1}{2a^{2}}\dot{\phi}^{2}+\frac{1}{2}\phi'^{2} + V(\phi) \right).
\end{eqnarray}

\begin{eqnarray}
T_{ii}&=&-g_{ii}\left(-\frac{1}{2a^{2}}\dot{\phi}^{2}+\frac{1}{2}\phi'^{2} + V(\phi) \right) .
\end{eqnarray}

\begin{eqnarray}
T_{55}&=&{\phi'}^{2} -\left(-\frac{1}{2a^{2}}\dot{\phi}^{2}+\frac{1}{2}\phi'^{2} + V(\phi) \right) \\
&=&\frac{1}{2}{\phi'}^{2}+\frac{1}{2a^{2}}\dot{\phi}^{2} - V(\phi) .
\end{eqnarray}

\begin{eqnarray}
T_{05}&=&\dot{\phi}\phi'. 
\end{eqnarray}

The equation of motion for the scalar field is given by
\begin{equation}
\nabla^{2}\phi-\frac{dV}{d\phi}=0, \qquad \nabla^{2} \equiv g^{AB}\nabla_A \nabla_B,
\end{equation}
following
\begin{eqnarray}
\nabla^{2}\phi &=& g^{AB}\nabla_A \partial_B \phi \\
&=& \nabla_A \partial^{A} \phi\\
&=& \frac{1}{\sqrt{-g}}\partial_A \left(\sqrt{-g} \partial^{A} \phi \right)\\
&=& \underbrace{\partial_A \ln\underbrace{\left(\sqrt{-g}\right)}_2}_3  \partial^{A} \phi+\underbrace{\partial_A\partial^{A} \phi}_{1}.
\end{eqnarray}
\begin{enumerate}
\item \begin{eqnarray}
\partial_A\partial^{A} \phi&=& \partial_A(g^{AB}\partial_B \phi) \\
&=& \partial_A g^{AB}\partial_B \phi+g^{AB}\partial_A\partial_B\phi \\
&=& \partial_0 g^{00}\dot{\phi}+g^{00}\ddot{\phi}+\partial_5 g^{55}{\phi'}+g^{55}{\phi''} \\
&=& 2\frac{\dot{a}}{a^{3}}\dot{\phi}-\frac{1}{a^{2}}\ddot{\phi}+{\phi''} .
\end{eqnarray}
\item \begin{eqnarray}
\sqrt{-g}&=&\frac{a^{4}r^{2}\sin \theta}{\sqrt{1-kr^{2}}}.
\end{eqnarray}
\item \begin{equation}
\partial_A \ln\left(\sqrt{-g}  \right)= 4 \partial_A\ln a+\partial_A \ln \left(\frac{r^{2}\sin \theta}{\sqrt{1-kr^{2}}} \right).
\end{equation}
\end{enumerate}
\begin{eqnarray}
\partial_0 \ln\left(\sqrt{-g}  \right) &=& 4 \partial_0\ln a \\
&=&4\frac{\dot{a}}{a}.
\end{eqnarray}
\begin{eqnarray}
\partial_5 \ln\left(\sqrt{-g}  \right) &=& 4 \partial_5\ln a \\
&=&4\frac{{a'}}{a}.
\end{eqnarray}
Hence
\begin{eqnarray}
 \nabla^{2}\phi&=&\left[\partial_0 \ln\left(\sqrt{-g}  \right)g^{00}\dot{\phi}+\partial_5 \ln\left(\sqrt{-g}  \right)g^{55}{\phi'}\right]+\left[2\frac{\dot{a}}{a^{3}}\dot{\phi}-\frac{1}{a^{2}}\ddot{\phi}+{\phi''}\right]\\
&=&\left[-4\frac{\dot{a}}{a^{3}}\dot{\phi}+4\frac{{a'}}{a}{\phi'}\right]+\left[2\frac{\dot{a}}{a^{3}}\dot{\phi}-\frac{1}{a^{2}}\ddot{\phi}+{\phi''}\right]\\
&=&\phi''-\frac{1}{a^{2}}\ddot{\phi}+4\frac{{a'}}{a}{\phi'}-2\frac{\dot{a}}{a^{3}}\dot{\phi},
\end{eqnarray}
Therefore, the equation of motion is given by
\begin{eqnarray}
0&=&\nabla^{2}\phi-\frac{dV}{d\phi}\\
&=&\phi''-\frac{1}{a^{2}}\ddot{\phi}+4\frac{{a'}}{a}{\phi'}-2\frac{\dot{a}}{a^{3}}\dot{\phi}-\frac{dV}{d\phi}.
\end{eqnarray}

By calculating the components of Einstein tensor we find 

\begin{footnotesize}
\begin{eqnarray}\nonumber
G_{00}&=&a^{2}\left\{3\left[ \frac{\dot{a}^{2}}{a^{4}}-\left( \frac{a''}{a}+\frac{a'^{2}}{a^{2}}\right)+\frac{k}{a^{2}}\right] +(1-kr^{2})\left[\frac{\bar{a}^{2}}{a^{4}}-2\frac{\bar{\bar{a}}}{a^{3}}-6\frac{\bar{a}}{a^{3}r}  \right]+2\frac{\bar{a}}{a^{3}r}  \right\}\\\nonumber 
G_{11}&=&-g_{11}\left\{\left[ \frac{1}{a^{2}}\left(2\frac{\ddot{a}}{a}-\frac{\dot{a}^{2}}{a^{2}}\right)-3\left( \frac{a''}{a}+\frac{a'^{2}}{a^{2}}\right)+\frac{k}{a^{2}}\right] -(1-kr^{2})\left[\frac{3\bar{a}^{2}}{a^{4}}+4\frac{\bar{a}}{a^{3}r}  \right]  \right\}\\\nonumber
G_{22}&=&-g_{22}\left\{\left[ \frac{1}{a^{2}}\left(2\frac{\ddot{a}}{a}-\frac{\dot{a}^{2}}{a^{2}}\right)-3\left( \frac{a''}{a}+\frac{a'^{2}}{a^{2}}\right)+\frac{k}{a^{2}}\right] -(1-kr^{2})\left[2\frac{\bar{\bar{a}}}{a^{3}}-\frac{\bar{a}^{2}}{a^{4}}+4\frac{\bar{a}}{a^{3}r}  \right]+2\frac{\bar{a}}{a^{3}r}   \right\}\\ \nonumber
G_{33}&=&\frac{g_{33}}{g_{22}}G_{22}\\ \nonumber
G_{55}&=&3\left[2\frac{a'^{2}}{a^{2}}- \frac{1}{a^{2}}\frac{\ddot{a}}{a}-\frac{k}{a^{2}}\right] +(1-kr^{2})\left[3\frac{\bar{\bar{a}}}{a^{3}}+9\frac{\bar{a}}{a^{3}r}  \right]-3\frac{\bar{a}}{a^{3}r} \\ \nonumber
G_{01}&=&2\left(2\frac{\bar{a}\dot{a}}{a^{2}}-\frac{\dot{\bar{a}}}{a} \right)\\ \nonumber
G_{05}&=&3\left(\frac{\dot{a}{a'}}{a^{2}}-\frac{\dot{{a}}'}{a} \right)\\ \nonumber
G_{15}&=&3\left(\frac{\bar{a}{a'}}{a^{2}}-\frac{{\bar{a}}'}{a} \right).
\end{eqnarray}
\end{footnotesize}
Supposing the separability of the scale/warp factor $a(t,r,y)=A(t,r)B(y)$, from $T_{01}=0$ we have

\begin{eqnarray}
0&=&2\left(2\frac{\bar{a}\dot{a}}{a^{2}}-\frac{\dot{\bar{a}}}{a} \right)\\
&=&2\frac{\bar{a}\dot{a}}{a}-{\dot{\bar{a}}}\\
&=&2\frac{\bar{A}B\dot{A}B}{AB}-\dot{\bar{A}}B\\
&=&2\frac{\bar{A}\dot{A}}{A}-\dot{\bar{A}}.
\end{eqnarray}
From the above equation we have two options, namely
\begin{eqnarray}
0&=&2\frac{\bar{A}\dot{A}}{A}-\dot{\bar{A}}\\
&=&2\frac{\bar{A}}{A}-\frac{\dot{\bar{A}}}{\dot{A}}\\
&=&\partial_r\ln A^{2}- \partial_r \ln \dot{A}\\
&=&\partial_r\ln \frac{A^{2}}{\dot{A}}\\\label{dota}
&\Rightarrow & \ln \frac{A^{2}}{\dot{A}}=T(t) \Leftrightarrow \dot{A}=A^{2}e^{-T},
\end{eqnarray}
or
\begin{eqnarray}
0&=&2\frac{\bar{A}\dot{A}}{A}-\dot{\bar{A}}\\
&=&2\frac{\dot{A}}{A}-\frac{\dot{\bar{A}}}{\bar{A}}\\
&=&\partial_t\ln \frac{A^{2}}{\bar{A}}\\\label{bara}
&\Rightarrow & \ln \frac{A^{2}}{\bar{A}}=R(r) \Leftrightarrow \bar{A}=A^{2}e^{-R}\\ \nonumber
&&\\ \label{dotbar}
&\therefore & \dot{A}=\bar{A}e^{R-T}.
\end{eqnarray}
By combining the equations \eqref{bara}, \eqref{dota} and \eqref{dotbar} follow some usefull relations,
\begin{eqnarray}
\dot{A}^{2}&=&A^{4}e^{-2T}\\
\bar{A}^{2}&=&A^{4}e^{-2R}\\
\dot{\bar{A}}&=&2\frac{\bar{A}\dot{A}}{A}=2A^{3}e^{-(T+R)}\\
\ddot{A}&=&\dot{A}\left(2\frac{\dot{A}}{A}-\dot{T} \right)=A^{2}e^{-T}\left(2Ae^{-T}-\dot{T}\right)\\
\bar{\bar{A}}&=&\bar{A}\left(2\frac{\bar{A}}{A}-\bar{R} \right)=A^{2}e^{-R}\left(2Ae^{-R}-\bar{R}\right).
\end{eqnarray}
Back to the energy-momentum tensor components, $T_{15}=0$ gives 
\begin{eqnarray}
0&=&3\left(\frac{\bar{a}{a'}}{a^{2}}-\frac{{\bar{a}}'}{a} \right)\\
&=&\frac{\bar{a}{a}'}{a}-{{\bar{a}}'}\\
&=&\frac{\bar{A}B{A}B'}{AB}-{\bar{A}}B'\\
&=&{\bar{A}}B'-{\bar{A}}B'.
\end{eqnarray}

Analogously, $T_{05}=\dot{\phi}\phi'$ gives

\begin{eqnarray}
\dot{\phi}\phi'&=&3\left(\frac{\dot{a}{a'}}{a^{2}}-\frac{\dot{{a}}'}{a} \right)\\
&=&3\left(\frac{\dot{A}B{A}B'}{A^{2}B^{2}}-\frac{\dot{A}B'}{AB} \right)\\
&=&3\left(\frac{\dot{A}B'}{AB}-\frac{\dot{A}B'}{AB} \right)\\
&=&0\\
&\Rightarrow& \dot{\phi}=0.
\end{eqnarray}
It imposes a scalar field depending only on the extra dimensional component.

Finally, the components of energy-momentum tensor becomes
\begin{eqnarray}
T_{\mu \nu}&=&-g_{\mu \nu}\left(\frac{1}{2}\phi'^{2} + V(\phi) \right) 
\end{eqnarray}
and
\begin{eqnarray}
T_{55}&=&\frac{1}{2}{\phi'}^{2} - V(\phi). 
\end{eqnarray}

Merging the energy-momentum tensor and the Einstein tensor we have, for the diagonal components 

\textbf{00:}

{\footnotesize
\begin{eqnarray}\nonumber
\frac{1}{2}\phi'^2+V(\phi) &=&3\left[ \frac{\dot{a}^{2}}{a^{4}}-\left( \frac{a''}{a}+\frac{a'^{2}}{a^{2}}\right)+\frac{k}{a^{2}}\right] +(1-kr^{2})\left[\frac{\bar{a}^{2}}{a^{4}}-2\frac{\bar{\bar{a}}}{a^{3}}-6\frac{\bar{a}}{a^{3}r}  \right]+2\frac{\bar{a}}{a^{3}r} \\ \nonumber
  &=&\frac{1}{B^{2}}\left[ -3\left( B{B''}+{B'^{2}}\right) +(1-kr^{2})\left(\frac{\bar{A}^{2}}{A^{4}}-4\frac{\bar{\bar{A}}}{A^{3}}-6\frac{\bar{A}}{A^{3}r}  \right)+2\frac{\bar{A}}{A^{3}r}+3\frac{\dot{A}^{2}}{A^{4}}+3\frac{k}{A^{2}}\right]\\ \nonumber
 &=&\frac{1}{B^{2}}\left\{ -3\left( B{B''}+{B'^{2}}\right) +\frac{1}{A^{2}}\left[(1-kr^{2})\left(\frac{\bar{A}^{2}}{A^{2}}-4\frac{\bar{\bar{A}}}{A}-6\frac{\bar{A}}{Ar}  \right)+2\frac{\bar{A}}{Ar}+3\frac{\dot{A}^{2}}{A^{2}}+3{k}\right]\right\},
\end{eqnarray}
}
\vspace{1cm}

then
\begin{align}\nonumber
C_0=&B^{2}\left( \frac{1}{2}\phi'+V(\phi) \right)+3\left( B{B''}+{B'^{2}}\right)\\ \nonumber
=&\frac{1}{A^{2}}\left[(1-kr^{2})\left(\frac{\bar{A}^{2}}{A^{2}}-4\frac{\bar{\bar{A}}}{A}-6\frac{\bar{A}}{Ar}  \right)+2\frac{\bar{A}}{Ar}+3\frac{\dot{A}^{2}}{A^{2}}+3{k}\right]
\end{align}
Where $C_0$ is a separability constant.

\textbf{11:}
{\footnotesize
\begin{eqnarray}\nonumber
\frac{1}{2}\phi'^2+V(\phi) &=&\left[ \frac{1}{a^{2}}\left(2\frac{\ddot{a}}{a}-\frac{\dot{a}^{2}}{a^{2}}\right)-3\left( \frac{a''}{a}+\frac{a'^{2}}{a^{2}}\right)+\frac{k}{a^{2}}\right] -(1-kr^{2})\left[\frac{3\bar{a}^{2}}{a^{4}}+4\frac{\bar{a}}{a^{3}r}  \right] \\ \nonumber
 &=&\left[ \frac{1}{B^{2}}\left(2\frac{\ddot{A}}{A^{3}}-\frac{\dot{A}^{2}}{A^{4}}\right)-\frac{3}{B^{2}}\left( B{B''}+{B'^{2}}\right)+\frac{1}{B^{2}}\frac{k}{A^{2}}\right] +\frac{1}{B^{2}}(1-kr^{2})\left[3\frac{\bar{A}^{2}}{A^{4}}+4\frac{\bar{A}}{A^{3}r}  \right]\\ \nonumber
  &=&\frac{1}{B^{2}}\left[ -3\left( B{B''}+{B'^{2}}\right) +(1-kr^{2})\left(3\frac{\bar{A}^{2}}{A^{4}}+4\frac{\bar{A}}{A^{3}r}  \right)+2\frac{\ddot{A}}{A^{3}}-\frac{\dot{A}^{2}}{A^{4}}+\frac{k}{A^{2}}\right]\\ \nonumber
 &=&\frac{1}{B^{2}}\left\{ -3\left( B{B''}+{B'^{2}}\right) +\frac{1}{A^{2}}\left[(1-kr^{2})\left(3\frac{\bar{A}^{2}}{A^{2}}+4\frac{\bar{A}}{Ar}  \right)+2\frac{\ddot{A}}{A}-\frac{\dot{A}^{2}}{A^{2}}+{k}\right]\right\},
\end{eqnarray}
}
thus
\begin{align}\nonumber
C_0=&B^{2}\left( \frac{1}{2}\phi'+V(\phi) \right)+3\left( B{B''}+{B'^{2}}\right)\\ \nonumber
=&\frac{1}{A^{2}}\left[(1-kr^{2})\left(3\frac{\bar{A}^{2}}{A^{2}}+4\frac{\bar{A}}{Ar}  \right)+2\frac{\ddot{A}}{A}-\frac{\dot{A}^{2}}{A^{2}}+{k}\right]
\end{align}

\vspace{1cm}

\textbf{22:}
{\scriptsize
\begin{eqnarray}\nonumber
\frac{1}{2}\phi'^2+V(\phi) &=&\left[ \frac{1}{a^{2}}\left(2\frac{\ddot{a}}{a}-\frac{\dot{a}^{2}}{a^{2}}\right)-3\left( \frac{a''}{a}+\frac{a'^{2}}{a^{2}}\right)+\frac{k}{a^{2}}\right] -(1-kr^{2})\left[2\frac{\bar{\bar{a}}}{a^{3}}-\frac{\bar{a}^{2}}{a^{4}}+4\frac{\bar{a}}{a^{3}r}  \right]+2\frac{\bar{a}}{a^{3}r}   \\ \nonumber
  &=&\frac{1}{B^{2}}\left[ -3\left( B{B''}+{B'^{2}}\right) +(1-kr^{2})\left(2\frac{\bar{\bar{A}}}{A^{3}}-\frac{\bar{A}^{2}}{A^{4}}+4\frac{\bar{A}}{A^{3}r}  \right)+2\frac{\ddot{A}}{A^{3}}-\frac{\dot{A}^{2}}{A^{4}}+\frac{k}{A^{2}}+2\frac{\bar{A}}{A^{3}r}\right]\\ \nonumber
 &=&\frac{1}{B^{2}}\left\{ -3\left( B{B''}+{B'^{2}}\right) +\frac{1}{A^{2}}\left[(1-kr^{2})\left(2\frac{\bar{\bar{A}}}{A}-\frac{\bar{A}^{2}}{A^{2}}+4\frac{\bar{A}}{Ar}  \right)+2\frac{\ddot{A}}{A}-\frac{\dot{A}^{2}}{A^{2}}+{k}+2\frac{\bar{A}}{Ar}\right]\right\},
\end{eqnarray}
}
with
\begin{align}\nonumber
C_0=&B^{2}\left( \frac{1}{2}\phi'+V(\phi) \right)+3\left( B{B''}+{B'^{2}}\right)\\\nonumber
=&\frac{1}{A^{2}}\left[(1-kr^{2})\left(2\frac{\bar{\bar{A}}}{A}-\frac{\bar{A}^{2}}{A^{2}}+4\frac{\bar{A}}{Ar}  \right)+2\frac{\ddot{A}}{A}-\frac{\dot{A}^{2}}{A^{2}}+{k}+2\frac{\bar{A}}{Ar}\right]
\end{align}
%
%
and\\
\textbf{55:}
{\footnotesize
\begin{eqnarray}\nonumber
\frac{1}{2}\phi'^2-V(\phi) &=&3\left[2\frac{a'^{2}}{a^{2}}- \frac{1}{a^{2}}\frac{\ddot{a}}{a}-\frac{k}{a^{2}}\right] +(1-kr^{2})\left[3\frac{\bar{\bar{a}}}{a^{3}}+9\frac{\bar{a}}{a^{3}r}  \right]-3\frac{\bar{a}}{a^{3}r} \\ \nonumber
  &=&3\left[2 \frac{B'^{2}}{B^{2}}-\frac{\ddot{A}}{A^{3}B^{2}} -\frac{k}{(AB)^{2}}\right] +(1-kr^{2})\left[3\frac{\bar{\bar{A}}}{A^{3}B^{2}}+9\frac{\bar{A}}{A^{3}B^{2}r}  \right]-3\frac{\bar{A}}{A^{3}B^{2}r}\\ \nonumber
 &=&\frac{1}{B^{2}}\left[ 6{B'^{2}}+(1-kr^{2})\left(3\frac{\bar{\bar{A}}}{A^{3}}+9\frac{\bar{A}}{A^{3}r}  \right)-3\frac{\ddot{A}}{A^{3}} -3\frac{k}{A^{2}}-3\frac{\bar{A}}{A^{3}r}\right]\\ \nonumber
 &=&-\frac{1}{B^{2}}\left\{ -6{B'^{2}} +\frac{3}{A^{2}}\left[(kr^{2}-1)\left(\frac{\bar{\bar{A}}}{A}+3\frac{\bar{A}}{Ar}  \right)+\frac{\ddot{A}}{A} +{k}+\frac{\bar{A}}{Ar}  \right]\right\}\\ \nonumber
\end{eqnarray}
}
resulting
\begin{equation}\nonumber
B^{2}\left( V(\phi)-\frac{1}{2}\phi' \right)+6{B'^{2}}=\frac{3}{A^{2}}\left[(kr^{2}-1)\left(\frac{\bar{\bar{A}}}{A}+3\frac{\bar{A}}{Ar}  \right)+\frac{\ddot{A}}{A} +{k}+\frac{\bar{A}}{Ar}  \right]=C_5
\end{equation}

\vspace{1cm}

The equations can be summarizes as follows: 

\vspace{1cm}

- Equations for $\phi$ or $y$

\begin{equation}
\phi''(y)+4\frac{{a'}}{a}{\phi'}(y)=\frac{dV}{d\phi}
\end{equation}

\begin{equation}
B(y)^{2}\left[ 3\left( \frac{B''(y)}{B(y)}+{\frac{B'(y)^{2}}{B(y)^{2}}}\right)+\frac{1}{2}\phi'(y)+V(\phi)\right]=C_0
\end{equation}

\begin{equation}
B(y)^{2}\left[ 6{\frac{B'(y)^{2}}{B(y)^{2}}}-\frac{1}{2}\phi'(y)+V(\phi)\right]=C_5
\end{equation}

\vspace{1cm}

- Equations for $t$ and $r$

\begin{equation}
\dot{A}(t,r)=A(t,r)^{2}e^{-T(t)}
\end{equation}

\begin{equation}\label{bA}
\bar{A}(t,r)=A(t,r)^{2}e^{-R(r)}
\end{equation}

\begin{equation}\label{e00}
\frac{1}{A^{2}}\left[(1-kr^{2})\left(\frac{\bar{A}^{2}}{A^{2}}-4\frac{\bar{\bar{A}}}{A}-6\frac{\bar{A}}{Ar}  \right)+2\frac{\bar{A}}{Ar}+3\frac{\dot{A}^{2}}{A^{2}}+3{k}\right]=C_0
\end{equation}

\begin{equation}\label{e11}
\frac{1}{A^{2}}\left[(1-kr^{2})\left(3\frac{\bar{A}^{2}}{A^{2}}+4\frac{\bar{A}}{Ar}  \right)+2\frac{\ddot{A}}{A}-\frac{\dot{A}^{2}}{A^{2}}+{k}\right]=C_0
\end{equation}

\begin{equation}\label{e22}
\frac{1}{A^{2}}\left[(1-kr^{2})\left(2\frac{\bar{\bar{A}}}{A}-\frac{\bar{A}^{2}}{A^{2}}+4\frac{\bar{A}}{Ar}  \right)+2\frac{\ddot{A}}{A}-\frac{\dot{A}^{2}}{A^{2}}+{k}+2\frac{\bar{A}}{Ar}\right]=C_0
\end{equation}

\begin{equation}\label{e55}
\frac{3}{A^{2}}\left[(kr^{2}-1)\left(\frac{\bar{\bar{A}}}{A}+3\frac{\bar{A}}{Ar}  \right)+\frac{\ddot{A}}{A} +{k}+\frac{\bar{A}}{Ar}  \right]=C_5
\end{equation}

\vspace{1cm}

Calculating \eqref{e11} - \eqref{e22} we find

\begin{eqnarray}
0&=&(1-kr^{2})\left(-2\frac{\bar{\bar{A}}}{A}+4\frac{\bar{A}^{2}}{A^{2}}  \right)-2\frac{\bar{A}}{Ar}\\
&=&(1-kr^{2})\left(-2{\bar{\bar{A}}}+4\frac{\bar{A}^{2}}{A}  \right)-2\frac{\bar{A}}{r}\\
&=&(1-kr^{2})\left[-2A^{2}e^{-R}(2Ae^{-R}-\bar{R})+4A^{3}e^{-2R}  \right]-2\frac{A^{2}}{r}e^{-R}\\
&=&(1-kr^{2})\left[2Ae^{-R}-\bar{R}-2Ae^{-2}  \right]+\frac{1}{r}\\
&=&-(1-kr^{2})\bar{R}+\frac{1}{r}\\
&\Rightarrow& \bar{R}(r)=\frac{1}{r(1-kr^{2})}.
\end{eqnarray}

From which we find the solution
\begin{equation}
R(r)=\ln \frac{c_1r}{\sqrt{1-kr^{2}}} \Rightarrow e^{-R(r)}=\frac{\sqrt{1-kr^{2}}}.{c_1r}.
\end{equation}

The equation $\bar{A}=A^{2}\,\frac{\sqrt{1-kr^{2}}}{c_1r}$, on the other hand, is solved by
%
%
%
%
%
%

{\large
\begin{equation}\label{A1}
A_k(t,r)=\frac{c_1}{c_1Y(t)+\ln \frac{\sqrt{1-kr^{2}}+1}{r}-\sqrt{1-kr^{2}}}
\end{equation}}
where $A_k(t,r)$ depends on $k=\pm 1$ or $k=0$, the FRW curvature parameter. Performing partial derivative with respect to $t$

\begin{equation}
\dot{A}_k(t,r)=-\frac{c_1^2\dot{Y}(t)}{\left( c_1Y(t)+\ln \frac{\sqrt{1-kr^{2}}+1}{r}-\sqrt{1-kr^{2}} \right)^2}\equiv -\frac{c_1^2\dot{Y}(t)}{\left( c_1Y(t)+f_k(r) \right)^2}
\end{equation}
with
\begin{eqnarray}\label{fk}
f_k(r)&=&\ln \frac{\sqrt{1-kr^{2}}+1}{r}-\sqrt{1-kr^{2}}.
\end{eqnarray}
Then

\begin{equation}
\frac{\dot{A}(t,r)}{A(t,r)^2}=-\dot{Y}(t)=e^{-T(t)}.
\end{equation}
We can simplify the field equations by using some previous results of our calculations

$$\begin{array}{ccccccc}
\frac{\bar{A}}{A^2} & = & e^{-R}, && \frac{\dot{A}}{A^2} & = & -\dot{Y} \\ 
\frac{\bar{\bar{A}}}{A^2} & = & e^{-R}(2Ae^{-R}-\bar{R}), && \frac{\ddot{A}}{A^2} & = & \dot{Y}\left(2A\dot{Y} -\frac{\ddot{Y}}{A^2}\right)
\end{array}, A=\frac{c_1}{c_1Y(t)+f(r)}. $$

Using this on Eq. \eqref{e00}

\begin{eqnarray*}\label{ae00}
C_0&=&\frac{1}{A^{2}}\left[(1-kr^{2})\left(\frac{\bar{A}^{2}}{A^{2}}-4\frac{\bar{\bar{A}}}{A}-6\frac{\bar{A}}{Ar}  \right)+2\frac{\bar{A}}{Ar}+3\frac{\dot{A}^{2}}{A^{2}}+3{k}\right]\\
C_0&=&(1-kr^{2})\left[e^{-2R}-4\frac{e^{-R}}{A}(2Ae^{-R}-\bar{R})-6\frac{e^{-R}}{Ar}  \right]+2\frac{e^{-R}}{Ar}+3{\dot{Y}^{2}}+\frac{3k}{A^2}\\
C_0&=&(1-kr^{2})e^{-R}\left[e^{-R}-4\frac{1}{A}(2Ae^{-R}-\bar{R})-6\frac{1}{Ar}  \right]+2\frac{e^{-R}}{Ar}+3{\dot{Y}^{2}}+\frac{3k}{A^2}\\
C_0&=&(1-kr^{2})e^{-R}\left[-7e^{-R}+\frac{2}{A}\left(2\bar{R}-\frac{3}{r} \right) \right]+\frac{1}{A}\left(2\frac{e^{-R}}{r}+\frac{3k}{A}\right)+3{\dot{Y}^{2}}\\
C_0&=&(1-kr^{2})e^{-R}\left[-7e^{-R}+\frac{2(c_1Y+f(r))}{c_1}\left(2\bar{R}-\frac{3}{r} \right) \right]+\frac{2(c_1Y+f(r))}{c_1}\frac{e^{-R}}{r}+\\
&&+\frac{(c_1Y+f(r))^2}{c_1^2}{3k}+3{\dot{Y}^{2}}\\
C_0&=&(1-kr^{2})e^{-R}\left[-7e^{-R}+\frac{2f(r)}{c_1}\left(2\bar{R}-\frac{3}{r} \right) \right]+\frac{2f(r)}{c_1}\frac{e^{-R}}{r}+3k\frac{f(r)^2}{c_1^2}+\\
&&+\left[2(1-kr^{2})e^{-R}\left(2\bar{R}-\frac{3}{r} \right)+2\frac{e^{-R}}{r}+\frac{6kf(r)}{c_1}\right]Y+3k{Y^2}+3{\dot{Y}^{2}}.\\
\end{eqnarray*}
Resulting on
\begin{equation}
\dot{Y}_k^2=-kY_k^2+w_k(r)Y_k+z_k(r)
\end{equation}
where $w_k(r)$ and $z_k(r)$ are:
\begin{small}
\begin{eqnarray} \nonumber
w_k(r)&=&-\frac{1}{3}\left[2(1-kr^{2})e^{-R}\left(2\bar{R}-\frac{3}{r} \right)+2\frac{e^{-R}}{r}+\frac{6kf(r)}{c_1}\right]\\ \nonumber
z_k(r)&=&-\frac{1}{3}\left\{(1-kr^{2})e^{-R}\left[\frac{2f(r)}{c_1}\left(2\bar{R}-\frac{3}{r} \right) -7e^{-R}\right]+\frac{2f(r)}{c_1}\frac{e^{-R}}{r}+3k\frac{f(r)^2}{c_1^2} -C_0\right\},
\end{eqnarray}
\end{small}
and $$w_0(r)=0.$$

According to $k=0,\pm1$ the solutions are
\begin{itemize}
\item[k=1]
\end{itemize}
\begin{equation}\label{y1}
Y^{\pm}_1(t)=\frac{1}{4}e^{\pm(t\mp \alpha_{-1})}\left[\left( e^{\mp(t\mp \alpha_{-1})} -w_1(r) \right)^2-4z_1(r) \right];
\end{equation}
\begin{itemize}
\item[k=0]
\end{itemize}
\begin{equation}\label{y0}
Y^{\pm}_0(t)=\alpha_0 \pm \sqrt{z_0(r)}t;
\end{equation}
\begin{itemize}
\item[k=-1]
\end{itemize}
\begin{equation}\label{ym1}
Y^{\pm}_{-1}(t)=\frac{1}{2}\left[w_{-1}(r) \pm\sqrt{w_{-1}^2(r)+4z_{-1}(r)}\sin(t+\alpha_{1}) \right];
\end{equation}
and
\begin{equation}\label{ym2}
Y^{\pm}_{-1}(t)=\frac{1}{2}\left[w_{-1}(r) \pm\sqrt{w_{-1}^2(r)-4z_{-1}(r)}\sin(t-\alpha_{1}) \right].
\end{equation}

Thus $A_k(t,r)$ is given by
\begin{equation}
A_k(t,r)=\frac{c_1}{c_1Y_k(t)+f_k(r)}
\end{equation}
where $f_k(r)$ is given by eq.\eqref{fk} and $Y_k(t)$ is given by eqs.\eqref{y1},\eqref{y0},\eqref{ym1} and \eqref{ym2}.

\vspace*{1cm}

From the equation \eqref{e11} follows

\begin{eqnarray*}\label{ae11}
C_0&=&\frac{1}{A^{2}}\left[(1-kr^{2})\left(3\frac{\bar{A}^{2}}{A^{2}}+4\frac{\bar{A}}{Ar}  \right)+2\frac{\ddot{A}}{A}-\frac{\dot{A}^{2}}{A^{2}}+{k}\right]\\
&=&(1-kr^{2})\left(3e^{-2R}+4\frac{e^{-R}}{Ar}  \right)+\frac{2}{A}\dot{Y}\left(2A\dot{Y}-\frac{\ddot{Y}}{\dot{Y}}\right)-{\dot{Y}^{2}}+\frac{k}{A^2}\\
&=&(1-kr^{2})e^{-R}\left(3e^{-R}+\frac{4}{Ar}  \right)+4\dot{Y}^2-2\frac{\ddot{Y}}{A}-{\dot{Y}^{2}}+\frac{k}{A^2}\\
&=&(1-kr^{2})e^{-R}\left(3e^{-R}+\frac{4}{c_1r}(c_1Y+f(r))  \right)-2\frac{\ddot{Y}}{c_1}(c_1Y+f(r))+\\
&&+\frac{k}{c_1^2}[c_1^2Y^2+2c_1f(r)Y+f(r)^2]+3{\dot{Y}^{2}}\\
&=&(1-kr^{2})e^{-R}\left(3e^{-R}+\frac{4}{c_1r}f(r)  \right)+\frac{k}{c_1^2}f(r)^2+kY^2\\
&&\left[(1-kr^{2})e^{-R}\frac{4}{r} +\frac{2k}{c_1}f(r)\right]Y-2\frac{f(r)}{c_1}\ddot{Y}-2Y\ddot{Y}+3{\dot{Y}^{2}}.
\end{eqnarray*}
Hence
\begin{equation}
2\ddot{Y}\left( Y+\frac{f(r)}{c_1}\right)=3\dot{Y}^2+kY^2+u(r)Y+v(r),
\end{equation}

where $u_k(r)$ and $v_k(r)$ are:

\begin{eqnarray} \nonumber
u_k(r)&=&\frac{4}{r}(1-kr^{2})e^{-R}+\frac{2k}{c_1}f(r)\\ \nonumber
v_k(r)&=&u_k(r)\frac{f(r)}{c_1}+3(1-kr^{2})e^{-2R}-\frac{k}{c_1^2}f(r)^2 -C_0
\end{eqnarray}

From the equation \ref{e55}
\begin{equation}\label{ae55}
\frac{3}{A^{2}}\left[(kr^{2}-1)\left(\frac{\bar{\bar{A}}}{A}+3\frac{\bar{A}}{Ar}  \right)+\frac{\ddot{A}}{A} +{k}+\frac{\bar{A}}{Ar}  \right]=C_5,
\end{equation}
follows
\begin{equation}
\ddot{Y}\left( Y+\frac{f(r)}{c_1}\right)=2\dot{Y}^2+kY^2+g(r)Y+h(r),
\end{equation}
where
\begin{align}
g(r)&=e^{-R}\left[(kr^2-1)\left(\frac{3}{r}-\bar{R}\right)+1\right]+k\frac{f(r)}{c_1}\\
h(r)&=g(r)\frac{f(r)}{c_1}-k\frac{f(r)^2}{c_1^2}-2\frac{1-(kr^2)^2}{c_1r}-\frac{C_5}{3},
\end{align}
From \ref{e22}
\begin{equation}\label{ae22}
\frac{1}{A^{2}}\left[(1-kr^{2})\left(2\frac{\bar{\bar{A}}}{A}-\frac{\bar{A}^{2}}{A^{2}}+4\frac{\bar{A}}{Ar}  \right)+2\frac{\ddot{A}}{A}-\frac{\dot{A}^{2}}{A^{2}}+{k}+2\frac{\bar{A}}{Ar}\right]=C_0,
\end{equation}
we have

\begin{equation}
2\ddot{Y}\left( Y+\frac{f(r)}{c_1}\right)=3\dot{Y}^2+kY^2+m(r)Y+n(r),
\end{equation}
where

\begin{align}
m(r)&=e^{-R}\left[-2(kr^2-1)\bar{R}+\frac{6}{r}-4kr\right]+2k\frac{f(r)}{c_1}\\
n(r)&=m(r)\frac{f(r)}{c_1}+3e^{-2R}(1-kr^2)-k\frac{f(r)^2}{c_1^2}-C_0.
\end{align}

Summarizing, the differential equation for $Y(t)$ are finally given by

\begin{align}
\dot{Y}_k^2&=-kY_k^2+w_k(r)Y_k+z_k(r)\\
2\ddot{Y}\left( Y+\frac{f(r)}{c_1}\right)&=3\dot{Y}^2+kY^2+u(r)Y+v(r)\\
\ddot{Y}\left( Y+\frac{f(r)}{c_1}\right)&=2\dot{Y}^2+kY^2+g(r)Y+h(r)\\
2\ddot{Y}\left( Y+\frac{f(r)}{c_1}\right)&=3\dot{Y}^2+kY^2+m(r)Y+n(r).
\end{align}

\bibliographystyle{unsrt}
\bibliography{Bibliography_arxiv}


\end{document}